\documentclass[letterpaper,12pt]{report}

\usepackage[top=1in, bottom=1in, left=1.5in, right=1in]{geometry}
\usepackage{setspace}
\usepackage{graphicx}
\usepackage{subfigure}
\usepackage{amssymb}
\usepackage{amsmath}
\usepackage{listings}
\usepackage{verbatim}
\usepackage{multirow}
\usepackage{docmute}
\usepackage{color}
\usepackage{url}
\usepackage{pdflscape}
\usepackage{caption}

\newcommand{\simgt}%
   {\,\hbox{\lower0.6ex\hbox{$\sim$}\llap{\raise0.6ex\hbox{$>$}}}\,}
\newcommand{\simlt}%
   {\,\hbox{\lower0.6ex\hbox{$\sim$}\llap{\raise0.6ex\hbox{$<$}}}\,}

\begin{document}

\pagestyle{empty} 

{
\centering 
{\Large\bfseries HALF-WAVE PLATES FOR THE SPIDER COSMIC MICROWAVE BACKGROUND POLARIMETER}\par
\vspace{7\baselineskip}
{by}\\[0.1\baselineskip]
{SEAN ALAN BRYAN\\[1.9\baselineskip]
Submitted in partial fulfillment of the requirements\\[0.5\baselineskip]
for the degree of Doctor of Philosophy\\[0.5\baselineskip]
}\par

\centering
\vspace{7\baselineskip}
Dissertation Adviser: Dr. John Ruhl \\[\baselineskip]

\centering
\vspace{7\baselineskip}
Department of Physics \\
CASE WESTERN RESERVE UNIVERSITY \\[\baselineskip]
May, 2014\par
\vfill
}

\clearpage

{

\centering
{\bfseries CASE WESTERN RESERVE UNIVERSITY \\ SCHOOL OF GRADUATE STUDIES}\par
\vspace{3\baselineskip}
\flushleft
{We hereby approve the thesis/dissertation of}\par
\vspace{1.5\baselineskip}
\rule{25em}{0.4pt}\\
[1\baselineskip]
candidate for the \rule{15em}{0.4pt} degree *.\\
[4\baselineskip]
(signed) \rule{21.2em}{0.4pt}\\
~~~~~~~~~~~~(chair of the committee) \\[1.5\baselineskip]
\textcolor{white}{(signed) }\rule{21.2em}{0.4pt}\\
~~~~~~~~~~~~\textcolor{white}{(chair of the committee)} \\[1.5\baselineskip]
\textcolor{white}{(signed) }\rule{21.2em}{0.4pt}\\
~~~~~~~~~~~~\textcolor{white}{(chair of the committee)} \\[1.5\baselineskip]
\textcolor{white}{(signed) }\rule{21.2em}{0.4pt}\\
~~~~~~~~~~~~\textcolor{white}{(chair of the committee)} \\[3\baselineskip]
(date) \rule{10em}{0.4pt}\\

\vspace{1\baselineskip}
*We also certify that written approval has been obtained for any proprietary material contained therein.
\vfill
}

\clearpage

\begin{center}
\vspace*{35\baselineskip}
\copyright~2014, Sean Bryan
\end{center}

\clearpage

For my family.

\clearpage

\setcounter{page}{1}
\pagenumbering{roman}
\pagestyle{plain}

\tableofcontents

\listoffigures

\listoftables

\chapter*{Acknowledgments}

{\doublespacing
Grad school at CWRU in Cleveland has been really great. I learned a lot, adopted two great cats, made awesome friends, and met and married Sarah.

Thank you Mom, Dad and Will for being my family. Thank you Gary and Christine for being Sarah's parents and now my family too. Thank you Sarah. Thank you to all my teachers and classmates in Saint Paul Public Schools for giving me such a solid educational start in life. Thank you to everyone in my life in college at the University of Minnesota.

Everyone on the Spider team is awesome. I always enjoy your hospitality when I come to visit Caltech, Princeton, and Toronto. Even with the long hours and hard work last summer in Texas, it was a blast because of the cool people I got to be around.

John Ruhl, thank you for being a great research adviser. I've learned so much about science and myself from you and the wonderful group you lead here. Thank you to everyone in the lab for being so fun to work with.

~

~

~}

\pagenumbering{roman}
\pagestyle{plain}
\setcounter{page}{7}

\chapter*{The Spider Team}

\begin{center}
\includegraphics[width=1.0\textwidth]{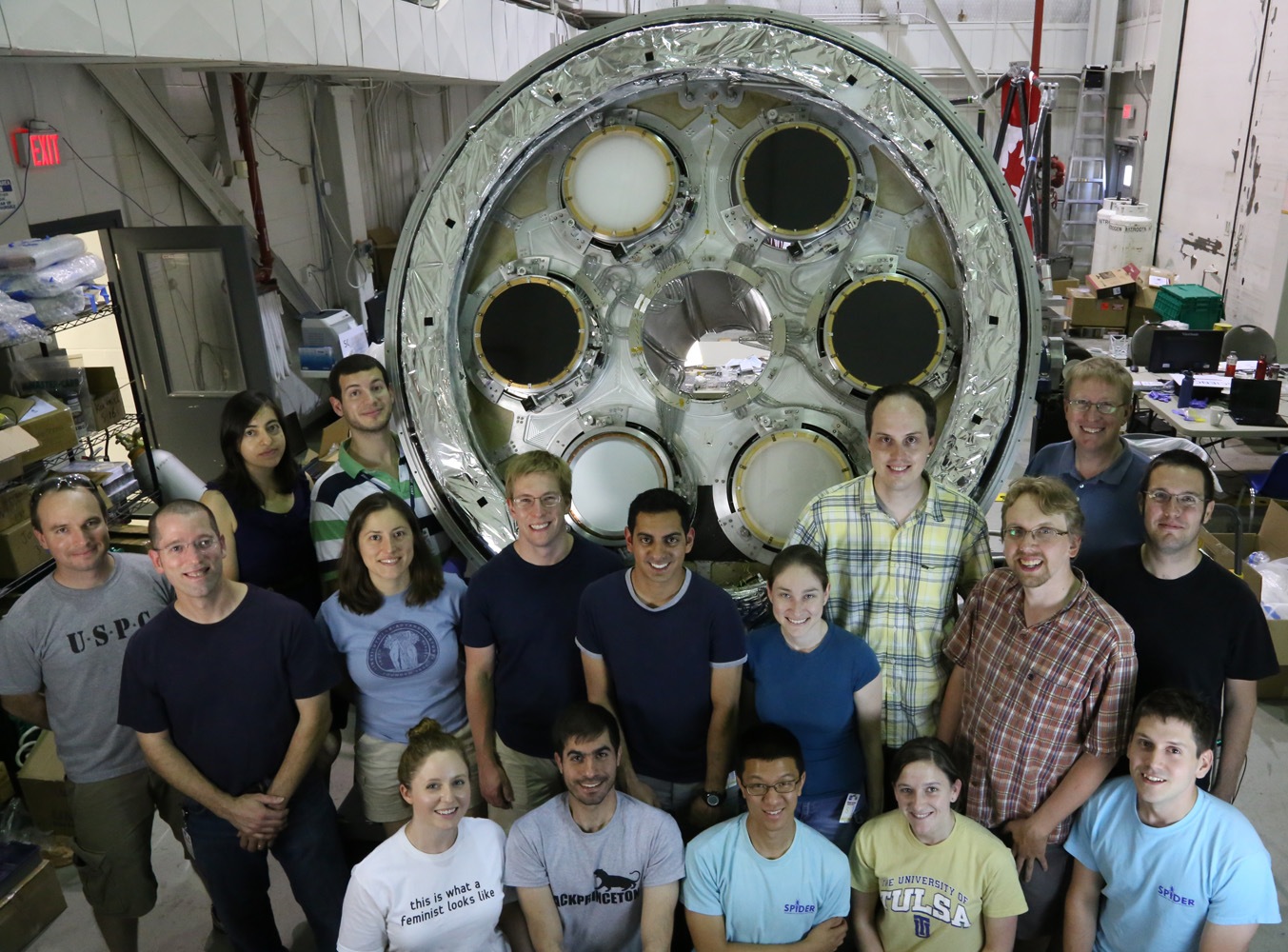} 
\end{center}

\noindent
P. A. R. Ade,$^1$ M. Amiri,$^2$ S. J. Benton,$^3$ R. F. Bihary,$^4$
J. J. Bock,$^{5,6}$ J. R. Bond,$^7$ J. A. Bonetti,$^6$ S. Bryan,$^4$
H. C. Chiang,$^8$ C. R. Contaldi,$^9$ B. P. Crill,$^{5,6}$
O. Dore,$^{5,6}$ M. Farhang,$^{7,3}$ J. P. Filippini,$^5$ L. M. Fissel,$^{10,3}$ A. A. Fraisse,$^{11}$  A. Gambrel,$^{11}$
N. N. Gandilo,$^{12}$ J. E. Gudmundsson,$^{11}$ M. Halpern,$^2$ M. Hasselfield,$^{13,2}$
G. Hilton,$^{14}$ W. Holmes,$^6$ V. V. Hristov,$^5$ K. Irwin,$^{15}$ W. C. Jones,$^{11}$
Z. K. Kermish,$^{11}$ C. J. MacTavish,$^{17}$ P. V. Mason,$^5$ L. Moncelsi,$^5$ T. E. Montroy,$^4$
T. A. Morford,$^5$ J. M. Nagy,$^4$ C. B. Netterfield,$^{3,12}$ A. S. Rahlin,$^{11}$
C. Reintsema,$^{14}$ J. E. Ruhl,$^4$ M. C. Runyan,$^6$
J. A. Shariff,$^{12}$ J. D. Soler,$^{17,12}$ A. Trangsrud,$^5$ C. Tucker,$^1$
R. S. Tucker,$^{5}$ A. D. Turner,$^6$ A. C. Weber,$^6$ D. Wiebe,$^2$ and E. Young,$^{11}$

\clearpage

{\footnotesize

\noindent
$^1$School of Physics and Astronomy, Cardiff University, Cardiff, UK \\
$^2$Department of Physics and Astronomy, University of British Columbia, Vancouver, BC, Canada \\
$^3$Department of Physics, University of Toronto, Toronto, ON, Canada \\
$^4$Department of Physics, Case Western Reserve University, Cleveland, OH, USA \\
$^5$Division of Physics, Mathematics \& Astronomy, California Institute of Technology, Pasadena, CA, USA \\
$^6$Jet Propulsion Laboratory, Pasadena, CA, USA \\
$^7$Canadian Institute for Theoretical Astrophysics, University of Toronto, Toronto, ON, Canada \\
$^8$School of Mathematics, Statistics \& Computer Science, University of KwaZulu-Natal, Durban, South Africa \\
$^9$Theoretical Physics, Blackett Laboratory, Imperial College, London, UK \\
$^{10}$CIERA- Northwestern University, Evanston, IL, USA \\
$^{11}$Department of Physics, Princeton University, Princeton, NJ, USA \\
$^{12}$Department of Astronomy and Astrophysics, University of Toronto, Toronto, ON, Canada \\
$^{13}$Department of Astrophysical Sciences, Princeton University, Princeton, NJ, U.S.A \\
$^{14}$National Institute of Standards and Technology, Boulder, CO, USA \\
$^{15}$Department of Physics, Stanford University, Stanford, CA, U.S.A. \\
$^{16}$Kavli Institute for Cosmology, University of Cambridge, Cambridge, U.K. \\
$^{17}$Institut d'Astrophysique Spatiale IAS, CNRS \& Université Paris-Sud, Orsay Cedex, France \\
}

\clearpage

\begin{center}
{\Large\bfseries Half-wave Plates for the Spider \\ Cosmic Microwave Background Polarimeter \\ }
\vspace{7 mm}
by \\
\vspace{7 mm}
 SEAN ALAN BRYAN
\end{center}

\vspace{14 mm}

\begin{center}
{\bfseries Abstract}
\end{center}

Spider is a balloon-borne array of six telescopes that will observe the Cosmic Microwave Background. The 2400 antenna-coupled bolometers in the instrument will make a polarization map of the CMB with $\sim$degree resolution at 150 GHz and 95 GHz. Polarization modulation is achieved via a cryogenic sapphire half-wave plate (HWP) skyward of the primary optic. In this thesis, the design, construction, and lab testing of the HWP system are discussed. The polarization modulation of these optical stacks is modeled using a physical optics calculation and Mueller matrices. Performance tests in both the lab and integrated in the flight cryostat show consistency with the model.

\pagenumbering{roman}
\pagestyle{plain}
\setcounter{page}{6}
\thispagestyle{empty}



\clearpage

\pagenumbering{arabic}
\setcounter{page}{1}
 
\doublespacing


\chapter{Introduction}

One of the key frontiers in modern cosmology is measuring the temperature and polarization anisotropies of the Cosmic Microwave Background (CMB) radiation. The anisotropies encode a wealth of information about inflation~\cite{baumann09}, re-ionization~\cite{zaldarriaga08}, and the large-scale structure of the universe \cite{smith08}. From the discovery of the CMB in 1965 by Penzias and Wilson~\cite{penzias65} through to the Planck satellite of today \cite{planck13a}, measurements of the CMB have been a key ingredient in establishing the standard model of cosmology.

The CMB is relic light from the early universe. The temperature of the universe before $t\sim380,000$ years after the Big Bang was high enough that the universe was filled with a plasma of coupled photons and charged particles. As the universe expanded and cooled, neutral hydrogen formed, capturing the free electrons. This event, called recombination, allowed the photons to travel without scattering electromagnetically. This free-streaming background of photons is the CMB that we observe today. Structure in the universe from that era is imprinted on the CMB as hot and cold spots, which means the CMB is the oldest direct picture of the universe we have. The CMB is sometimes called the ``smoking gun'' of the Hot Big Bang scenario.

The uniformity of the CMB temperature across the sky actually presents a challenge to the Hot Big Bang scenario as it was originally conceived. The largest volume of the universe that could have been in causal contact at the time of recombination has an apparent angular size of today of $\sim2^\circ$. However, the CMB temperature is highly uniform across the entire sky, to better than one part in $10^4$. This paradox is called the Horizon Problem. Inflation, conceived by Guth \cite{guth81} and independently by Starobinsky \cite{starobinsky80}, postulates that there was an early time in the universe's history when the universe expanded exponentially, growing a small causally-connected volume in the pre-inflation era into a large enough volume to explain the uniformity of the CMB. Quantum fluctuations in the field driving the inflationary expansion seeded density perturbations that later grew through gravitational collapse into the fluctuations seen in the CMB, and much later grew into the large-scale structure of the universe today. The quantum fluctuations also perturbed the rest of the metric to create a background of gravitational waves that left an imprint on the polarization of the CMB. Precision measurements of the CMB temperature and polarization are therefore a powerful probe to learn about the physics of inflation.

\section{CMB Polarization}
\label{cmb_polarization}

\subsection{Physical Origin}
Once the dipole due to the proper motion of the earth is subtracted, the CMB has temperature anisotropies at the $\sim 100~\mu$K level, about 4 parts in $10^5$ \cite{hu97}. These were generated by temperature variations in the plasma, Doppler shifts due to motion of the plasma along our line of sight, and redshifting of photons as they climb out of potential wells formed by overdensities and underdensities of Dark Matter. The plasma also generated linearly-polarized light through Thomson scattering. In Thomson scattering, photons scatter off electrons, and the differential cross section is
\begin{equation}
\frac{d \sigma}{d \Omega} = \frac{3 \sigma_T}{8 \pi} {|\hat{\epsilon}' \cdot \hat{\epsilon}|}^2,
\end{equation}
where $\hat{\epsilon}'$ and $\hat{\epsilon}$ are unit vectors in the direction of the scattered and incident photons respectively. The incident electromagnetic wave causes the electron to move. Accelerating charge radiates, so light is re-emitted with a directional dependence governed by the Thomson cross section. 

\begin{figure}
\begin{center}
\includegraphics[width=0.95\textwidth]{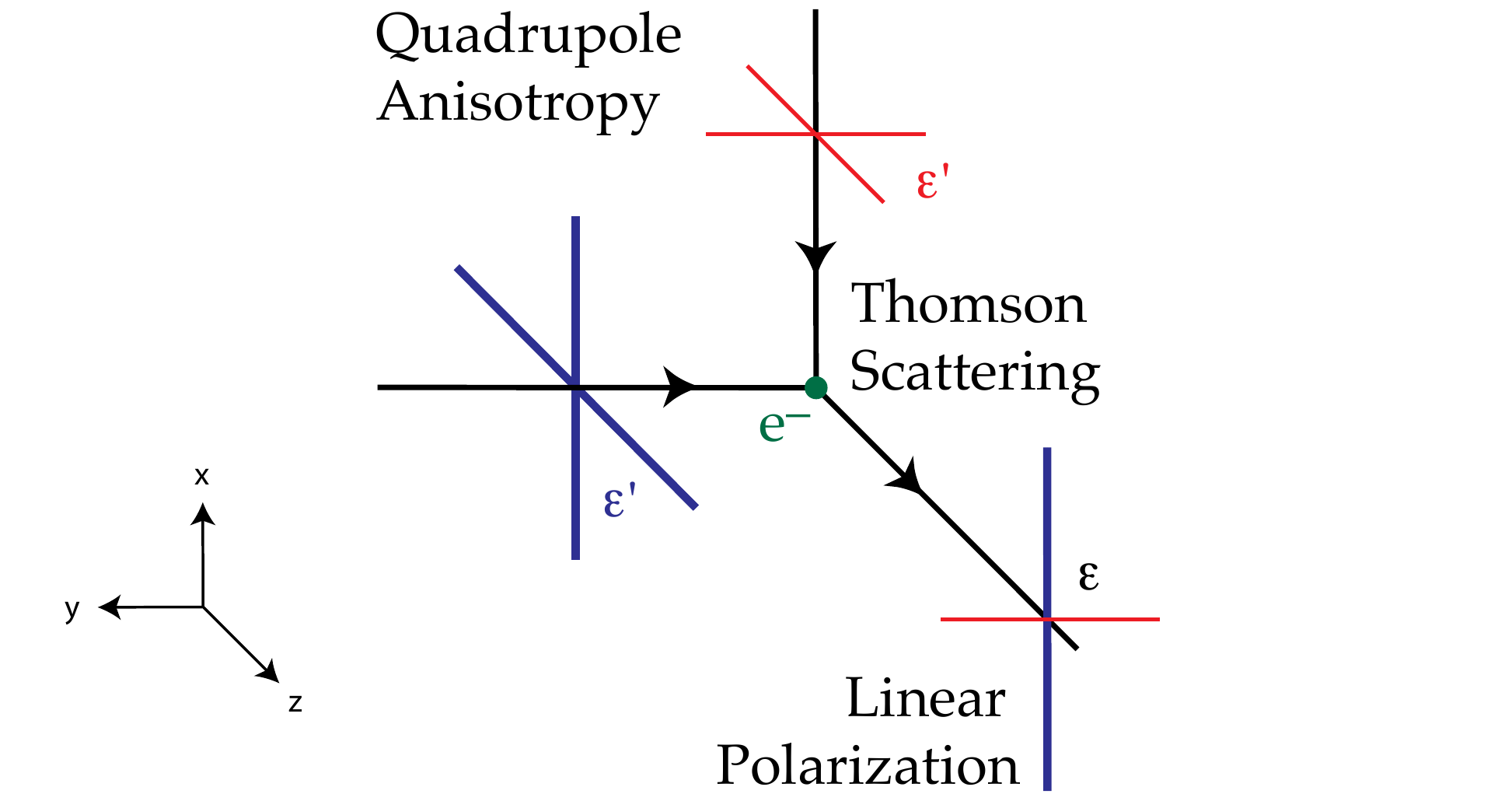}
\caption[Illustration of the generation of linear polarization in the CMB.]{A local quadrupole generates linear polarization. The light incident on the electron is from a cooler source in the vertical direction, and a warmer source in the horizontal direction. This causes the scattered light to be partially polarized. Figure from Hu and White \cite{hu97}. \label{linear_pol}}
\end{center}
\end{figure}

The CMB photons we observe today were emitted along a line of sight to us from the spherical surface of last scattering. Light incident on a small volume of plasma from the last-scattering surface causes the electrons in that plasma patch to move and radiate out to us in the line-of-sight direction. If the plasma is surrounded by an isotropic field of photons, the E-field amplitudes incident along the last scattering surface will be equal, and the radiation in the line-of-sight direction will be unpolarized. As illustrated in Figure~\ref{linear_pol}, if the plasma is surrounded by a field of photons with a quadrupolar brightness distribution (generated by density perturbations or gravity waves), the E-field amplitudes coming from one direction will be higher than those waves from the orthogonal direction. This causes the radiation in the line-of-sight direction towards us to be partially polarized.

Since they naturally describe partially polarized incoherent light, we use the Stokes Parameters to characterize the polarization state of light of the CMB. For light traveling in the $z$ direction, these are defined in \cite{hecht74} as
\begin{eqnarray}
I &=& <a_x^2> + <a_y^2> \\
Q &=& <a_x^2> - <a_y^2> \\
U &=& < 2 a_x a_y \cos{(\theta_x - \theta_y)} > \\
V &=& < 2 a_x a_y \sin{(\theta_x - \theta_y)} > ,
\end{eqnarray}
where
\begin{eqnarray}
E_x &=& a_x(t) \cos{(kz - \omega t - \theta_x(t))} \\
E_y &=& a_y(t) \cos{(kz - \omega t - \theta_y(t))}
\end{eqnarray}
are the $x$ and $y$ components of the electric fields, and the $<>$ operator indicates averaging over a timescale much longer than $1/\omega$. The physical interpretations of the Stokes Parameters are $I$ $\sim$ intensity, $Q,U$ $\sim$ linear polarization, and $V$ $\sim$ circular polarization. For example $(I,Q,U,V) = (1,0,0,0)$ is completely unpolarized light, $(1,1,0,0)$ is completely linearly polarized light, and $(1,0.5,0,0)$ is partially linearly polarized light. Thomson scattering only produces linear polarization, so $V=0$ for the CMB. This means that light from the CMB can be characterized as a pseudo-vector with polarization fraction $\sqrt{Q^2 + U^2} / I$ and polarization angle $\frac{1}{2}$atan2${(U,Q)}$. Here $\mathrm{atan2}$ is the two-argument arctangent function that returns a signed angle, indicating the rotation direction. It is defined as
\begin{equation}
\mathrm{atan2}(y,x) \equiv 2 \tan^{-1} {\frac{y}{\sqrt{x^2 + y^2} + x}}.
\end{equation}
A full-sky map of the polarization anisotropy of the CMB can therefore be visualized as a pseudo-vector field on the surface of the sphere.

At angular scales larger than $~\sim1^\circ$, the two dominant sources of polarization anisotropies are density perturbations and gravity waves. These two kinds of perturbations differ in their transformation properties under global parity flips. For a field defined in a plane, a global parity flip is equivalent to looking at the field in a mirror. This picture can be generalized to a sphere, since a sphere is locally flat around any point. The density perturbations generate a polarization field $\vec{P}(\hat{n})$, where $\hat{n} = (\theta,\phi)$ is a unit vector pointing to a direction on the sky. Since density is a scalar quantity, this polarization field must be invariant under a global parity flip. It follows that the field $\vec{P}(\hat{n})$ must have a vanishing curl. By analogy with electromagnetism, polarization patterns with a vanishing curl are called E-modes.

Gravity waves also generate local quadrupoles in temperature, and therefore generate polarization. Considering a spherical patch of plasma, a passing gravity wave would induce an elliptical distortion. This would compress and heat the plasma along one direction, and rarefy and cool the plasma along the orthogonal direction. This would induce a quadrupolar brightness distribution, and therefore would radiate partially polarized light as illustrated in Figure~\ref{linear_pol}. Since gravity waves are tensors, the gravity wave perturbations need not be invariant under a global parity flip, so the polarization field they generate need not be invariant. This implies that the $\vec{P}(\hat{n})$ generated by this mechanism can contain a non-zero curl. Again by analogy with electromagnetism, polarization patterns with curl are called B-modes. The only physical mechanism that can generate B-modes at the surface of last scattering is a background of gravity waves. Because of their global transformation properties, density perturbations (scalars) produce only E-modes, and gravity waves (tensors) produce roughly equal amounts of E- and B-modes. Detecting B-modes at large angular scales is therefore a detection of the primordial gravity waves in the universe at the time of last scattering.

\subsection{Angular Power Spectra}

Angular power spectra can be calculated from maps of the CMB temperature and polarization for comparison with theory and constraining cosmological parameters. The spherical harmonic functions $Y_l^m$ are the basis for estimating the angular power spectrum of the CMB, since the maps are on the surface of a sphere. For the temperature maps, $C_l^{\mathrm{TT}}$ are the spherical harmonic coefficients when the temperature map is decomposed onto spherical harmonics and the $a_{lm}$'s are combined according to
\begin{equation}
C_l^{\mathrm{TT}} = <a_{lm} a_{lm}^{*}>,
\end{equation}
where the brackets denote an average over $m$. This decomposition works if the ordinary spin-0 spherical harmonics are used, since temperature is a scalar quantity.

Because polarization maps transform under global parity flips and rotations as spin-2 objects, polarization maps may be expressed using the spin-2 spherical harmonic functions as a basis \cite{balbi06}. The maps are expressed as
\begin{equation}
Q(\hat{n})+iU(\hat{n}) = \sum_{l>0} \sum_{m=-l}^l (a_{lm}^E + i a_{lm}^B) Y_l^m(\hat{n}), 
\end{equation}
where $Y_l^m$ are the spin-2 spherical harmonics, not the ordinary spin-0 spherical harmonics. In the convention used in \cite{page07}, the $C_l$ are calculated with
\begin{equation}
C_l^{\mathrm{XY}} = <a_{lm}^X a_{lm}^{Y*}>.
\end{equation}
Here X and Y are T (temperature), E (E-modes), or B (B-modes). The possible spectra to calculate from a given observation are the temperature spectrum $C_l^{\mathrm{TT}}$, the temperature-polarization cross spectra $C_l^{\mathrm{TE}}$ and $C_l^{\mathrm{TB}}$, and the polarization spectra $C_l^{\mathrm{EB}}$, $C_l^{\mathrm{EE}}$ and $C_l^{\mathrm{BB}}$.

All of these measured angular power spectra can be compared with calculated values of the theoretical angular power spectrum. The TT, EE, and BB spectra can all be non-zero, since they are auto-correlations. The review in Hu and White \cite{hu97} shows that because of how E- and B-modes transform under parity, the TE spectrum is non-zero but TB and EB are identically zero. TB and EB can appear to be non-zero if there is a global rotation of all of the polarization directions from cosmic birefringence, or due to incorrect calibration of the detector sensitivity angles \cite{keating13}.

\begin{figure}
\begin{center}
\includegraphics[width=0.62\textwidth]{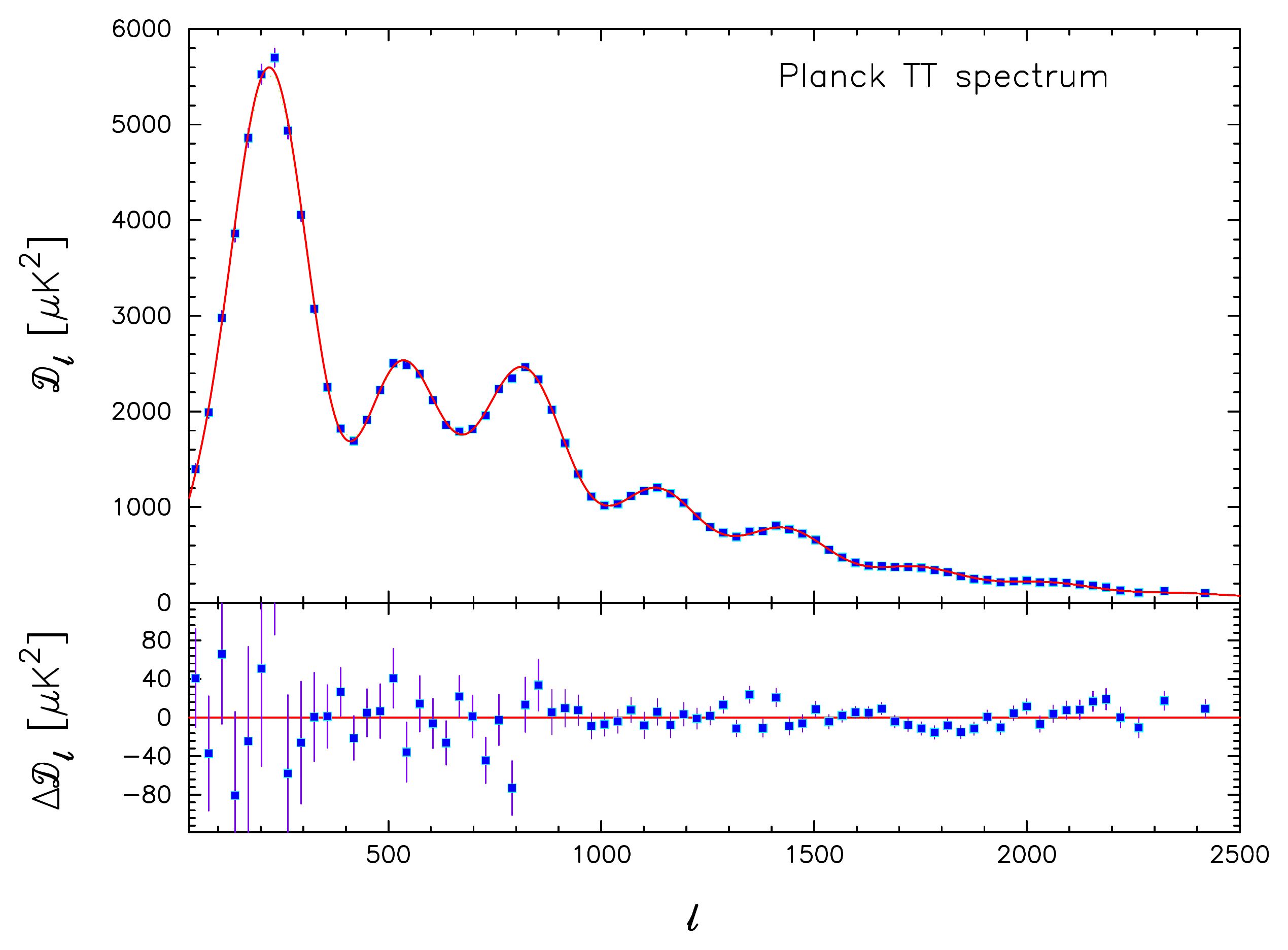}
\includegraphics[width=0.67\textwidth]{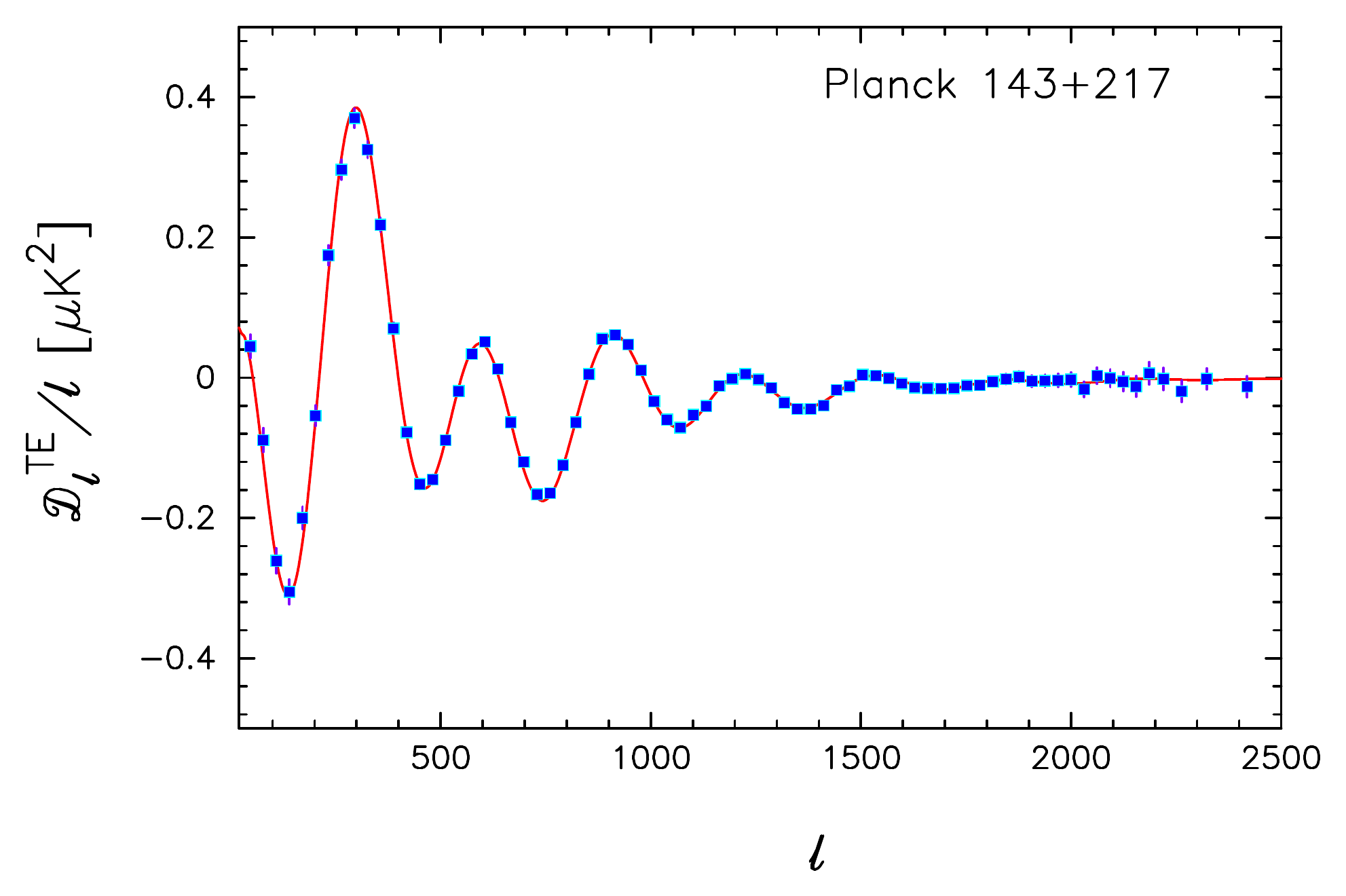}
\includegraphics[width=0.65\textwidth]{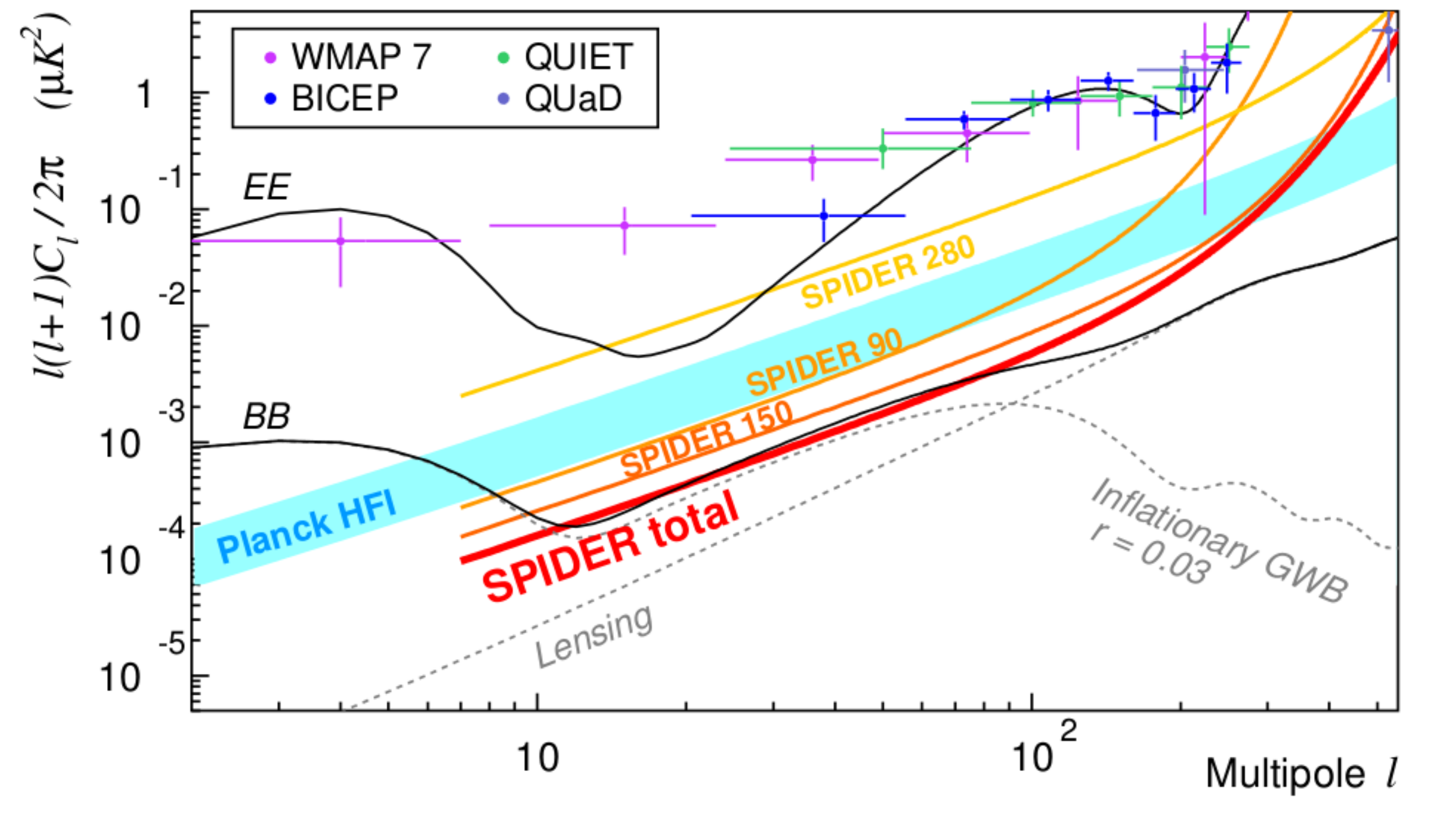}
\caption[Planck TT and TE spectra, EE spectra, and expected Spider results.]{Planck TT and TE spectra, EE spectra from several experiments, and expected Spider BB sensitivity per multipole $l$. This projection assumes a first flight with 3x 95 GHz receivers, and 3x 150 GHz receivers, and a second flight with 2 receivers each at 95 GHz, 150 GHz, and 280 GHz. From Planck \cite{planck13b} and Fraisse et al. \cite{fraisse13} .\label{spider}}
\end{center}
\end{figure}


\section{Inflation}
\label{inflation}

\subsection{Horizon Problem}

The CMB is remarkably homogeneous, suggesting that the entire currently observable universe was once in causal contact and thermal equilibrium. It turns out that in the Hot Big Bang scenario, there never was such a time! This paradox is the Horizon Problem. Ryden \cite{ryden03} reviews the problem as follows. The horizon distance at the time of last scattering is calculated to be $d_{\mathrm{hor}}(t_{\mathrm{ls}}) \approx 0.27~\mathrm{Mpc}$ in the Standard Model, and the angular-diameter distance to the surface of last scattering is calculated to be $d_A \approx$ 13 Mpc. This means that a causally-connected region of space at the time of last scattering has an apparent angular size today of
\begin{equation}\label{last_scattering_horizon_angle}
\theta_{\mathrm{apparent}} = \frac{2 d_{\mathrm{hor}}(t_{\mathrm{ls}})}{d_A} \approx \frac{0.53~\mathrm{Mpc}}{13~\mathrm{Mpc}} \approx 2^\circ.
\end{equation}
This means that in the Hot Big Bang model, the CMB should not appear isotropic on scales larger than $2^\circ$. However, the entire sky is uniformly $2.7$ K, to about four parts in $10^5$. This means there must have been some event in the early universe that caused the currently observable universe to be in causal contact.

Inflation solves this problem by postulating a period in which the universe was briefly dominated for a time $t_{\mathrm{inflation}}$ by a component with an equation of state $w<1/3$. For a cosmological constant, $w=-1$, and the scale factor grows exponentially with time during this period:
\begin{equation}
a(t) \propto e^{H t}.
\end{equation}
In grand unified theories (GUT) of particle physics, in order to solve the monopole problem (i.e. the non-detection of monopoles today, despite their calculated abundant production in the early universe in grand unified theories of particle physics), inflation must have started after the temperature of the universe was at the GUT scale, $\sim10^{16}$ GeV. Since the universe was radiation-dominated at the GUT time, the horizon distance just before inflation was $d_{\mathrm{hor}}(t_{\mathrm{before}}) = 2ct_{\mathrm{before}}$. Ryden shows that just after inflation, the horizon size at the end of inflation expanded to 
\begin{equation}
d_{\mathrm{hor}}(t_{\mathrm{before}} + t_{\mathrm{inflation}}) = e^{N} c (2 t_{\mathrm{before}} + H_{\mathrm{before}}^{-1}) \approx e^N 3 c t_{\mathrm{before}},
\end{equation}
where $N \equiv H t_{\mathrm{inflation}}$ is the number of e-folds of inflation. After inflation, this horizon keeps growing according to the usual radiation-driven, matter-driven, then cosmological-constant-driven expansion history of the universe from then to the present day. For inflation at the GUT scale, $N \geq 60$ is required to solve the horizon problem.

\subsection{Perturbations}
Inflation was conceived to solve the monopole problem and the horizon problem, but it makes other testable predictions. If inflation is driven by a scalar field $\phi$ with a potential $V(\phi)$, the value of this field will vary spatially due to quantum fluctuations. Inflation expands these virtual quantum fluctuations to scales larger than the horizon, which turns them into real macroscopic perturbations. At the end of inflation, the scalar field will decay into dark matter, baryons, photons, and all of the other Standard Model particles. This means that the fluctuations in the inflaton field will decay into density fluctuations in the baryon-photon plasma and dark matter. Measuring these fluctuations in cosmological observables is a way to directly probe the physics of inflation.

Measurements of the large-scale temperature anisotropies of the CMB give a measurement of the overall amplitude of the primordial density perturbations. If inflation was the mechanism that generated these perturbations, the review by Liddle and Lyth \cite{liddle00} shows that this in turn is a constraint on the quantity $(V/\epsilon)^{\frac{1}{4}}$, where
\begin{equation}
\epsilon \equiv \frac{m_{\mathrm{p}}^2}{16 \pi} \left( \frac{V'}{V}\right)^2
\end{equation}
is one of the slow-roll parameters. The current value of this constraint is given in \cite{baumann09} as $(V/\epsilon)^{\frac{1}{4}} \approx 6.30 \times 10^{16}$ GeV.

\begin{figure}
\begin{center}
\includegraphics[width=1.0\textwidth]{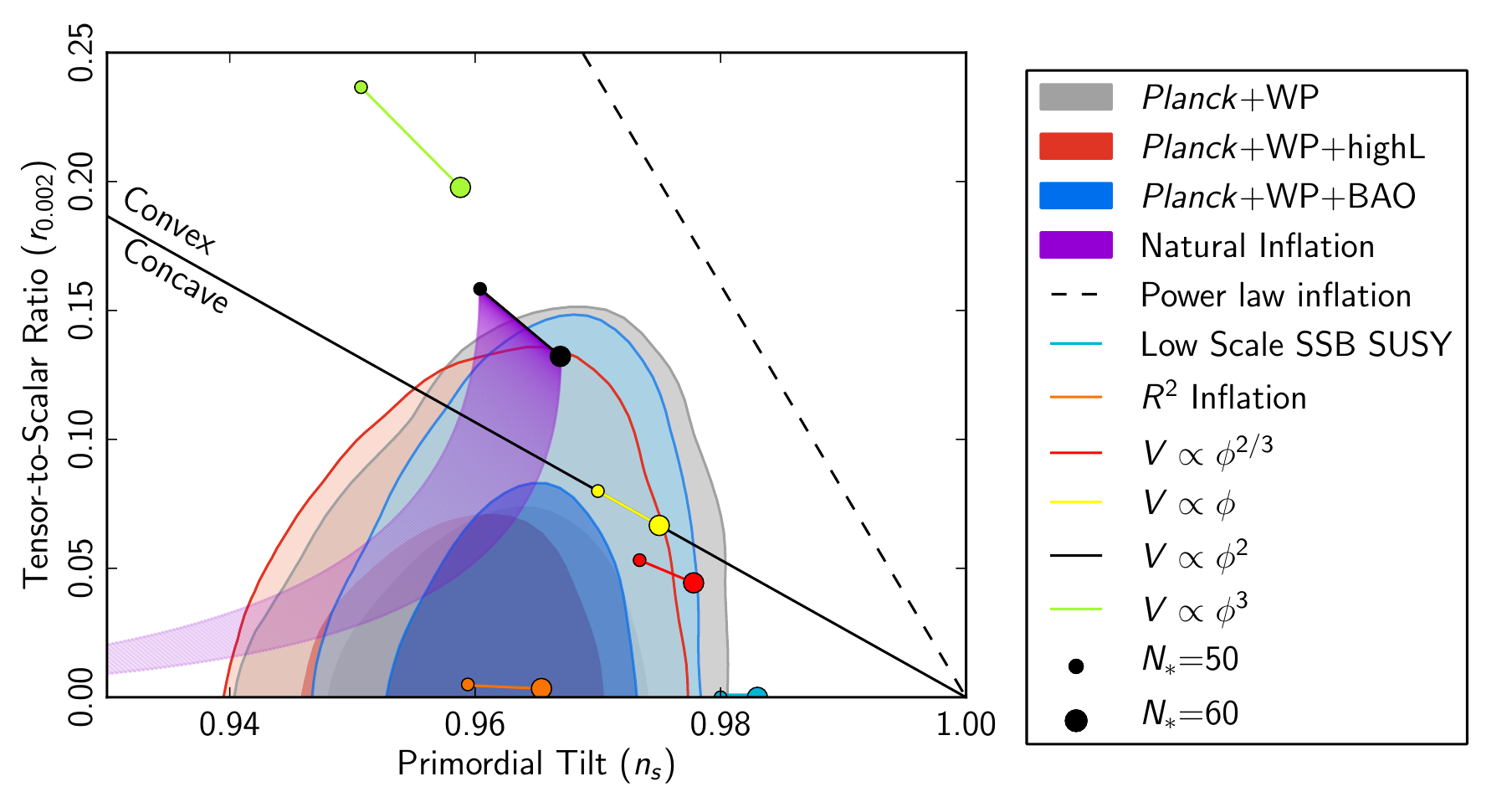}
\caption[$n_s$ and $r$ constraints on inflationary models]{Both $n_s$ and $r$ can constrain inflationary models. The plot above shows the constraints from Planck \cite{planck13b} and other experiments, overlaid with predictions for several inflationary models. The vertical axis shows constraints on $r_{0.002}$, which is the tensor-to-scalar ratio $r$ measured at $k = 0.002~\mathrm{Mpc}^{-1}$. \label{ns_r}}
\end{center}
\end{figure}

Just as inflation causes density perturbations, it also perturbs the entire metric $\tilde{g}_{\mu \nu} = g_{\mu \nu} + h_{\mu \nu}$, where $g_{\mu \nu}$ is the background Friedmann-Robertson-Walker metric. Taking $h$ to be small means the perturbations are linear. Following the review in Peacock \cite{peacock99}, the RMS amplitude of the gravity wave (tensor) perturbations are $h_{\mathrm{rms}} \sim H/m_{\mathrm{p}}$, small enough for linear theory. Assuming inflation is driven by a simple scalar field, quantum field theory can be used to calculate the ratio $r$ of the amplitude of the tensor perturbations $\Delta_\mathrm{T}$ to the amplitude of the scalar perturbations $\Delta_\mathrm{S}$. The result is
\begin{equation}
\frac{\Delta_\mathrm{T}^2}{\Delta_\mathrm{S}^2} \equiv r \approx 12.4 \epsilon.
\end{equation}
Combining this result with the bound on $(V/\epsilon)^{\frac{1}{4}}$ yields a relationship between $r$ and the energy scale of inflation $V^{\frac{1}{4}}$, given in \cite{baumann09} as
\begin{equation}
E_{inflation} \equiv V^{\frac{1}{4}} = 1.06 \times 10^{16} ~\mathrm{GeV} ~ \left( \frac{r}{0.01} \right)^{\frac{1}{4}}.
\end{equation}
This means that a measurement of $r$ would be a measurement of the energy scale of inflation.

Measurements of the spectrum of the primordial density perturbations also constrain inflation. Continuing the calculation presented in \cite{peacock99}, in inflation the scalar spectral index $n_s$ is calculated to deviate from scale invariance ($n_s = 1$) by
\begin{equation}
1 - n_s = 6 \epsilon - 2 \eta,
\end{equation}
where
\begin{equation}
\eta \equiv \frac{m_{\mathrm{p}}^2}{8 \pi} \left( \frac{V''}{V}\right)
\end{equation}
is the other slow-roll parameter. For typical polynomial potentials, $\eta \sim \epsilon$, so inflation generically predicts that $r$ and $n_s$ will be related by
\begin{equation}
r \approx 3 (1-n_s).
\end{equation}
Planck has measured that $(1 - n_s) = 0.0376 \pm 0.0075$ \cite{planck13a}, so single-field power-law-potential inflation generically predicts $r \sim 0.1$. It is of course possible to come up with a model of inflation with a different relationship among the primordial perturbation, $r$ and $n_s$, which could result in a vanishingly small $r$. Predictions for $r$ and $n_s$ in two such models are also shown in Figure~\ref{ns_r}.

The best current upper limits on $r$ come indirectly from the large-angle CMB temperature spectrum. If $r$ were large enough, the tensor perturbations would induce temperature anisotropies at large angular scales in the CMB, so the current non-detection of excess anisotropies at large angular scales sets a limit of $r < 0.1$ at $95\%$ confidence \cite{planck13a}. Since those measurements have low enough instrument noise now that they are cosmic-variance limited, current upper limits on $r$ will not improve significantly with improved temperature measurements. However, the BB polarization spectrum directly gives a measurement of $r$. Figure \ref{ns_r} shows the current observational constraints on $n_s$ and $r$ along with predictions from several models of inflation. According to simulations \cite{fraisse13}, after two flights the BB measurements of Spider will detect or set an upper limit on $r$ of 0.03 at $99\%$ confidence, which would detect or rule out the simple $V \sim \phi^N$ models of inflation shown in the figure.

\section{Observing B-modes in the CMB}

\subsection{Primordial B-mode Signal}

The gravity wave background generated by inflation induces a B-mode polarization pattern at the surface of last scattering, a signal that would peak at roughly $l=100$ as shown in Figure~\ref{spider}. This is the signal Spider is hunting for. The peak appears at large enough angular scales that the roughly half-degree beam sizes in Spider are enough to resolve the feature. As discussed further in Section~\ref{spider_scan}, the feature appears on small enough angular scales that we can concentrate the sensitivity of the instrument on a relatively small observing region, roughly $10\%$ of the sky. Also, the scan speed of the instrument will put this signal at roughly 1 Hz in the detector timestreams, which is a high enough frequency that the detector drifts will not be a problem, and low enough to be well below the time constant of the detectors.

Later in the history of the universe, between redshifts of 10 and 5, the neutral hydrogen in the universe was reionized, creating a diffuse population of charged particles. Light from the CMB Thomson-scattered from the electrons in this plasma, creating E- and B-mode polarization patterns that we could observe today. This signal appears on very large angular scales, roughly $l<10$. Searching for this signal would require mapping a larger fraction of the sky than is available to Spider observing during its Antarctic flight. Also, as discussed in Section~\ref{foreground_section}, foreground contamination is higher at large angular scales, which would present an additional challenge to searching for this signal.


\subsection{Lensing B-mode Signal}

Gravitational lensing by large-scale structure distorts the primordial temperature and polarization anisotropies of the CMB as they travel to us from the surface of last scattering. As reviewed in \cite{smith08}, lensing takes the primordial maps emitted by the surface of last scattering $[I(\hat{n}),Q(\hat{n}),U(\hat{n})]$ and deflects the rays as they travel to us to form the lensed maps $[I(\hat{n} + \nabla \varphi(\hat{n})),Q(\hat{n} + \nabla \varphi(\hat{n})),U(\hat{n} + \nabla \varphi(\hat{n}))]$ that we observe. The angular power spectrum of this lensing deflection field $\varphi(\hat{n})$ is approximately
\begin{equation}
C_l^{\varphi \varphi} = \frac{8 \pi^2}{l^3} \int_0^{z_{rec}} \frac{dz}{H(z)} D(z) \left( \frac{D(z_{rec}) - D(z)}{D(z_{rec})  D(z)} \right)^2 P_\psi(z,k=l/D(z)),
\end{equation}
where $z_{rec}$ is the redshift of recombination, $D$ is the comoving distance to redshift $z$ and $P_\psi$ is the power spectrum of the gravitational potential generated by large-scale structure. This relation shows that measuring the lensing deflection field is a measurement of both the growth of structure in the recent $z \simlt 5$ universe ($P_\psi)$ and the expansion rate of the recent universe (encoded in $H(z)$ and $D(z)$).

A natural basis for analyzing polarization data in search of the lensing signal is to use an estimator optimized to measure the deflection field $\varphi(\hat{n})$, its angular power spectrum $C_l^{\varphi \varphi}$, or its correlation with other measurements. However, thinking instead in the usual E-mode and B-mode decomposition, lensing induces a B-mode signal by distorting the primordial E-mode signal. This lensing B-mode signal is shown in Figure~\ref{spider} overplotted in the same panel that shows the expected B-mode sensitivity of Spider. The signal is expected to peak near $l\approx1000$, and its expected amplitude is calculable from existing measurements of large-scale structure. The signal was recently detected by the SPTpol instrument \cite{hanson13}. On angular scales smaller than $l\approx90$, B-modes generated by lensing are larger than the primordial $r=0.03$ B-mode signal that Spider is searching for. At all scales, the lensing signal is below the sensitivity of Spider, so we believe that confusion between lensing and primordial B-modes will not limit our results. Future experiments searching for primordial B-modes well below $r\sim 0.01$ may need to ``delens'' their maps by measuring the lensing deflection field in their maps and removing its effects before hunting for the primordial $r$ signal in their data.

\subsection{Dust Foreground Removal}
\label{foreground_section}

\begin{figure}
\begin{center}
\includegraphics[width=0.55\textwidth]{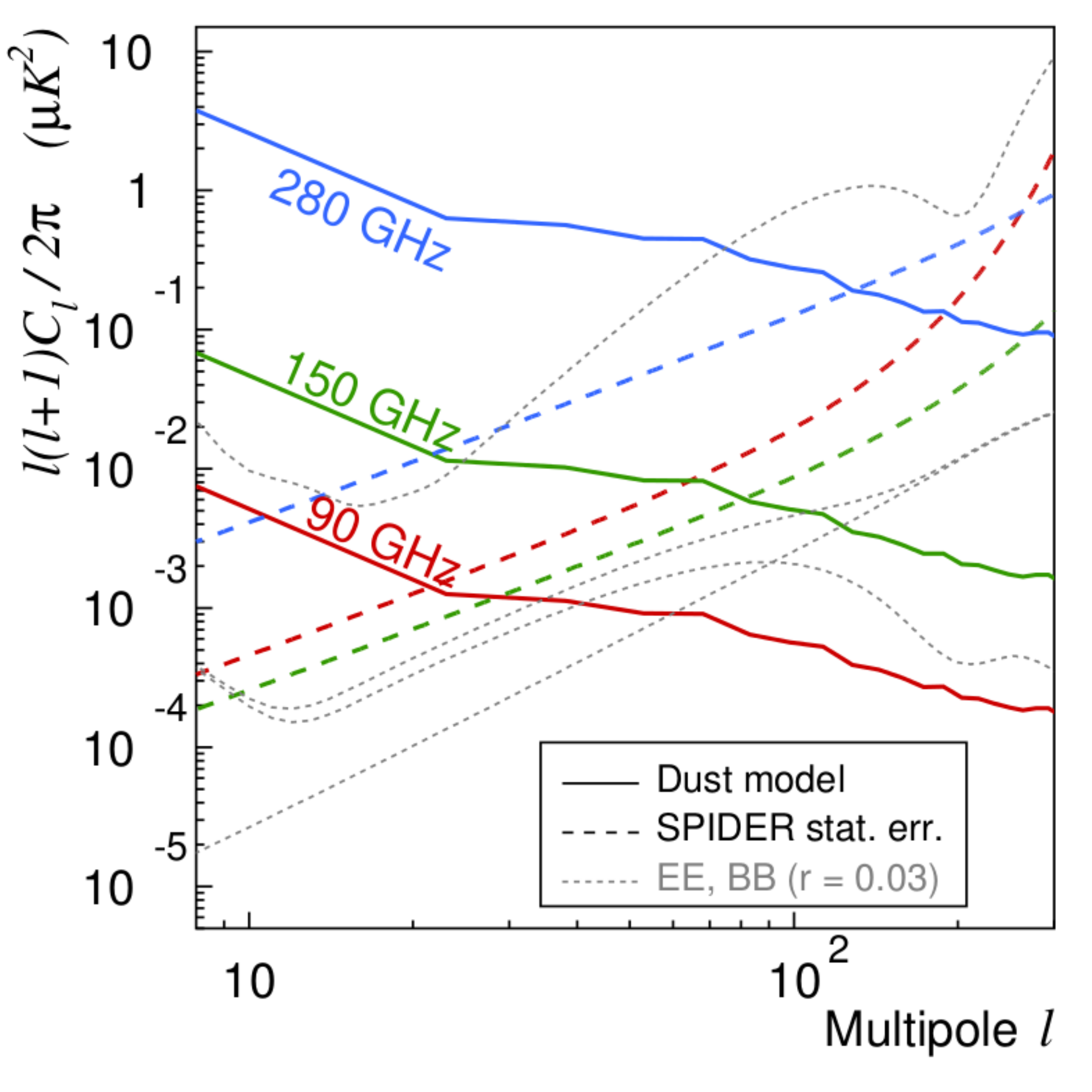}
\caption[Estimated foreground levels for Spider.]{Estimated foreground levels for Spider. Over most of the angular scales of our observations, we expect to be dominated by dust foregrounds. Combining the maps made at both of our observing frequencies, as well as using data from Planck, will let us remove this contamination and after our second flight we will reach a final inflationary B-mode sensitivity of $r=0.03$ at $99\%$ confidence. Figure from Fraisse et al. \cite{fraisse13}. \label{foreground_estimate}}
\end{center}
\end{figure}

The dominant foreground in Spider's frequency bands and observing region is expected to be polarized dust emission from our galaxy. Since other instruments have not yet made low-noise polarized maps at Spider's observing frequencies in this part of the sky, there is not existing data on the exact nature of this or other foregrounds. Fraisse et al. \cite{fraisse13} extrapolated data from earlier measurements to estimate the level of polarized foreground contamination in our observing region, which is shown in Figure~\ref{foreground_estimate}. At large angular scales ($l=10$), the expected dust foreground contamination is about 20 times larger than the instrument noise level at 95 GHz, and about 300 times larger at 150 GHz. The dust foreground drops to the level of the instrument noise by $l=20$ for 95 GHz, and $l=100$ for 150 GHz.

Combining the observations made at the different observing frequencies in Spider will be key to reducing the impact of foreground contamination. Modeling suggests that over our observing bands, dust contamination scales with observing frequency as a power law. The CMB fluctuations are known to have a blackbody spectrum scaling with observing frequency as $\frac{dB}{dT} (\nu, \mathrm{2.725~K})$. This means that the total observed temperature and polarization maps $S \equiv [I(\hat{n}),Q(\hat{n}),U(\hat{n})]$ can be modeled as
\begin{eqnarray}
S_{obs~(95~GHz)} &=& \frac{dB}{dT} (\mathrm{95~GHz}, \mathrm{2.725~K}) S_{CMB} + \left( \frac{\mathrm{95~GHz}}{\mathrm{95~GHz}} \right)^\beta S_{dust} \nonumber \\
S_{obs~(150~GHz)} &=& \frac{dB}{dT} (\mathrm{150~GHz}, \mathrm{2.725~K}) S_{CMB} + \left( \frac{\mathrm{150~GHz}}{\mathrm{95~GHz}} \right)^\beta S_{dust} \nonumber \\
I_{obs~(Planck~217~GHz)} &=& \frac{dB}{dT} (\mathrm{217~GHz}, \mathrm{2.725~K}) I_{CMB} + \left( \frac{\mathrm{217~GHz}}{\mathrm{95~GHz}} \right)^\beta I_{dust} . ~~~~~~
\end{eqnarray}
This system of equations relates the seven observations (95 GHz Spider $[I,Q,U]$, 150 GHz Spider $[I,Q,U]$, and 217 GHz Planck $I$) to the seven unknowns (CMB $[I,Q,U]$, Dust $[I,Q,U]$, and the Dust spectral index $\beta$). This means we can solve this system to get best estimates for maps of the CMB and foregrounds separately. In simulations this has been shown to reduce Spider's sensitivity to $r$ slightly below its theoretical best level, but if we take data at 280 GHz in a second flight the total sensitivity after the foreground removal procedure is still projected to be $r=0.03$ at $99\%$ confidence.

\chapter{The Spider Instrument}\label{spider_chapter}

Spider is a high-sensitivity microwave telescope array optimized for the low optical loading of the environment available from a scientific balloon payload. The instrument will be launched from McMurdo Station, Antarctica. By eliminating the detector loading and resulting photon noise from observing through the atmosphere, and due to its large number of polarized detectors, Spider will be able to make very deep maps of the 95 GHz and 150 GHz polarized sky on angular scales from $\sim$1-20 degrees with only a few weeks of observing time.

\section{The Balloon Platform}
\label{spider_scan}
The Spider instrument will operate as a stratospheric balloon payload in Antarctica. From altitudes of roughly 120,000 feet, very little of the earth's atmosphere remains between Spider and the microwave sky. Ground-based instruments suffer a large noise penalty due to photon loading from looking through the entire atmosphere. Because of the dramatic reduction in atmospheric loading, the detector noise in Spider will be significantly lower in flight, allowing the instrument to make deep maps of the microwave sky. The flight will be in the austral summer of 2014-2015 over the Antarctic continent. Circular stratospheric wind patterns centered on the South Pole can keep balloon payloads over the continent for weeks at a time. While this is much shorter than the months or years of observing time for a ground-based telescope, the reduced detector noise resulting from the low optical loading makes up for it. This will allow Spider to make competitive measurements with a far more compact dataset. The Columbia Scientific Balloon Facility (CSBF), part of NASA, provides the ballooning support for Spider.

Spider will execute azimuth scans across the sky with a sinusoidal velocity profile, a peak speed of 6 degrees per second, a peak-to-peak scan amplitude of 90 degrees, and a peak acceleration at the scan turnaround of 0.8 deg/s/s. The scan is fast enough to put the B-mode peak at $l\sim80$ at roughly 1 Hz in the detector timestreams, well below the high-frequency cutoff caused by the 1-10 ms time constants of our detectors, and well above the expected low frequency cutoff from 1/f noise at 20-40 mHz.

The instrument weighs roughly 5,000 lbs in total, so controlling the rapid scans of the instrument to the required precision of roughly an arcminute is a technical challenge. The instrument is supported by a structure called the gondola frame constructed from carbon fiber rods. The gondola is described in more detail in \cite{soler13}. This design allows the gondola to be light and strong. The torque to execute the scans is provided by a motor turning against a reaction wheel, and also using a pivot motor at the top of the instrument to torque against the balloon itself. The scans are servoed in real time using tiltmeters and gyros that monitor the motion of the instrument. The servoing ensures that the motion is smooth and that each scan has a well-controlled speed profile. Pinhole sun sensors that monitor the location of the sun, star cameras that use the relative locations of stars to determine the direction and rotation of the gondola, and differential GPS sensors all will be used after the flight to reconstruct the direction the instrument was pointing at all times during the flight. This will be crucial in converting the detector timestreams into maps of the microwave sky.

\begin{landscape}
\begin{figure}
\begin{center}
\includegraphics[width=1.5\textwidth]{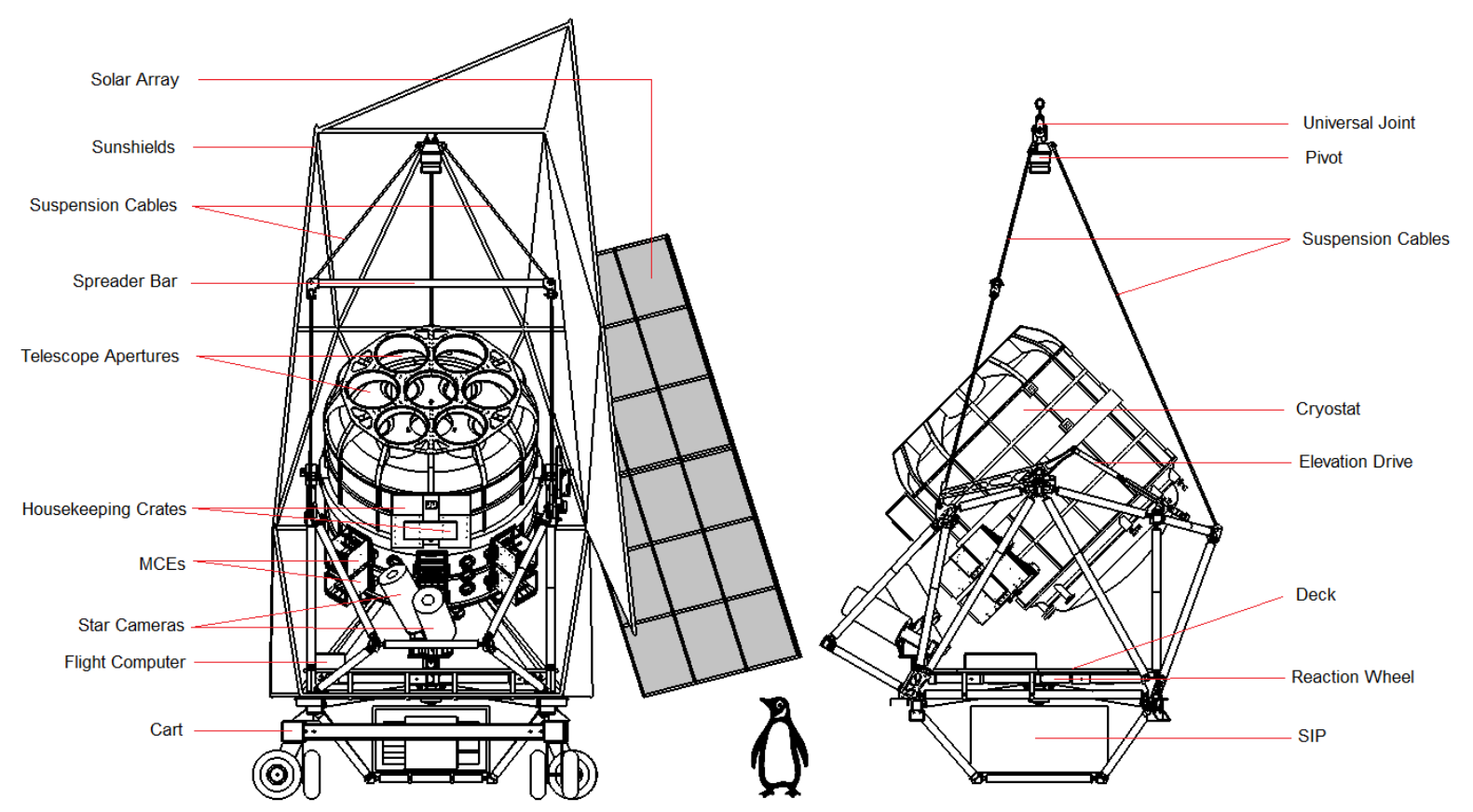}
\caption[Drawing of the Spider balloon payload.]{Drawing of the Spider balloon payload. A 1 m tall penguin is shown for scale. From \cite{soler13}. \label{spider_drawing_section}}
\end{center}
\end{figure}
\end{landscape}

\section{Cryogenics}

\begin{figure}
\begin{center}
\includegraphics[width=1.0\textwidth]{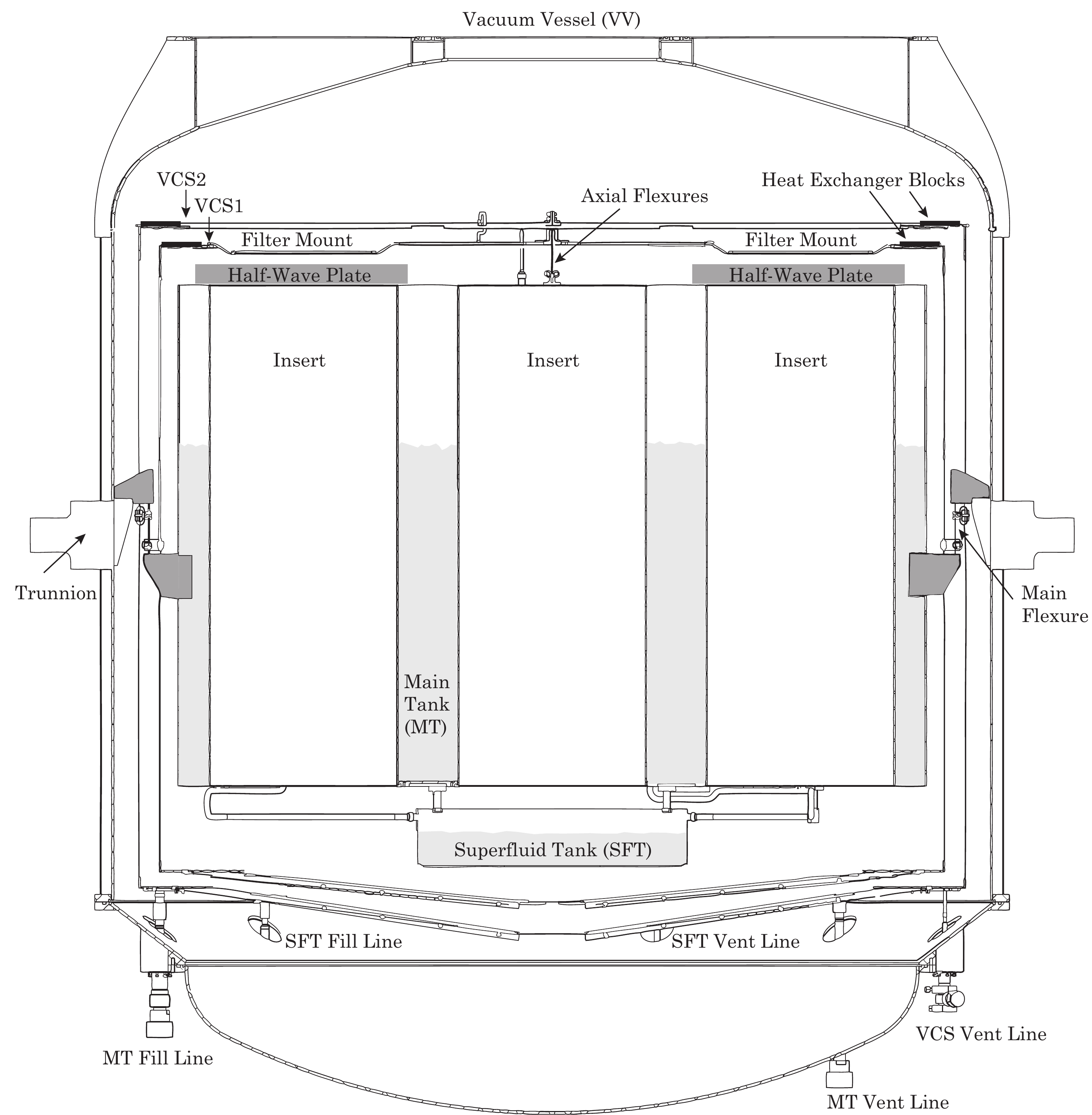}
\caption[Cross section view of the flight cryostat.]{Cross section view of the flight cryostat. The cryostat is attached to the gondola frame with the trunnion. The vacuum vessel is pumped out to eliminate thermal loading from convection. Vapor cooled shields at intermediate temperatures (VCS2 and VCS1) intercept thermal radiation before it gets to the main tank of liquid helium. The bottom of the cryostat holds the superfluid tank and associated fill and vent lines, providing cooling power at 2 K. The telescopes are mounted in the insert ports, which each look skyward through the HWPs and filter stacks. Figure from \cite{gudmundsson10}.\label{theo_cross_section}}
\end{center}
\end{figure}

Spider is an array of six refracting telescopes housed in a single liquid-helium cryostat. A cross-sectional drawing of the cryostat is shown in Figure~\ref{theo_cross_section}. The cryostat holds 1284 L of liquid helium, which is expected to keep the instrument cold for roughly 20 days. This long hold time is enabled by two vapor-cooled radiation shields at intermediate temperatures intercepting radiation from the ambient temperature vacuum vessel. The helium gas boiling off from the main tank is directed through plumbing and heat exchangers, and is used to cool the intermediate temperature shields before being vented out of the cryostat.

The cryostat also contains a superfluid liquid-helium reservoir. The superfluid tank is replenished via capillary lines connected to the main liquid bath. The inner diameter and length of the capillary tubes are fine enough to limit the flow rate of liquid helium, keeping the heat load on the superfluid tank low. In the lab environment, a pump is used to evaporatively cool the superfluid tank to below 2 K. In flight, the superfluid tank will be opened to ambient pressure, which will be low enough to pump on the superfluid tank and keep it at temperature. The tank provides cooling power to helium-3 refrigerators that cool the detectors to 250 mK. The tank also cools the telescope tubes to 2 K, which reduces their thermal radiation onto the detectors. The cryostat is described in more detail in \cite{gudmundsson10}.

\section{Detectors and Optics}

Spider employs a detector technology designed to allow us to deploy thousands of detectors. Spider uses phased-array antenna feeds to couple the radiation from free space into the detector system, instead of the more conventional feed horn technology. Each phased-array antenna consists of several slot antennas each with an individual microstrip transmission line. After being coupled into the antennas, the transmission lines carry the radiation through passband-defining filters, and on to the detector. The individual transmission lines are designed to combine radiation from each slot antenna in-phase at the detector. The phased-arrays are polarization selective, with cross-polar response at roughly the $1\%$ level. Each detector pixel consists of two interleaved phased-array antennas, each sensitive to a single polarization. The phased-arrays are fabricated on the same wafer as the rest of the detector system using standard photolithography techniques, which simplifies construction.

The detectors are superconducting Transition Edge Sensor (TES) bolometers. The focal plane is cooled below the transition temperature of the superconductors. The TESs are voltage-biased to keep them heated to a temperature midway along their superconducting transition. In this state, a small change in optical power is nearly exactly counterbalanced by a corresponding change in the electrical power dissipated in the device, a process called electrothermal feedback. One drawback of this kind of device is that there is a limit to the ability of electrothermal feedback to respond to incident optical power. Too much optical power will saturate the device. This limited dynamic range presents a challenge to testing, because the amount of optical input power from the 300 K lab environment is far higher than the optical loading during flight from the cold microwave sky. To enable operation under both loading conditions, each bolometer has an aluminum TES that can be biased onto its transition under high loading, in series with a titanium TES that will be used in flight loading conditions.

\begin{figure}
\begin{center}
\includegraphics[width=0.65\textwidth]{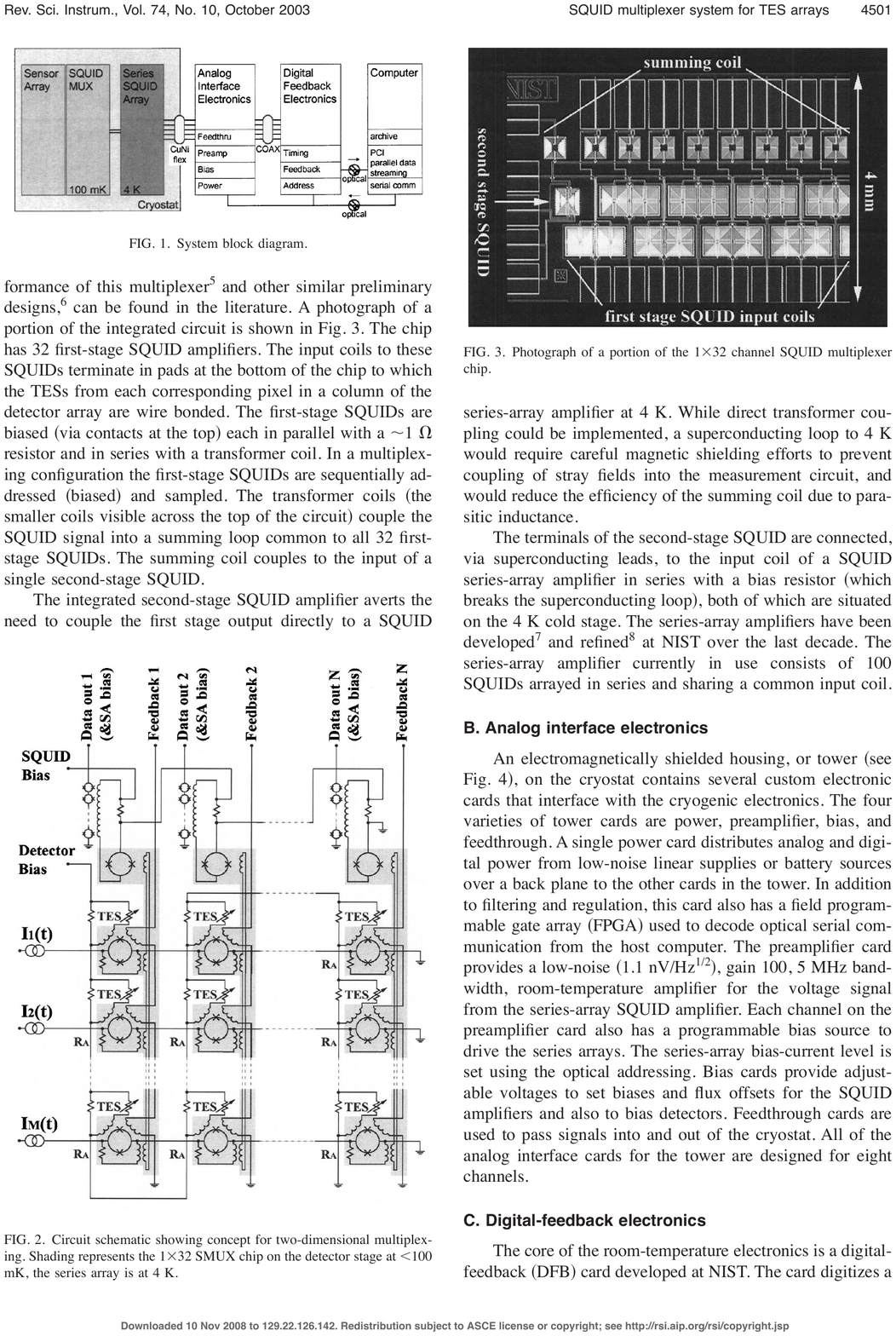}
\caption[Overview diagram of the multiplexed detector readout scheme.]{Overview diagram of the multiplexed detector readout scheme. From \cite{reintsema03}. \label{tmux_circuit}}
\end{center}
\end{figure}

The change in electrical power is sensed by measuring the current flowing through the TES using a Superconducting Quantum Interference Device (SQUID). Coils lithographed on the same chip as the SQUID are used to convert the current to magnetic field, and that field is detected with the SQUID. The SQUID readout system is constructed such that 32 detectors are read out by a single SQUID amplifier chain. The circuit is shown in Figure~\ref{tmux_circuit}. Each detector has its own SQUID. In a given column, all of these first stage SQUIDs except one are left in the unbiased state. The signals from all of the SQUIDs in a column are summed together with a coil. Since only the biased first stage SQUID contributes any signal to the sum, no information is lost by summing the signals and reading them out with a second-stage SQUID. Finally, the signals are amplified by a series-array SQUID, which boosts the signal enough that it can be measured outside the cryostat. Each of the other first stage SQUIDs is turned on one by one, and the rest of the amplifier chain gets the signal from that detector out of the cryostat. This allows a single amplifier chain to read out a column of 32 detectors. Each telescope in Spider uses 16 columns of 32 detectors. Since wiring for SQUIDs and detectors in a given row is shared across all columns, this grid approach significantly reduces the cryogenic wiring requirements. The antennas, detectors, and SQUID readout for Spider are discussed in more detail in \cite{trangsrud11}.

The detectors are fabricated on tiles. A 150 GHz tile contains an 8-by-8 array of polarized detector pairs, and a 95 GHz tile contains a 6-by-6 array of polarized detector pairs. Each of the 6 focal planes is populated with 4 tiles, for a total of 24 detector tiles in the instrument. The focal plane is coupled to the sky through a two-lens refracting telescope. The assembly of the lenses, detectors, and telescope tube is called an insert, which mounts in one of the six insert ports in the flight cryostat. Each insert is optimized to operate at a single passband, either 95 GHz or 150 GHz. The mechanical structure of the insert is supported with carbon fiber rods for lightweighting and strength. The lenses are machined from cast HDPE, and anti-reflection coated with porous PTFE sheets manufactured by Porex. The lenses are cooled to 4 K by the liquid helium bath to reduce their thermal radiation onto the detectors.

The sidelobes of the phased-array antennas in the focal plane are sensitive to radiation from the telescope tube itself. To reduce the amount of excess loading and photon noise this causes, and to prevent this stray coupling from receiving light from the sky, the telescope tube is blackened and cooled using the superfluid tank to below 2 K. Spillover onto the edges of primary lens is controlled with a blackened aperture stop cooled to below 2 K located on the cold side of the lens.

\begin{figure}
\begin{center}
\includegraphics[width=0.73\textwidth]{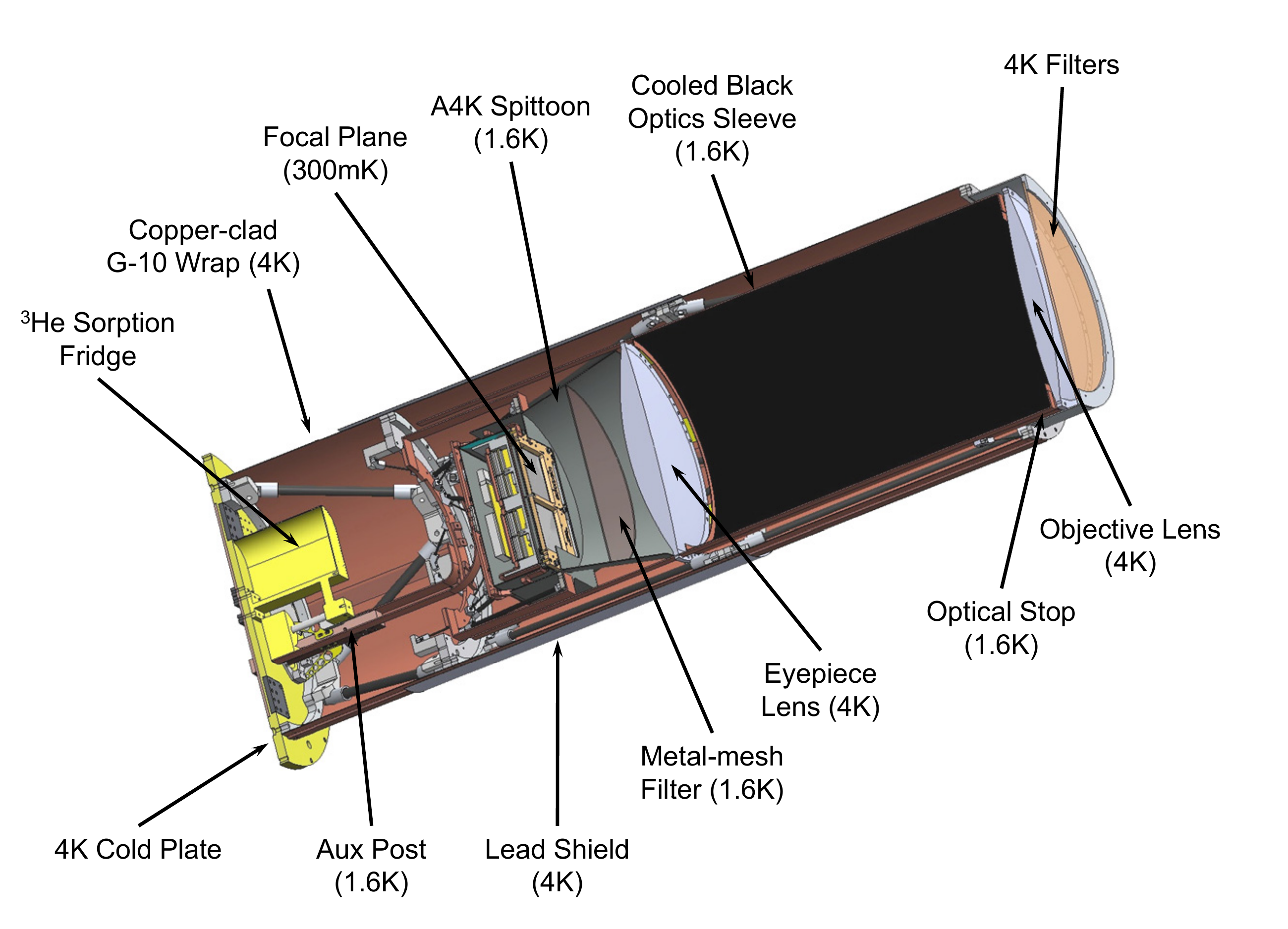}\includegraphics[width=0.26\textwidth]{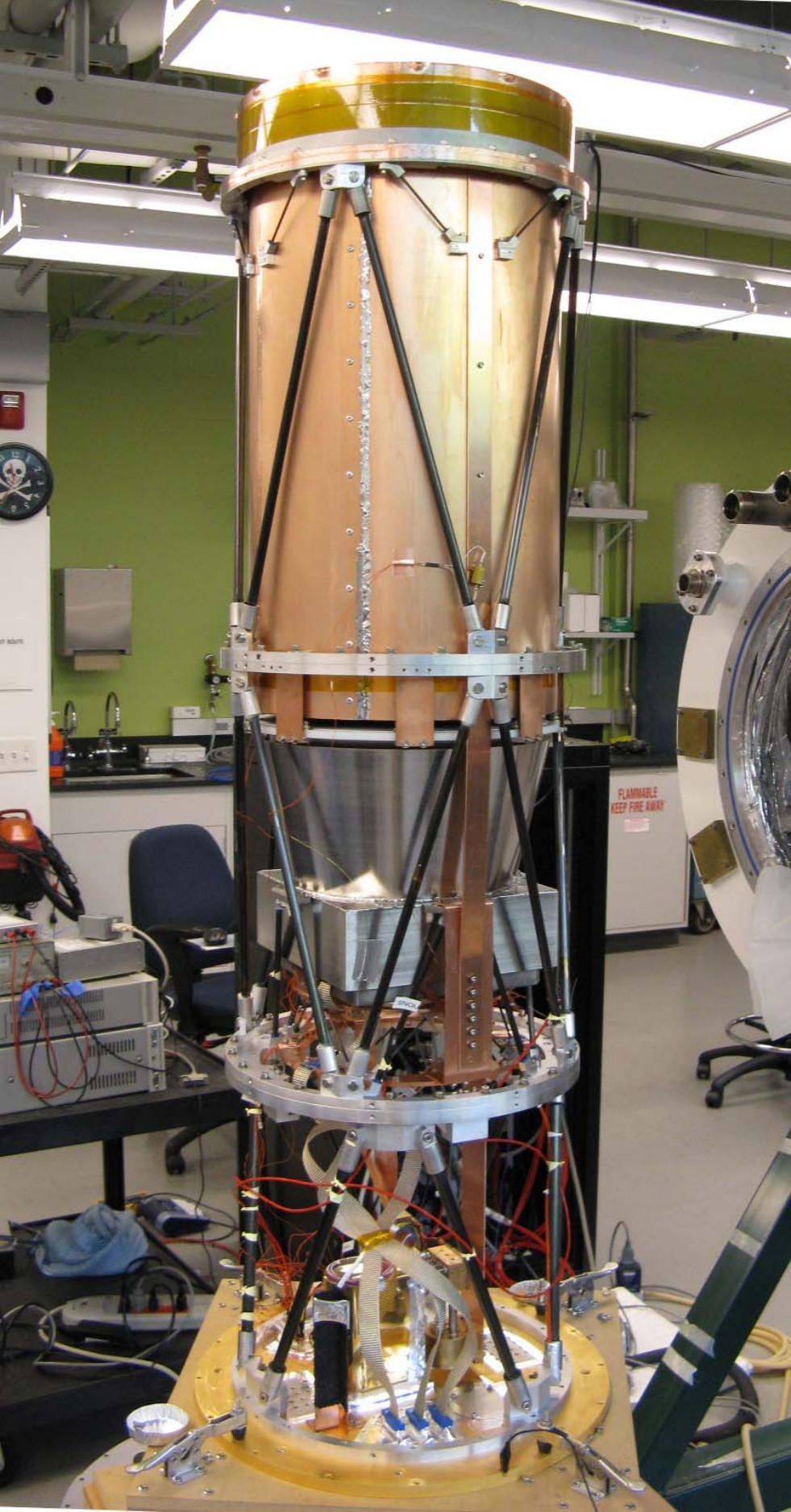}
\caption[Cross section view of a Spider telescope.]{Cross section view of a Spider telescope. From \cite{runyan10}.\label{insert_cross_section}}
\end{center}
\end{figure}

The first flight of the Spider instrument will have three 95 GHz receivers, and three 150 GHz receivers. The detector count and other specifications are shown in Table~\ref{bolo_count}.

\begin{table}
\begin{center}
\noindent\makebox[\textwidth]{%
\footnotesize
\begin{tabular}{c|c|c|c|c|c|}
  \textbf{Band} & \textbf{Beam} & \textbf{Pixel count}          & \textbf{Good} & \textbf{Typical}                  & \textbf{Total}\\
  \textbf{Center} & \textbf{FWHM} & ($\times N_{telescopes}$)          & \textbf{Detector} & \textbf{Detector}                  & \textbf{Instrument}\\
    &  &  ($\times$ dual polarization)       &  \textbf{Yield} & \textbf{Sensitivity}                 & \textbf{Sensitivity} \\ \hline \hline
   95 GHz & 49 arcminutes & $144\times3\times2$ = 864       & $83.3\%$ & $134~\mu\mathrm{K}_{CMB}\sqrt{\mathrm{s}}$ & $5.0~\mu\mathrm{K}_{CMB}\sqrt{\mathrm{s}}$ \\ \hline
      150 GHz & 29 arcminutes & $256\times3\times2$ = 1536    & $84.7\%$      & $130~\mu\mathrm{K}_{CMB}\sqrt{\mathrm{s}}$ & $3.6~\mu\mathrm{K}_{CMB}\sqrt{\mathrm{s}}$ \\ \hline

\end{tabular}}
\caption[Spider detector count and other specifications, based on measured performance at CSBF.]{Spider detector count and other specifications, based on measured performance at CSBF. Sensitivities are noise-equivalent temperature (NET). The total instrument sensitivity is calculated by combining the measured NET of each detector tile. The ``typical detector sensitivity'' is the total sensitivity scaled by $\sqrt{N}$ to give a sense of typical device performance. Data courtesy of Jeffrey Filippini and the rest of the Spider team. Simulations in Fraisse et al. assumed the same beam FWHM, but assumed instrument sensitivities of $5.5~\mu\mathrm{K}_{CMB}\sqrt{\mathrm{s}}$ at 95 GHz, and $4.2~\mu\mathrm{K}_{CMB}\sqrt{\mathrm{s}}$ at 150 GHz \cite{fraisse13}. \label{bolo_count}}
\end{center}
\end{table}


\chapter{Modeling the Spider Beam}

\label{zemax_chapter}

For many optical systems, ray tracing is a way to quantitatively model the performance of the optical elements. The ray approximation holds when the physical size of the optics and the detectors is far larger than the wavelength of light. However, for Spider the optics are $\sim0.3$ m in diameter, which is only $\sim100$ times larger than the observing wavelengths. The detectors are small, only $\sim3$ wavelengths in size. This means that a ray trace can still guide the design of the optics, and will be a good qualitative simulation. However, to fully understand the system it is necessary to model the electromagnetic fields themselves as they propagate through the system. Specifically, this model is necessary to calculate the amount of stray light radiating onto the detectors from the finite temperature aperture stops in the system, and also to model the far-field beam maps measured through the full optical system.

To model the Spider optical system, we use the Zemax EE software program, since it can both trace rays and numerically model field propagation. It is also helpful to check analytically the field propagation results from Zemax using the Gaussian-beam formalism. 

\section{Gaussian-Beam Formalism}

This brief review follows Goldsmith \cite{goldsmith98}. Spider is a receiver, but thinking in the time-reversed sense, many microwave feeds launch electric-field distributions that have an approximately Gaussian profile. With the opening of the feed in the $xy$ plane and the feed looking in the $z$ direction, the electric-field amplitude (suppressing two phase factors) is approximately
\begin{equation}\label{gaussbeam-def}
E(x,y,z) = \sqrt{ \frac{2}{ \pi w(z)^{2}} } \exp{\left( -\frac{(x^{2} + y^{2})}{w(z)^{2}} - \frac{i \pi (x^{2} + y^{2})}{\lambda R(z)}  \right)},
\end{equation}
where the beam width $w(z)$ and phase-front curvature $R(z)$ vary as the beam propagates through the different surfaces in an optical system. Since this is a traveling-wave solution, the magnetic-field distribution follows directly from the electric-field distribution. What makes the Gaussian-beam approximation convenient is that the evolution of $w$ and $R$ through free space, through lenses, or in reflection from a curved mirror, can be calculated straightforwardly using ray transfer matrices. It is also a useful approximation since many optical systems do not dramatically distort a Gaussian field distribution as it travels through the optics.

To start the calculation, it is necessary to determine the Gaussian beam that is launched by the antenna. The array of slot antennas that make up a Spider antenna can be well approximated as a uniform-phase square-patch antenna with a physical width $a$. At the focal plane, this electric field distribution $\phi$ (normalized such that $\iint \phi^{*} \phi dx dy$ = 1) is
\begin{equation}
\phi(x,y) = 
\begin{cases}
\frac{1}{a} & -\frac{a}{2} < (x,y) < \frac{a}{2} \\
0 & \mathrm{elsewhere}.
\end{cases}
\end{equation}
To calculate the corresponding Gaussian beam for this antenna, following Goldsmith \cite{goldsmith98} we consider the coupling coefficient $c$ between this distribution $\phi$ and a Gaussian beam $\psi$ with a width $w_{g}$ and $R = \infty$. The coupling integral is
\begin{equation}\label{field_coupling}
c \equiv \iint \phi^{*} \psi dx dy = \int_{-\frac{a}{2}}^{\frac{a}{2}} \int_{-\frac{a}{2}}^{\frac{a}{2}} \frac{dx dy}{a} \sqrt{ \frac{2}{ \pi w_{g}^{2}} } \exp{\left( - \frac{(x^{2} + y^{2})}{w_{g}^{2}} \right) }.
\end{equation}
The parameter $w_{g}$ is varied until the coupling is maximized, i.e. $\frac{dc}{dw_{g}} = 0$. Solving this equation reduces to
\begin{equation}
\frac{\sqrt{\pi}}{2} \frac{1}{a/w_{g}} \mathrm{erf}\left(\frac{1}{2} \frac{a}{w_{g}} \right) = \exp{\left( - \frac{1}{4} \left( \frac{a}{w_{g}}\right)^{2} \right)}.
\end{equation}
The numerical solution of this equation is
\begin{equation}\label{gauss-equiv}
w_{g} = a \times 0.5051.
\end{equation}
Evaluating the integral in Equation~\ref{field_coupling} with this width shows that $c^{2} = 79\%$ of the power of the antenna goes into this Gaussian mode, which means that the Gaussian-beam formalism is appropriate for approximately modeling an optical system fed with this antenna.

To propagate a Gaussian beam through an optical system, first the width and curvature parameters launched by the detector are combined into the complex beam parameter $q$, defined as
\begin{equation}\label{q-def}
\frac{1}{q} \equiv \frac{1}{R} - \frac{i \lambda}{\pi w^{2}}.
\end{equation}
For a Spider detector, $w = w_{g}$ from Equation~\ref{gauss-equiv} and $R=\infty$. Then the ray transfer matrices $\mathsf{H}_{i}$ from the focal plane through any of the lenses, free space, and mirror reflections up to Nth surface of interest are multiplied together to yield the combined ray transfer matrix defined as
\begin{equation}
\left[ \begin{array}{cc} A & B  \\ C & D \end{array} \right] \equiv \mathsf{H}_{N} \ldots \mathsf{H}_{2} \mathsf{H}_{1}.
\end{equation}
Several useful ray transfer matrices are shown in Table~\ref{ray-trans-mat}. This allows the complex beam parameter
\begin{equation}
q_{out} = \frac{A q_{in} + B}{C q_{in} + D}
\end{equation}
at that surface to be calculated. After inverting Equation~\ref{q-def} to obtain $w$ and $R$ at the surface of interest, the electric field distribution can then be calculated using Equation~\ref{gaussbeam-def}. The power distribution $P(x,y)$ at that surface is then calculated by taking the absolute value $E^{*}(x,y) \times E(x,y)$ of the field distribution.

\begin{table} 
\begin{center}
\noindent\makebox[\textwidth]{%
\small
\begin{tabular}{|c|c|}
\hline
\textit{Optical} & \textit{Ray Transfer} \\
\textit{Element} & \textit{Matrix} \\ \hline \hline
 &  \\
Distance $L$ in a uniform medium of any index & $\left[ \begin{array}{cc} 1 & L  \\ 0 & 1 \end{array} \right]$ \\
 &  \\ \hline
    &  \\
Spherical interface from index $n_{1}$ to $n_{2}$ of radius $R$. & $\left[ \begin{array}{cc} 1 & 0  \\ \frac{n_{2} - n_{1}}{n_{2} R} & \frac{n_{1}}{n_{2}} \end{array} \right]$ \\
$R>0$ if concave towards incident light &  \\ 
  &  \\ \hline
  &  \\
Slab of thickness $L$ of index $n_{2}$ material surrounded by material with $n_{1}$ & $\left[ \begin{array}{cc} 1 & L \frac{n_{1}}{n_{2}}  \\ 0 & 1 \end{array} \right]$ \\
 &  \\ \hline

\end{tabular}}
\caption[Selected ray transfer matrices.]{Selected ray transfer matrices. After a table in Goldsmith \cite{goldsmith98}. \label{ray-trans-mat}}
\end{center}
\end{table}

\subsection{Effective Lens Curvature}

The lenses in Spider are defined according to a axially-symmetric conic surface profile. Following the convention used in Zemax, the height-vs-radius (i.e. the ``sag'') of a lens surface in Spider is
\begin{equation}
z(r) = \frac{c r^2}{1 + \sqrt{1 - (1+k) c^2 r^2}},
\end{equation}
where $c$ is the inverse curvature of the lens and $k$ is the conic constant. However, the ray transfer matrices for Gaussian beams only can handle spherical surfaces. Simply assuming that the lens is spherical with a curvature radius $1/c$ does not take into account the significant impact of the conic constant on the lens shape.

To approximately handle this effect for a Gaussian-beam calculation, we calculate an effective curvature $c_{eff}$ by requiring that at the outer radius of the lens $r_{edge}$, the corresponding spherical lens have the same sag as the real conic lens. The sag $z_{edge}^{real}$ of the real lens is
\begin{equation}
z_{edge}^{real} = \frac{c r_{edge}^2}{1 + \sqrt{1 - (1+k) c^2 r_{edge}^2}}.
\end{equation}
Equating this with the sag of a spherical lens yields an equation for $c_{eff}$,
\begin{equation}
\frac{c_{eff} r_{edge}^2}{1 + \sqrt{1 - c_{eff}^2 r_{edge}^2}} = z_{edge}^{real}.
\end{equation}
Solving this equation for $c_{eff}$ yields
\begin{equation}
c_{eff} = \frac{2 z_{edge}^{real}}{(z_{edge}^{real})^2 + r_{edge}^2}. \label{ceff}
\end{equation}
So, for example, the sky-side surface of the secondary lens (i.e. the top surface of the lower lens shown in Figure~\ref{ray_trace}) has a real curvature parameter $c = 1/(-3643.06~\textrm{mm})$ and a conic constant $k = 532.204$ (i.e. the surface is an oblate ellipsoid). The clear diameter of the lenses in Spider is 290 mm, so $r_{edge} = $ 145 mm. Evaluating Equation~\ref{ceff} for this lens surface yields an effective curvature of $c_{eff} = 1/(-2541.46~\textrm{mm})$, a significant correction factor.

\section{Zemax Physical Optics}
The Gaussian-beam formalism has several important limitations. As just noted, Gaussian beams cannot model optical systems with aspheric lenses. Also, for Spider $21\%$ of the power launched by an antenna does not go into a Gaussian beam pattern, so it would be desirable to model the propagation of that power through the optical system as well. The Gaussian-beam formalism cannot model the effect of aperture stops on an optical system. A truncated Gaussian is no longer a Gaussian beam, so ignoring the stops is the only way to use the formalism. One possible way around this would be to consider the propagation of higher-order electric field modes in addition to the Gaussian fundamental mode. Finally, a critical limitation for a telescope like Spider with a large field of view is that these approaches cannot model an optical system off-axis.

The physical optics propagation feature of Zemax EE solves all of these problems. The user can specify an arbitrary initial electric field distribution on the focal plane, and that distribution is numerically propagated through an optical system (including aperture stops) using an FFT-based algorithm. Zemax can model off-axis detectors, which is useful because of the large focal plane in Spider. Also unlike the Gaussian-beam formalism, Zemax can easily calculate the electric field propagation through aspheric lenses, such as those used in Spider. The disadvantage of Zemax is a small speed penalty, and the complexity of the program. Because of the approximations it makes, a Gaussian-beam model requires only a few analytic calculations, whereas simulating the full electric field distributions for Spider with Zemax requires several minutes of desktop computer time. Still, Zemax is a complex enough program that it is worthwhile to use Gaussian beams as an approximate check on Zemax's results, and to build intuition.

\section{Spider Optical System}

\begin{figure}
\begin{center}
\includegraphics[width=0.99\textwidth]{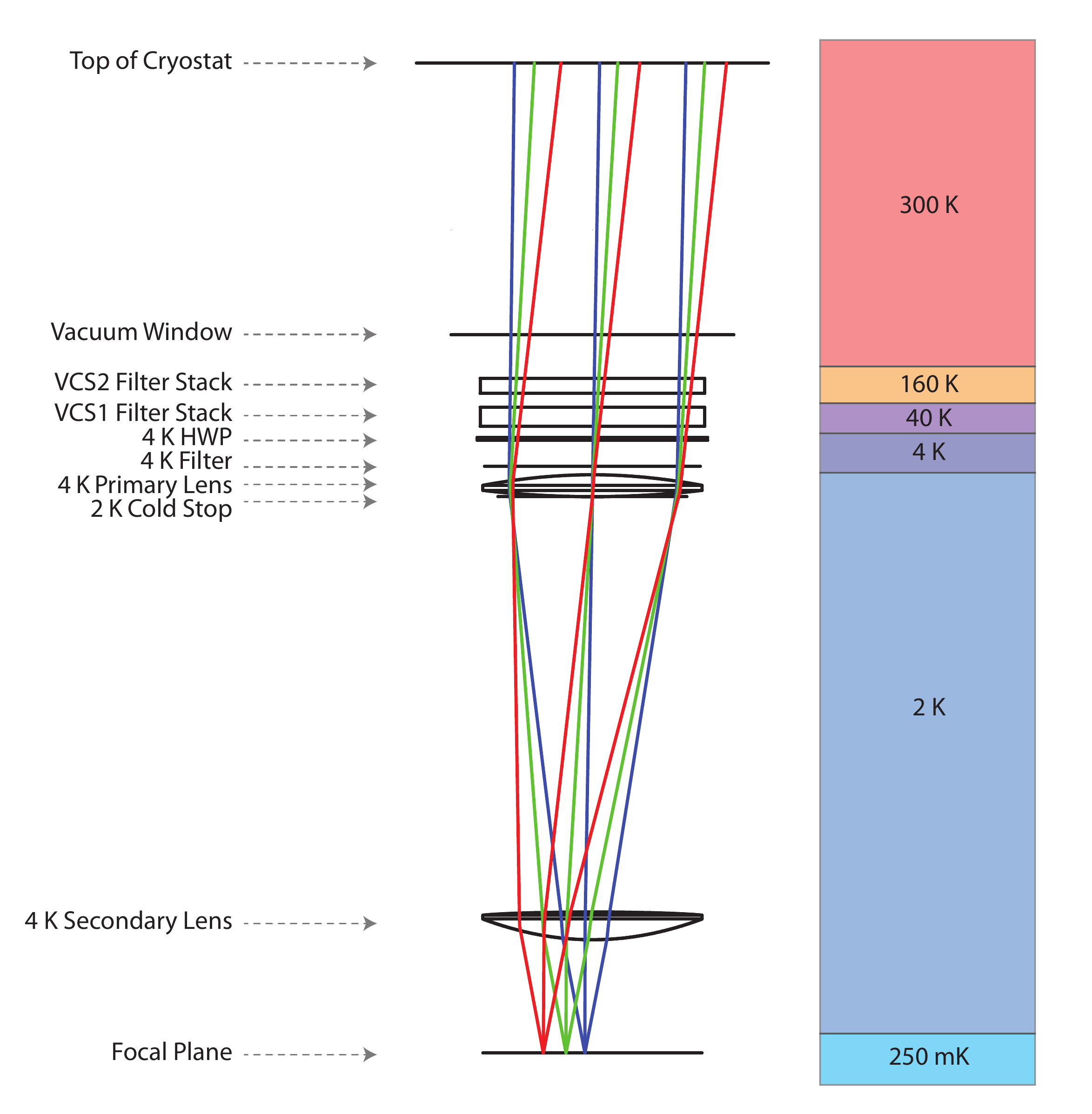}
\caption[Diagram of the Spider optical system. ]{Diagram of the Spider optical system, showing the clear apertures of all of the optical elements to scale. The 150 GHz detectors on the focal plane launch an $f/2.6$ beam ($f/2.3$ at 95 GHz) onto the secondary lens. A 2 K aperture stop is located before the primary lens. The two lenses are actually at 4 K, but there is a 2 K blackened optics sleeve surrounding the telescope. There is a filter at 4 K, then the HWP, followed by a filter stack at each of the vapor-cooled shields (VCS) to prevent excess thermal radiation from reaching the 4 K parts of the cryostat. There is then a thin UHMWPE vacuum window, followed by a recessed baffle that continues to the edge of the cryostat. Not shown is the baffle attached to the cryostat that guards against stray light. The blue, green, and red rays respectively are $f/2.6$ cones of rays traced from detectors at the inside corner, center, and outside corner of one of the four detector tiles. \label{ray_trace}}
\end{center}
\end{figure}

A Spider 150 GHz antenna is a 6.95 mm square patch. Using Equation~\ref{gauss-equiv}, this means that the equivalent Gaussian beam has a width of $6.95~\mathrm{mm} \times 0.5051 = 3.51$ mm. In the far field, the $1/e$ radius of a Gaussian beam expands at a full-width angle of $(2 \lambda)/(\pi w)$ radians. (This radius encloses $86.5\%$ of the beam power \cite{goldsmith98}.) This angle can be converted to an effective $f-$number for the feed,
\begin{equation}
f_{square~patch} = \left( 2 \times \tan{\frac{\lambda}{\pi (a \times 0.5051)}} \right)^{-1}.
\end{equation}
A 150 GHz detector is therefore approximately $f/2.6$. The 95 GHz is $f/2.3$ since it is a 9.27 mm patch, slightly smaller than a frequency-scaled 150 GHz detector. These feed $f-$numbers both mate well with $f-$number of the Spider telescope as shown in Figure~\ref{ray_trace}.

\begin{figure}
\begin{center}
\includegraphics[width=0.7\textwidth]{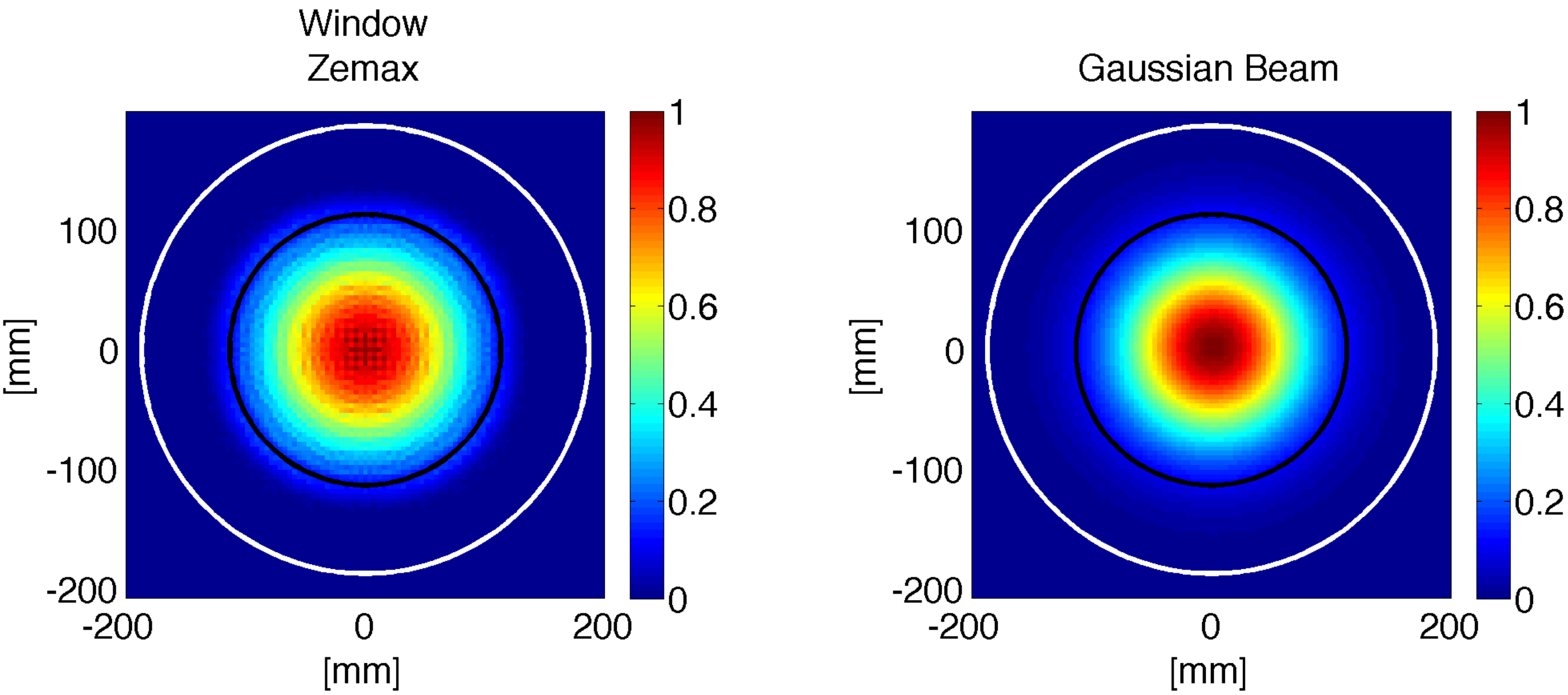}
\includegraphics[width=0.7\textwidth]{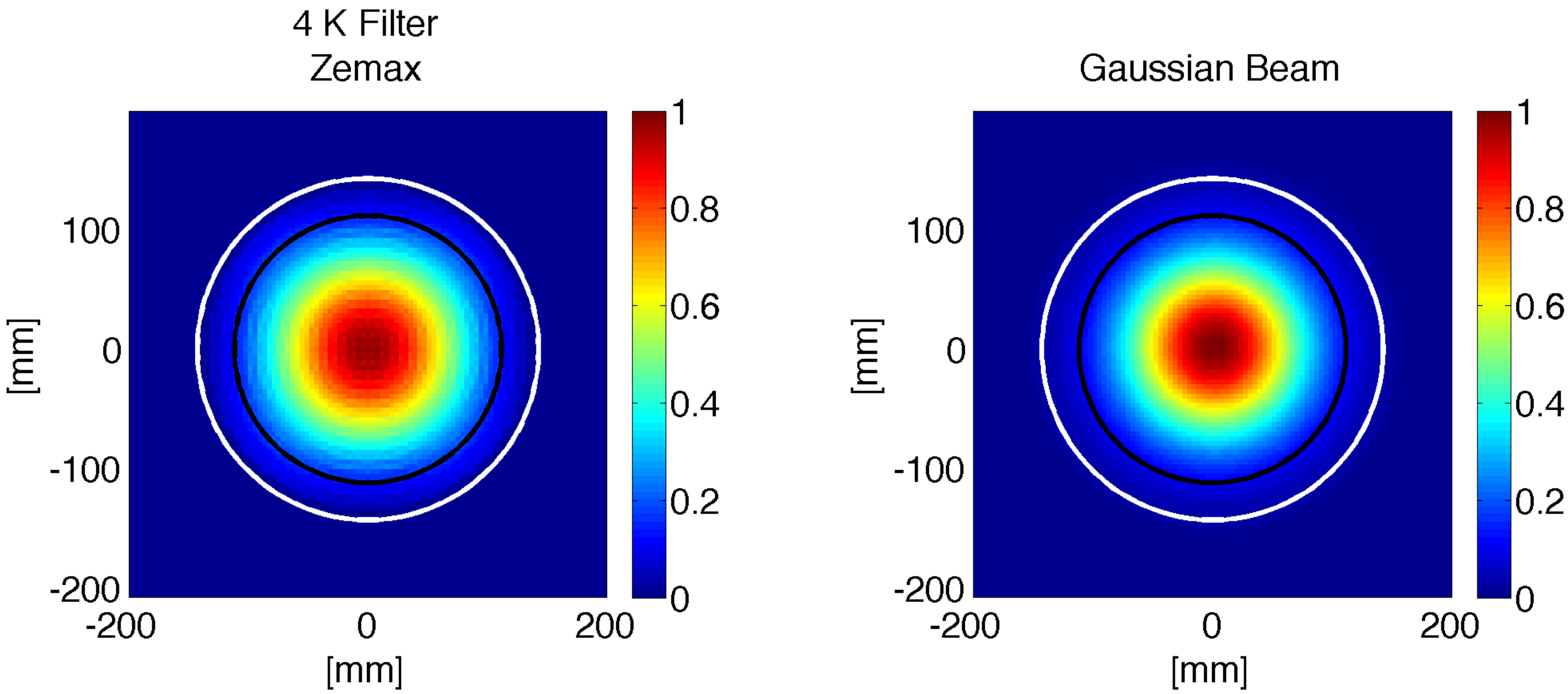}
\includegraphics[width=0.7\textwidth]{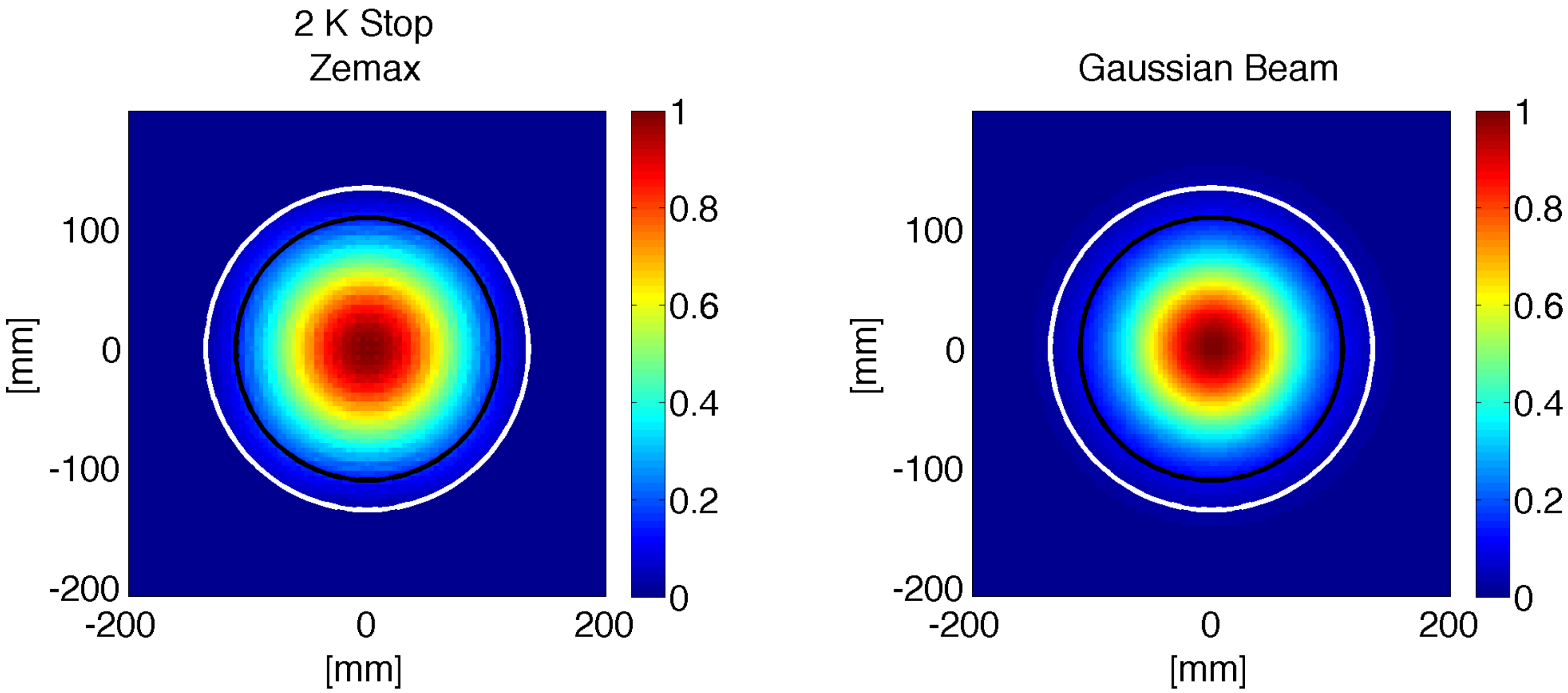}
\includegraphics[width=0.7\textwidth]{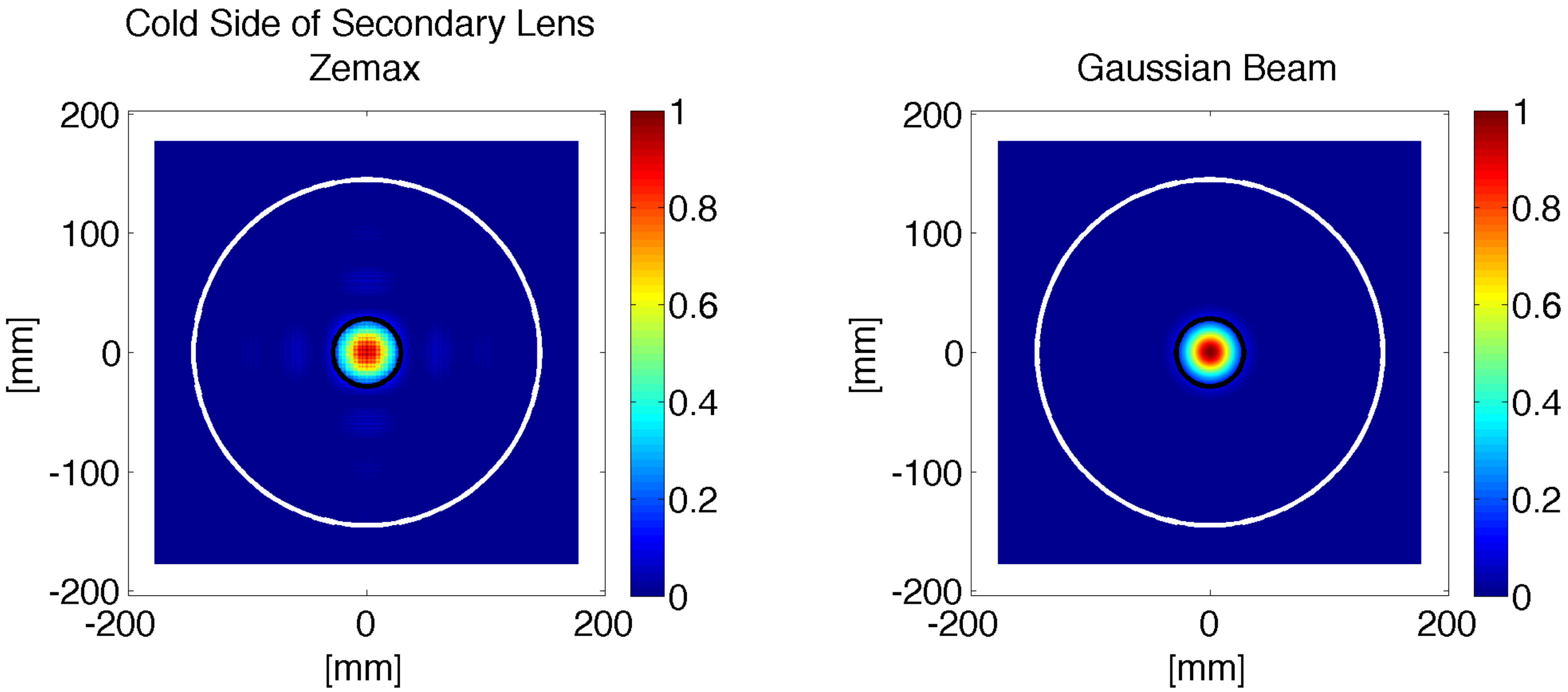}
\caption[Zemax and Gaussian-beam power distribution simulations of an on-axis Spider detector at 150 GHz.]{Zemax physical optics (left) and analytic Gaussian -beam (right) power distribution simulations of an on-axis Spider detector at 150 GHz. A white circle shows the clear diameter of the surface, and a black circle shows the size of an $f/2.6$ ray bundle traced using the same ABCD matrix used in the Gaussian-beam calculation. \label{zemax_and_gaussbeam}}
\label{fig:cont}
\end{center}
\end{figure}

Figure~\ref{zemax_and_gaussbeam} shows a Zemax simulation of a Spider 150 GHz detector placed at the center of the focal plane as its power pattern propagates up to the cold side of the secondary lens, then up to the cold stop, through the primary lens onto the 4 K filter, and finally up to the UHMWPE vacuum window. The Gaussian-beam calculation is shown for comparison. Both methods assume monochromatic radiation, 2 mm wavelength for the 150 GHz calculation, and 3 mm for the 95 GHz calculation. The two calculations nearly agree, except in the calculated spillover onto each aperture in the system. This is because Zemax properly truncates the electric field distribution when it encounters an aperture, unlike the Gaussian-beam formalism which cannot do this. Also, four of the non-Gaussian near field sidelobes launched by the square antenna are visible in the Zemax simulation at the secondary lens. These are absorbed by the 2 K stop and cooled optics sleeve, and do contribute to the loading on the detectors. This effect is handled by the Zemax simulation.

\begin{figure}
\begin{center}
\includegraphics[width=0.7\textwidth]{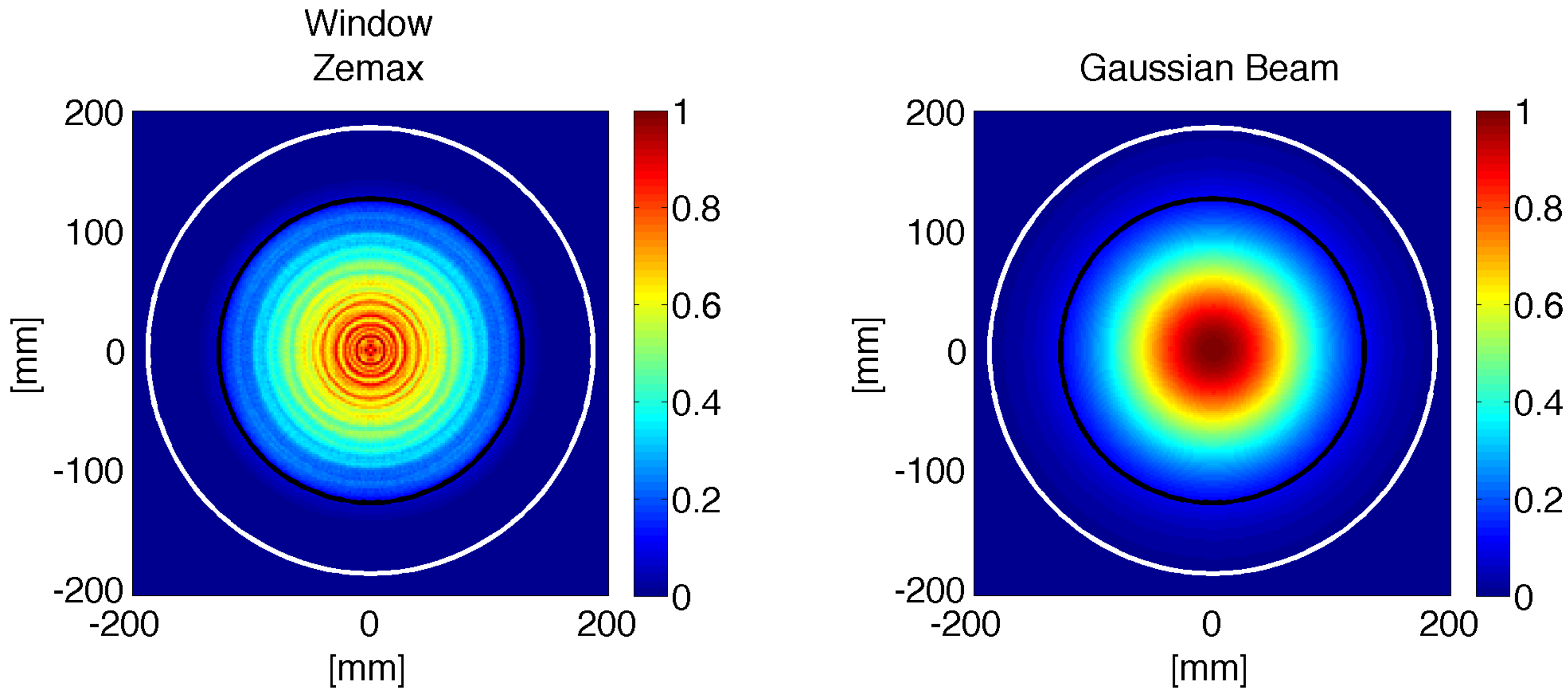}
\includegraphics[width=0.7\textwidth]{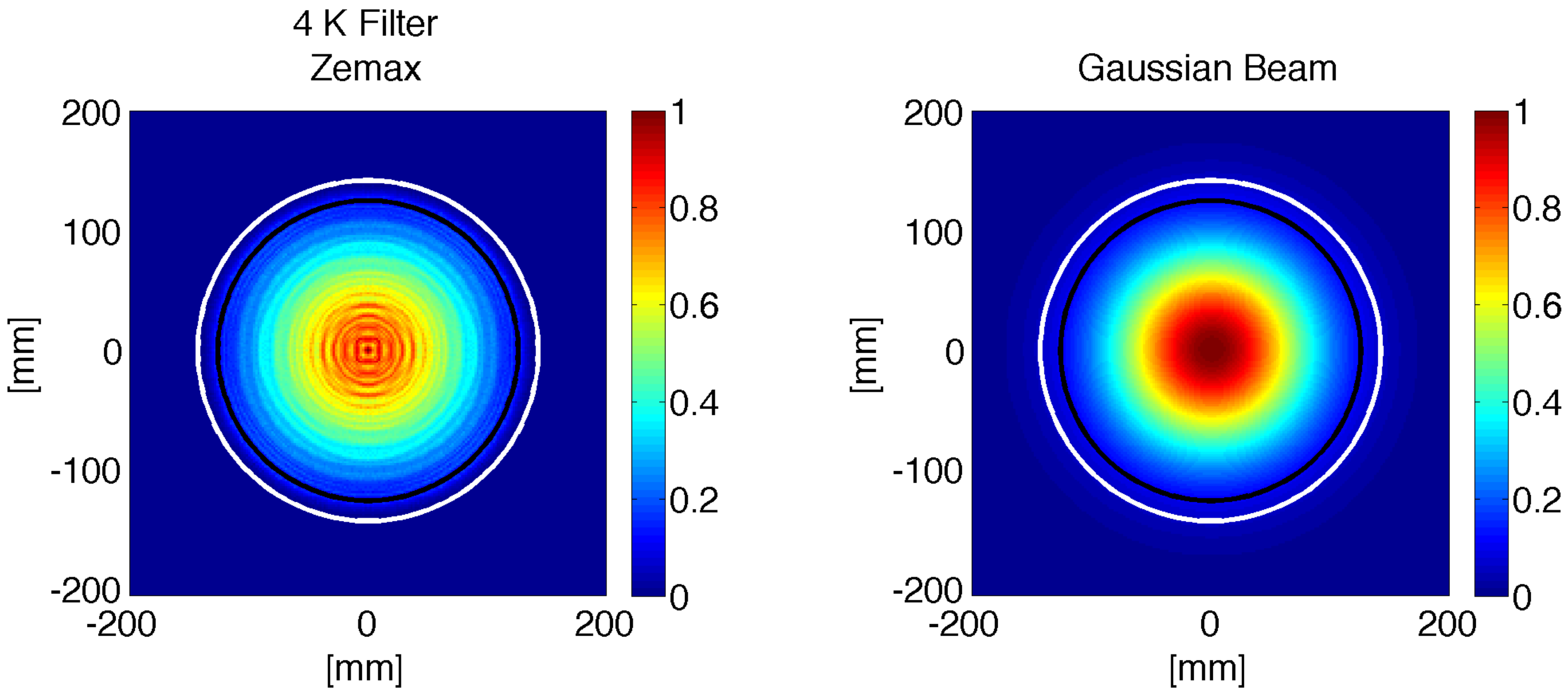}
\includegraphics[width=0.7\textwidth]{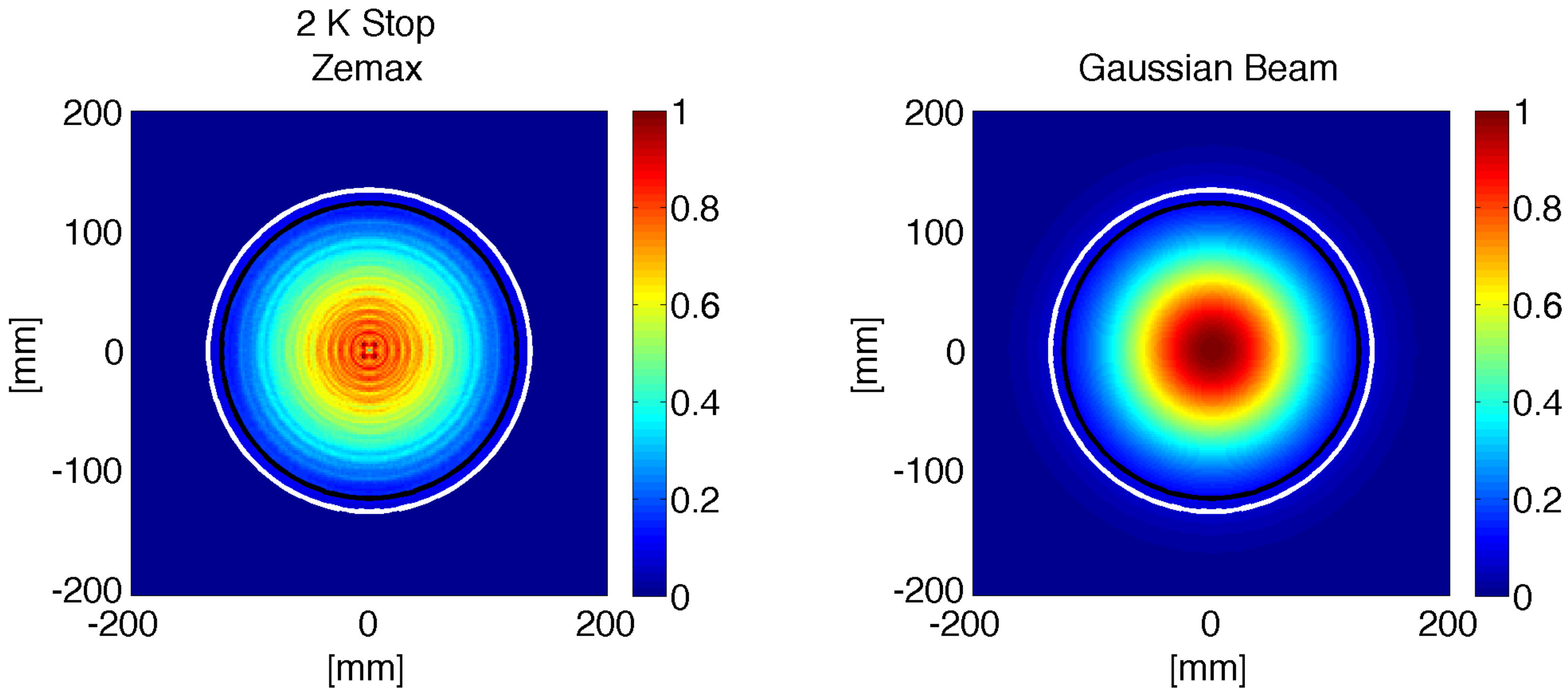}
\includegraphics[width=0.7\textwidth]{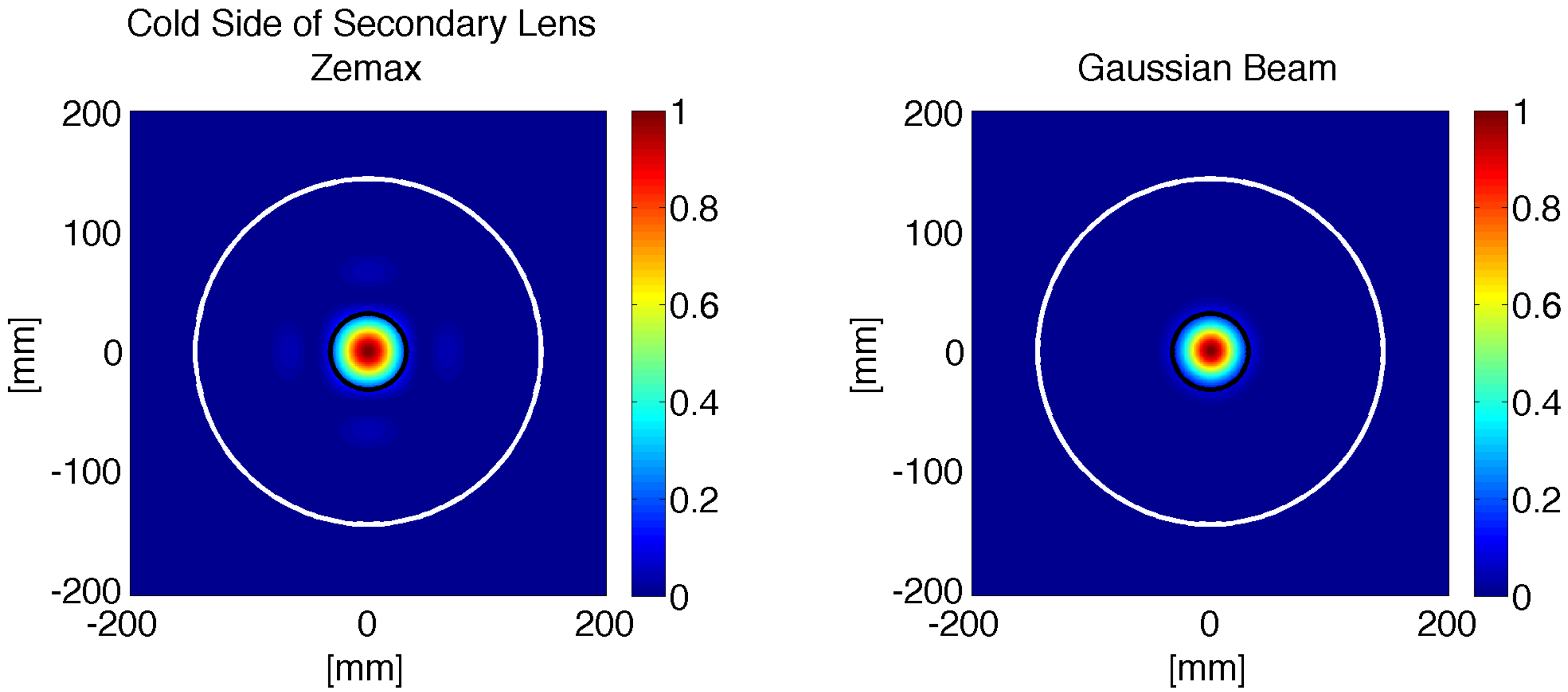}
\caption[Zemax and Gaussian-beam power distribution simulations of an on-axis Spider detector at 95 GHz.]{Same as Figure~\ref{zemax_and_gaussbeam} but for 95 GHz. The size of the detector feed was changed to the 95 GHz size, and a $f/2.3$ ray bundle was traced.}
\label{fig:cont}
\end{center}
\end{figure}

\section{Far-field Beam Maps}

The sidelobes of the far field response can be measured using a bright broadband noise source. This source uses Johnson noise in a resistor followed by a chain of amplifiers and frequency doublers to produce incoherent radiation launched by a feed horn. A clean-up polarizer is placed after the feed, and the entire device can be rotated.

\begin{figure}
\begin{center}
\includegraphics[width=0.95\textwidth]{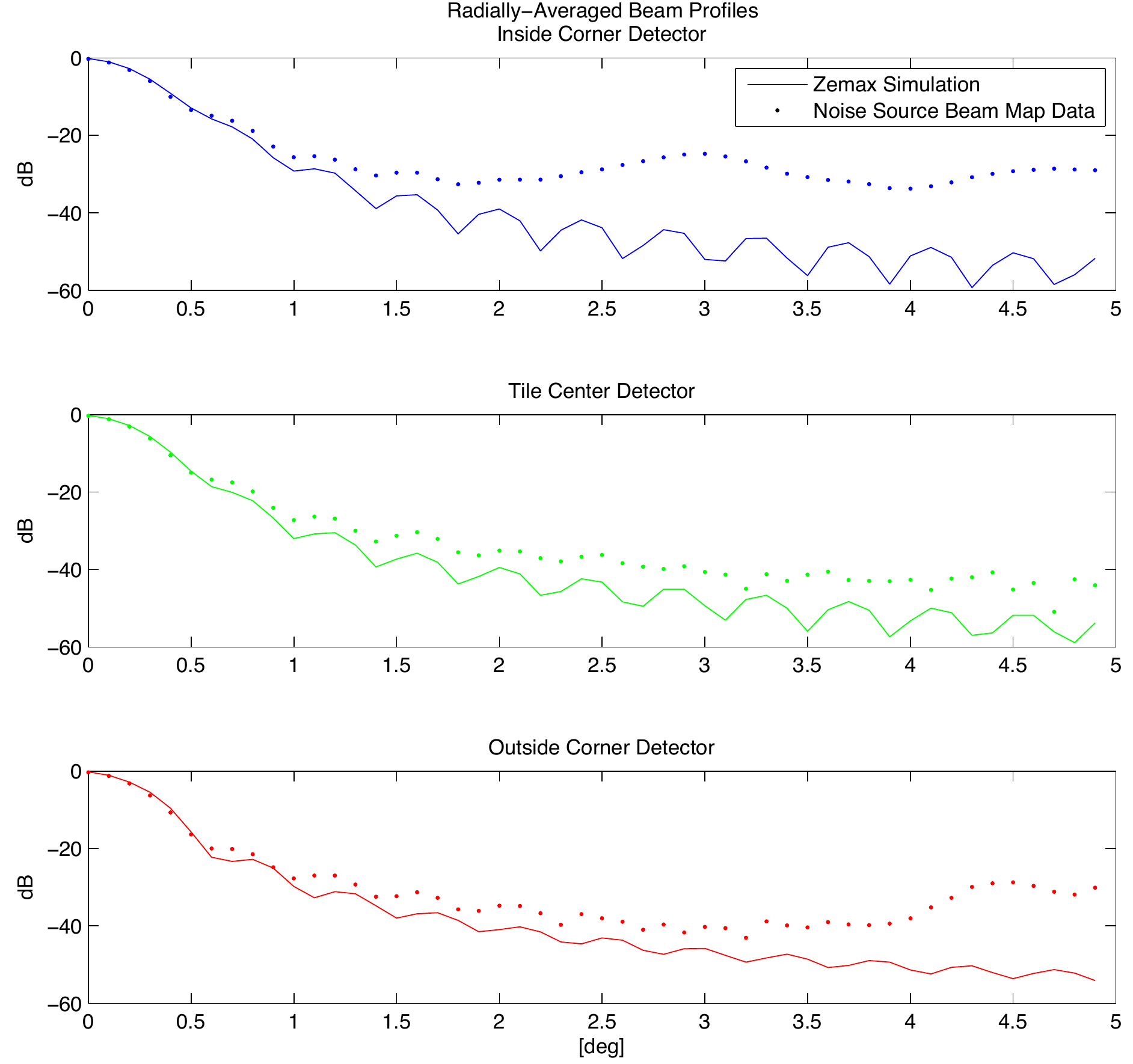}
\caption[Radially-averaged beam profiles of 150 GHz detectors.]{Radially-averaged beam profiles of 150 GHz detectors. A Zemax simulation assuming single-frequency light is overplotted. The source was bright enough that the data are not limited by noise. Data and radial averaging code courtesy of Rebecca Tucker. \label{radial_profiles}}
\end{center}
\end{figure}

Radially-averaged beam maps from three selected detectors taken with a Spider 150 GHz receiver at the Caltech highbay using the broadband noise source are shown in Figure~\ref{radial_profiles}. One detector was near the ``inside corner'' of one of the four detector tiles (i.e. considering just one of the four tiles, the detector in that tile closest to the center of the focal plane), another was near the ``tile center'' (i.e. a detector at the center of a tile and therefore some distance away from the focal plane center) and the third was an ``ouside corner'' detector (i.e. a detector maximally distant from the center of the focal plane). The source was bright enough that the beam profile measurements were not limited by detector noise. A radially-averaged Zemax simulation of approximately the same places on the focal plane are overplotted. The model matches the data well out to 1 degree, roughly the -30 dB point. The main beam profiles of the data and model agree well. Between 1 and 5 degrees the data profile plateaus at roughly -30 to -40 dB, whereas the simulation continues to drop to -50 dB. The effect of the ghost beam is visible in the inside corner detector's profile as a bump at $\sim$3 degrees, which is its expected location. (Ghost beam modeling is shown in Chapter~\ref{ghost_chapter}, and ghost beam map images are shown in Figure~\ref{ghost_amplitude_images}.) The inside and outside corner detector profiles show an unexplained bump at $\sim$4.5 degrees that is common to many detectors at different locations on the focal plane. The simulation did not include scattering from the cryostat vacuum window or any other optical elements, or multiple reflections inside the room. Both effects could create diffuse large angle beam coupling and could possibly explain the difference between the data and the simulation.

\clearpage

\section{Excess Optical Loading}

For a Spider 150 GHz detector, both calculations and measurements indicate that $80\%$ of the power pattern of the antenna makes it through the optical system onto the sky. The other $20\%$ is stopped down by the surfaces in the optical system, each of which is at finite temperature. The apertures around each of these surfaces will radiate onto the detectors causing an additional loading contribution above the power from the microwave sky and loading due to loss in each optical element. Here, we use the Zemax physical optics simulation of Spider's beam to calculate how much of the beam terminates on each surface in the optical system, and from that calculate the resulting excess optical loading.

The total power absorbed by the detector from a blackbody of temperature $T_{load}$ radiating onto the entire power pattern of a single-moded polarized $A \Omega = \lambda^{2}$ antenna is
\begin{equation}
P_{opt}(T_{load}) = \eta \int_{\nu_{min}}^{\nu_{max}} d\nu \frac{h \nu}{\exp{\left(\frac{h \nu}{k T_{load}}\right)} - 1},
\end{equation}
where $\nu_{min}$ and $\nu_{max}$ are the minimum and maximum photon frequencies of the detector passband, and $\eta$ is the optical efficiency \cite{krauss66}. For simplicity, this integral assumes a uniform frequency response spectrum of the detector in its passband. For this calculation, we assume perfect optical efficiency ($\eta$ = 1) of the detectors, and assume there is no loss in any of the filters, optics, HWP, or window. Based on lab measurements, the cold stop and the area outside both of the lenses are assumed to be 1.7 K, the 4 K filter is assumed to be 4 K, VCS1 is assumed to be 40 K, VCS2 is assumed to be 160 K, and room temperature surfaces are assumed to be 295 K.

The relative normalizations of the calculated Zemax field distributions are used to calculate the percentage of the power pattern that spilled over onto each aperture. When Zemax propagates the field distribution between two surfaces, if some power is stopped down by the second surface it zeros out those pixels. This reduces the calculated $P~=~\sum_{pixels}(E^{*} \times E)\Delta x \Delta y$ of that surface by the amount of power that spilled over onto that surface's aperture stop. This means that the fraction of power stopped down at a particular surface is $(P_{\mbox{\em last surface}} - P_{\mbox{\em current surface}})/P_{\mbox {\em initial}}$, where $P_{\mbox {\em initial}}$ is the power at the focal plane. To calculate the power that this particular surface, which is at a temperature $T_{\mbox{\em current surface}}$,  will radiate onto the detectors, the fractional spillover is multiplied by $P_{opt}(T_{\mbox{\em current surface}})$.

This calculation is accurate if the optical system consists of lenses, mirrors and blackened knife-edge aperture stops. Since most of the spillover happens inside the 2 K space which has a blackened cold optics sleeve, and the cold stop is a blackened knife edge, this is a good approximation for Spider. However, surfaces like the filters, HWP, and window are not blackened outside their clear aperture. In fact, they are shiny metal, probably reflective, which Zemax cannot treat at all in its physical optics propagation calculation. For the outside corner pixels, in the Zemax calculation there is a significant amount of spillover onto warmer non-blackened surfaces. It is difficult to quantitatively bound the effect this will have on the outside corner pixels in Spider, but it seems safe to assume that pixels closer to the center of the focal plane should not be affected.

\begin{figure}
\begin{center}
\subfigure{\includegraphics[width=0.77\textwidth]{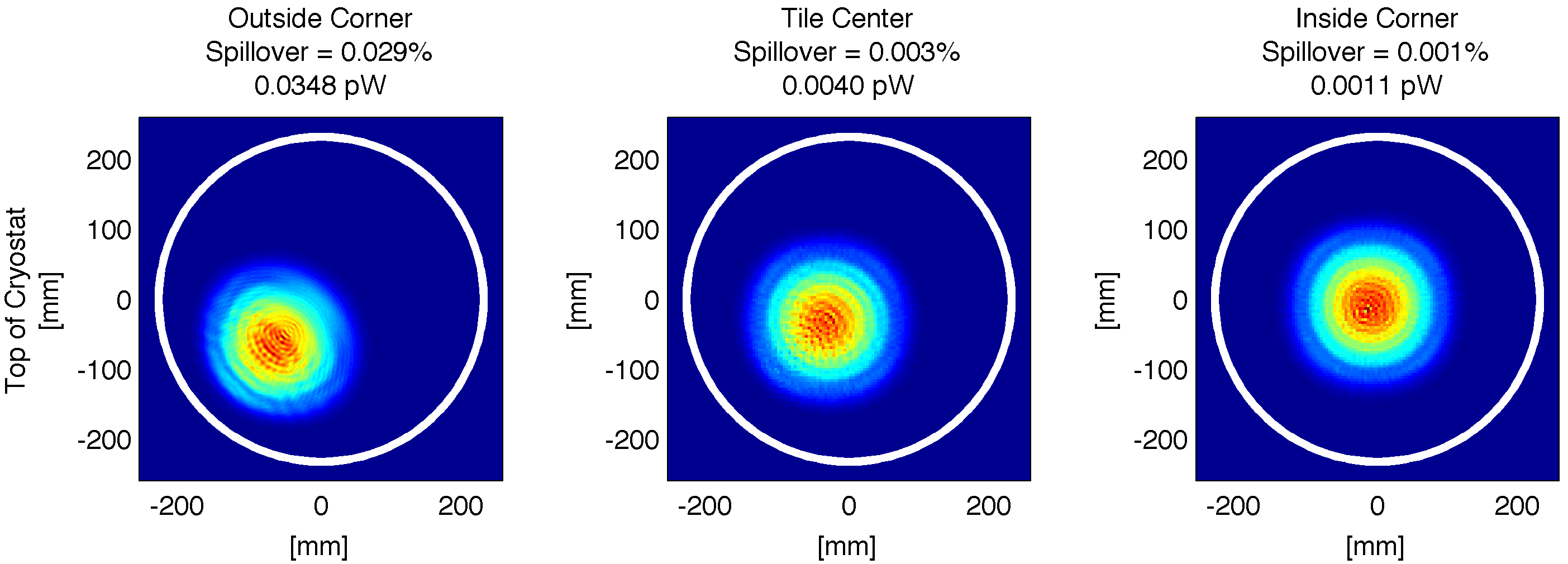}}
\subfigure{\includegraphics[width=0.77\textwidth]{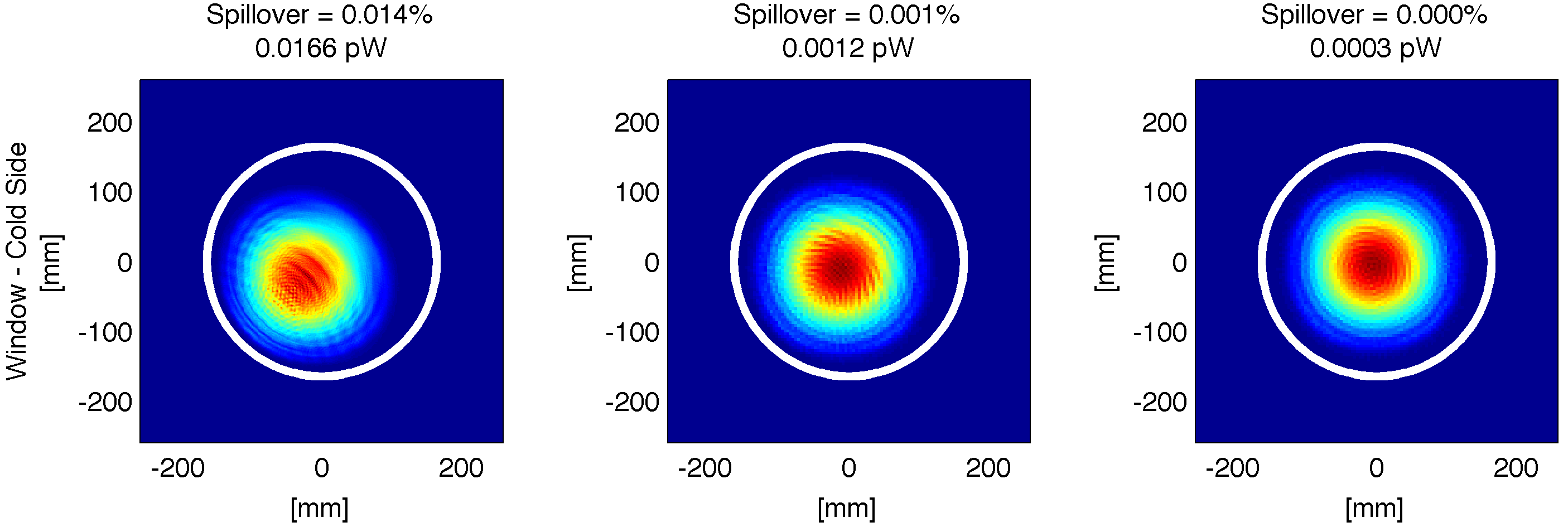}}
\subfigure{\includegraphics[width=0.77\textwidth]{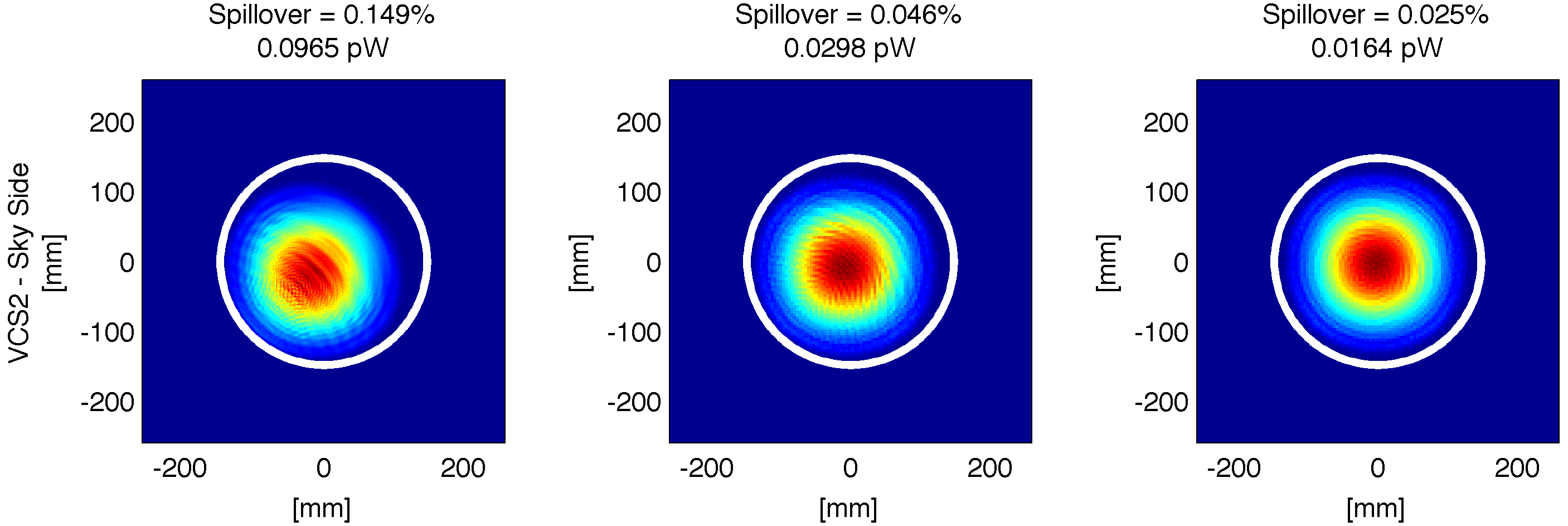}}
\subfigure{\includegraphics[width=0.77\textwidth]{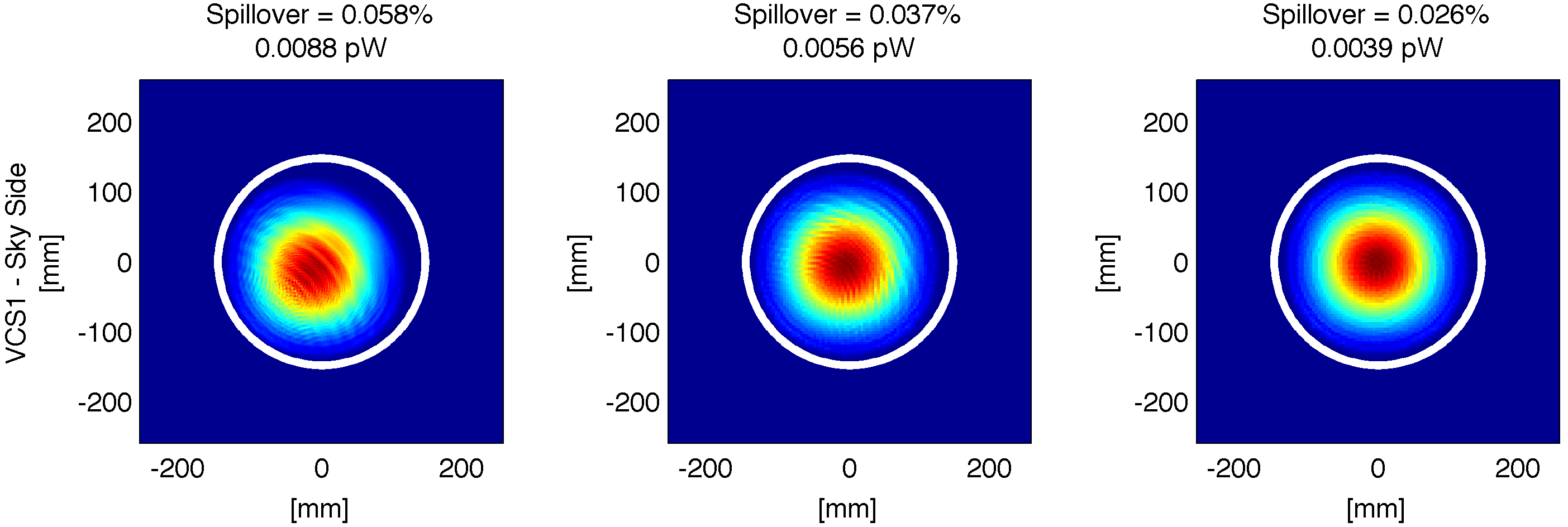}}
\subfigure{\includegraphics[width=0.77\textwidth]{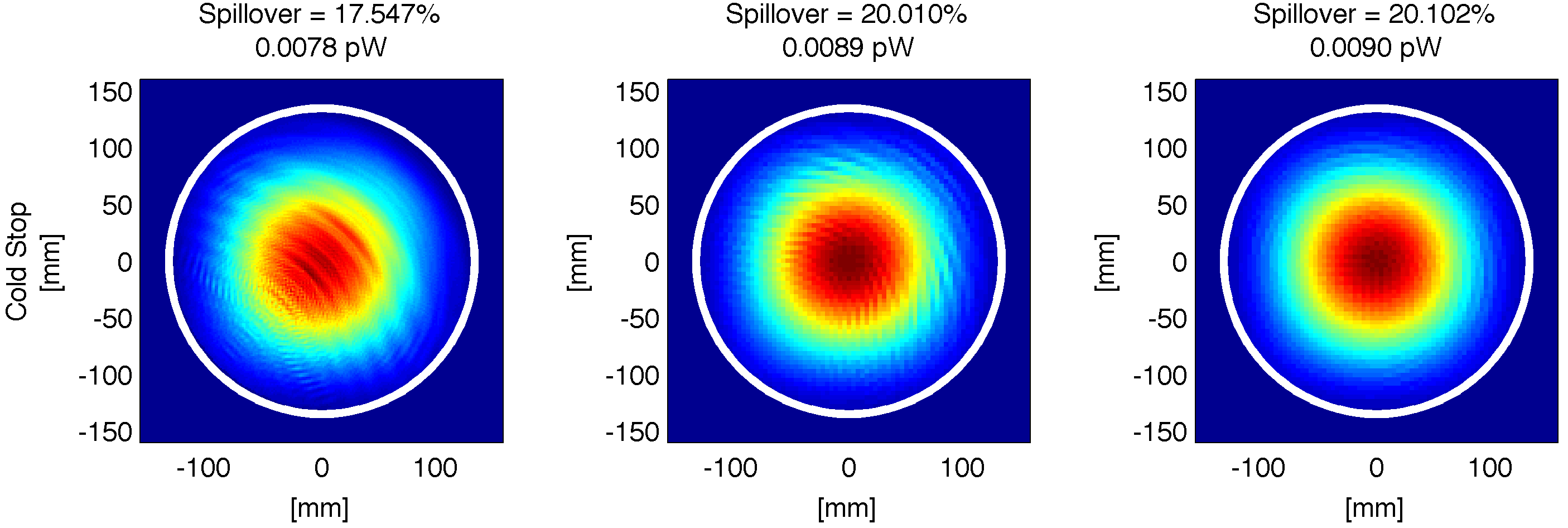}}
\caption[Power distributions and excess loading at selected surfaces for 150 GHz.]{Power distributions and excess loading at selected surfaces calculated for 150 GHz pixels at the outside corner (left plots), center, and inside corner (right plots) of a detector tile in the focal plane.  \label{loading}}
\end{center}
\end{figure}

\begin{figure}
\begin{center}
\subfigure{\includegraphics[width=0.77\textwidth]{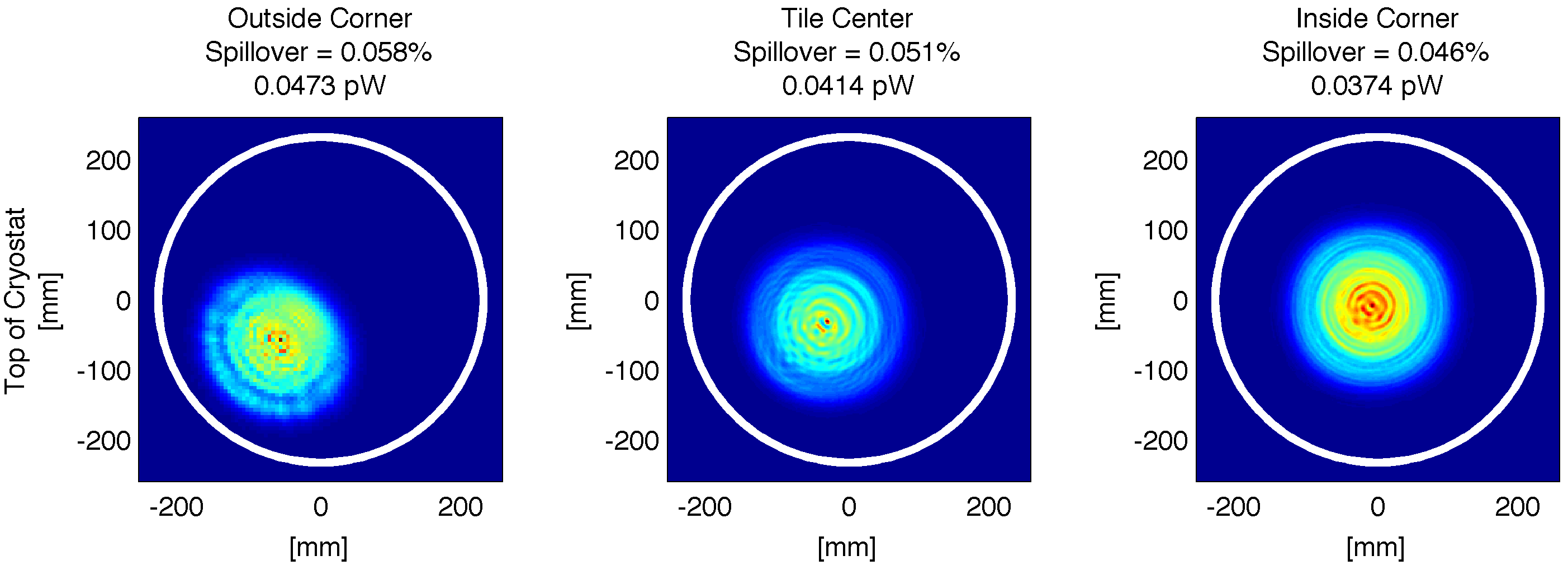}}
\subfigure{\includegraphics[width=0.77\textwidth]{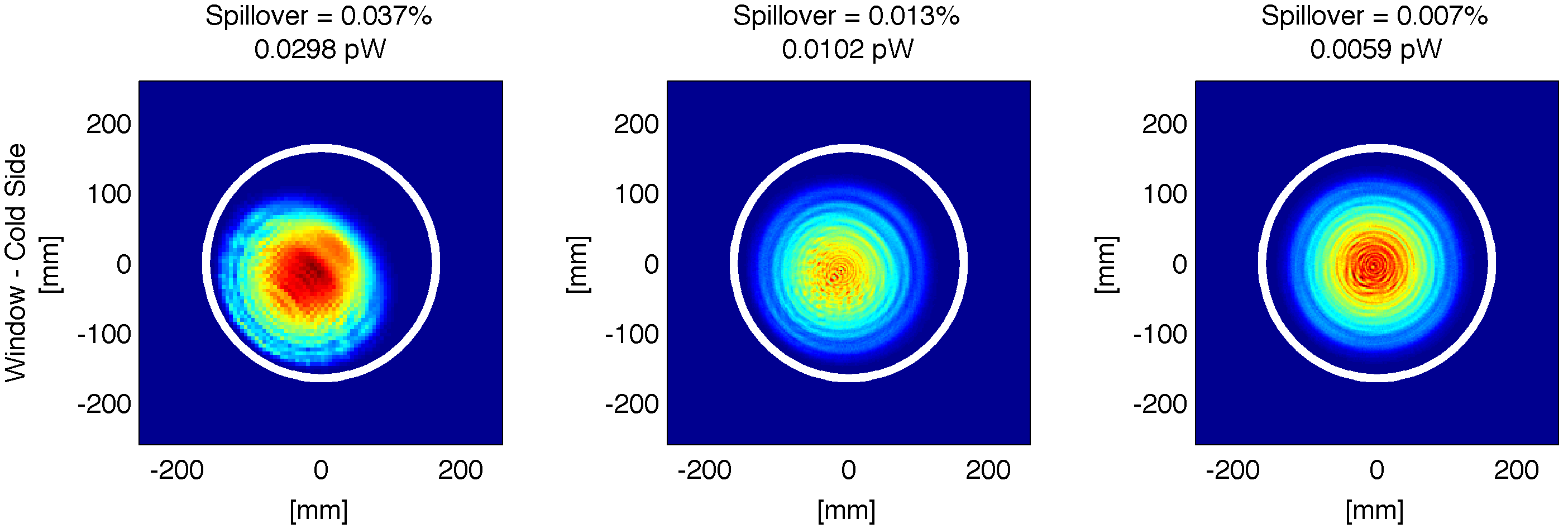}}
\subfigure{\includegraphics[width=0.77\textwidth]{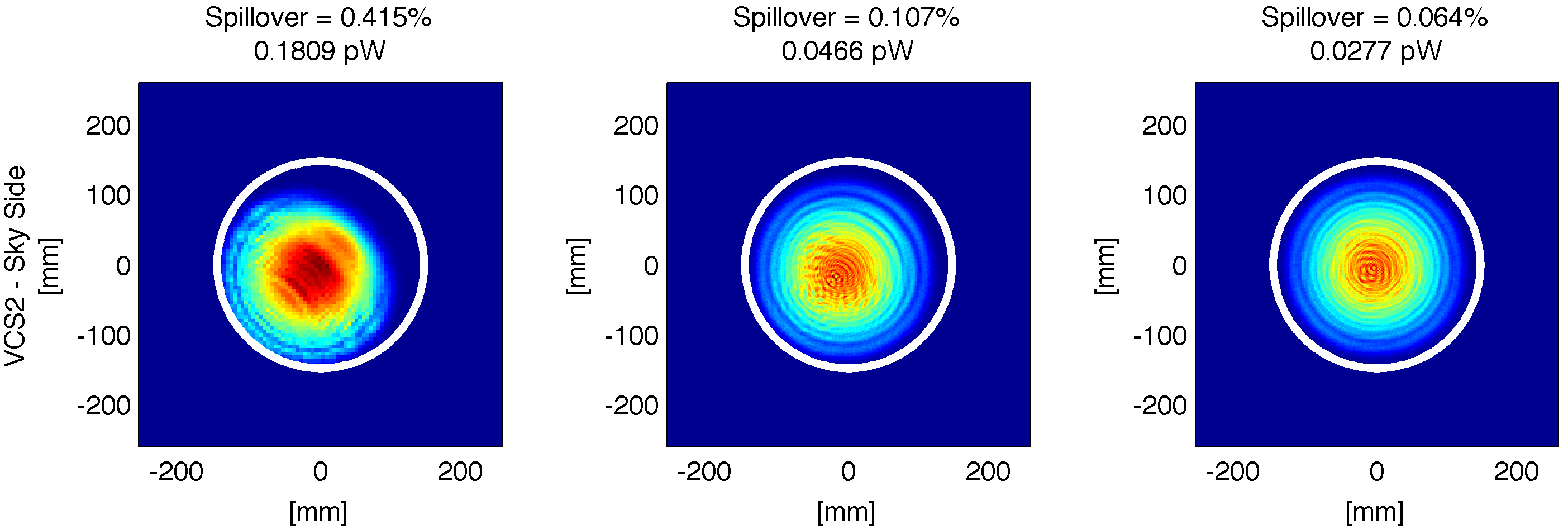}}
\subfigure{\includegraphics[width=0.77\textwidth]{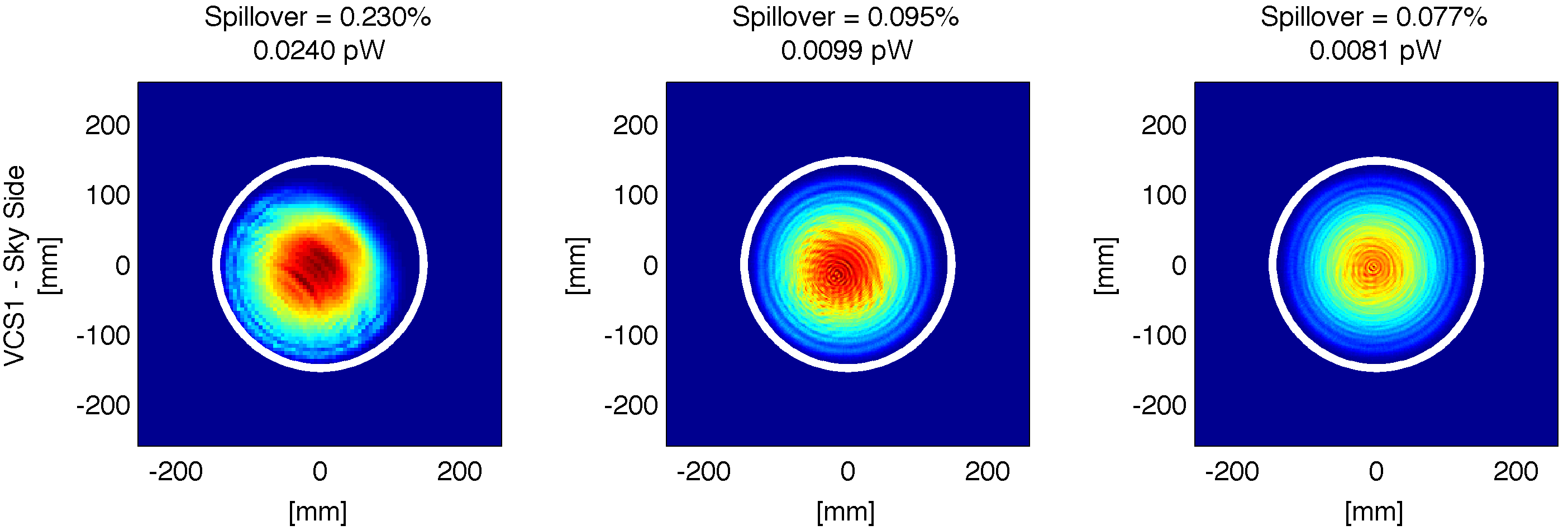}}
\subfigure{\includegraphics[width=0.77\textwidth]{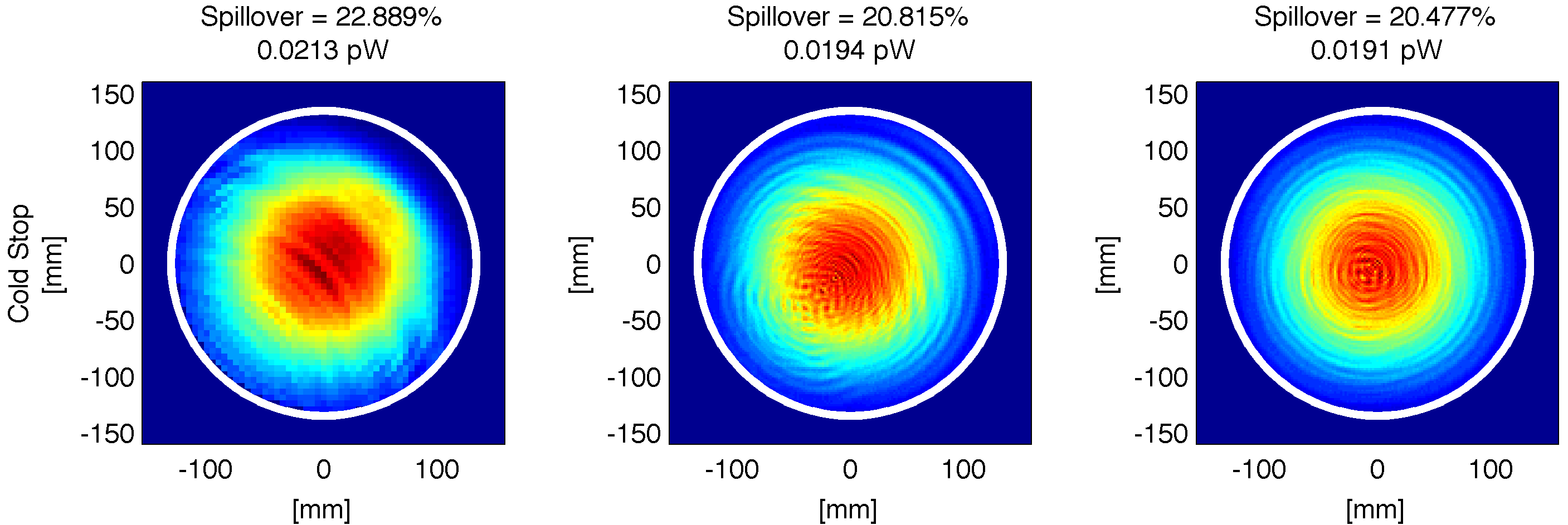}}
\caption[Power distributions and excess loading at selected surfaces for 95 GHz.]{Same as Figure~\ref{loading} but for 95 GHz.}
\end{center}
\end{figure}

As shown in Table~\ref{spillover_fraction_table}, adding up all the loading at 150 GHz from all of the surfaces yields 0.16 pW, 0.05 pW, and 0.03 pW for the outside corner, center, and inside corner pixels respectively. At 95 GHz the excess loading is calculated to be slightly higher, 0.31 pW, 0.13 pW, and 0.06 pW. This is dominated by the spillover onto VCS2. The higher spillover at 95 GHz is caused by the slightly faster feed of the 95 GHz pixels. No filter efficiencies or detector optical efficiency factors have been applied to this calculation, so the actual excess loading will be significantly lower.

\begin{table}
    \begin{tabular}{c|c|c|c|c|}
                     & \textbf{~~~Outside~~~}   & \textbf{~~~Tile~~~}   & \textbf{~~~Inside~~~}    &   \\ 
    \textit{Surface} & \textbf{~~~Corner~~~}   & \textbf{~~~Center~~~}   & \textbf{~~~Corner~~~}   & \textit{Temperature}    \\ \hline \hline
    Cold Stop        & 17.547\%                     & 20.010\%                     & 20.102\%                     & 1.7 K       \\ \hline
    4 K Filter       & \textcolor{white}{0}0.027\%  & \textcolor{white}{0}0.015\%  & \textcolor{white}{0}0.013\%  & \textcolor{white}{00}4 K         \\ \hline
    VCS1             & \textcolor{white}{0}0.058\%  & \textcolor{white}{0}0.037\%  & \textcolor{white}{0}0.026\%  & \textcolor{white}{0}40 K        \\ \hline
    VCS2             & \textcolor{white}{0}0.149\%  & \textcolor{white}{0}0.046\%  & \textcolor{white}{0}0.025\%  & 160 K       \\ \hline
    Window           & \textcolor{white}{0}0.014\%  & \textcolor{white}{0}0.001\%  & \textcolor{white}{0}0.000\%  & 295 K       \\ \hline
    Top of Cryostat  & \textcolor{white}{0}0.029\%  & \textcolor{white}{0}0.003\%  & \textcolor{white}{0}0.001\%  & 295 K       \\ \hline \hline
    \textit{Total}   & 17.824\%                     & 20.112\%                     & 20.167\%      & ~           \\
    \end{tabular}
  \\
  \\
  \\
  \\
    \begin{tabular}{c|c|c|c|c|}
                     & \textbf{~~~Outside~~~}   & \textbf{~~~Tile~~~}   & \textbf{~~~Inside~~~}       \\ 
    \textit{Surface} & \textbf{~~~Corner~~~}   & \textbf{~~~Center~~~}   & \textbf{~~~Corner~~~}       \\ \hline \hline
    Cold Stop        & 0.0078 pW                 &  0.0089 pW             & 0.0090 pW \\ \hline
    4 K Filter       & 0.0002 pW                 &  0.0001 pW             & 0.0001 pW \\ \hline
    VCS1             & 0.0088 pW                 &  0.0056 pW             & 0.0039 pW \\ \hline
    VCS2             & 0.0965 pW                 &  0.0298 pW             & 0.0164 pW \\ \hline
    Window           & 0.0166 pW                 &  0.0012 pW             & 0.0003 pW \\ \hline
    Top of Cryostat  & 0.0348 pW                 &  0.0040 pW             & 0.0011 pW \\ \hline \hline
    \textit{Total}   & 0.1647 pW                 &  0.0496 pW             & 0.0308 pW \\
    \end{tabular}
    \caption[Fractional spillover calculation for 150 GHz.]{Fractional spillover calculation for 150 GHz. The top table shows the calculated fraction of the beam that is stopped down at several critical surfaces. The bottom table shows the loading in pW resulting from the thermal emission of that surface. The total loading is shown on the bottom row. \label{spillover_fraction_table}}
\end{table}

\begin{table}
    \begin{tabular}{c|c|c|c|c|}
                     & \textbf{~~~Outside~~~}   & \textbf{~~~Tile~~~}   & \textbf{~~~Inside~~~}    &   \\ 
    \textit{Surface} & \textbf{~~~Corner~~~}   & \textbf{~~~Center~~~}   & \textbf{~~~Corner~~~}   & \textit{Temperature}    \\ \hline \hline
    Cold Stop        & 22.889\%                     & 20.815\%                     & 20.477\%                     & 1.7 K       \\ \hline
    4 K Filter       & \textcolor{white}{0}0.027\%  & \textcolor{white}{0}0.015\%  & \textcolor{white}{0}0.013\%  & \textcolor{white}{00}4 K         \\ \hline
    VCS1             & \textcolor{white}{0}0.230\%  & \textcolor{white}{0}0.095\%  & \textcolor{white}{0}0.077\%  & \textcolor{white}{0}40 K        \\ \hline
    VCS2             & \textcolor{white}{0}0.415\%  & \textcolor{white}{0}0.107\%  & \textcolor{white}{0}0.064\%  & 160 K       \\ \hline
    Window           & \textcolor{white}{0}0.037\%  & \textcolor{white}{0}0.013\%  & \textcolor{white}{0}0.007\%  & 295 K       \\ \hline
    Top of Cryostat  & \textcolor{white}{0}0.058\%  & \textcolor{white}{0}0.051\%  & \textcolor{white}{0}0.046\%  & 295 K       \\ \hline \hline
    \textit{Total}   & 23.656\%                     & 21.096\%                     & 20.684\%      & ~           \\
    \end{tabular}
  \\
  \\
  \\
  \\
    \begin{tabular}{c|c|c|c|c|}
                     & \textbf{~~~Outside~~~}   & \textbf{~~~Tile~~~}   & \textbf{~~~Inside~~~}       \\ 
    \textit{Surface} & \textbf{~~~Corner~~~}   & \textbf{~~~Center~~~}   & \textbf{~~~Corner~~~}       \\ \hline \hline
    Cold Stop        & 0.0213 pW                 & 0.0194 pW              & 0.0191 pW \\ \hline
    4 K Filter       & 0.0002 pW                 & 0.0006 pW              & 0.0006 pW \\ \hline
    VCS1             & 0.0240 pW                 & 0.0099 pW              & 0.0081 pW \\ \hline
    VCS2             & 0.1809 pW                 & 0.0466 pW              & 0.0277 pW \\ \hline
    Window           & 0.0298 pW                 & 0.0102 pW              & 0.0059 pW \\ \hline
    Top of Cryostat  & 0.0473 pW                 & 0.0414 pW              & 0.0374 pW \\ \hline \hline
    \textit{Total}   & 0.3053 pW                 & 0.1281 pW              & 0.0567 pW \\
    \end{tabular}
    \caption[Fractional spillover calculation for 95 GHz.]{Same as Table~\ref{spillover_fraction_table} but for 95 GHz.}
\end{table}

Before optical efficiency, an estimate of Spider's total in-band optical loading during flight is 1.8 pW. This comes from 0.3 pW from the CMB, 0.2 pW from atmospheric emission, 0.6 pW from thermal emission of the optical elements, 0.1 pW from the stray light baffle mounted outside the cryostat, and 0.6 pW from thermal emission of the vacuum window. In principle, this will be different for the 95 GHz and 150 GHz bands, but the uncertainty in the estimates is larger than any differences there. This estimate did not include the aperture stop effects calculated in this section, so that loading should be added on. This means the worst case total loading would be 1.8~pW~+~0.3~pW~=~2.1~pW, and would occur for the outside corner pixels in the 95 GHz telescopes. After detector optical efficiency, which is measured to be typically around $40\%$, this means that the total power dissipated on the TES during flight should be about 0.8 pW. For comparison, the measured saturation power of the TES bolometers ranges between 1.5 pW and 3.0 pW, which means we have a safety factor of roughly 2 to 4.

\section{Loading and Detector Noise}

Optical loading is the key factor that determines the detector noise in Spider. This occurs directly because of fluctuations in the thermal photons incident on the detector. Also, loading is the main driver in designing the thermal link in the bolometers, which determines the thermal fluctuation noise in the detector itself.

Thermal fluctuation noise in the detectors depends on the temperature of the detectors, and their thermal link to the refrigerator. The titanium superconducting TESs that we will use during flight have a superconducting transition temperature of approximately 500 mK. The base temperature of the refrigerator is approximately 250 mK. This means that if we want the TES saturation power to have a safety factor of two on the expected 0.8 pW of loading in flight, the thermal conductivity of the link between the TES and fridge should be roughly $G = (2\times0.8~\mathrm{pW}) / (500 - 250~\mathrm{mK}) = 6.4~\mathrm{pW}/\mathrm{K}$. As presented in \cite{lueker10}, the thermal fluctuation noise caused by this is
\begin{eqnarray}
NET_{TFN} &=& \frac{\sqrt{4kT_{detector}^2 G}}{\sqrt{2} \eta \int d\nu \frac{dB(\nu,T_{CMB})}{dT} \frac{c^2}{\nu^2} } \\
&=& 77~\mu\mathrm{K}_{CMB}\sqrt{\mathrm{s}}~\mathrm{at~95~GHz} \nonumber \\
&=& 71~\mu\mathrm{K}_{CMB}\sqrt{\mathrm{s}}~\mathrm{at~150~GHz}. \nonumber
\end{eqnarray}
Here $T_{detector}$ is the $500$ mK temperature of the detector biased into its superconducting transition, $B$ is the blackbody function, and $\eta$ is the optical efficiency of the detector, measured to be approximately $40\%$ in Spider. A correction factor relating to the non-equilibrium nature of the fluctuations has been omitted, which causes the calculation here to be a slightly overestimate of the noise. Minimizing this noise term is why the detectors were designed to have a saturation power only a factor of two to four above the expected loading.

Fluctuations in the thermal photons landing on the detector also causes noise. Also presented in \cite{lueker10}, the photon noise is
\begin{eqnarray}\label{tfn_approx}
NET_{photon} &=& \frac{\sqrt{\int d\nu P(\nu) h \nu}}{\sqrt{2} \eta \int d\nu \frac{dB(\nu,T_{CMB})}{dT} \frac{c^2}{\nu^2} } \\
&=& 77~\mu\mathrm{K}_{CMB}\sqrt{\mathrm{s}}~\mathrm{at~95~GHz} \nonumber \\
&=& 107~\mu\mathrm{K}_{CMB}\sqrt{\mathrm{s}}~\mathrm{at~150~GHz}, \nonumber
\end{eqnarray}
where $P(\nu)$ is the spectrum of light incident on the polarized detector. In reality, $P(\nu)$ is the sum of all of the blackbody spectra at all of the different temperature surfaces contributing to the total loading. For simplicity in Equation~\ref{tfn_approx}, this was approximated by the equivalent single blackbody spectrum that radiates 2.1 pW onto the polarized detector, which has a temperature of 8.14 K at 150 GHz, 9.70 K at 95 GHz, and a functional form of
\begin{equation}
P(\nu) = \frac{h \nu}{\exp{\left(\frac{h \nu}{k T_{load}}\right)} - 1}.
\end{equation}
Adding the thermal fluctuation and photon noise terms in quadrature yields a total noise of 121 $\mu\mathrm{K}_{CMB}\sqrt{\mathrm{s}}$ at 95 GHz and 128 $\mu\mathrm{K}_{CMB}\sqrt{\mathrm{s}}$ at 150 GHz. These estimates represent a best-possible noise level of bolometric detectors operating with Spider's photon loading, cryogenic system, and TES material choice. The measured noise performance in Spider, 134 $\mu\mathrm{K}_{CMB}\sqrt{\mathrm{s}}$ at 95 GHz, 130 $\mu\mathrm{K}_{CMB}\sqrt{\mathrm{s}}$ at 150 GHz, and shown in Table~\ref{bolo_count} is nearly this good.


\chapter{Advantages of a HWP}

Many CMB polarization
experiments including Spider, EBEX \cite{sagiv10},
POLARBEAR \cite{lee08}, ABS \cite{essinger-hileman09} and others
use a half-wave plate (HWP) to modulate the polarization state of
light from the sky before measurement by the detectors.
HWPs may be periodically stepped (eg between maps of the same part of the sky)
to reduce the effect of beam asymmetries and instrumental polarization of optical elements below the HWP. Alternatively, 
they can be continuously rotated to modulate the signal and to also reject atmospheric variations and 1/f noise~\cite{johnson06}.

\section{Polarimetry}

Spider employs pairs of orthogonally polarized detectors. In instrument coordinates, the sum of the two detectors of a pair measures the $I$ Stokes parameter, and the difference of the two measures $Q$. In order to estimate $U$ with the same pair of detectors, some kind of modulation is required. The sky rotates by several degrees for an Antarctic balloon payload with an observing strategy similar to Spider, so that provides some modulation. However, turning $Q$ sensitivity completely into $U$ sensitivity requires a rotation of $45^\circ$, so there is room to improve significantly by adopting another modulation scheme.

Rotating the instrument itself about its optical axis is one way to allow the same detector to measure both $Q$ and $U$. A disadvantage is that while the desired polarization sensitivity rotates with the instrument, so do many undesired systematic effects in the instrument. If the systematics are measured with sufficient precision, their contaminating effects could be carefully simulated and subtracted from the science data using a technique called deprojection \cite{aikin13}.

\begin{figure}
\begin{center}
\includegraphics[width=0.57\textwidth]{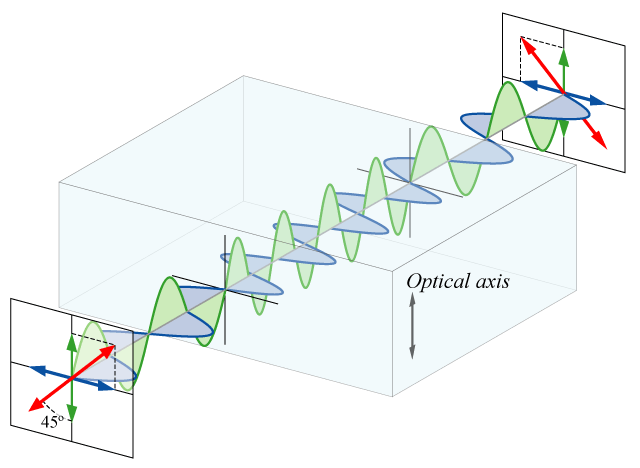}
\caption[Illustration of an ideal HWP rotating polarized light.]{Illustration of an ideal HWP rotating light polarized at an angle of $45^\circ$ relative to the crystal axes of the sapphire. In coordinates oriented along the crystal axes of the sapphire, the half-wave phase delay applied by the birefringent sapphire to the $x$ component of the electric field causes the total field to rotate by $90^{\circ}$ when the $x$ and $y$ components are recombined at the other side of the plate. In general, for light polarized at an angle $\theta$ relative to the crystal axes, an ideal HWP rotates the polarization by an angle $2\theta$. (Illustration from \cite{wikipedia}.) \label{hwp_cartoon}}
\end{center}
\end{figure}

For Spider, we chose to rotate the polarization sensitivity of the detectors using a rotatable HWP. The Spider HWP is constructed from a plate of birefringent sapphire. In a coordinate system oriented to align with the crystal axis of the sapphire, horizontally- and vertically-polarized light waves experience a different speed of light in the crystal structure of sapphire, which causes a phase delay to accumulate between the two states as they propagate. This has the effect of rotating linearly-polarized light as it travels through the plate. The amount of rotation depends on the relative orientation $\theta$ between the polarized light and the HWP. An ideal HWP rotates linear polarization by an angle of $2\theta$. Figure~\ref{hwp_cartoon} illustrates light polarized at an angle $\theta=45^\circ$ being rotated $2\theta=90^\circ$ by a HWP (i.e. $U$ polarized light being rotated into $-U$ polarized light). Since the HWP in Spider is mounted skyward of the primary optic, beam systematics originating in the telescope or detectors will remain unchanged as the HWP rotates to modulate the polarization sensitivity. This means that comparing maps made at different HWP angles will straightforwardly enable separating out the polarized signal from beam systematics.

\section{Beam Systematics}

\begin{figure}
\begin{center}
\includegraphics[width=0.95\textwidth]{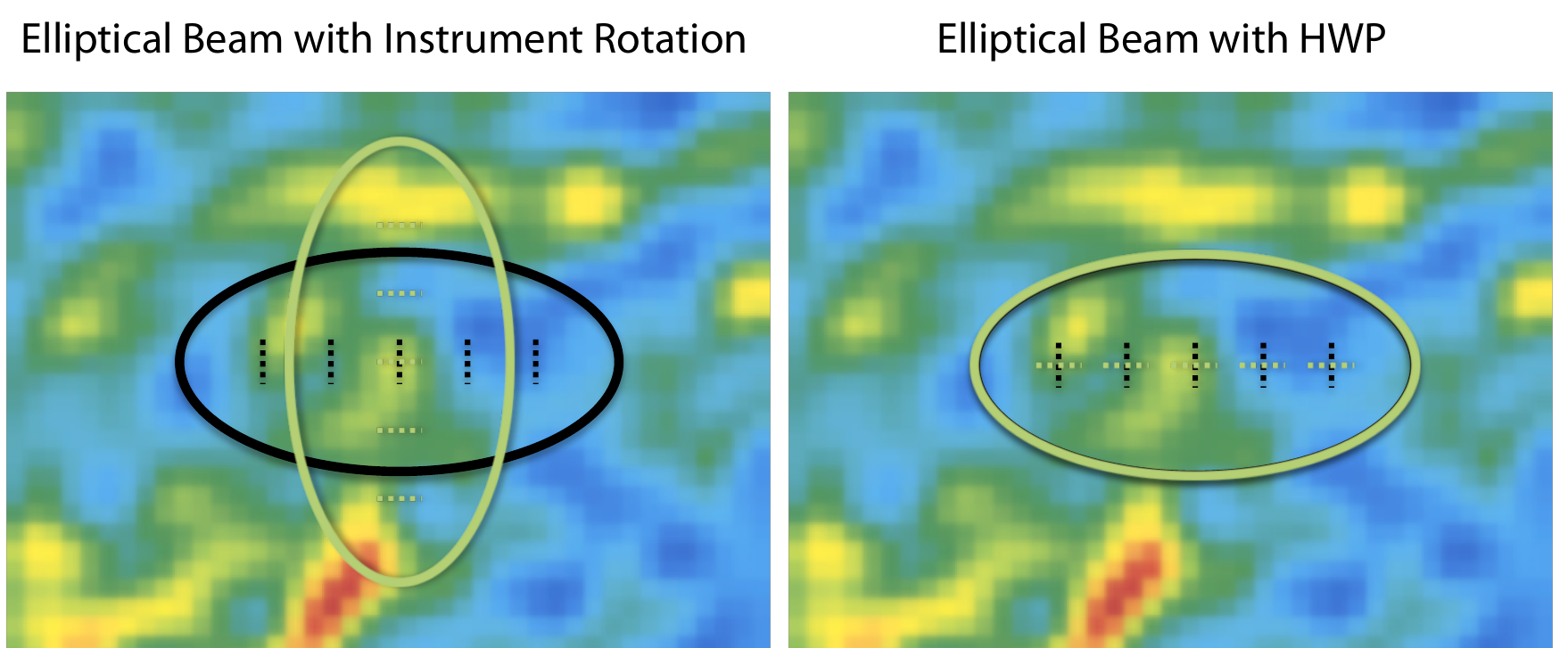}
\caption[Illustration of the effect of non-ideal beam shape on an instrument with and without a HWP. ]{Illustration of the effect of non-ideal beam shape on an instrument with and without a HWP. The left panel shows a light-green ellipse representing the beam of an instrument sensitive to horizontal polarization without a HWP. The black ellipse shows that when the instrument rotates to measure polarization, it also couples to intensity differently. The right panel shows that since a HWP rotates polarization sensitivity without rotating the elliptical beam, the impact of non-ideal beam shape is greatly reduced. \label{elliptical_beam_cartoon}}
\end{center}
\end{figure}

By design, the beam pattern of a single detector is intended to be circularly symmetric. However, in practice it may be slightly asymmetric, possibly having a slight elliptical shape. As illustrated in Figure~\ref{elliptical_beam_cartoon}, this causes a problem for polarimetry that relies on instrument rotation. Without a HWP, the instrument could observe with an instrument angle of $0$, where the detector measures $d_{0}\sim(I + Q)$, and later observe with an instrument angle of $90^{\circ}$, where the detector measures $d_{90}\sim(I - Q)$. These two measurements would then be subtracted to obtain an estimate of $Q$. However, the orientation of the elliptical beam would rotate with the instrument, so the $I$ on the sky that the beam encloses would be different at the two instrument angles. This would corrupt the estimate of $Q \sim d_{0} - d_{90}$ with a coupling to intensity fluctuations. Since the intensity fluctuations of the CMB are $\sim100$ times larger in amplitude than the polarization fluctuations, this places very strict performance requirements on the beam shape performance of CMB polarimeters without a HWP.

\begin{figure}
\begin{center}
\includegraphics[width=0.95\textwidth]{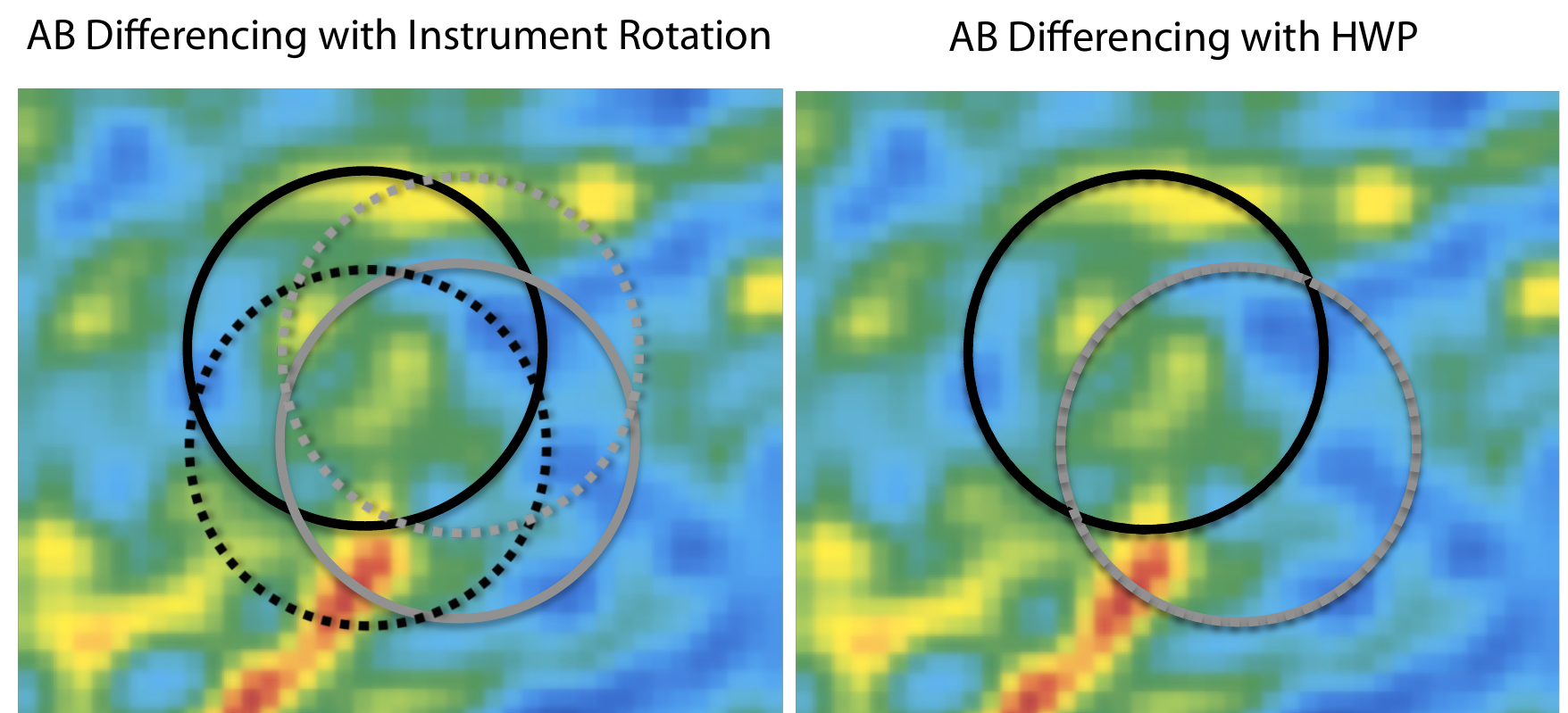}
\caption[Illustration of the effect of AB detector differencing systematics with and without a HWP.]{Illustration of the effect of AB detector differencing systematics with and without a HWP. The solid black and grey circles indicate the beams of an A and B polarized detector respectively. When the detectors are differenced to estimate $Q$, there will also be a coupling to intensity because the A and B beams are not located in exactly the same place on the sky. The left panel shows the beam locations with the instrument rotated to $0^\circ$ (solid lines) and $90^\circ$ (dotted lines). Both the polarization sensitivity and the coupling to intensity change with instrument rotation. However, the right panel shows the beam locations with the HWP at $0^\circ$ (solid lines) and $45^\circ$ (dotted lines, not visible because they are co-located with the solid lines). Since the HWP is able to change the polarization sensitivity without changing the intensity coupling, the intensity coupling caused by the fact that the A and B beams are not co-located can be disentangled from true polarization on the sky.\label{differencing_cartoon}}
\end{center}
\end{figure}

Spider has two polarization sensitive detectors at each point in the focal plane. The A detector responds to horizontal polarization, i.e. $(I + Q)$ and the B detector responds to vertical polarization, i.e. $(I - Q)$. One promising analysis strategy for Spider is to difference the A and B timestreams to obtain an estimate of $Q$. However, as illustrated in Figure~\ref{differencing_cartoon}, even if the beam patterns of both detectors are completely circularly symmetric, they may not be perfectly co-located on the sky due to slight differences between the antenna or feed properties of the two detectors. This means that each detector's beam would enclose a different patch of the intensity distribution on the sky, and this would corrupt the differenced timestreams with coupling to intensity. Moreover, if the instrument or the sky rotates the polarization sensitivity of the detector pair, the two beams would also move and the coupling to the intensity distribution would change. This would make it difficult to disentangle the contributions of polarization and intensity onto the differenced timestream.

The HWP cures these two classes of systematics, since it allows for polarization modulation without changing the coupling to intensity. As the HWP rotates by an angle $\theta_{hwp}$, it rotates the sensitivity of an individual or detector pair that measures $Q$ polarization by $2 \theta_{hwp}$. However, the beam pattern remains unchanged as the HWP rotates, which means that it is possible to isolate true polarization on the sky.


\chapter{HWP Fabrication}

The Spider HWPs are constructed from 330 mm diameter birefringent single-crystal sapphire plates grown by Crystal Systems in Salem, MA. The thickness of the plate needs to be chosen such that the phase delay accumulated between the two polarizations is exactly one half wavelength at the band center. In Chapter~\ref{index_measurements}, the two indices of refraction of sapphire at 5 K are measured to be $n_s$ = 3.336 and $n_f$ = 3.019. For the 150 GHz HWPs, we chose a sapphire thickness of 3.16 mm. At 5 K, that is a half-wave retarder at a frequency of $(1/2) \times (c/ (3.336-3.019)) / (3.16~\mathrm{mm}) = 149.6~\mathrm{GHz}$. For the 95 GHz HWPs, we chose a sapphire thickness of 4.93 mm, which is a half-wave retarder at 95.6 GHz.

The high index of refraction of the sapphire would cause large reflections at both surfaces of the plate. Calculated using the methods outlined in Section~\ref{hwp_mueller}, the total unpolarized reflection of the plate across the passband would be $42.6\%$ at 150 GHz and $42.4\%$ at 95 GHz. To reduce these large reflections, we apply a quarter-wave anti-reflection (AR) coating to both sides. A quarter-wave AR coating reduces total reflection because the waves reflected from the front and back of the coating layer have a relative path length difference of a half-wavelength, meaning the two waves destructively interfere. The cancellation will be perfect if the amplitudes of the two reflected waves are exactly matched. This occurs if the index of refraction of the material is chosen to be $\sqrt{n}$, where $n$ is the index of refraction of the material to which the coating is applied \cite{hecht74}. The optimal AR coating material for sapphire would therefore have an index of refraction in the range $\sqrt{3.019} = 1.738$ to $\sqrt{3.336} = 1.826$. We were not able to find a suitable material with exactly that index of refraction. With the materials we did choose, again calculating using the methods outlined in Section~\ref{hwp_mueller}, the AR-coated HWPs should have reflection of $2.3\%$ at 150 GHz and $3.2\%$ at 95 GHz. The measured performance in Chapter~\ref{texas_results} is consistent with low reflection.

For the 150 GHz HWPs we use Cirlex as the AR coat material, manufactured by the Fralock corporation. We chose a thickness of 0.254 mm (i.e. 0.010'') because that was a stock thickness. Lau et al. \cite{lau06} measured the optical properties of Cirlex. At room temperature, averaged in the mm-waves from 100 GHz to 400 GHz, they measured the index of refraction of Cirlex to be $n_{300~\mathrm{K}}^{\mathrm{mm-wave}} = \sqrt{3.37}$. At room temperature averaged over the far IR, 0.3-3.0 THz, they measured $n_{300~\mathrm{K}}^{\mathrm{far-IR}} = \sqrt{3.6}$. They then cooled the sample and measured its far IR optical properties again, obtaining a value of $n_{5~\mathrm{K}}^{\mathrm{far-IR}} = \sqrt{4.0}$. Assuming the index of refraction scales the same way in the mm-wave band as it does in the far-IR, an estimate of the mm-wave index of refraction of Cirlex at 5 K would be
\begin{equation}
n_{5~\mathrm{K}}^{\mathrm{mm-wave}} \approx n_{300~\mathrm{K}}^{\mathrm{mm-wave}} \times \frac{n_{5~\mathrm{K}}^{\mathrm{far-IR}}}{n_{300~\mathrm{K}}^{\mathrm{far-IR}}} = 1.935.
\end{equation}
With this index of refraction, our 0.254 mm sheets are quarter-wave coatings at 152.5 GHz.

As described in this Chapter, we bond the Cirlex to the sapphire with a 5.8 $\mu$m layer of HDPE as an adhesive layer. Lamb \cite{lamb96} quotes a measurement of the index of refraction of HDPE at 77 K from 26-40 GHz as $n=1.51$. At room temperature, Lamb \cite{lamb96} quotes measurements from 160-970 GHz showing that the index is almost totally independent of frequency, so using the 26-40 GHz cryogenic measurement should be appropriate even for the Spider 150 GHz band. When calculating the spectral properties of the 150 GHz HWPs in Chapter~\ref{hwp_mueller}, this layer is included in the model.

We AR coat 95 GHz HWPs with fused quartz wafers manufactured by Mark Optics. Lamb \cite{lamb96} quotes a measurement of the index of refraction of fused quartz at 245 GHz as $n=1.951$ at room temperature. Since quartz thermally contracts very little upon cooling (unlike sapphire and Cirlex), if fused quartz follows the Lorentz-Lorenz formula its index of refraction at liquid helium temperatures will be almost unchanged. We chose a fused quartz thickness of 0.427 mm, which is a quarter-wave coating at 90.0 GHz. 

The thicknesses of each material used in the HWPs is shown in Table~\ref{thickness_150}, and the 95 GHz thicknesses are shown in Table~\ref{thickness_95}. The manufacturing tolerance on all of these thicknesses is 10-20 microns. This matters more for the AR coatings than the sapphire, but it only results in $\sim1\%$ level variations in the calculated optical properties.

\begin{table}
\begin{center}
    \begin{tabular}{c|c|c|c|c|}
            & \textbf{4 K Index} & \textbf{Nominal} & \textbf{Thickness} & \textbf{Nominal} \\
            & \textbf{of Refraction} & \textbf{Thickness} & \textbf{Tolerance} & \textbf{Frequency} \\ \hline \hline

    Cirlex  & 1.935 & 0.254 mm          & $\pm$0.013 mm            & 152.5 GHz         \\ \hline
    HDPE    & 1.51 & 0.006 mm          & $\pm$0.001 mm            & N/A               \\ \hline
    Sapphire & 3.336,3.019 & 3.160 mm          & $\pm$0.020 mm            & 149.6 GHz         \\ \hline
    HDPE     & 1.51 & 0.006 mm          & $\pm$0.001 mm            & N/A               \\ \hline
    Cirlex   & 1.935 & 0.254 mm          & $\pm$0.013 mm            & 152.5 GHz         \\ \hline
    \end{tabular}
    \caption[Thicknesses of each optical layer in the 150 GHz HWPs.]{Thicknesses of each optical layer in the 150 GHz HWPs, and a measurement (sapphire) or best estimate from literature data (AR coat materials) of the index of refraction.
\label{thickness_150}}
\end{center}
\end{table}

\begin{table}
\begin{center}
    \begin{tabular}{c|c|c|c|c|}
    ~        & \textbf{4 K Index} & \textbf{Nominal} & \textbf{Thickness} & \textbf{Nominal } \\
    ~        & \textbf{of Refraction} &\textbf{Thickness} & \textbf{Tolerance} & \textbf{Frequency} \\ \hline \hline
    Quartz   & 1.951 & 0.427 mm          & $\pm$0.020 mm            & 90.0 GHz         \\ \hline
    Sapphire & 3.336,3.019 & 4.930 mm          & $\pm$0.020 mm            & 95.6 GHz         \\ \hline
    Quartz   & 1.951 & 0.427 mm          & $\pm$0.020 mm            & 90.0 GHz         \\ \hline
    \end{tabular}
    \caption[Thicknesses of each optical layer in the 95 GHz HWPs.]{Same as Table~\ref{thickness_150} but for the 95 GHz HWPs.
\label{thickness_95}}
\end{center}
\end{table}

\section{150 GHz HWPs}
\subsection{Bonding}

The Cirlex needs to be securely bonded to the sapphire in order to keep air gaps from forming, which would cause reflections that would severely degrade the performance of the AR coat. However, the bond needs to be strong to survive the stress of differential thermal contraction upon cooling to liquid helium temperature. In the limit of good gluing, the AR/sapphire/AR stack will shrink according to the thermal contraction of the sapphire, since it is so much thicker and stronger than the AR coats. The $\Delta L / L$ of sapphire from room to 4 K is 0.010''~/~13'' for the full diameter \cite{ekin07}. The Cirlex AR coats, if they were not bonded to anything, would naturally shrink by a different amount. We do not have a good literature source for the thermal contraction of Cirlex down to 4 K. We mounted a Cirlex sheet in an aluminum frame and cooled it to 77 K, and observed that the Cirlex shrinks less than the frame holding it. Here we estimate its thermal contraction as half that of aluminum. That means its $\Delta L$ is 0.030''. However, the sapphire will only shrink by 0.010'', leaving stress inside the Cirlex because of the extra 0.020'' of shrinkage it ``wants'' to undergo. Using the room temperature value Young's Modulus of Cirlex from the manufacturer's data sheet, we can use the definition of Young's Modulus,
\begin{equation}
(\mathrm{Young's~Modulus}) = \frac{(\mathrm{Internal~Stress~in~psi})}{\Delta L/L},
\end{equation}
to estimate how that 0.020'' translates into stress in psi. Since Cirlex has a room temperature Young's Modulus of $330\times10^{3}$ psi, the internal stress will be 507 psi. This will be a tensile stress, since the Cirlex is being prevented from shrinking. Imagine a small patch of the 0.010''-thick Cirlex, one inch square in area, near the outer edge of the stack. The 507 psi internal stress corresponds to 5 lbs of sideways force on that square inch near the edge. The bond will need to accommodate approximately that much force to prevent the stack from coming apart upon cooling.

We use a hot press process to bond the Cirlex to the sapphire. The concept is to create $\sim$micron and smaller sized pores on the surface of the sapphire and the Cirlex by roughening both surfaces. Then we place a thin layer of HDPE material between the sapphire and cirlex. When we apply high pressure and temperature, the heat melts the HDPE bond layer, and the high pressure forces the molten HDPE into all of the pores. As the stack cools back down, the HDPE solidifies inside the pores on both surfaces and adheres to them, creating a bond. The bond appears to be robust at room temperature, and survives several thermal cycles to cryogenic temperatures before beginning to fail. So, this appears to be a good 150 GHz HWP fabrication solution for Spider.

\subsection{Fabrication Process}

The first step in bonding is to prepare the surfaces by making them rougher. Sanding the sapphire by hand for several minutes with a 60 grit Norton/Saint-Gobain 66260306360 diamond sanding pad increases its surface roughness from 1.1~$\mu$m RMS to 1.5~$\mu$m RMS (measured with an optical profilometer at the MORE Center at CWRU), and also creates visible non-Gaussian scratches and pits. Sanding the Cirlex by hand for several minutes with 240 grit aluminum oxide sandpaper increases its surface roughness from 0.3~$\mu$m RMS to 2.1~$\mu$m RMS (measured with a stylus profilometer at the MORE Center at CWRU), and also creates even more visible scratches and pits on its surface. After the surfaces are roughened, they are rinsed with isopropyl alcohol, cleaned with trichloroethylene, then rinsed again with isopropyl alcohol to remove large dust particles from sanding as well as dissolve any oils or other impurities that may be on the surfaces.

The sapphire and Cirlex are then taken into a clean room to prepare them for bonding. Preparing the bond in a regular room environment leaves dust particles in the bond. This creates visible pockets of weak bonding that have been shown to grow in size upon repeated thermal cycling, eventually causing the entire bond to fail. The clean room facility we used does not have an established rating, but a particle counter regularly registers fewer than 10 particles per cubic foot of air. (This counter registers roughly half a million particle counts per cubic foot of lab or office air.) In the clean room, the sapphire is cleaned again with isopropyl alcohol and visually inspected and dusted with a Kimwipe until it is free of visible dust particles. Then, both sides of the HDPE bond material are cleaned with isopropyl alcohol and dusted with a Kimwipe until they are free of visible dust particles. The bond layer is a ULINE S-7317 trash liner, which is a 5.8 micron thick layer of HDPE. It is stretched tight over the sapphire until it is free of wrinkles using pieces of tape attached outside the bond area, then it is dusted with a Kimwipe. Finally, the Cirlex is recleaned with isopropyl alcohol and dusted, then placed on top of the sapphire-HDPE stack.

\begin{figure}
\begin{center}
\includegraphics[width=0.8\textwidth]{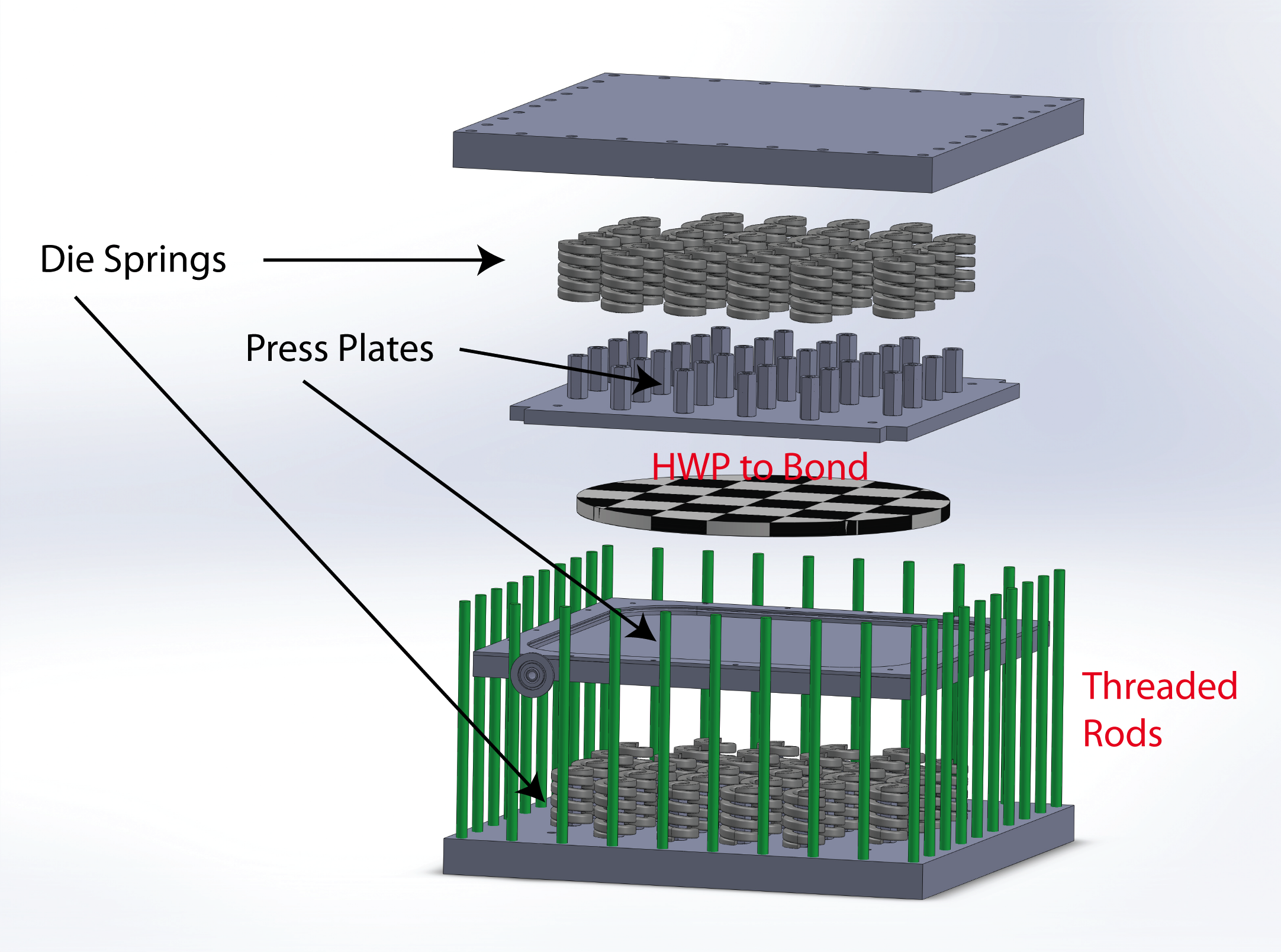}
\includegraphics[width=0.8\textwidth]{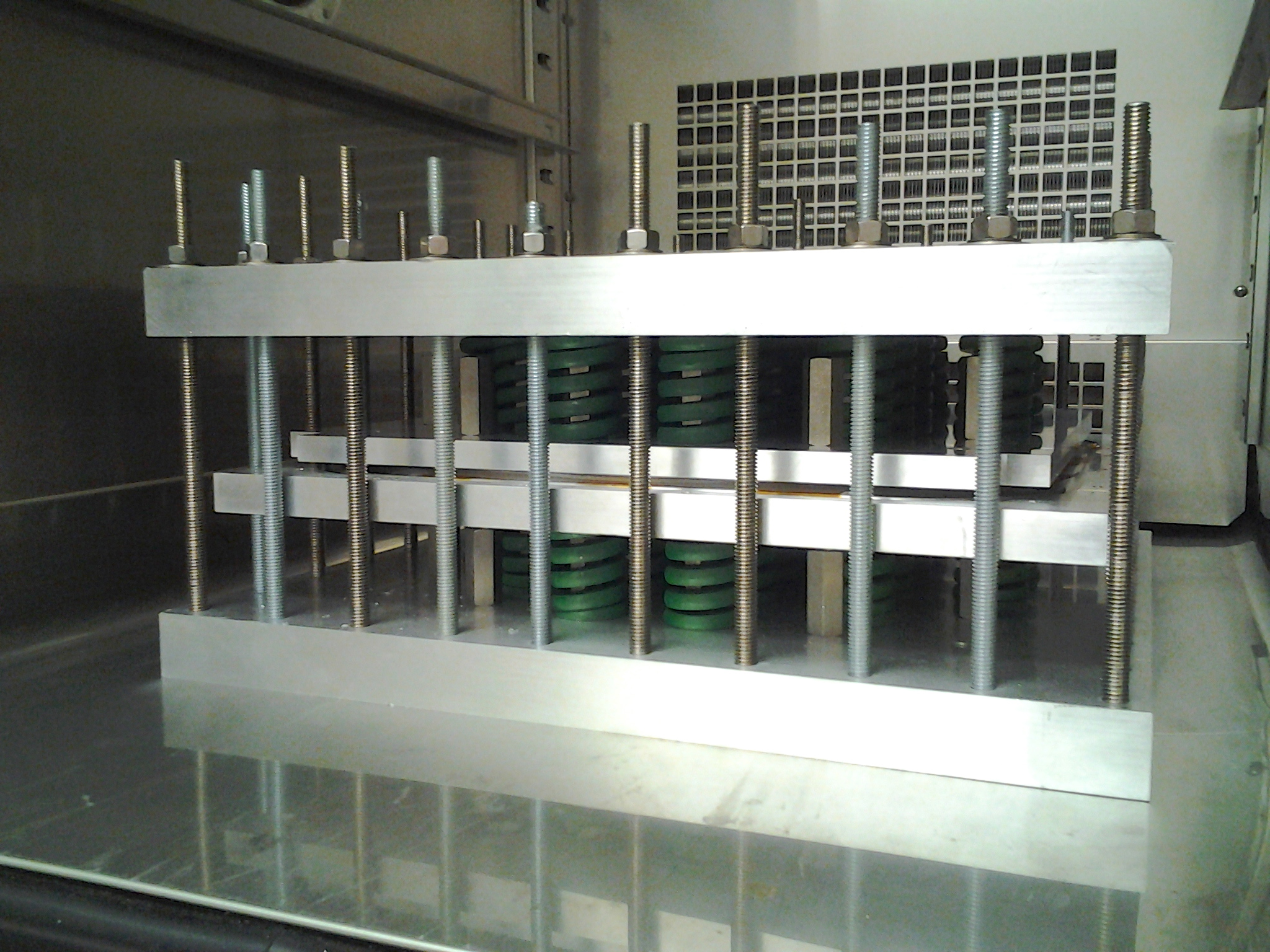}
\caption[Exploded view drawing of the HWP press, and a photo of the press in the oven before use.]{Exploded view drawing of the HWP press, and a photo of the press in the oven before use.
\label{spring_press_photo}}
\end{center}
\end{figure}

The optical stack is prepared inside the press that applies the pressure during the bond. The press is shown in Figure~\ref{spring_press_photo}. The optical stack is prepared with three layers of $1/16"$-thick synthetic felt, McMaster 8877K62, on each side to cushion it from the aluminum press plates. The press plates were machined to a planarity tolerance of $\pm0.001$ inches, and have a machined surface. The press plates are pushed together from outside with a total of 48 McMaster 9595K17 die springs. The springs in turn are pushed by outer plates, $17.50"~\times~17.50"\times~1.25"$ aluminum 6061-T6, that are pulled together with 36 3/8-16 threaded rods arranged on their perimeter. Nuts on the threaded rods push together two press plates when the nuts are tightened. A McMaster 5429T13 oilite thrust bearing is used under each nut, so the nuts turn easily with a wrench even under high pressure. This press design allows for a grid of springs to apply more even pressure. Also, the outer plates bow outward from the force of the springs, but the inner press plates remain planar to protect the optical stack.

In the press, each $2"$ long spring delivers a force of 1128 pounds when compressed by $0.300"$, according to the supplier. When bonding, we compress the springs by $0.160"$. Since the supplier says the spring is linear, this means that each spring exerts a force of 602 pounds, and all 48 exert a combined force of 28,896 pounds. All this force is spread out over both sides of the $13"$-diameter HWP stack, which is an area of 265.5 square inches. This means the bond takes place under a pressure of 108.8 psi, which is 7.4 atmospheres.

The outer plates each support all of the 28,896 pound load. This will bow the plates outward, and cause internal stress. The plates are made of aluminum 6061-T6, a strong alloy. From the Machinery's Handbook, a square plate supported at its edges without clamping with a uniform load spread across it (a rough approximation of this system) will bow outwards at its center by a distance
\begin{eqnarray}
d &=& \frac{0.0443 \times (\mathrm{total~load~in~pounds}) \times (\mathrm{length~of~plate})^2}{(\mathrm{material~elasticity~modulus~when~in~tension})\times  (\mathrm{plate~thickness})^3} \\
&=& \frac{0.443 \times (28896~\mathrm{lbs}) \times (17.5~\mathrm{in})^2}{(29 \times 10^6~\mathrm{psi}) \times (1.25~\mathrm{in})^3} \nonumber \\
&=& 7.0~\mathrm{thou}. \nonumber
\end{eqnarray}
This means that the center springs will be compressed $7.0/160 = 4.4\%$ less than the springs near the edge where the outer plate is supported, which is a load non-uniformity that the inner press plates should be able to handle without deforming significantly. The outer plates will also experience internal stress from the load. Again according to the Machinery's Handbook, the highest stress anywhere across the outer plate will be
\begin{eqnarray}
S &=& \frac{0.28 \times (\mathrm{total~load~in~pounds})}{(\mathrm{plate~thickness})^2} \\
&=& 5,174~\mathrm{psi}. \nonumber
\end{eqnarray}
Since the yield stress of Aluminum 6061-T6 is 35,000 psi, this is a safe design and the press should be reusable because it is always operated well below the yield stress of the material.

While under pressure, the press and HWP stack are baked in an ESPEC ESL-3CA oven. This oven circulates temperature-controlled air, settable from -$35^\circ$ C to $150^\circ$ C with an error of $\pm 0.1^\circ$ C. This applies heat to both press plates on both sides of the HWP. We start the bake cycle by slowly ramping up to $140^\circ$ C over 6 hours to avoid thermal shock. This is hot enough to easily melt the HDPE bond layer material. The oven spends 12 hours at temperature to ensure the melted HDPE flows into both surfaces enough to form a good bond. Then, the oven spends 6 hours ramping back down to room temperature. The temperature of the aluminum press plates tracks the air temperature with very little lag.

One AR coating is bonded onto the sapphire at a time. This means the coating applied first is actually rebaked while the second coating is being bonded. This has not been shown to cause any problems, and in fact we have shown that a bond with air pockets can be repaired by simply putting the HWP back in the press, applying pressure, and baking the entire stack again.

\begin{figure}
\begin{center}
\includegraphics[width=0.48\textwidth]{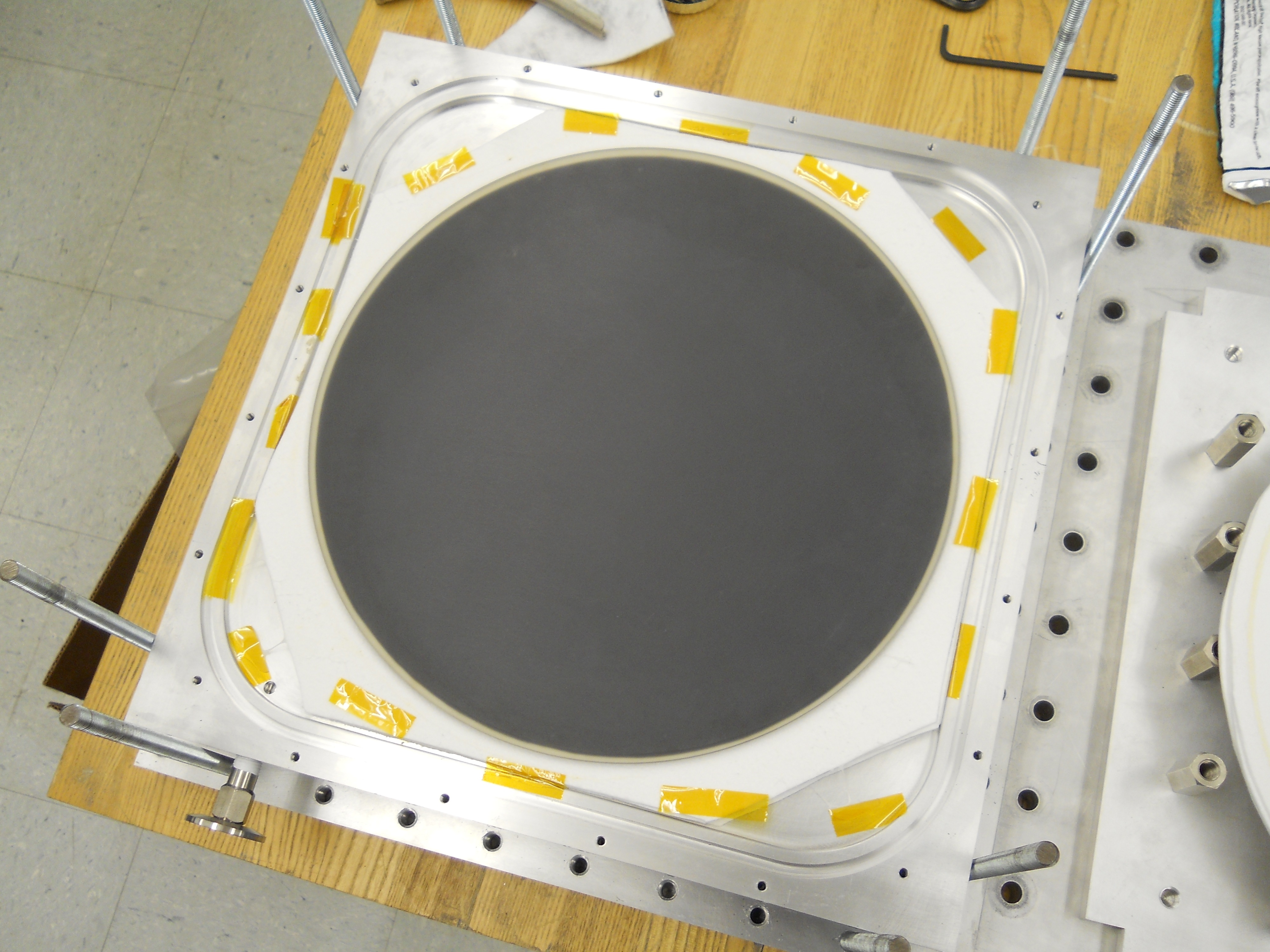}
\includegraphics[width=0.48\textwidth]{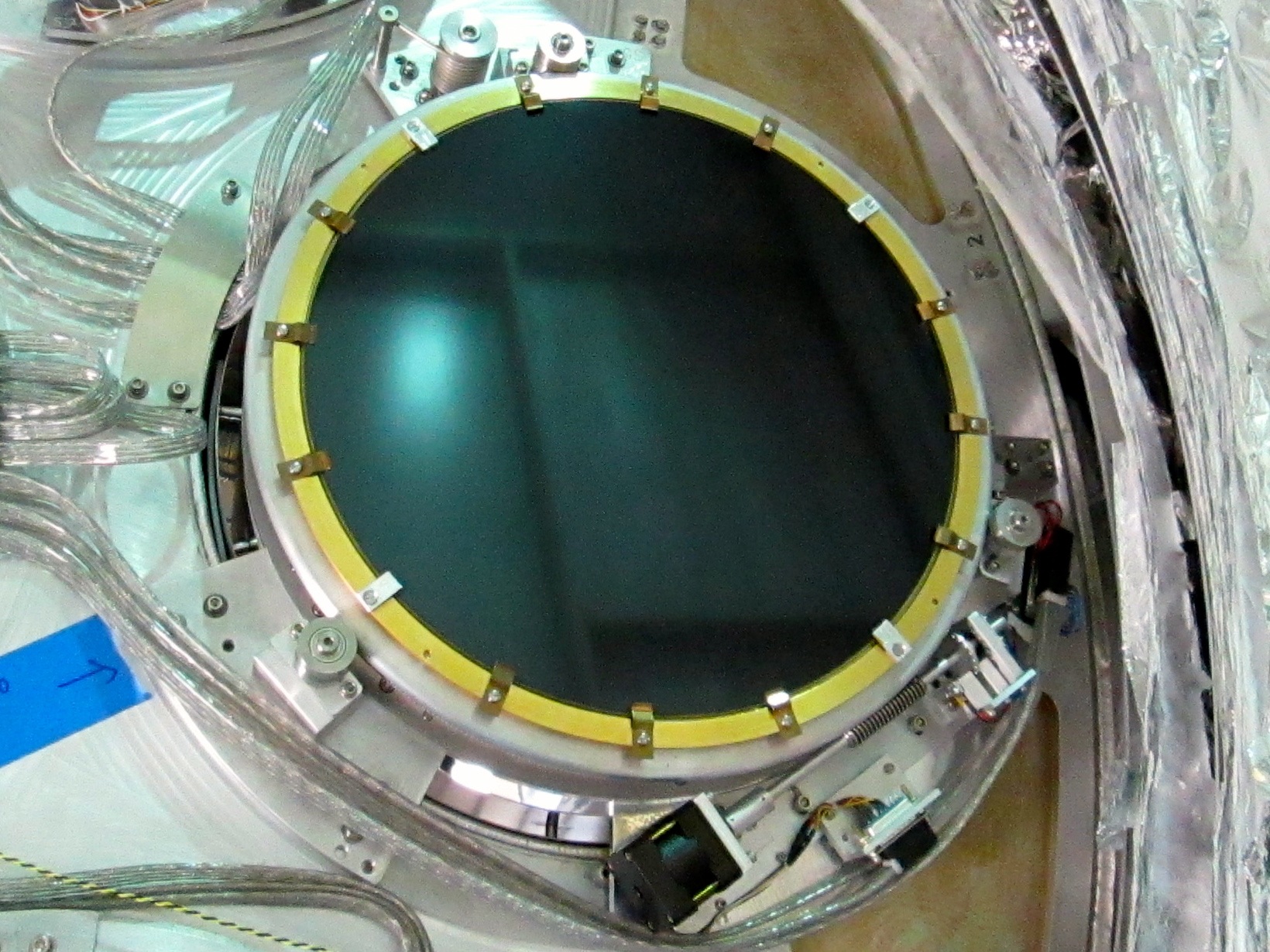}
\caption[Photographs of the Cirlex AR coat bonding results.]{Photographs of the Cirlex AR coat bonding results. The left panel shows a sapphire with a Cirlex AR coat bonded onto the side away from the camera. This allows the bond quality to be viewed through the sapphire, and there are no visible features in this bond. The right panel shows a completed 150 GHz HWP installed in the flight cryostat. \label{cirlex_photos}}
\end{center}
\end{figure}

\section{95 GHz HWPs}

When using the $\sim 50\%$ thicker sapphire and Cirlex necessary for the 95 GHz HWP optical stack, we found that the bond process described above for the 150 GHz HWPs did not survive even a single thermal cycle to liquid nitrogen temperatures. Instead, we anti-reflection coated the 95 GHz HWPs using a wafer of fused quartz fabricated by Mark Optics. The wafer is bonded to the sapphire at the center to prevent too large of an air gap from forming between the quartz and the sapphire. A gap larger than 50 $\mu$m would significantly impair the performance of the AR coating and cause large reflections. Without bonding, it is easy for such a gap to form because the fused quartz wafers are heavy enough and thin enough to sag under their own weight. Also, the inward radial force on the wafer from differential thermal contraction between the HWP mount and the fused quartz wafer may be enough to cause the wafer to bow outward away from the sapphire. Finally, the quartz wafers are not perfectly planar and have waves across their surface tens of $\mu$m in amplitude and $\sim$inches in horizontal size. Unfortunately, the fused quartz wafers cannot be bonded to the sapphire across their entire surface, because they are too fragile to survive the stress of differential thermal contraction. However, a small bond at the center of the wafer provides some control over the air gaps. Keeping the bond diameter small in practice keeps the stress from differential thermal contraction between the sapphire and quartz from causing the bond to fail.

A roughly one-inch diameter patch near the center of the sapphire and the fused quartz wafer was roughened. Sandpaper was used for the quartz, and a diamond sanding pad for the sapphire. The surfaces were cleaned with trichloroethylene then rinsed with isopropyl alcohol. A thin layer of Lord AP-134 adhesion promoter was applied to both surfaces and allowed to dry for one hour. While it was drying, the parts were moved into the clean room. Then, a small drop of Eccobond 24 adhesive prepared with its Part B catalyst was applied to the sapphire, and the quartz wafer was then placed onto the sapphire. The weight of a lead brick was used to clamp the bond together while it dried for 24 hours. The final bond is estimated to be much thinner than 25 $\mu$m, which is calculated to have a minimal impact on the AR coat performance.

\section{Mounting the HWP}

\begin{figure}
\begin{center}
\includegraphics[width=0.95\textwidth]{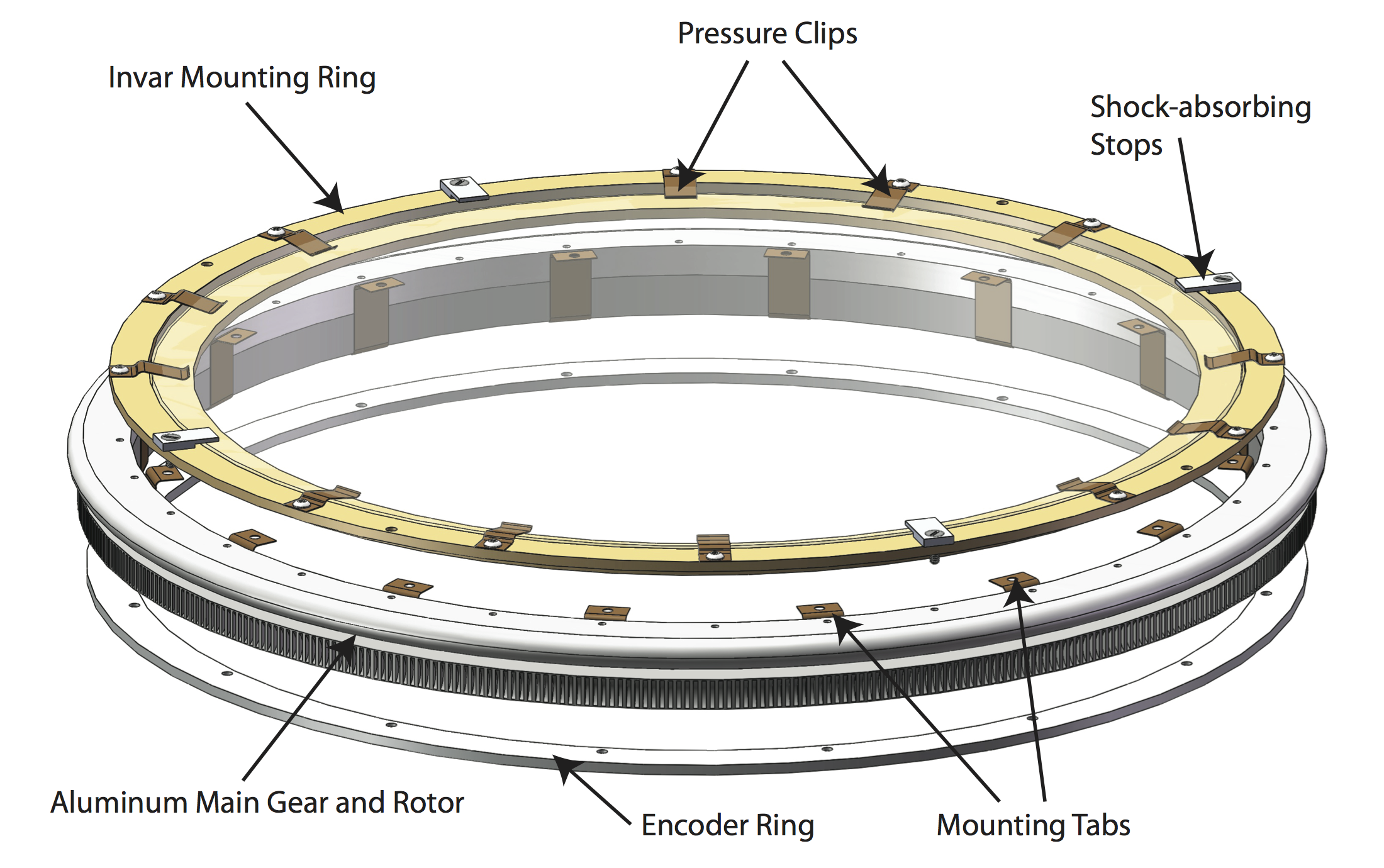}
\caption[Exploded view of an Invar HWP mount and aluminum main gear.]{Exploded view of an Invar HWP mount and aluminum main gear. The HWP stack is held down by 12 pressure clips and 4 shock-absorbing stops to ensure that the HWP does not come out of the mount in the event of a mechanical shock. 16 flexible mounting tabs allow for differential thermal radial contraction between the aluminum main gear and the Invar HWP mount. \label{exploded_hwp_mount}}
\end{center}
\end{figure}

The HWP optical stacks are mounted as shown in Figure~\ref{exploded_hwp_mount}. The mount needs to hold the HWP securely to prevent movement during shocks that could be experienced during shipping and launch. To achieve this, we use 12 phosphor bronze clips to apply pressure to hold the HWP in an Invar mounting plate with a clear diameter of 12''. To prevent the Invar mounting plates from corroding, we had them plated with 25 micro-inches of gold by GCG Corporation in Glendale, CA. The clips are made of 0.016''-thick shim stock and are 0.375'' wide. The part of the clip that applies pressure is approximately 0.550'' long, which means that at 0.050'' deflection all 12 clips are calculated to apply 25 lbs of force. This is consistent with lab measurements, and is enough to hold the 2.5 lb 150 GHz HWP even in the event of a 10g shock. The fused quartz wafers that AR coat the 95 GHz HWPs are more fragile, so we only apply 15 lbs of force using 0.030'' of clip deflection. This will protect against 4g shock with the 3.7 lb 95 GHz HWP. As added protection against shock, we added 4 limiting stops. These thick aluminum tabs are wrapped with enough Kapton tape to almost touch the HWP. The purpose of the limiting stops is to provide a cushion of Kapton tape that will stop the HWP if a hard upward shock is applied to it, instead of allowing that shock to stress the phosphor bronze mounting clips and deform them.

The Invar mounting plate has low thermal contraction, and is expected to shrink by approximately 0.006'' in diameter upon cooling to 4 K  \cite{ekin07}. Since it is the thickest and stiffest part of the optical stack, the sapphire will define the overall thermal contraction of the HWP. The sapphire is expected to shrink by approximately 0.010'', which is fairly well-matched to the Invar holder.

The Invar-mounted HWP is attached to an aluminum main gear and bearing with 16 flexible phosphor bronze mounting tabs. The main gear was fabricated by GoTo \& Tracking Systems in Arlington, TX. The aluminum main gear is expected to shrink by up to 0.060'' \cite{ekin07}, so the flexible mounting tabs are necessary to take up the differential thermal contraction between the main gear and the Invar mounting ring. The tabs are made from 0.008''-thick phosphor bronze shim stock, and the flexible part is $1/2$'' wide and $1$'' tall. The tabs are rigid enough in the lateral direction to mount the HWP securely, but flexible enough radially to repeatedly comply with the differential thermal contractions.


\chapter{Rotation Mechanism}

\section{Design Goals}

Polarization modulation with the HWP is accomplished by rotating it relative to the detector polarization sensitivity angle. Each rotation mechanism needs to be able to rotate its HWP to any desired angle with an absolute accuracy of $\pm 0.1$ degrees to keep HWP angle error from dominating the $\pm1^\circ$ polarization angle error budget of the experiment \cite{fraisse13}. They also need to hold the HWPs at 4 K in the flight cryostat to cool them. Cooling reduces the loss of the sapphire, quartz, and Cirlex and therefore reduces their thermal emission onto the detectors. Mounting the HWP inside the cryostat also makes reflections from the HWP terminate on cold surfaces instead of on a warm surface outside the cryostat. 

Since Spider will use the HWPs in a step-and-integrate mode, the mechanisms must prevent the HWPs from rotating while the instrument is observing. To facilitate lab characterization and reduce downtime during flight, the mechanisms must turn smoothly at a minimum of one degree per second while cold. The mechanisms must generate only a small amount of heat to conserve liquid helium in the main tank during flight.

\begin{figure}
\begin{center}
\includegraphics[width=1.0\textwidth]{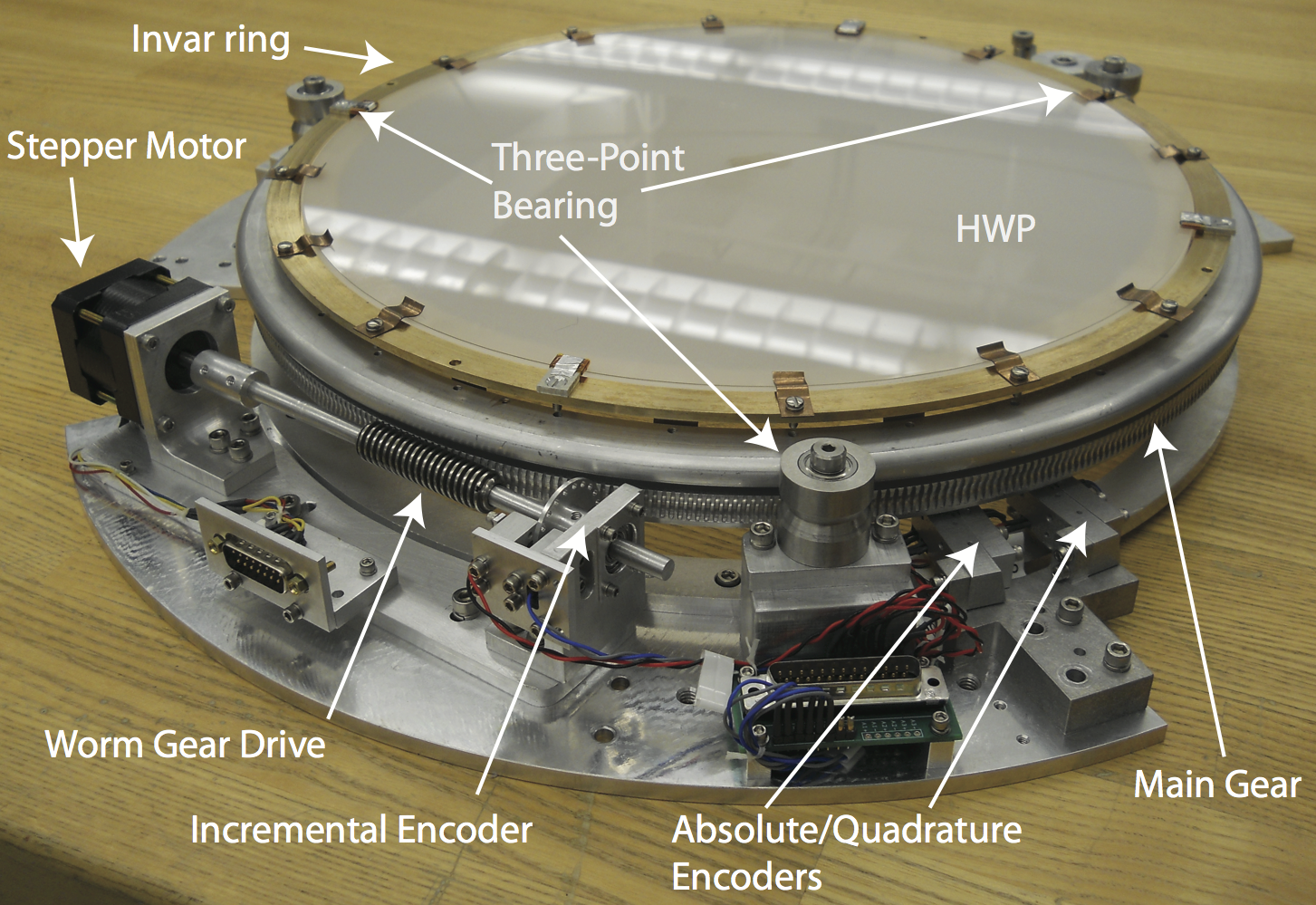}
\caption[Overview of one of the HWP rotation mechanisms with a 95 GHz HWP installed.]{Overview of one of the HWP rotation mechanisms with a 95 GHz HWP installed. The HWP optical stack is held in a 12'' clear diameter Invar mounting ring attached to the main gear. The rotor turns on a three-point bearing and is driven by a cryogenic stepper motor turning a worm gear. Optical encoders verify that the rotor is at the desired angle. \label{hwp_mechanism_overview}}
\end{center}
\end{figure}

To meet these design goals, the HWP rotation mechanism consists of a rotor mounted in a three-point mechanical bearing rotated by a worm gear connected to a stepper motor. The mechanical bearing provides a low-friction platform for our step-and-integrate observing strategy, and the worm gear prevents the bearing from moving more than $0.1^{\circ}$ even when the stepper motor is not energized. We custom-built cryogenic optical encoders to monitor the bearing angle. A photograph of a rotation mechanism is shown in Figure~\ref{hwp_mechanism_overview}.

Our motor solution differs from the methods used by some other experiments in the field. The torque for the HWP rotation mechanisms in Maxipol \cite{johnson06} , EBEX \cite{reichborn10},  BLAST-Pol \cite{fissel10}, and PILOT \cite{salatino10} is provided via a rotating shaft fed through the vacuum wall of the cryostat. Since Spider is an array of six telescopes in a single cryostat, having six independent rotating shaft feed-throughs is undesirable. We therefore chose to use cryogenic stepper motors. This means that the only connections outside the cryostat for a single mechanism are four high-current ($\sim1$ A) wires, and 14 low current angle encoder readout wires. 

To measure the rotation angle of our HWP bearing, Spider uses optical encoders mounted inside the cryostat. This also differs from the approach taken in other instruments. The HWP encoder in BLAST-Pol consists of a leaf spring making contact with a potentiometer located near the bearing. Maxipol, EBEX, and PILOT all have optical encoders. Maxipol has a commercial absolute optical encoder outside the cryostat. EBEX has a relative optical encoder on the main bearing. PILOT has a 3 bit absolute encoder on its main bearing which allows the bearing to step between 8 discrete angles. In contrast with Spider and EBEX which use optical encoders located completely inside the cryostat, PILOT has optical fibers running to and from outside the cryostat. This allows for room-temperature light sources and detectors but requires fiber optic vacuum feed throughs at the cryostat wall. 

\section{Bearings}

To hold the HWP in place yet allow it to rotate smoothly, we use three V-groove guide bearings evenly spaced around the circumference of the rotor. This allows for a large clear aperture, but keeps the individual bearings small to minimize the effect of thermal contraction on the moving bearing parts. Each of the three guide bearings consists of a stainless steel V-groove turning on two VXB SR4Z-Kit8526 ball bearings inside. We chose the lowest performance grade of ball bearing, ABEC-1, because it has the largest mechanical clearances available, which is an advantage when faced with thermal contraction. To take up thermal contraction as the aluminum main gear shrinks, we spring-loaded one of the guide bearings with a McMaster-Carr 9287K136 torsion spring to push it radially inwards as the mechanism cools. This spring is rated to deliver 40 in lbs of torque at $180^{\circ}$ of rotation. Since it is only deflected by $90^{\circ}$ and applies its torque at a radius of 2 inches, it maintains a nearly constant radial force of approximately 10 lbs.

The stainless steel ball bearings come from the manufacturer with a light oil coating for lubrication and to prevent corrosion, but the oil coating will cause the bearings to seize at cryogenic temperatures. To remove the oil, we cleaned the ball bearings with acetone and used compressed air to blast each part dry. The acetone/dry cycle is repeated, and the bearings were rinsed with isopropyl alcohol, then blasted dry. The bearings were then cleaned in a beaker of isopropyl alcohol in an ultrasonic cleaner for 20 minutes. Finally, the bearings were baked in a $150^{\circ}$ F oven for a few hours to dry them completely and remove any stray moisture left over from the isopropyl solution. The ball bearings were then stored in a sealed plastic bag with a desiccant pouch, and have not corroded after several months of storage. To prevent oil or moisture from getting on the ball bearings and possibly causing corrosion, the ball bearings were always handled while wearing latex gloves until they were installed inside the full assembly.

Since the rotor is thermally connected to the cold plate only via three point contacts, it was necessary to verify that it is sufficiently heat sunk to the 4 K mounting surface. We measured that in the temperature range from 10 K to 30 K, the thermal conductance between the main gear and the 4 K plate is approximately 2 mW/K. We tested a prototype a 95 GHz HWP mounted in an aluminum mount without any glue bonding between layers. We measured that there is a strong heat link between the outer edge of the top AR coat and the aluminum mount, greater than 10 mW/K. However, due to the low cryogenic thermal conductivity of fused quartz, there can be a substantial thermal gradient between the edge and center of the AR coats. With a heater and thermometer attached near the center of the top AR coat, we measured a thermal conductance of approximately 0.2 mW/K. Since the literature value of the bulk thermal conductivity of fused quartz is 0.1 W/(m K) in this temperature range \cite{lakeshore12}, we expect a thermal conductance of 0.9 mW/K for a 0.195''-thick disc with the heater near the center and the thermometer $1/4''$ slightly off center. This is broadly consistent with the measured performance.

The finite thermal conductivity of the fused quartz 95 GHz AR coats means that thermal radiation will heat them slightly above the temperature of the 4 K main tank. When mounted and cooled in the flight cryostat, the HWPs are estimated to be less than $0.5\%$ absorptive, with most of the loss happening in the AR coats. The heat load on the main helium tank in the flight cryostat with 5 telescopes is 2 W, compared with 1 W in earlier runs. Conservatively assuming that all of that extra 1 W is optical loading through the apertures yields an estimate of 200~mW on each telescope, or 200 mW $\times~0.5\% \times (1/2) = $ 0.5 mW absorbed at each AR coat in each HWP. Based on the measured performance of the quartz AR coats used in the 95 GHz HWPs, this means that the worst case is the AR coat is heated by 0.5~mW~/~0.2~mW/K = 2.5 K above liquid helium temperatures. 

For the flight 95 GHz HWPs, we glue the fused quartz to the sapphire with a roughly 1-inch diameter glue bond near the center. This should heat sink the center of the fused quartz discs to the high thermal conductivity sapphire, but there still can be thermal gradients on the unglued parts of the fused quartz. For the flight 150 GHz HWPs, we bond Cirlex to the sapphire, so it should be very well heat sunk to the sapphire all the way across the plate. Since these are improvements of the thermal conductivity of the system, the measurements of the prototype 95 GHz HWP represent a worst-case scenario for the flight system.

\section{Motor}
\label{steppermotor}

We use a Mycom PS445-01A stepper motor to rotate the HWP mechanism. Stepper motors are rotated through small discrete angle steps by alternately energizing the two sets of coils in the motor \cite{acarnely07}. This means the angle of the rotor can be precisely set by the motor driver without the need for realtime feedback from the angle encoders. The motor is rated at 1.2 A drive current; we nominally drive at 0.8 A, and a minimum of 0.1-0.2 A is needed to overcome friction in the rest of the rotation mechanism when cold. 200 steps rotate the motor shaft by an entire revolution. Since the main gear has 463 gear teeth, and each revolution of the stepper motor and worm gear advances the main gear by one tooth, this means that in principle the rotation angle can be set in increments of $(360^{\circ}/463~\mathrm{teeth})~\times~(1~\mathrm{tooth}/200~\mathrm{motor~steps})~=~0.004^{\circ}$. In practice, the bearing's actual precision is fundamentally limited to about $\pm 0.05^{\circ}$ to $\pm 0.10^{\circ}$ by the mechanical clearance left between the worm gear and the main gear to prevent interference upon thermal contraction.

\begin{figure}
\begin{center}
\includegraphics[width=0.9\textwidth]{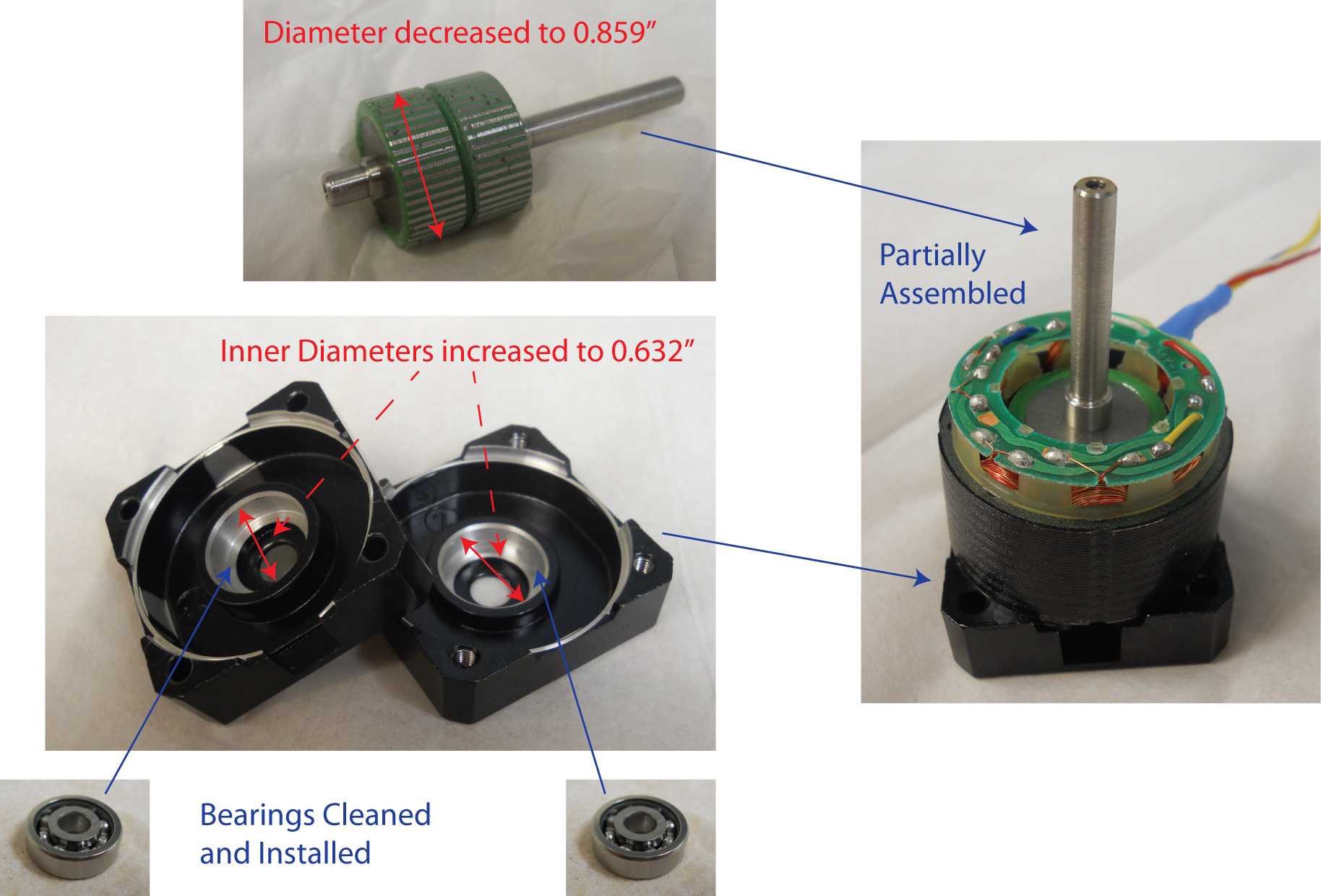}
\caption[Modifications to a Mycom PS445-01A stepper motor for cryogenic operation. ]{Modifications to a Mycom PS445-01A stepper motor for cryogenic operation. To accommodate differential thermal contraction, we increased the clearance between the rotor (shown in the top left) and stator, as well as the clearance between the rotor bearings (bottom left) and the motor housing (center left). We also cleaned the bearings to remove oils that become sticky at cryogenic temperatures. The right panel shows a partially-assembled motor. The bottom half of the motor housing, the stator, and the rotor are visible. \label{motor_mods}}
\end{center}
\end{figure}

We use VXB S625Z-Kit8525 ball bearings in the stepper motor. As it arrives from the manufacturer, the motor has an aluminum housing that will shrink onto the stainless steel ball bearings as the motor cools, binding them. The coil assembly also shrinks onto the permanent magnet rotor. To prevent this from causing mechanical interference, we increased the clearance in both parts of the motor. To do this, the permanent magnet rotor was turned down to a 0.859'' diameter, a reduction of 0.006''. The motor drive shaft was polished to fit more easily inside the ball bearing assemblies, allowing for a gentler assembly process that reduces the risk of warping the bearing or motor shaft. The parts of the aluminum motor housing holding the ball bearing were reamed to a 0.632'' diameter, an increase of 0.002''. These modifications are illustrated in Figure~\ref{motor_mods}.

After making all of these modifications, the motor parts were cleaned in acetone and isopropyl alcohol. To remove small magnetic particles that may have stuck to the permanent magnet rotor, the rotor was rotated in a hand drill while wiping with a Kimwipe, and then turned while blasting with compressed air. The rotors were rinsed again with isopropyl alcohol and all of the motor parts were baked in a $150^{\circ}$~F oven for a few hours to dry completely.

To prevent corrosion of the stainless steel bearing parts, we store the motors and assembled rotation mechanisms in airtight plastic bags with desiccant. This precaution is probably not critical. During test and integration, motors have been open to the room environment for months without any visible corrosion or appreciable degradation of their subsequent cold performance. This robustness to the room environment is important for our application, and is a significant improvement over a commercial cryogenic stepper motor that we tested early in our development process that required a clean room environment.

\subsection{Motor Drivers}

We initially chose to use an AllMotion EZHR17EN stepper motor driver to provide the drive current. This driver provides full steps and discrete microstep drive waveforms, but not the continuous sine wave waveform of some other stepper motor drivers. This driver has the advantage of low cost and small size, and provides ample current to drive our motors. It also has been successfully used in several scientific balloon flights for other applications. It uses pulse-width modulation to provide adjustable drive current. Unfortunately, the $\sim$MHz pulse-width modulation switching frequency of the drive current produced a large amount of RF electrical pickup in the angle encoders, rendering them unreadable. To solve this problem, we switched to a Phytron MCC-2 LIN stepper motor driver. Like the AllMotion, this driver also provides full steps and microsteps, and does not provide continuous sine wave waveforms. This driver does not use a pulse-width modulation scheme to adjust its drive current, and therefore does not produce significant electrical pickup in the encoders. The Phytron driver is somewhat larger, heavier and more expensive. The Phytron system we deployed to drive 6 stepper motors fits in a 17''$\times$8''$\times$3'' enclosure, weighs 15 lbs, and cost about \$3,500. A system to drive 6 stepper motors based on AllMotion products would likely fit in a 8''$\times$6''$\times$3'' enclosure, weigh about 2 lbs, and would cost around \$1,500. To our knowledge the Phytron MCC-2 LIN has not yet been used in any balloon flights, but it operates well in lab vacuum chamber testing, and in environmental chamber testing.

\subsection{Cryogenic Performance}

To verify that our modifications were sufficient to allow the motors to turn when cold, we dunk tested each motor individually in liquid nitrogen. We mounted the motor on an aluminum beam that could be dunked in a small liquid nitrogen bucket, and attached a 1-inch diameter spool on the motor shaft. Fishing line was wound around the spool and used to lift a weight bucket, with two pulleys carrying the line. To determine the maximum torque that the motor can deliver, we added weights to the bucket until the motor failed to lift it. We tested each motor at room temperature, and again with the motor immersed in liquid nitrogen. For these tests the motor speed was fixed at 0.5 shaft revolutions per second and we used a drive current of 0.8 A. To remove the water that condensed on the motors as they warmed up after these tests, each motor was disassembled, completely re-cleaned, and assembled again before installation in a rotation mechanism.

As shown in Table~\ref{motor_torque_testing}, each modified motor delivered between 7 and 9 inch ounces of torque at room temperature, compared with the unmodified motor's performance of $15.0 \pm 0.4$ inch ounces. The drop in torque is likely because the cryogenic modifications increased distance between the permanent magnet rotor and the coils. Immersed in liquid nitrogen, the motors delivered between 8 and 11 inch ounces of torque. This increase upon cooling is expected because the distance between the permanent magnet rotor and coils should decrease upon cooling, increasing the torque. This also means that we left enough clearance to prevent thermal contraction from causing extra friction in the ball bearings.

\begin{table}
\begin{center}
\begin{tabular}{ c | c | c| c| c|}
                                    &                                        &                                      & & \textbf{Min. Current}\\
\textbf{\textit{Motor}} & \textbf{293 K Torque} & \textbf{77 K Torque} & & \textbf{in Mechanism}\\
\hline
\hline
\textit{Unmodified Motor} & $15.0 \pm 0.4$ in oz& n/a & & n/a \\
\hline
\hline
Spider \#1 & $\textcolor{white}{0}7.2 \pm 0.4$ in oz & $\textcolor{white}{0}8.7 \pm 0.2$ in oz & & 0.2 A - 0.1 A\\
\hline
Spider \#2 & $\textcolor{white}{0}7.4 \pm 0.4$ in oz & $\textcolor{white}{0}8.7 \pm 0.2$ in oz & & \textcolor{white}{000...}$\leq$\textcolor{white}{.}0.1 A \\
\hline
Spider \#3 & $\textcolor{white}{0}8.4 \pm 0.2$ in oz & $10.4 \pm 0.4$ in oz & & \textcolor{white}{000...}$\leq$\textcolor{white}{.}0.1 A \\
\hline
Spider \#4 & $\textcolor{white}{0}8.4 \pm 0.2$ in oz & $10.0 \pm 0.4$ in oz & & \textcolor{white}{000...}$\leq$\textcolor{white}{.}0.1 A \\
\hline
Spider \#5 & $\textcolor{white}{0}7.8 \pm 0.4$ in oz & $\textcolor{white}{0}9.7 \pm 0.4$ in oz & & \textcolor{white}{000...}$\leq$\textcolor{white}{.}0.1 A \\
\hline
Spider \#6 & $\textcolor{white}{0}8.8 \pm 0.4$ in oz & $10.4 \pm 0.4$ in oz & & \textcolor{white}{000...}$\leq$\textcolor{white}{.}0.1 A \\
\hline
Spider \#7 & $\textcolor{white}{0}8.1 \pm 0.4$ in oz & $\textcolor{white}{0}9.7 \pm 0.2$ in oz & & 0.2 A - 0.1 A\\
\hline
Spider Spare A & $\textcolor{white}{0}8.6 \pm 0.2$ in oz & $\textcolor{white}{0}9.9 \pm 0.4$ in oz & & n/a \\
\hline
Spider Spare B & $\textcolor{white}{0}8.4 \pm 0.2$ in oz & $\textcolor{white}{0}9.3 \pm 0.2$ in oz & & n/a \\
\hline
\end{tabular}
\end{center}
\caption[Stepper motor torque measurements.]{Stepper motor torque measurements. The two center columns show the measured torque of each stepper motor at room temperature and immersed in liquid nitrogen. The right column shows the minimum drive current to turn at $\sim10$ K when the motor was integrated into its rotation mechanism. (The drive current can only be set in 0.1 A increments.) For comparison, the top row shows the torque from an unmodified stepper motor. Quoted errors on the torque measurements are the range of torque loads over which the motor was marginally able to turn. \label{motor_torque_testing}}
\end{table}

We cooled each of the seven rotation mechanisms (six for the Spider flight cryostat, and one spare) in a pulse tube test cryostat for testing before installation in the Spider flight cryostat. As shown in Table~\ref{motor_torque_testing}, we measured the minimum drive current necessary to turn at 1 degree per second at $\sim10$~K, which was always 0.2~A or lower. For all of the mechanisms, the minimum current to turn when cold does not change much over a speed range from 0.5 degrees per second up to 5 degrees per second. Since we plan to operate at 0.8~A, the minimum current measurements show that we have a torque safety factor of 4 or more.

We also measured the cryogenic heat dissipation of the mechanism by turning two mechanisms simultaneously in the cryostat for 40 minutes with 0.6 A drive current at 1 degree per second. The cold head of the pulse tube rose to an equilibrium temperature indicating that an additional 0.8~W of loading was present. This load includes heating from the operation of the LEDs in the angle encoders, which is approximately 0.17 W for two mechanisms as shown in Section~\ref{enc_hardware}. We plan on driving at 0.8 A. If the motor coils are purely inductive, then the power dissipation is purely due to the motion of the motor and is therefore independent of drive current. If the motor coils are also resistive, there will be a $I^2 R$ component to the power dissipation. In the absolute worst case, scaling the power of a single motor by $(0.8~\mathrm{A})^2 / (0.6~\mathrm{A})^2$ gives an upper bound on the power of 0.7 W. At this power level, an individual mechanism turning at 1 degree per second would boil off only 7 mL of liquid helium in a $22.5^{\circ}$ turn.

\subsection{Reliability}
We built two rotation mechanisms for the Keck instrument \cite{ogburn12} that successfully operated when deployed at the South Pole. Each HWP was rotated by $45^{\circ}$ every four days throughout the entire 2011 observing season. This was an important proof of concept for the Spider rotation mechanisms, and also showed that our understanding of the Spider mechanism was sufficient to allow us to redesign and construct new rotation mechanisms for the different mechanical constraints of the Keck instrument cryostats. The rotation mechanisms are durable and have a long operational lifetime. We lab-tested five of the Spider rotation mechanisms for longevity by rotating them each through more than 700 turns of $22.5^{\circ}$ below 20 K and observed no degradation of their mechanical performance.

\clearpage

\section{Angle Encoders}

\subsection{Hardware}\label{enc_hardware}

\begin{figure}
\begin{center}
\includegraphics[width=0.85\textwidth]{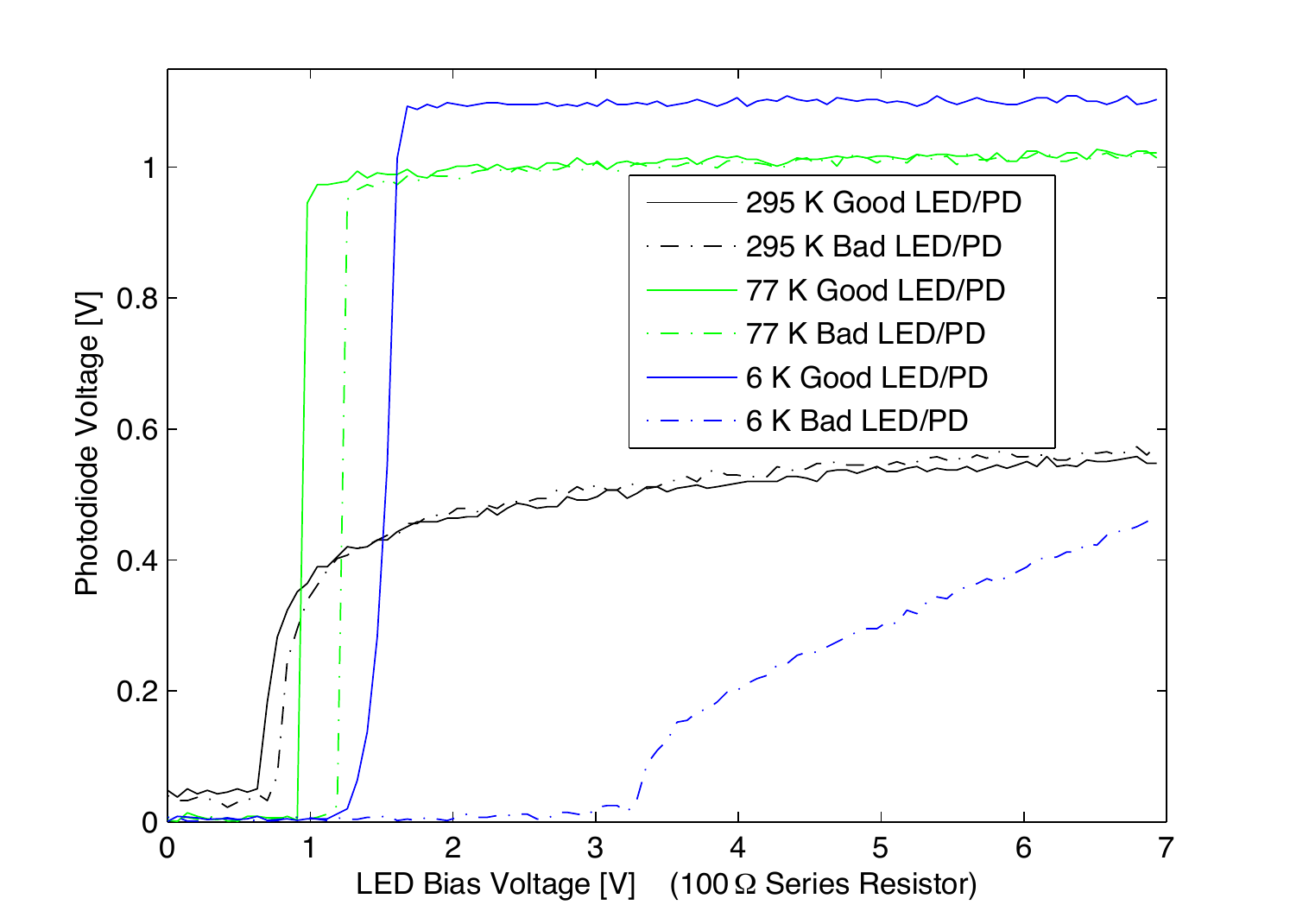}
\caption[Comparison of an operational and defective LED/photodiode pair. ]{Comparison of an operational and defective LED/photodiode pair.  In the operational pair (solid lines), by the time the LED bias voltage exceeds 2 V the photodiode is saturated with LED light. The defective pair (dot-dashed lines) operates normally at room and liquid nitrogen temperatures, but at liquid helium temperatures the photodiode fails to saturate with light for any of the LED bias voltages. (Data courtesy of Ben Saliwanchik and Johanna Nagy.) \label{bad_ledpd}}
\end{center}
\end{figure}

Our design goal is to be able to monitor the HWP angle with a precision of $\pm 0.1$ degrees. We built a system of several optical encoders to measure the bearing angle while it is turning. For each encoder, we used an Industrial Fiberoptics IF-E91A light emitting diode (LED), and a Vishay BPV23NF photodiode detector, operating in a band centered at 940 nm. Although the manufacturers do not guarantee cryogenic performance, these components are inexpensive, readily available, and typically work well at liquid helium temperatures. We screen for units that work well at 4 K by individually dip-testing each LED/photodiode pair in liquid helium before using the pair in a rotation mechanism. For the dip test, the LED was DC voltage-biased with a programmable power supply and a 100~$\Omega$ series resistor. A typical dip test result of a functional pair and a defective pair is shown in Figure~\ref{bad_ledpd}. 74 out of 84 tested pairs functioned well at 6~K. So far in our testing, no LED or photodiode that has passed this initial screening has subsequently failed.

To monitor the stepper motor shaft, we placed an incremental encoder wheel on the motor shaft as shown in Figure~\ref{hwp_mechanism_overview}. The LED shines light towards the photodiode detector, and the beam is alternately blocked and passed as 20 equally-spaced holes rotate through the beam, creating a chopped light signal on the detector.  Each period of the shaft encoder signal corresponds to a main bearing rotation of $(360^{\circ}/463~\mathrm{main~gear~teeth}) \times (1~\mathrm{tooth}/20~\mathrm{shaft~encoder~periods}) = 0.04^{\circ}$. This is sufficient angle resolution for a single turn, but roundoff error from using this encoder alone could accumulate over many turns and soon exceed our angle error goal of $\pm0.1^{\circ}$.

\begin{figure}
\begin{center}
\subfigure{\raisebox{.33 in}{\includegraphics[width=0.55\textwidth]{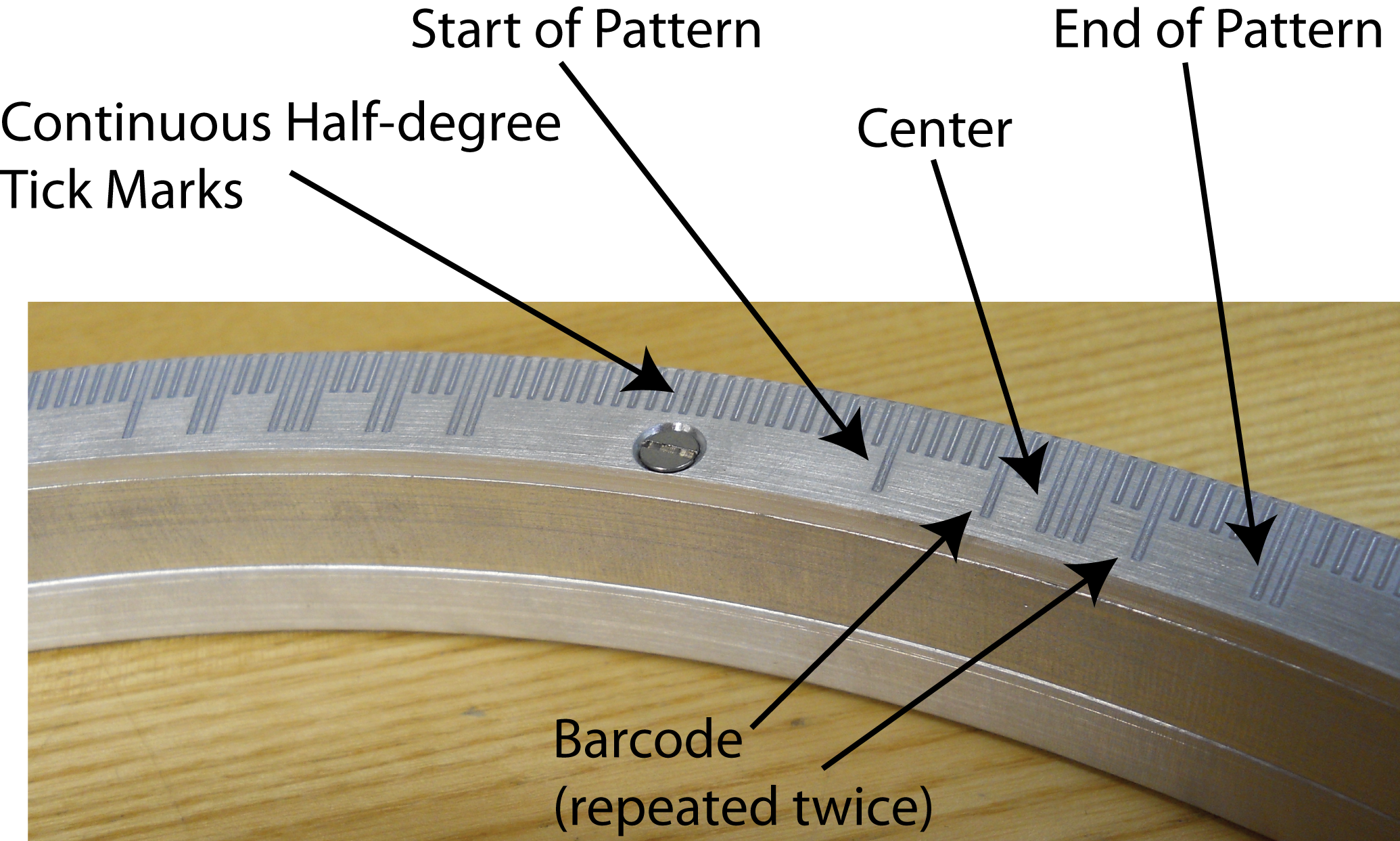}}}
\subfigure{\includegraphics[width=0.85\textwidth]{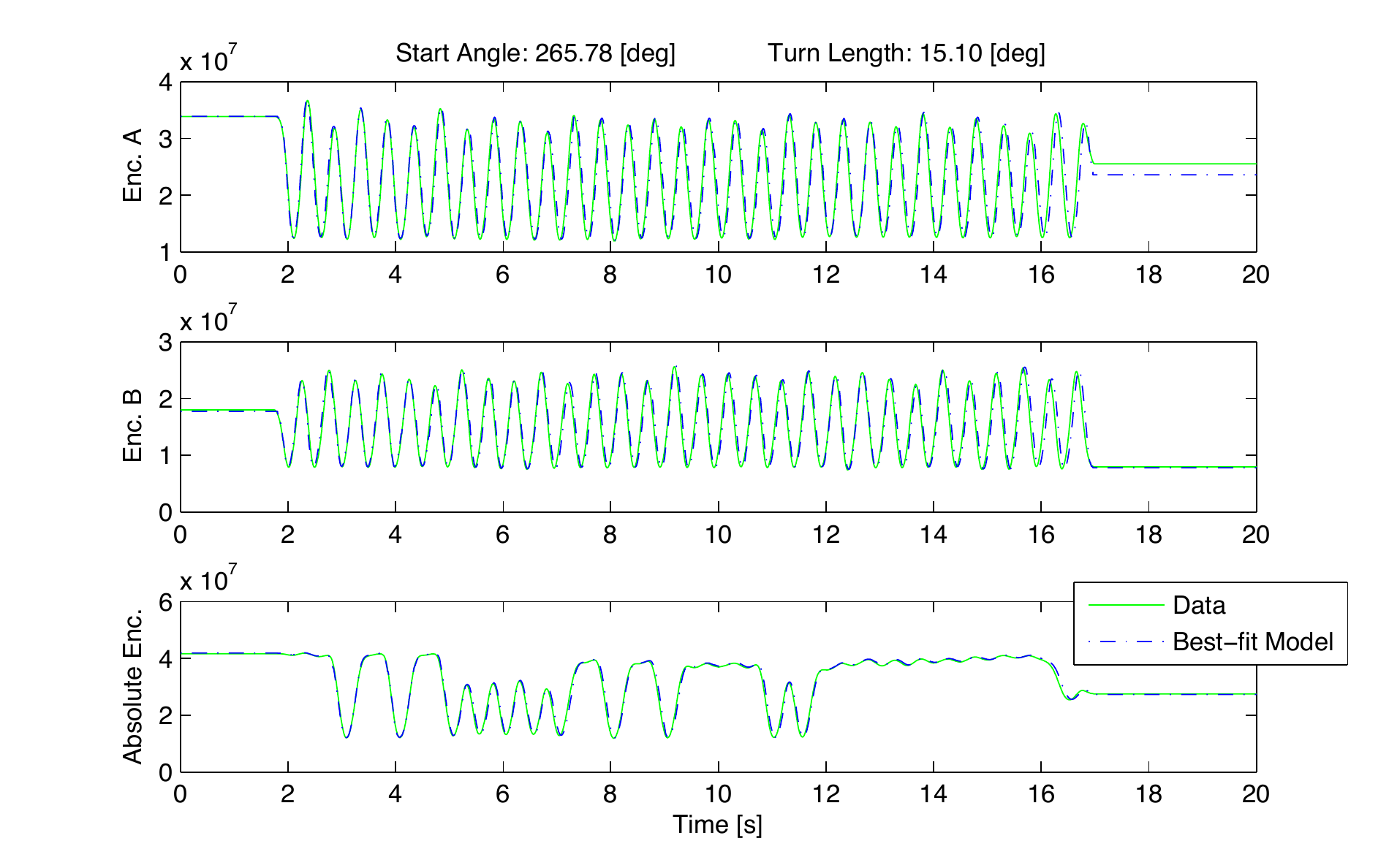}}
\caption[Main encoder.]{Main encoder. The left panel is a photograph showing both the continuous and barcode tracks of the encoder attached to the main gear. The right panel is a plot of encoder data (in raw ADC units) from a 15-degree turn taken with a mechanism operating cold in the Spider flight cryostat. The best-fit model timestream is overplotted, which allows a precise estimate of the start and stop angles. \label{turn_timestream}}
\end{center}
\end{figure}

To prevent angle error from accumulating, we mounted an absolute encoder directly on the main bearing. To conserve space, instead of a chopper wheel we use a reflective encoder attached below the main gear. The encoder pattern is shown in Figure~\ref{turn_timestream}. Tick marks laser-etched in aluminum are less reflective than bare aluminum, presumably because of the higher surface roughness of the laser-etched ticks. Laser Design and Services Company in Willoughby,~OH etched a continuous track of tick marks every $0.5^{\circ}$ around the main bearing. Each tick is $0.030$'' wide. A second track was etched containing a unique barcode pattern every $22.5^{\circ}$, shown in Figure~\ref{turn_timestream}. The barcodes allow the absolute start and stop angle of a partial turn longer than $22.5^{\circ}$ to be determined solely from that turn's encoder data. Each barcode pattern starts with a single tick and ends with two ticks in a row, indicating the rotation direction. At the center of each pattern are three ticks in a row to clearly indicate a reference point in the pattern. A unique binary pattern of ticks and skipped ticks is placed between the center reference and the end of the pattern, and repeated in reverse order between the center reference and the start of the pattern. This barcode scheme is somewhat redundant, but it is easy to identify both by eye and with a computer algorithm. After implementing it, we found that our two-track scheme is conceptually similar to the Virtual Absolute encoders made by Gurley Precision Instruments. 

\begin{figure}
\begin{center}
\subfigure{\includegraphics[width=0.45 \textwidth]{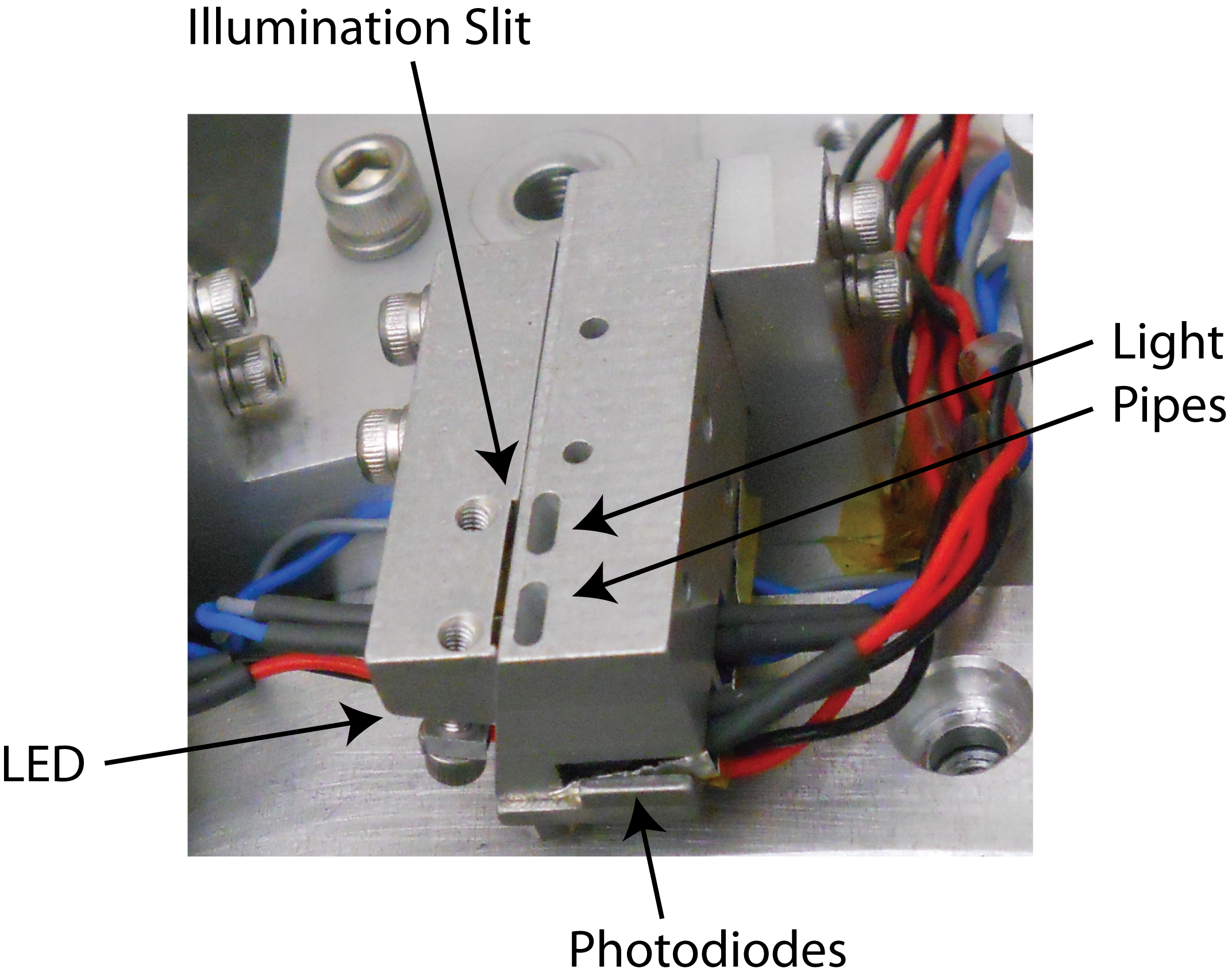}}
\subfigure{\raisebox{.33 in}{\hspace{0.03 \textwidth} \includegraphics[width=0.50 \textwidth]{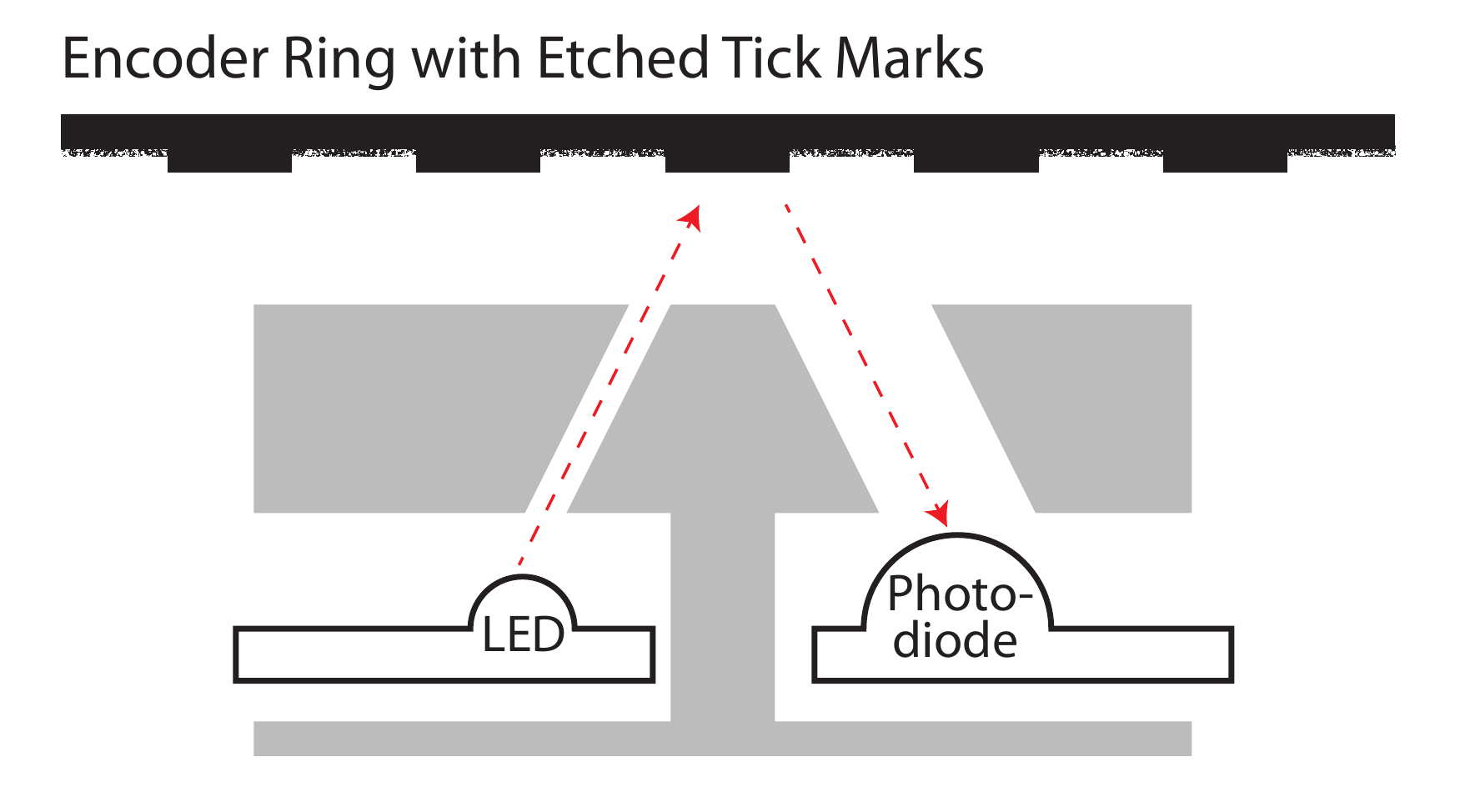}}}
\caption[LED and photodiode mount to read out the reflective encoder on the bottom of the main bearing. ]{LED and photodiode mount to read out the reflective encoder on the bottom of the main bearing. The LED shines up through the narrow illumination slit and reflects off both tracks of the encoder. One light pipe for each track of the encoder carries light down to the photodiodes for detection.\label{encoder_led_pd_mount}}
\end{center}
\end{figure}

We use LED/photodiode pairs to sense the encoder ticks on the main bearing as they pass by. The mounting assembly is shown in Figure~\ref{encoder_led_pd_mount}. The LED shines up through an illumination slit onto both encoder tracks. The slit is narrow enough that light is only bouncing off one encoder tick at a time. Two photodiode detectors are mounted below two light pipes, with one for each track of the encoder. An additional LED/photodiode pair also views the continuous track, but is staggered by a quarter-tick. These two form a quadrature encoder readout. This makes the angular sensitivity more uniform because as the bearing turns one of the two continuous encoder signals is always transitioning between high and low, and it also indicates the direction of the turn. 

To reduce stray pickup from the stepper motor drive current, the encoders are read out with a lock-in amplifier circuit. The switching is accomplished using a Vishay SI2318DS-T1-E3 MOSFET in series with each LED driven by a digital logic signal. Each LED is chopped at 1.5 kHz. The rotor LEDs are voltage biased at 6 V with a 100~$\Omega$ series resistor. Since it shines directly onto the photodiode, a 1 k$\Omega$ series resistor is used for the shaft encoder LED to prevent it from saturating the photodiode. When cooled, the voltage drop across this model of LED is approximately 2~V, which implies a calculated power dissipation at 4 K of approximately 80~mW for each main encoder LED, and 8 mW for the shaft encoder LED. The LEDs are chopped at a 50\% duty cycle, which means the total power dissipation for a single mechanism is calculated to be 84 mW. 

The two wires carrying the current from a photodiode are connected across a $30~\mathrm{k}\Omega$ resistor at room temperature, and the readout electronics measure the voltage across that resistor. For lab testing we built an analog lock-in circuit. Each channel used an Analog Devices AD620 instrumentation amplifier as a preamp for an AD630 balanced modulator/demodulator followed by a low-pass active filter at 500 Hz. We chose to not deploy this readout system for flight, since the thermometer readout electronics for the Spider flight cryostat have a digital lock-in system that can demodulate the photodiode signals. Biasing the LED creates a 10-100 mV demodulated signal in the photodiode detector with very low noise. The system is readout noise limited with both the analog ($\sim$0.5 mV noise) and digital ($\sim$0.1 mV noise) lock-in electronics. Since the LED voltage drop varies with temperature and between different LEDs, robustly biasing at lower current to reduce the cryogenic heat load would require a more sophisticated bias circuit.

\subsection{Software}

The analog encoder signals are interpreted in software to precisely estimate the start and stop angle of a turn. For Spider, a typical operation during observations will be to make a $22.5^{\circ}$ turn, and the encoder data will be interpreted afterwards to measure the exact angle. One approach for this post-processing would be to digitize all of the encoder signals by finding whether each encoder was at a ``high'' or ``low'' voltage level at each encoder voltage sample. Counting the rising and falling edges would yield an estimate of the relative angle between the start and stop of a turn. The shaft encoder would be used to interpolate between the half-degree encoder ticks. This method has some difficulties. Because of the $\pm0.05^{\circ}$ mechanical clearance between the main and worm gears, at a turn start or stop the shaft encoder can tick without moving the main gear, which would cause error in the turn length estimate. Also, interpreting the digitized barcode encoder to yield a precise absolute angle estimate would require some kind of pattern recognition to robustly identify the middle of a barcode. The algorithm could fail if not enough of a barcode appeared during a short turn to properly trigger the pattern recognition scheme.

Instead of digitizing the encoders to ``high'' and ``low'', we rely on the repeatability of the encoder analog voltage levels to interpret the signals. To do this, we use data taken earlier from a single long continuous turn to create a template of all the encoder voltages as a function of absolute rotation angle. We use this template to make a model of the encoder voltages of an arbitrary short turn. This allows us to take a parameter estimation approach when analyzing the real data from a short turn, as we can vary the start angle, stop angle, and other parameters of our model until it matches the data. This yields highly precise estimates of the absolute rotation angle.

To form an encoder template, our algorithm uses data from a continuous rotation longer than $360^{\circ}$. In principle, interpreting the encoder signals based on rising and falling edges would yield the best sensitivity. However, due to variations in the aluminum surface around the encoder ring, the ``high'' and ``low'' signal levels vary by up to a factor of two around the ring. However,  finding maxima and minima in the signal is still a clear way to identify the encoder ticks. 

The algorithm looks for sign changes in the numerical derivative of one of the half-degree-tick encoders to find all of the maxima and minima of the signal. The algorithm then fits a parabola near each estimated maximum or minimum to improve the estimate of the tick location. These maxima and minima are etched to be exactly a quarter-degree apart. The algorithm assumes the turn is smooth and uses linear interpolation between these ticks to get a relative angle estimate at each encoder sample. The algorithm then asks for user input to find the barcode for the zero angle in the absolute encoder timestream. This completes the template, which consists of the two half-degree tick encoders voltages, and the absolute encoder voltage, as a function of absolute bearing angle. The algorithm to generate a template (implemented in Matlab) takes roughly 5 minutes of user interaction to verify that the template is accurate. A template only needs to be generated once for each rotation mechanism.

The analog encoder voltages are extremely repeatable, and the stepper motor turns the bearing very smoothly. This means there are only a few parameters necessary to model the encoder timestream of a partial turn. Each turn has a start time $t_{start}$ and a stop time $t_{stop}$. The motor driver takes a time $\tau$ to gradually bring the bearing up to speed, turns at a constant angular velocity $\omega_{bearing}$, and gradually reduces the speed at the end of the turn.  The direction of the turn (clockwise or counterclockwise) is indicated by the sign of $\omega_{bearing}$. The angular velocity model for a single turn is therefore
\begin{equation} \label{alpha_of_t}
\omega(t) = 
\begin{cases} 
      0 & t < t_{start} \\
      \frac{t - t_{start}}{\tau} \omega_{bearing} & t_{start} \leq t < (t_{start}+\tau) \\
      \omega_{bearing} & (t_{start}+\tau) \leq t \leq (t_{stop} - \tau) \\
      \frac{t_{stop} - t}{\tau} \omega_{bearing} & (t_{stop} - \tau) < t \leq t_{stop} \\
      0 & t > t_{stop}
   \end{cases}
\end{equation}
Analytically integrating this yields a model for the bearing angle as a function of time. The model picks up one more parameter after integration, namely the start angle $\theta_{start}$ of the turn. The stop angle of the turn is a derived parameter.  To generate a model timestream for a given set of parameters, we first use the integral of Equation~\ref{alpha_of_t} to calculate the model's absolute rotation angle for each time sample. We then use a nearest-neighbor lookup from the encoder voltage template to generate the model of the encoder voltages for each time sample. 

We take a least-squares fitting approach to use the encoder model to estimate the start and stop angles of a short turn. We find the first and last voltage sample that is significantly different from the DC encoder level to estimate $t_{start}$ and $t_{stop}$, and use a grid search to obtain the best-fit values of the other turn parameters $\{ \tau, \omega_{bearing}, \theta_{start} \}$. To get a good guess for the starting angle to set the range of the grid search, the algorithm assumes nominal values of the other parameters, and sweeps over all possible starting angles. This is repeated assuming a reverse-direction turn. The starting angle and turn direction with the lowest $\chi^{2}$ from this initial search are used as the initial guess for the grid search. Running on a dual-core laptop, the algorithm (implemented in C) takes approximately a minute to estimate the start and stop angles of a $22.5^{\circ}$ turn. A sample timestream and best-fit model is shown in Figure~\ref{turn_timestream}.

We verified the precision and accuracy of our angle encoding method by checking it on the bench against a Teledyne Gurley 8225-6000-DQSD optical relative encoder we connected to the main bearing. The manufacturer certifies that this encoder is accurate to $\pm 0.03^{\circ}$ which is sufficient to verify whether or not we have met our design goal of $\pm 0.1^{\circ}$ angle accuracy. We tested four of our seven mechanisms, and they performed equally well within measurement errors. Each half-degree tick in our encoder is measured to be $0.50^{\circ}$ apart according to the reference encoder, with scatter below the $\pm0.03^\circ$ error level of the reference encoder. When we commanded a set of 10 $22.5^{\circ}$ turns from the motor controller, the resulting turns from the mechanism with the highest measured scatter were measured by the reference encoder to be $22.56^{\circ} \pm 0.06^{\circ}$ long (i.e. mean $\pm$ RMS scatter in the set of measured turn lengths). The turn lengths estimated by our encoder differed from the reference encoder by $0.01^{\circ} \pm 0.05^{\circ}$. We also commanded a set of 14 forward and reverse turns of lengths varying from $22.5^{\circ}$ to $157.5^{\circ}$. In the mechanism with the highest measured scatter, the reference encoder estimates differed from our encoder by $0.00^{\circ} \pm 0.10^{\circ}$. Our encoder measured that in each pair of forward and reverse turns, the mechanism's stop angle differed from its starting angle by $0.01^{\circ} \pm 0.05^{\circ}$. This set of tests verifies that on the bench, we can reliably control and measure the absolute angle of the bearing to at worst $\pm 0.1^{\circ}$. Upon cooling, the encoder voltage template does not change appreciably, only differing by an overall gain and DC offset caused by thermal effects in the LED and photodiode. We interpret this as evidence that the encoders perform equally well inside a cryostat.


\chapter{Sapphire Optical Properties Measurements}
\label{index_measurements}

The material in this chapter has been published as a SPIE Proceedings article \cite{bryan10a}.

The optical action of a HWP in Spider comes from a plate of birefringent sapphire. A HWP made from a single slab of birefringent material is designed so the optical path length difference $(n_s~-~n_f)~\times~d$ between waves polarized along the two crystal axes is exactly a half wavelength. This phase delay causes linearly-polarized light to rotate as it passes through the plate. In order to choose the thickness $d$ of the plate, we needed to measure $n_s - n_f$, and the values of both indices are needed to optimize the AR coat design. To do this, we used a Fourier Transform Spectrometer (FTS) to measure the transmission as a function of photon frequency of samples of 3.05 mm thick sapphire at room temperature, and also a sample cooled to 5 K. We then fit the spectra to a model to determine the indices of refraction. In this chapter we also show spectra of a prototype 150 GHz AR-coated HWP that was made with a quartz AR coat.

\section{Measuring the Spectral Response}

\begin{figure*}
\begin{center}
\includegraphics[width=0.80\textwidth]{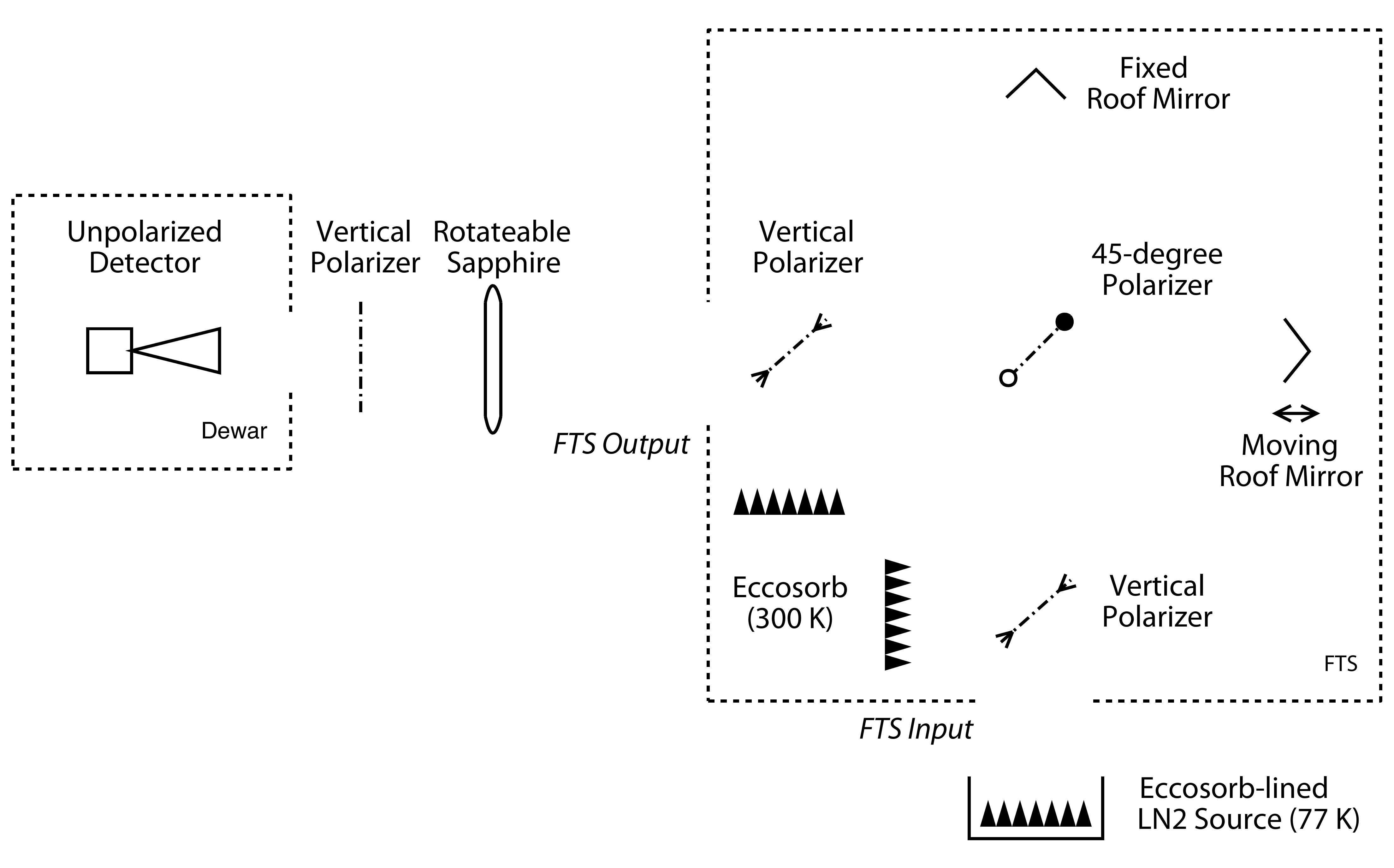}
\caption[Instrument configuration for the broadband spectra of the sapphire.]{Instrument configuration for the broadband spectra of room temperature sapphire. We used a polarized Fourier Transform Spectrometer to shine polarized light through the sapphire slab. An analyzer polarizer at the other side of the sapphire is aligned with the output of the FTS. We used a broadband sub-kelvin bolometer as the detector to get HWP spectra at .75 GHz resolution from 100 GHz to 240 GHz. For the cryogenic spectra of sapphire, we removed the final vertical polarizer, moved the sapphire into the dewar, and used a rotatable polarizer at the output of the FTS to take spectra at several angles near both crystal axes of the sapphire. \label{fts_setup}}
\end{center}
\end{figure*}

The frequency response of the Spider HWPs was characterized by taking broadband transmission spectra of sapphire using a polarized Martin-Pupplet FTS \cite{lesurf90}. A block diagram of the apparatus is shown in Figure~\ref{fts_setup}.  Since the light travels down and back each arm of the interferometer, moving one mirror by a half-wavelength changes the optical path difference between the two arms by a full wavelength. This means that the mirror movement required to generate a full period in the detector timestream is $\Delta x_{mirror} = \frac{\lambda}{2}$. If the mirror is moving at a constant speed $v_{mirror}$, then $\Delta x_{mirror} = \Delta t ~v_{mirror}$. This signal will appear in the detector timestream at a frequency $f_{audio} = 1/ \Delta t$. Combining these relations yields the relationship between the audio frequency $f_{audio}$ of the signal in the detector timestream and its corresponding photon frequency $\nu$,
\begin{equation} \label{freq_conv}
\left( \frac{c}{2 v_{mirror}}\right) f_{audio} = \nu.
\end{equation}
The Fourier transform of the timestream is the product of the source, detector and HWP spectra, with the frequency axis determined by Equation~\ref{freq_conv}. We analyze the data with a discrete cosine transform to eliminate noise biasing. The frequency resolution of the FTS is determined by the maximum distance the mirror moves from the white light fringe. The data presented here was taken with a total mirror travel of .2 m, and therefore a frequency resolution of $c / (2 \times .2~\mathrm{m}) = .75$ GHz.

For the source at the input port of the FTS, we use an Eccosorb-lined liquid nitrogen bath, which emits a nearly Rayleigh-Jeans spectrum in the mm-wave band. A wire grid polarizer was placed at the output of the FTS, followed by the sapphire HWP on a rotatable bearing, which allowed a polarized spectrum to be measured at each rotation angle $\theta$ of the material. Another wire grid polarizer aligned with the output polarizer on the FTS was placed between the HWP and the detector. This allows measurements of the sapphire rotating the polarization state of the light passing through it. A broadband sub-kelvin bolometer sensitive from 100 GHz to 240 GHz mounted in a lab cryostat was used as the detector. 

The transmission spectra of a 100 mm diameter sample of A-cut sapphire at room temperature are plotted in Figure~\ref{sapphire_spectra}. The slab has an air-medium interface on each side, both of which cause reflections. Considering the sapphire when it is rotated such that one of its crystal axes is aligned with the detector polarization angle, at certain frequencies the interference between the two reflected waves adds constructively, causing a minimum in the transmission spectrum. At other frequencies, the interference is destructive and leads to a peak in the transmission spectrum. This shows up in the color plot in Figure~\ref{sapphire_spectra} as vertical bands. The frequency spacing between peaks depends on the index of refraction the polarized wave experiences as it travels through the material. For waves traveling along the slow axis of the sapphire, the spacing corresponds to the slow index of refraction of sapphire $n_s$, while for waves polarized along the fast axis, the spacing corresponds to the fast index $n_f$. We use spectra taken with the light traveling at many angles through the material, and fit them to a model. This appears in the color plot in Figure~\ref{sapphire_spectra} as a variation in the vertical band spacing at different angles. The fitting procedure implicitly uses the peak spacing and amplitude to estimate the two indices of refraction of the sapphire.

\section{Fitting the Spectra to Measure the Indices of Refraction}

To fit the observed spectra as a function of sample angle $S_{obs}(\nu,\theta)$, the transmission $T_{xx}$ through the sapphire slab and aligned polarizers is calculated using a physical optics model similar to the one described in Savini et al. \cite{savini06}. The model extends the 2-by-2 matrix formalism reviewed in Hecht and Zajac \cite{hecht74} for modeling multiple layers of isotropic materials to a 4-by-4 matrix formalism for multiple layers of birefringent materials. The model uses the electromagnetic boundary conditions at each of the air-material interfaces to map the incident electric and magnetic fields onto the transmitted fields. This fully treats multiple reflections and interference effects. The model can handle lossy materials, but here we assume that all materials are lossless.

The model for the HWP transmission is multiplied by the detector response spectrum $F(\nu)$ and an overall normalization factor $a$ to obtain a model for the set of observed FTS spectra,
\begin{equation}
S_{calc}(\nu , \theta) = a \times F(\nu) \times T_{xx}(\nu, \theta - \theta_{0} \mid  \{ n_s, n_f\}),
\end{equation}
where $\theta_0$ is the angle of the crystal axes relative to the angle of the incident polarized light. The data is then fit for the parameters $\{ n_s, n_f, \theta_0, a\}$ using a Monte-Carlo Markov Chain \cite{chib95}, which allows for non-gaussian likelihood, and gives a straightforward estimate of covariances in the parameter estimates. The covariance between parameter estimates is less than 1\% for the room-temperature measurements reported here, so it is ignored below. The phase of the modulation with HWP angle determines the angle of the crystal axis, and the spacing in frequency between spectrum peaks determines the indices.

\begin{figure}
\begin{center}
\subfigure{\includegraphics[width=0.37\textwidth]{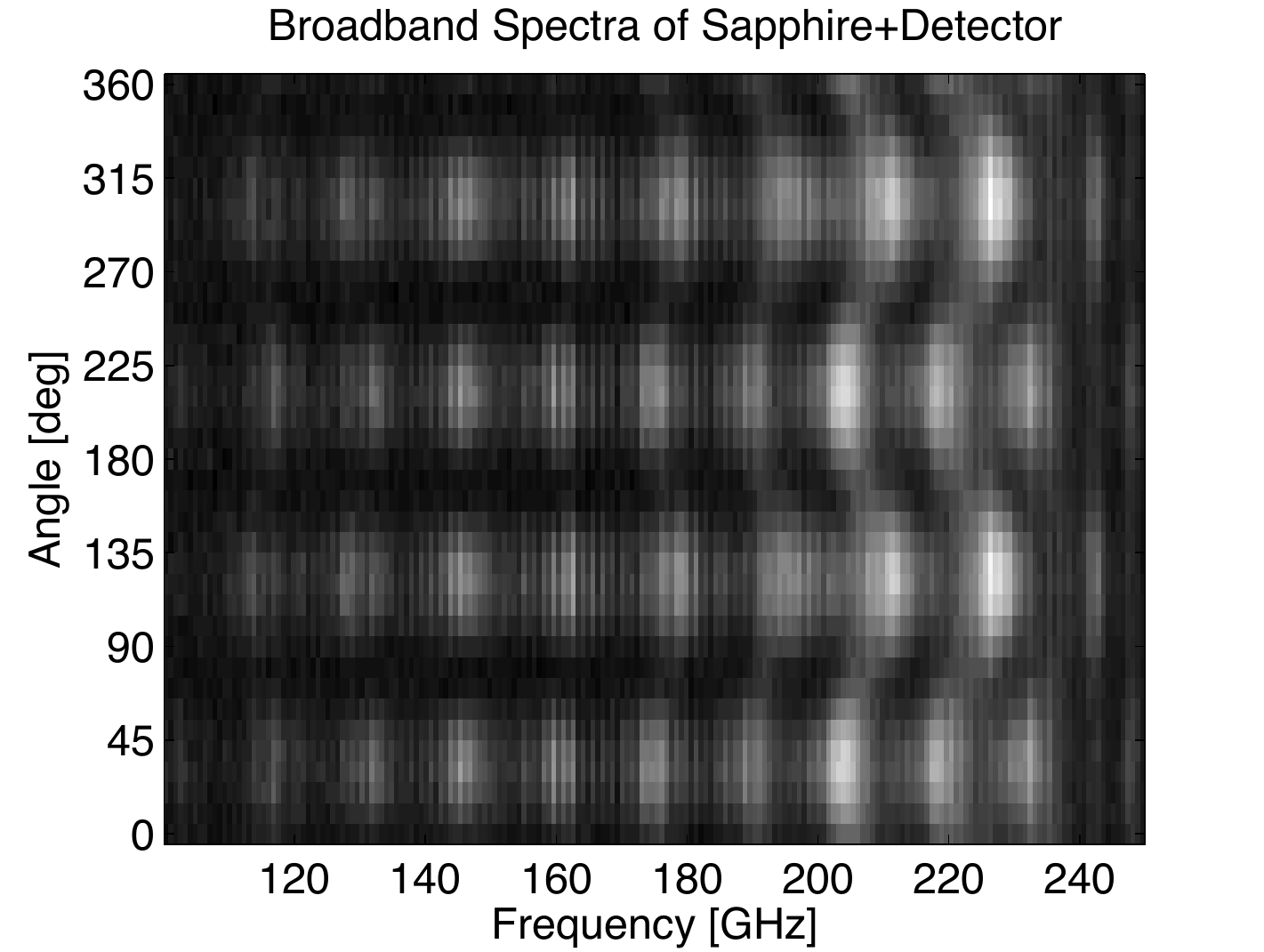}} 
\subfigure{\includegraphics[width=0.62\textwidth]{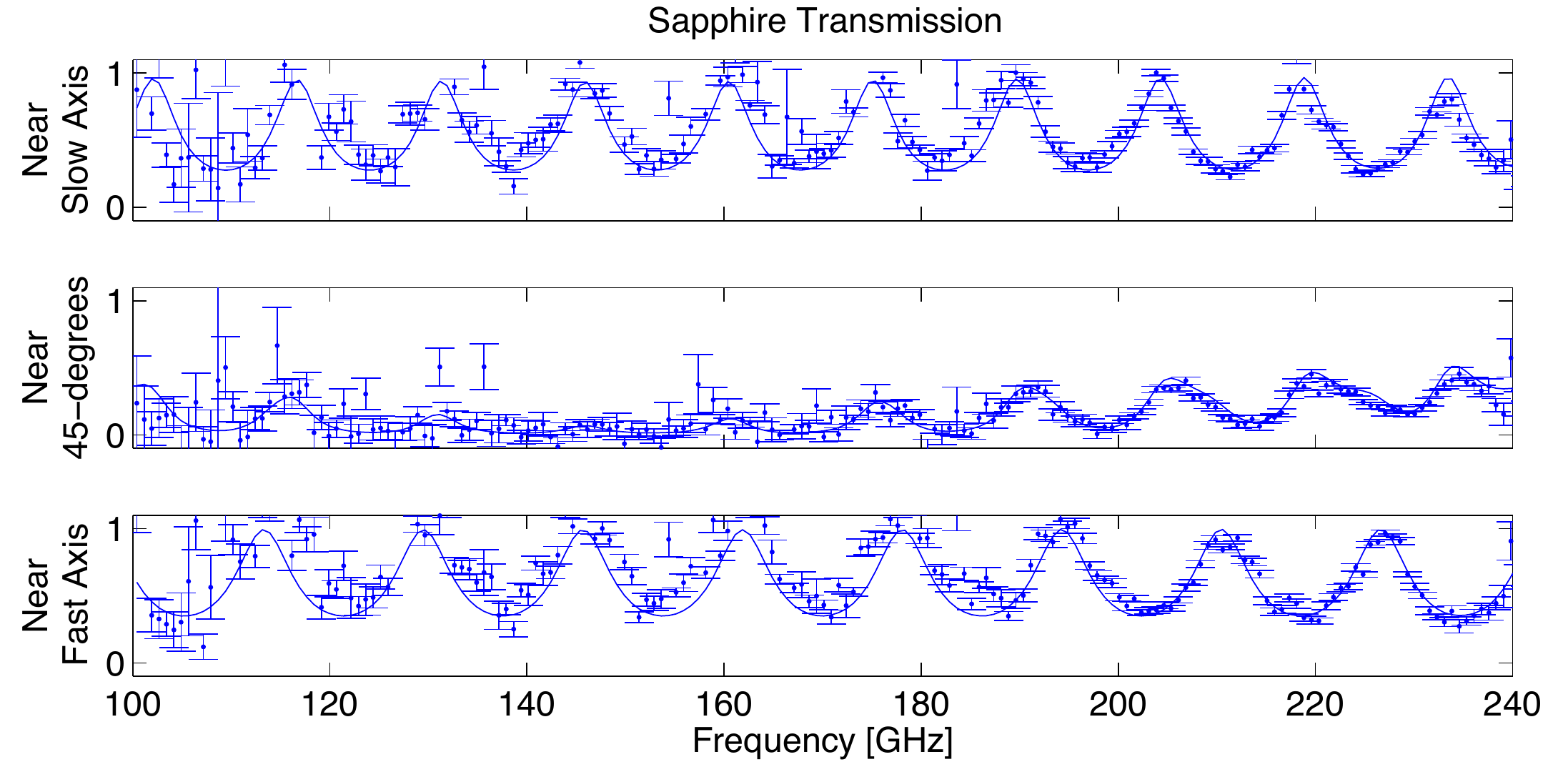}} 
\caption[Broadband millimeter-wave spectra of a 3.05 mm thick sapphire slab at room temperature.]{Broadband millimeter-wave spectra of a 3.05 mm thick sapphire slab at room temperature. The left panel shows the observed spectra as function of angle, where the intrinsic detector response has not been removed. Polarized spectra were measured with the sapphire oriented at $10^{\circ}$ intervals. The variation with angle is caused by the birefringent sapphire rotating the incident linearly-polarized light from the FTS. The right panel shows spectra taken near both of the crystal axes of the birefringent sapphire, and a spectrum taken at an intermediate angle. The intrinsic detector response was removed from this plot, and the best-fit spectrum is shown as a smooth curve. The error bars on the data points are from detector noise only. The frequency spacing between the peaks in the spectrum taken at an angle near the slow axis of the sapphire is different than the peak spacing for the spectrum taken near the fast axes. This allows a precise estimate of both indices of refraction of sapphire. \label{sapphire_spectra}}
\end{center}
\end{figure}

The room temperature sapphire spectra of the 100 mm diameter sample and curves from the best-fit model are shown in Figure~\ref{sapphire_spectra}. The best-fit indices of refraction are shown in Table~\ref{uniformity}. The values are almost 1\% below the values listed in Lamb \cite{lamb96}, but index differences $(n_{s}-n_{f})$ are in agreement. 
Also listed in the table are index values derived from temperature measurements of a 330~mm diameter sapphire used in the AR-coated Spider prototype discussed below.  For that larger sapphire, measurements were made of a roughly 50~mm diameter patch at the center, and three similar patches centered at a radius of $\sim 80$~mm to test uniformity of the material.  
As shown in the table, the best fit indices for all 4 locations on the 330~mm sample agreed within errors, and agreed with the values from the 100 mm sample. 

\begin{table}
\begin{center}
\begin{tabular}{ c | c | c | c | c | c |}
 & $\mathbf{n_s}$ & $\mathbf{n_f}$ & $\mathbf{n_{s} - n_{f}}$ \\
\hline
\hline
\textbf{Room Temp.} & & &\\
\hline
Lamb \cite{lamb96}&$3.403\phantom{4} \pm .003\phantom{0}$& $3.069\phantom{4} \pm .003\phantom{0}$& $.334\phantom{4} \pm .004\phantom{0}$   \\
\hline
100 mm Diameter &$3.3736 \pm .0002$& $3.0385 \pm .0002$& $.3350 \pm .0004$\\
\hline
330 mm Diameter & & &\\
\hline
\textit{Center}&$3.3742 \pm .0003$& $3.0373 \pm .0003$& $.3369 \pm .0004$   \\
\hline
\textit{Location 1}&$3.372\phantom{4} \pm .002\phantom{4}$& $3.031\phantom{4} \pm .002\phantom{4}$& $.341\phantom{4} \pm .003\phantom{4}$  \\
\hline
\textit{Location 2}&$3.371\phantom{4} \pm .002\phantom{4}$& $3.033\phantom{4} \pm .003\phantom{4}$& $.338\phantom{4} \pm .004\phantom{4}$   \\
\hline
\textit{Location 3}&$3.370\phantom{4} \pm .002\phantom{4}$& $3.030\phantom{4} \pm .003\phantom{4}$& $.340\phantom{4} \pm .004\phantom{4}$   \\
\hline
\hline
\textbf{LHe Temp.} & & &\\
\hline
Loewenstein \cite{loewenstein73} (1.5 K)&$3.361\phantom{4 \pm .0004}$& $3.047\phantom{4 \pm .0004}$& $.314\phantom{4 \pm .0004}$   \\
extrapolated \cite{johnson04} to 150 GHz & & & \\
\hline
5 K, 100 mm Diameter &$3.336\phantom{4} \pm .003\phantom{4}$& $3.019\phantom{4} \pm .003\phantom{4}$& $.317\phantom{4} \pm .004\phantom{4}$   \\
\hline
\hline
\textbf{Est. Systematics} &$\phantom{3.3734} \pm .003\phantom{4}$&$\phantom{3.3734} \pm .002\phantom{4}$& \\
\hline
\end{tabular}
\caption[Room temperature and cryogenic indices of refraction of A-cut birefringent sapphire near 150 GHz]{Room temperature and cryogenic indices of refraction of A-cut birefringent sapphire near 150 GHz. Statistical error from detector noise is shown next to each index value, and our estimated systematic uncertainties are shown in the last row. The indices of the 330 mm diameter sample were measured by taking spectra through its center, and at three other locations. The large sample has uniform optical properties at room temperature within statistical error. \label{uniformity}}
\end{center}
\end{table}

\begin{table}
\begin{center}
\begin{tabular}{ c | c | c|}
\textbf{Effect} & \textbf{Index Error} & Comment \\
\hline
\hline
Tilted Sample &  $+\phantom{.00}0\%$ &$\pm 2.5^{\circ}$ tilt.\\
 & $-.008\%$ &  \\
 \hline
Expanding Beam & $\pm.03\% $&Uncertainty in  correction \\
& & for \textit{f}/3.3 optics. \\
\hline
Polarizer alignment & $\pm .01\%$& $\pm 2^{\circ}$ rotation\\
\hline
Mirror Speed & $\pm .07\%$& $\pm2~\mu$m/s for 3 mm/s motion.\\
\hline
\hline
\textit{Quadrature Sum} & $\pm.08\%$& Detector noise is $\pm .006\%$ at 300 K,\\
& & \phantom{00000......}and $\pm .09\%$ at 5 K. \\
\end{tabular}
\caption[Estimated systematic error budget.]{Estimated systematic error budget. All of the systematics considered cause a fractional change in the observed indices. The increase in optical path due to a tilted sample and observing through a finite aperture can only bias the observed indices higher than the true value. The fitting code was run again with the polarizers in the model rotated by $\pm2^{\circ}$ to estimate the effect of mis-aligned polarizers on the results. The mirror motion varies at the $\pm2~\mu$m/s level due to the stage encoder resolution and motor drive feedback error. Added in quadrature, these effects are comparable to detector noise for the cold measurements, and dominate over detector noise for the warm measurements. \label{systematics}}
\end{center}
\end{table}

Previous measurements of the indices of sapphire at liquid helium temperatures, such as those of Loewenstein et al. \cite{loewenstein73} from .9 THz to 9 THz at 1.5 K, demonstrate that cryogenic indices are shifted from their room temperature values. 
Johnson \cite{johnson04} extrapolated these shifts from the high frequencies of Loewenstein et. al. down to 150 GHz. 
To check this extrapolation, the 100 mm diameter sapphire sample was cooled down to 5~K, and the spectra were fit to obtain its indices of refraction. Rather than rotating the sample at 5~K, it was left in a fixed position and spectra were taken with the FTS polarized at angles very near the slow and fast indices of the material. The measured cold indices, listed in Table~\ref{uniformity}, differ by nearly 1\% from the extrapolated values calculated by Johnson, but the index difference is in agreement. 


To estimate the systematic uncertainties of our method, four effects were considered:
\begin{itemize}
\item Tilt in the sample mount.  The optical path of a ray traveling through a tilted HWP is larger than the optical path at normal incidence;  any tilt of the HWP biases our observed indices above their true values.
\item The non-parallel beam.  The HWP is positioned in the converging \textit{f}/3.3 beam between the FTS and the detector.  Similar to the tilt effect listed above, the average non-normal incidence affects the measured indices.  This effect is corrected for, and a conservative 50\% uncertainty is assigned in the correction due to the uncertain illumination profile.
\item Polarizer misalignment.  To estimate the effect of a misalignment of the two polarizers bracketing the HWP, the fitting code was run again with mis-aligned polarizers in the physical optics model to see how the derived indices changed.
\item FTS mirror speed.  The FTS position linear encoder indicates the velocity varies at the $\pm2~\mu$m/s level for the 3~mm/sec speed used while taking the data here.  Since the peak spacing affects the indices, this couples directly to an uncertainty in the derived indices.
\end{itemize}
The systematic uncertainties associated with these effects, along with their quadrature sum, are listed in Table~\ref{systematics}.  Mirror speed is the dominant effect, and leads to a systematic uncertainty that is greater than the statistical uncertainties for our room temperature measurements, comparable to the statistical uncertainty for the 5~K measurement, and not large enough to explain the $1\%$ disagreement with previous published values.

\begin{figure}
\begin{center}
\subfigure{\includegraphics[width=0.41\textwidth]{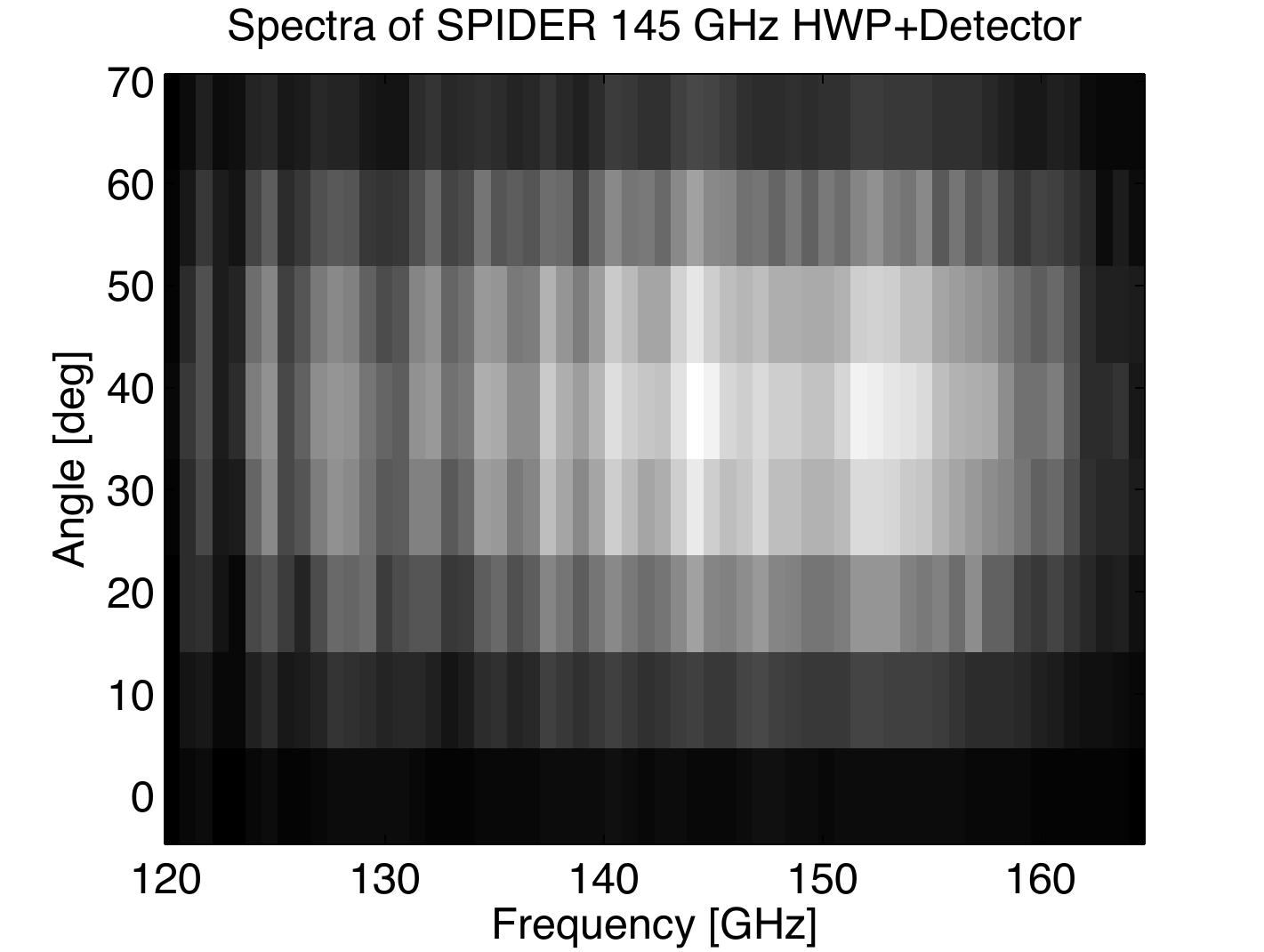}} 
\subfigure{\includegraphics[width=0.58\textwidth]{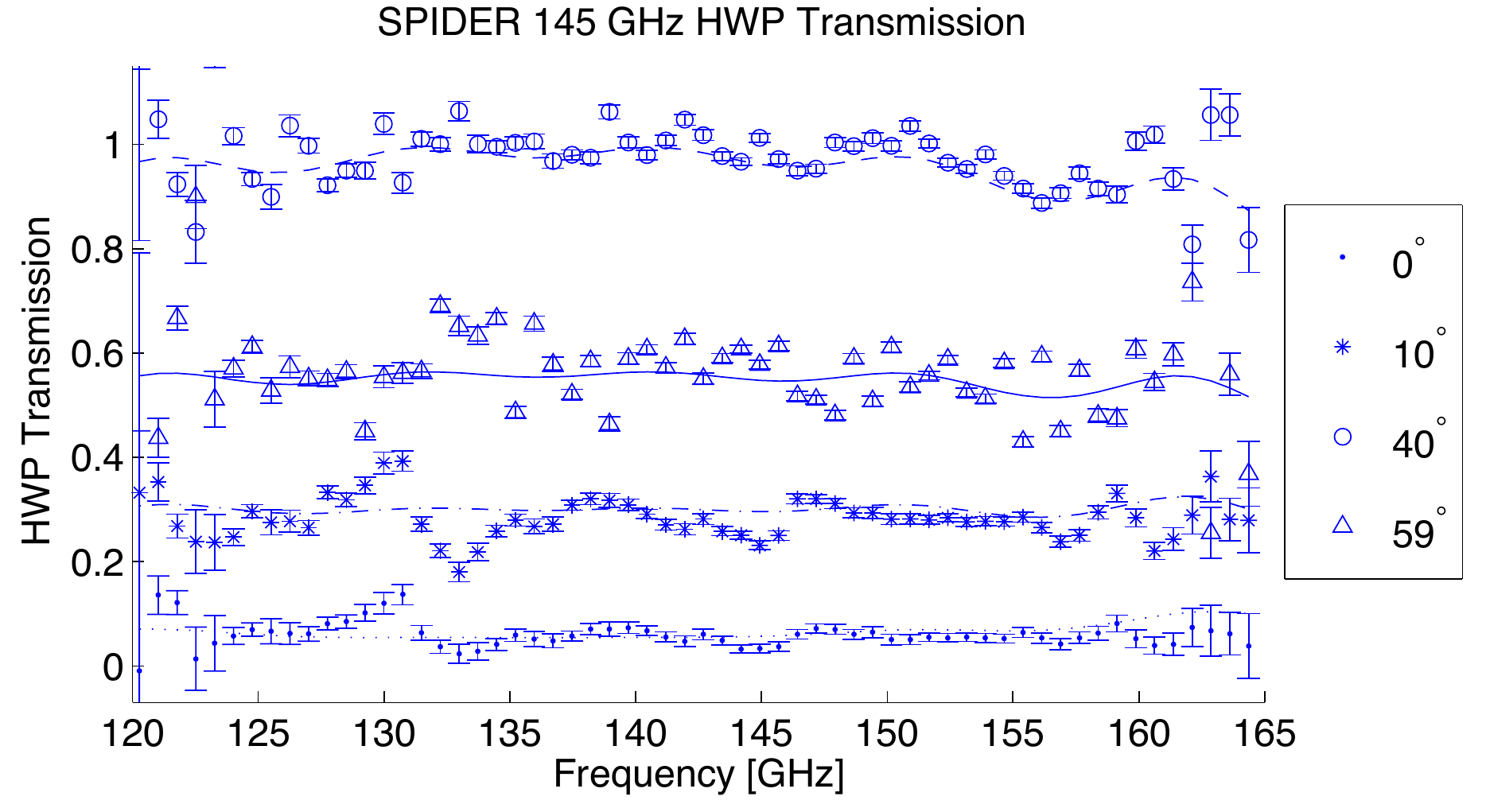}} 
\caption[Spectra taken with a single polarized detector of the cryogenic AR-coated HWP mounted in the prototype Spider receiver.]{Spectra taken with a single polarized detector of the cryogenic AR-coated HWP mounted in the prototype Spider receiver. The left panel shows the spectra of the combined detector and HWP taken at eight HWP angles between 0 and $66^\circ$. The right panel shows a HWP transmission model (smooth curves) and data scaled by an estimated detector-only spectrum. Since we did not cool down the instrument a second time to measure the detector-only spectrum, as described in the text we combined the HWP+detector spectra from all HWP angles to estimate the detector-only spectrum. \label{test_cryostat_spectra}}
\end{center}
\end{figure}

A prototype AR-coated HWP and its cryogenic rotation mechanism were mounted in a prototype Spider receiver \cite{runyan10}, and polarized FTS spectra were taken with the HWP rotated to eight angles. The spectra from one polarized detector are plotted in Figure~\ref{test_cryostat_spectra}. In this run there was only data from the combined HWP and detector. This means that in this dataset, separating out the spectral properties of each is a challenge, but can be attempted by looking for the spectrum of the response that is independent of HWP angle and assigning that to the detector. To attempt to isolate the detector response, first the entire dataset was divided by a fiducial HWP model. If this model is correct, dividing through leaves only the intrinsic response of the detector in the dataset. The indices were kept the same the same, but an MCMC was used to vary the crystal axis angle $\theta_{0}$ in the HWP model until the remaining spectrum for each angle in the dataset was similar. This allows data from the eight angles to be combined to produce an estimate of the detector response. The raw spectra are then divided by this estimated detector response to produce the estimated HWP-only spectra plotted in Figure~\ref{test_cryostat_spectra}. The observed polarization modulation agrees broadly with the fiducial model, but the scatter is larger than the noise estimate. This may be due to an incorrect estimate of the detector-only spectrum. Running the instrument in the same configuration but without the HWP would directly give a detector-only spectrum, and would allow a more straightforward determination of the in-band HWP properties. Still, these results show polarization modulation that is broadly consistent with the optics model and the broadband lab testing.

\chapter{Polarimetric Model of a Non-ideal HWP}

The material in this chapter has been published as an Applied Optics article \cite{bryan10}.

An ideal HWP with one of its crystal axes oriented at an angle $\theta_{hwp}$ to the plane of polarization of incident light rotates that polarization plane by  $2\theta_{hwp}$ as the light passes through it.  Real HWPs made from birefringent materials have several important non-idealities. Since the phase delay is only a half-wave at a single frequency, the exact angle through which a HWP rotates the polarization state is frequency-dependent.  Additionally, reflections from the material interfaces reduce transmission and induce non-ideal rotation;  to minimize these effects we use AR coatings on the surfaces of the HWP.  However,  AR coatings are frequency dependent and do not fully eliminate these non-ideal behaviors.  

These effects have been treated in a variety of ways by other authors.  O'Dea et al. \cite{odea07} and Brown et al. \cite{brown09} used a Mueller matrix formalism to parameterize a HWP and polarized detector, but did not connect their parameterization with a physical model of a HWP. Savini et al. \cite{savini06} described in detail a physical model for stacks of dielectric birefringent materials. Their model works directly with the electric fields, which allows it to handle the input and output polarization state of the light, multiple reflections from dielectric interfaces, and the finite bandwidth over which the stack is a half-wave retarder. 
Matsumura \cite{matsumura06} modeled a multiple-layer HWP using a similar approach.  We employ the methods of Savini et al. \cite{savini06} here using Jones and Mueller matrix methods to derive exact couplings from sources (of arbitrary but known spectra) on the sky to a detector, through a non-ideal HWP made of any number of dielectric layers. 

After the initial calculation of some band-averaged parameters, our Mueller matrix method is analytic and
does not require repetitive matrix multiplications or repetitive integration over frequency.  Therefore, 
modeling detector timestreams is far faster than repeated use of a Jones-formalism code. 
As an example, in 10 seconds of computer time in Matlab on a laptop, the direct Jones method can simulate 300 Spider detector samples, while the Mueller matrix method can simulate 24,000, a speedup by a factor of $\sim75$. A version of the Mueller matrix code written in C for use in the instrument simulation code for Spider can simulate 10 million detector samples in 10 seconds on the same machine.

\section{The HWP Mueller Matrix}
\label{hwp_mueller}

Our goal is to derive a model of the detector output $d$ of a HWP plus polarized-detector system for arbitrary orientations of the instrument and HWP relative to the coordinates defining the Stokes parameters of the incoming radiation, including the effects of reflections at the various dielectric interfaces and the effect of band averaging. We will accomplish this by calculating the Mueller matrices of the different parts of the system. First we will calculate the Mueller matrix of the HWP $\mathsf{M}_{HWP}$ (given in Equation~\ref{j_single_plate_hwp}). We will use rotation matrices to account for the relative orientations of the instrument, HWP, and detectors. The polarized detector is modeled as a partial linear polarizer with Mueller matrix $\mathsf{M}_{pol}$ (given in Equation~\ref{m_pol}) followed by a total power detector that is sensitive only to the $I$ Stokes parameter. The total Mueller matrix $\mathsf{M}$ is then calculated by combining the Mueller matrices of the partial polarizer, HWP, and the various rotation matrices.  The detector output model $d$ for incident light with a Stokes vector $S = [I, Q, U, V]$ is then given by summing over the top row the product of $\mathsf{M}$ and $S$~\cite{jones08},
\begin{equation} \label{eqn_one}
d = I  M_{II} + Q  M_{IQ} + U  M_{IU} + V  M_{IV} ,
\end{equation}
where the $M_{XX}$ are elements of the top row of the total Mueller matrix.

Since physical models of the action of a birefringent HWP are constructed using electric fields, we start the calculation in the Jones formalism. A Jones matrix is a 2-by-2 matrix of complex numbers that describes the action of an optical system on the $x$- and $y$-components of the electric field of an incident plane wave. A general retarder has the Jones matrix
\begin{equation}\label{j_hwp}
\mathsf{J}_{ret}(f) = \left[ \begin{array}{cc} a(f) & \epsilon_1(f)  \\ \epsilon_2(f) & b(f) e^{i \phi(f)} \end{array} \right],
\end{equation}
where 
$a(f)$, $b(f)$, and $\phi(f)$ are real, and because we chose the $xy$ coordinate system to be rotationally aligned with the optical axes, $\epsilon_1(f)$ and $\epsilon_2(f)$ are small and complex~\cite{odea07}. 

For a HWP made from a single layer of birefringent dielectric material, as we use in Spider, $x$- and $y$-polarized states defined in the crystal axis basis cannot couple into each other. This means that in the HWP Jones matrix of Equation~\ref{j_hwp}, $\epsilon_1(f)~=~\epsilon_2(f)~=~0$, so the Jones matrix of a single-plate HWP is
\begin{equation}\label{j_single_plate_hwp}
\mathsf{J}_{ret}(f) = \left[ \begin{array}{cc} a(f) & 0  \\ 0 & b(f) e^{i \phi(f)} \end{array} \right].
\end{equation}
This leaves only three parameters $a(f)$, $b(f)$ and $\phi(f)$ that are necessary to completely characterize the HWP. The transmission coefficients $a(f)$ and $b(f)$ vary with frequency because of the frequency dependence of the AR coating and the interference of multiple reflections inside the birefringent layer. The relative phase delay $\phi(f)$ also varies with frequency because the path length difference for polarization states traveling along the slow and fast crystal axes is $(n_s - n_f) d = \frac{\phi(f)}{2 \pi} \frac{c}{f}  $.

To convert this to a Mueller matrix, we follow Jones et. al.~\cite{jones08} and use
\begin{equation}\label{jones2mueller}
M_{ij} = \frac{1}{2} \mathrm{trace} ( \sigma_i \mathsf{J} \sigma_j \mathsf{J}^\dagger )
\end{equation}
from Born and Wolf \cite{born}, where $\sigma_i$ are the Pauli matrices,
\begin{eqnarray}
\begin{array}{cc}
\sigma_1 = \left[ \begin{array}{cc} 1 & 0  \\ 0 & 1 \end{array} \right] &
\sigma_2 = \left[ \begin{array}{cc} 1 & 0  \\ 0 & -1 \end{array} \right] \\
 & \\
\sigma_3 = \left[ \begin{array}{cc} 0 & 1  \\ 1 & 0 \end{array} \right] &
\sigma_4 = \left[ \begin{array}{cc} 0 & -i  \\ i & 0 \end{array} \right]. \\
\end{array}
\end{eqnarray}
This yields the Mueller matrix as a function of frequency $\mathsf{M}_{ret}(f)$ of the single-plate HWP,
\begin{eqnarray}\label{m_ret}
\mathsf{M}_{ret}(f) = \left[ \begin{array}{cccc} \frac{1}{2}(a^2 + b^2) & \frac{1}{2}(a^2 - b^2) & 0 & 0 \\ \frac{1}{2}(a^2 - b^2) & \frac{1}{2}(a^2 + b^2) & 0 & 0  \\ 0 & 0 & a b \cos(\phi) & - a b \sin(\phi) \\ 0 & 0 & a b \sin(\phi) & a b \cos(\phi) \end{array} \right],
\end{eqnarray}
where $a$, $b$, and $\phi$ are all functions of frequency. This reduces to the result given in Tinbergen \cite{tinbergen96} for an ideal retarder ($a = b = 1$). 

Since Mueller matrices can be band-averaged, we integrate this single-frequency Mueller matrix against a CMB or foreground spectrum $S(f)$, as well as the detector passband $F(f)$. This gives the band-averaged Mueller matrix
\begin{equation}\label{band_average}
\mathsf{M}_{HWP} = \frac{\int df \mathsf{M}_{ret}(f)  S(f) F(f)}{\int df S(f) F(f)}.
\end{equation}
The band-averaging means that $\mathsf{M}_{HWP}$ will not be of the same functional form as $\mathsf{M}_{ret}(f)$ and will have four (rather than three) independent non-zero elements,
\begin{equation}\label{m_hwp}
\mathsf{M}_{HWP} \equiv \left[ \begin{array}{cccc} T & \rho & 0 & 0 \\ \rho & T & 0 & 0  \\ 0 & 0 & c & -s \\ 0 & 0 & s & c \end{array} \right].
\end{equation}

To calculate the numerical values of this Mueller matrix from first principles, we use a physical optics model similar to the one in Savini et al. \cite{savini06}. The model extends the 2-by-2 matrix formalism described by Hecht and Zajac~\cite{hecht74} for modeling multiple dielectric layers of isotropic materials to a 4-by-4 matrix formalism for multiple layers of potentially birefringent materials. The model uses the electromagnetic boundary conditions at the interface between each layer of material to map the incident electric and magnetic fields onto the transmitted fields. This fully treats multiple reflections and interference effects, and can also handle lossy materials. Since the model can handle multiple layers of material, the frequency dependence of the AR coatings is also included in this calculation.

We use this model to calculate the elements of the HWP Jones matrix shown in Equation~\ref{j_single_plate_hwp} by
calculating the transmitted electric field amplitude $\vec{E}_{out}$
at a frequency $f$ with an incident electric field amplitude $\vec{E}_{in}$, for both
$x$- and $y$-polarized incident waves. When the model is run with input $x-$polarization, the resulting transmitted field is the first column of the Jones matrix in Equation~\ref{j_single_plate_hwp}
\begin{equation}\label{hwp_jones_calculation}
\left[ \begin{array}{c} a(f) \\ 0 \end{array} \right] = \vec{E}_{out} \left(\left[ \begin{array}{c} 1 \\ 0 \end{array} \right],f \right), 
\end{equation}
and running the model with input $y-$polarization yields the second column of the Jonex matrix
\begin{equation}
\left[ \begin{array}{c} 0 \\ b(f) e^{i \phi(f)} \end{array} \right] = \vec{E}_{out} \left(\left[ \begin{array}{c} 0 \\ 1 \end{array} \right],f \right).
\end{equation}
At each frequency, this Jones matrix is then converted to a Mueller matrix using Equation~\ref{jones2mueller}.

\begin{figure}
\begin{center}
\includegraphics[width=1.0\textwidth]{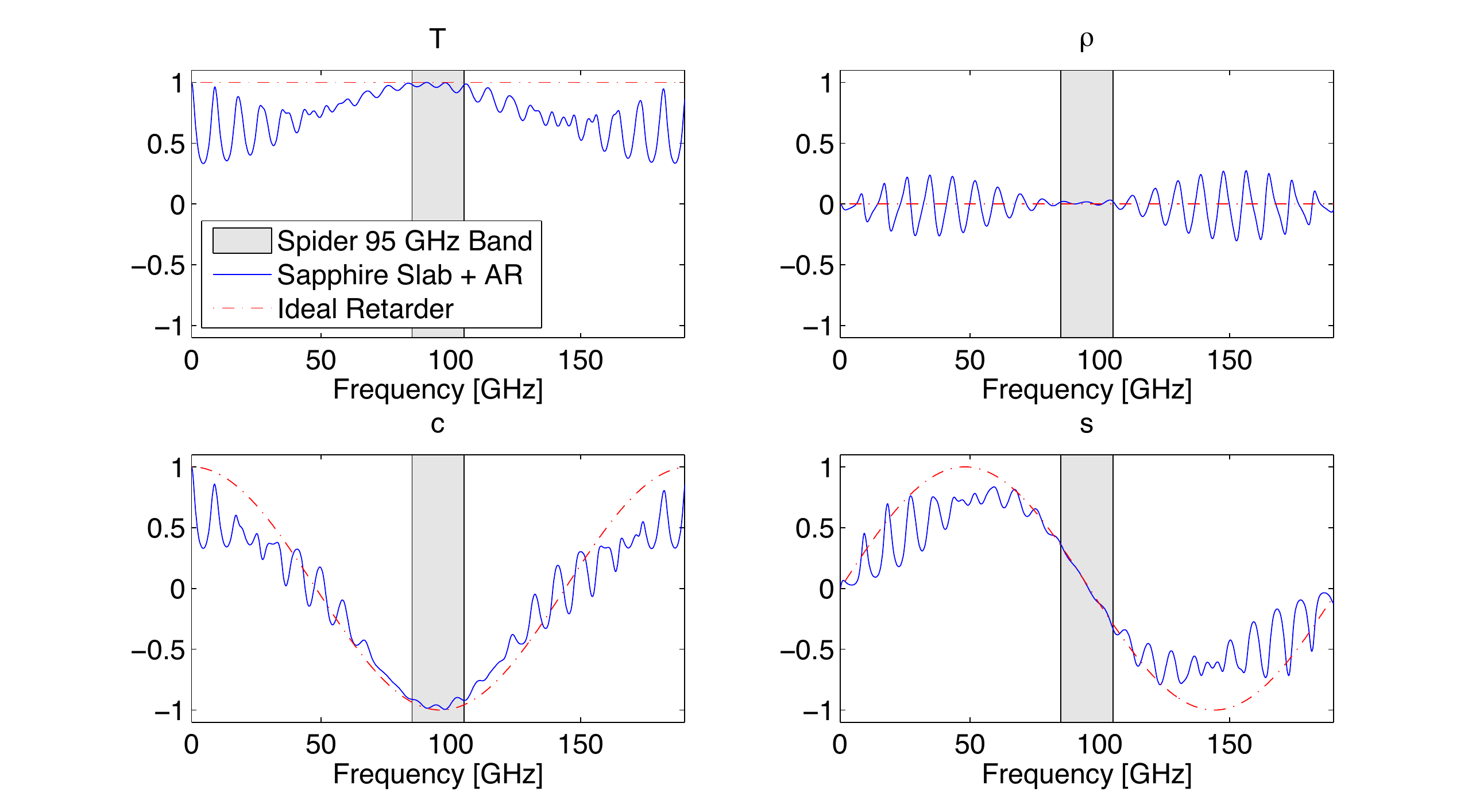}
\caption[Calculated Mueller matrix elements as a function of frequency of a Spider 95 GHz HWP. ]{Same as Figure~\ref{hwp_mueller} but for the 95 GHz HWP. \label{hwp_mueller_90}}
\end{center}
\end{figure}

For the specific case of the Spider HWPs, we used the material thicknesses in Table~\ref{thickness_150} to calculate the elements of this Mueller matrix in Equation~\ref{m_hwp} as a function of frequency, and the results are shown in Figures~\ref{hwp_mueller} and \ref{hwp_mueller_90}. Since the CMB is not expected to be circularly-polarized, in the case where the subsequent detector and optical system does not induce sensitivity to circular polarization the $s$ parameter will not be relevant for CMB polarimetry.

\begin{figure}
\begin{center}
\includegraphics[width=1.0\textwidth]{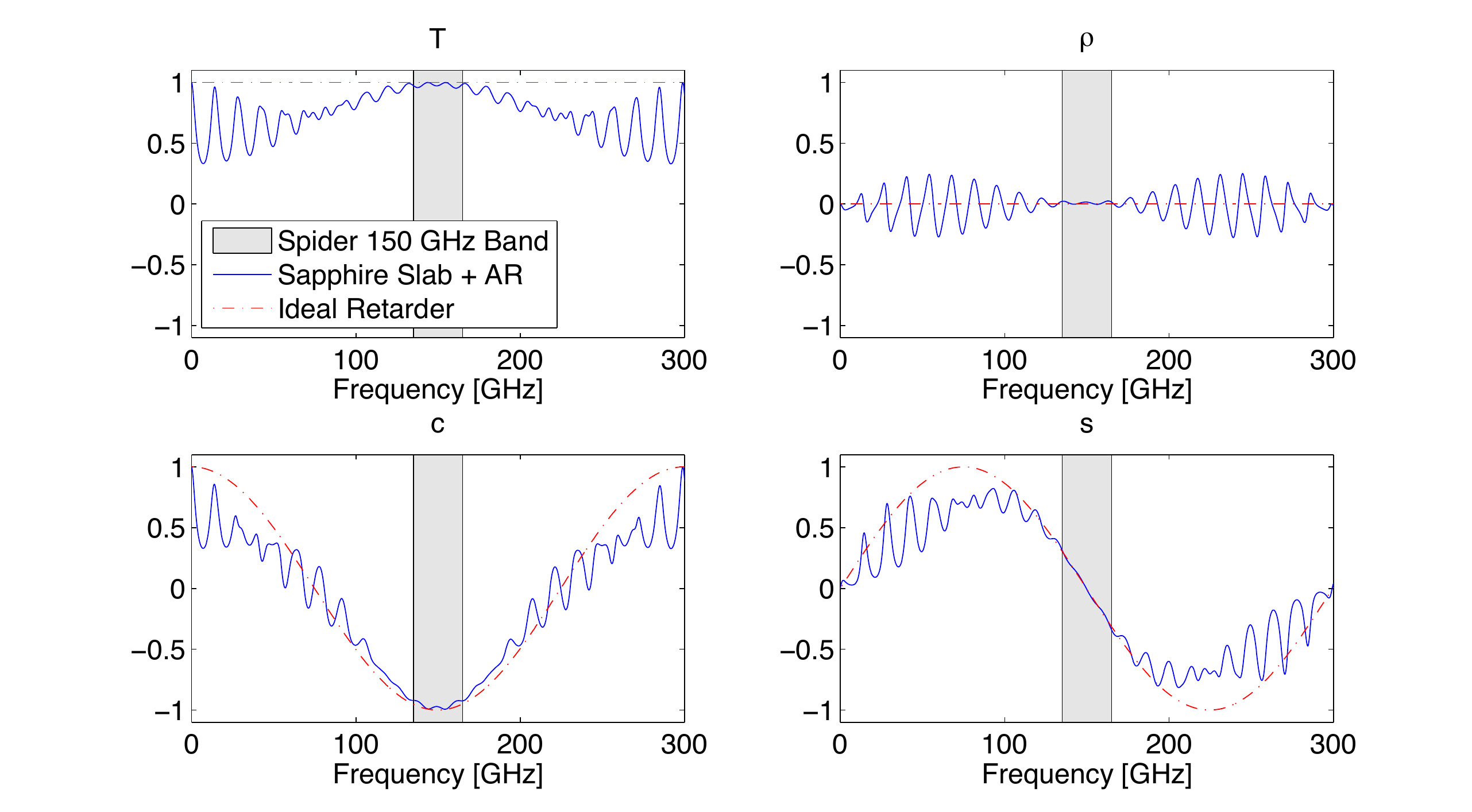}
\caption[Calculated Mueller matrix elements as a function of frequency of a Spider 150 GHz HWP. ]{Calculated Mueller matrix elements (see Equation~\ref{m_hwp} for where these elements lie in the matrix) as a function of frequency of a Spider 150 GHz HWP (solid blue lines). The dot-dash red lines show the Mueller matrix elements of a HWP retarder with a perfect AR coating for comparison. The grey box shows the Spider 150 GHz band. \label{hwp_mueller}}
\end{center}
\end{figure}

We then use Equation~\ref{band_average} to calculate the band-averaged matrix elements of the Spider HWPs, 
assuming a top-hat detector spectrum from 135 GHz to 165 GHz. For the source spectra we use
\begin{align}
S(f) \propto 
\begin{cases} 
  1 & \mathrm{Flat} \\
  \frac{dB}{dT} (f,2.725~\mathrm{K})  & \textrm{CMB~\cite{fixsen03}} \\
  f^{1.67} B(f,9.6~\mathrm{K}) \\
  ~~~+ 0.0935 f^{2.7} B(f,16.2~\mathrm{K}) & \textrm{Dust~\cite{finkbeiner99}} \\
  f^{-1} & \textrm{Synchrotron~\cite{bennett03}} \\
  f^{-.14} & \textrm{Free-free~\cite{oster61}},
\end{cases}
\end{align}
where $B(f,T)$ is the blackbody function, 
as estimates of the CMB and astrophysical foregrounds.  The results are shown in Tables~\ref{hwp_params} and \ref{hwp_params_90}. 

\begin{table}
\begin{center}
\begin{tabular}{ c | c | c | c | c|}
 & $T$ & $\mathbf{\rho}$ & $c$  & $s$\\
\hline
\hline
\textbf{Flat} & 0.97688 & 0.00958 & -0.96001 & -0.00639 \\
\hline
\textbf{CMB} & 0.97683& 0.00962 & -0.95992 & -0.01622 \\
\hline
\textbf{Dust} & 0.97651 & 0.00984 & -0.95915 & -0.04149 \\
\hline
\textbf{Synchrotron} & 0.97693 & 0.00953 & -0.96010 & \phantom{-}0.00395\\
\hline
\textbf{Free-free} & 0.97689 & 0.00958 & -0.96002 & -0.00494 \\
\hline
\textrm{(Ideal HWP)} & 1 & 0 & -1 & 0\\
\hline
\end{tabular}
\caption[Calculated Mueller matrix elements for the Spider 150 GHz HWPs.]{Calculated Mueller matrix elements for the Spider 150 GHz HWPs. The first row shows the HWP parameters averaged within the Spider 150 GHz passband. The CMB, Dust, Synchrotron, and Free-free rows all are band-averaged against the source spectra within the passband. The last row shows the parameter values of an ideal HWP for comparison. \label{hwp_params}}
\end{center}
\end{table}

\begin{table}
\begin{center}
\begin{tabular}{ c | c | c | c | c|}
 & $T$ & $\mathbf{\rho}$ & $c$  & $s$\\
\hline
\hline
\textbf{Flat} & 0.96784 & 0.01021 & -0.94920 & \phantom{-}0.02756 \\
\hline
\textbf{CMB} & 0.96667 & 0.01038 & -0.94851 & \phantom{-}0.01013 \\
\hline
\textbf{Dust} & 0.96509 & 0.01057& -0.94730 & -0.01156 \\
\hline
\textbf{Synchrotron} & 0.96857 & 0.01028 & -0.94956 & \phantom{-}0.03884\\
\hline
\textbf{Free-free} & 0.96794 & 0.01030 & -0.94925 & \phantom{-}0.02914 \\
\hline
\textrm{(Ideal HWP)} & 1 & 0 & -1 & 0\\
\hline
\end{tabular}
\caption[Calculated Mueller matrix elements for the Spider 95 GHz HWPs.]{Same as Table~\ref{hwp_params} but for the 95 GHz HWP. \label{hwp_params_90}}
\end{center}
\end{table}

In addition to being large amplitude, the $s$ parameter depends significantly on which source spectrum is used;  this is not surprising given the form of $s$ across the band, shown in Figure~\ref{hwp_params}, which shows that tailoring $s$ to be near zero requires near symmetric placement and weighting across the band, since $s$ is large but asymmetric.

The variations of $T$, $\rho$, and $c$ between the CMB and foreground sources are not as worrisome.  The parameters vary among the different sources at the $0.1\%$ level. For comparison, Fraisse et al. \cite{fraisse13} shows that to reach our target B-mode sensitivity at the largest angular scales in our map, the goal for foreground subtraction is to reduce it by roughly $95\%$-$99\%$. Since the HWP response varies among the different source spectra at the $0.1\%$ level, the HWP performance is more than sufficient for Spider to analyze its data using HWP parameters that do not depend on source spectra. For experiments targeting $r \ll 0.01$, the spectral dependence of the HWP may impact their foreground removal, and they may need a more sophisticated approach.

\section{Rotating the Instrument and HWP}

Jones et. al.~\cite{jones08} modeled a polarization-sensitive detector as a rotatable instrument with a partial-polarizer followed by a total power detector.   The Jones matrix of a vertical partial polarizer is
\begin{equation}\label{j_pol_a}
\mathsf{J}_{pol} = \left[ \begin{array}{cc} \eta & 0  \\ 0 & \delta \end{array} \right],
\end{equation}
which can be turned into a corresponding Mueller matrix
\begin{eqnarray}\label{m_pol}
\mathsf{M}_{pol} = \left[ \begin{array}{cccc} \frac{1}{2}(\eta^2 + \delta^2) & \frac{1}{2}(\eta^2 - \delta^2) & 0 & 0 \\ \frac{1}{2}(\eta^2 - \delta^2) & \frac{1}{2}(\eta^2 + \delta^2) & 0 & 0  \\ 0 & 0 & \eta \delta & 0 \\ 0 & 0 & 0 & \eta \delta \end{array} \right].
\end{eqnarray}
For an ideal polarized detector, $\eta = 1$ and $\delta = 0$. For a detector with crosspol, $\delta$ will be nonzero. In Spider we expect roughly $1\%$ crosspol in power. Since the Jones matrix formalism works with electric fields not power, this means we expect $\delta = 0.1$. The requirement $\eta^2 + \delta^2 = 1$ yields $\eta = 0.995$.  Given the instrument rotation matrix~\cite{tinbergen96}
\begin{equation}
\mathsf{M}_{\psi} = \left[ \begin{array}{cccc} 1 & 0 & 0 & 0 \\ 0 & \textcolor{white}{-}\cos{2 \psi_{inst}} & \sin{2 \psi_{inst}} & 0  \\ 0 & -\sin{2 \psi_{inst}} & \cos{2 \psi_{inst}} & 0 \\ 0 & 0 & 0 & 1 \end{array} \right],
\end{equation}
the detected radiation is given by the coupling to $I$ in the product $\mathsf{M}_{pol}~\mathsf{M}_{\psi}$;  for an 
ideal polarizer and an arbitrary instrument angle $\psi_{inst}$, the detector signal is
\begin{equation}
d = \frac{1}{2} \left(I + Q \cos(2 \psi_{inst}) + U \sin(2 \psi_{inst}) \right).
\end{equation}

\begin{figure}
\begin{center}
\includegraphics[width=0.65\textwidth]{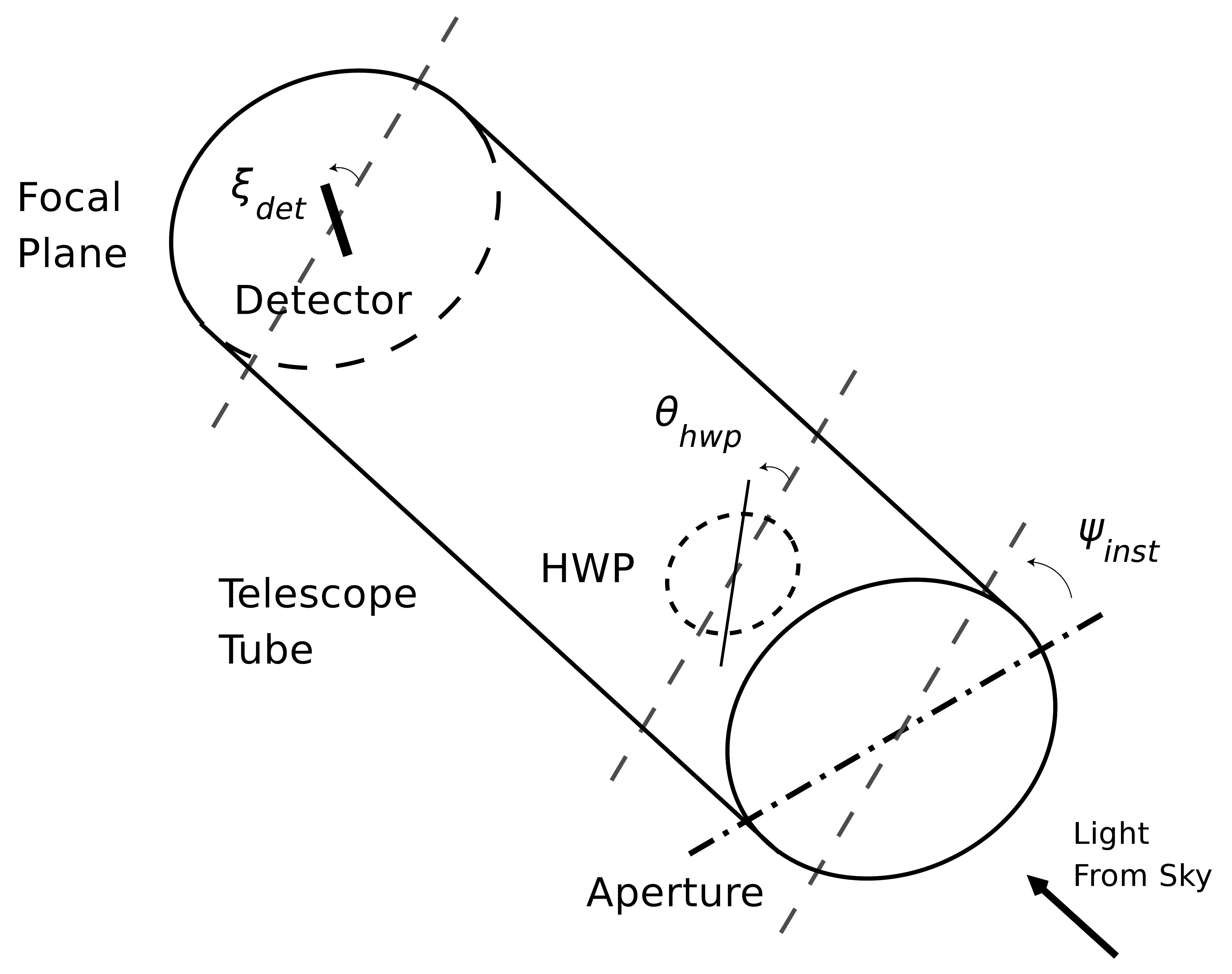}
\caption[Our definitions of the detector, HWP, and instrument angles.]{Our definitions of the detector, HWP, and instrument angles. The instrument angle is defined relative to a fixed reference on the sky that determines the absolute orientations of $Q$ and $U$, and the detector and HWP angles are defined relative to the instrument. \label{instrument_and_hwp_angles}}
\end{center}
\end{figure}

The addition of a HWP to the instrument can be modeled by a product of the Mueller matrices of the HWP, detector, and rotation matrices for the relative orientation of the HWP to the instrument and detector. Figure~\ref{instrument_and_hwp_angles} illustrates the definition of the angles used for the rotation matrices;  while only two angles are needed to define the single-detector problem, we use three here to make the problem more straightforward to visualize, and to easily accommodate calculations of focal planes with detectors at multiple orientation angles.  Here we consider only a single detector with $\xi_{det} = 0$. The matrix product that models the HWP, detector, and their orientations to the instrument is therefore
\begin{equation}\label{mat_multiply}
\mathsf{M} = \mathsf{M}_{pol}~\mathsf{M}_\xi~\mathsf{M}_{-\theta}~\mathsf{M}_{HWP}~\mathsf{M}_{\theta}~\mathsf{M}_{\psi},
\end{equation}
where $\mathsf{M}_{\theta}$ is the rotation matrix by the HWP angle $\theta_{hwp}$, and $\mathsf{M}_{\xi}$ is the rotation matrix by the detector orientation angle $\xi_{det}$.

The top row of the resulting Mueller matrix for a single-plate HWP is
\begin{eqnarray}\label{hwp_matrix_elements}
M_{II} &=& \frac{1}{2} \left[ T (\eta^2 + \delta^2) + \rho \cos(2 \theta_{hwp}) (\eta^2 - \delta^2) \right] \nonumber \label{start_of_big_equation} \\
M_{IQ} &=&  \textcolor{white}{-}\mathcal{F} \sin(2 \psi_{inst})  +  \mathcal{G} \cos(2 \psi_{inst}) \nonumber \\
M_{IU} &=&  - \mathcal{F} \cos(2 \psi_{inst})  + \mathcal{G} \sin(2 \psi_{inst}) \nonumber \\
M_{IV} &=& \frac{1}{2} s \sin(2 \theta_{hwp}) (\eta^2 - \delta^2),
\end{eqnarray}
where $\mathcal{F}$ and $\mathcal{G}$ are
\begin{eqnarray}
\mathcal{F} &\equiv&   - \frac{1}{4} (T-c) \sin(4 \theta_{hwp}) (\eta^2 - \delta^2) \nonumber \\
&-& \frac{1}{2} \rho \sin(2 \theta_{hwp})(\eta^2 + \delta^2) \\
\mathcal{G} &\equiv& \frac{1}{4} \left[T + c + (T-c) \cos(4 \theta_{hwp}) \right] (\eta^2 - \delta^2) \nonumber \\
&+& \frac{1}{2} \rho \cos(2 \theta_{hwp}) (\eta^2 + \delta^2). \label{end_of_big_equation}
\end{eqnarray}
Nominal calculated values for $T$, $\rho$, $c$, and $s$ are given in Tables~\ref{hwp_params} and \ref{hwp_params_90}. This formula reduces to the result given in Jones et. al. \cite{jones08} in the limit of no HWP ($T = c = 1,~\rho = s = 0$), and in the case of an ideal HWP ($T=1,~c=-1,~\rho = s = 0$) reduces to
\begin{eqnarray}
M^{ideal}_{II} &=& \frac{1}{2} (\eta^2 + \delta^2) \nonumber \\
M^{ideal}_{IQ} &=&   \frac{1}{2} \cos(2 ( \psi_{inst} + 2 \theta_{hwp})) (\eta^2 - \delta^2) \nonumber \\
M^{ideal}_{IU} &=&  \frac{1}{2} \sin(2 ( \psi_{inst} + 2 \theta_{hwp})) (\eta^2 - \delta^2) \nonumber \\
M^{ideal}_{IV} &=& 0.
\end{eqnarray}

Armed with the band-averaged Mueller matrix representations of the HWP plus detector system given by Equation~\ref{mat_multiply}, we can use Equation~\ref{eqn_one} to calculate the detector output as a function of input Stokes parameters.


\begin{figure}
\begin{center}
\subfigure{\includegraphics[width=0.48\textwidth]{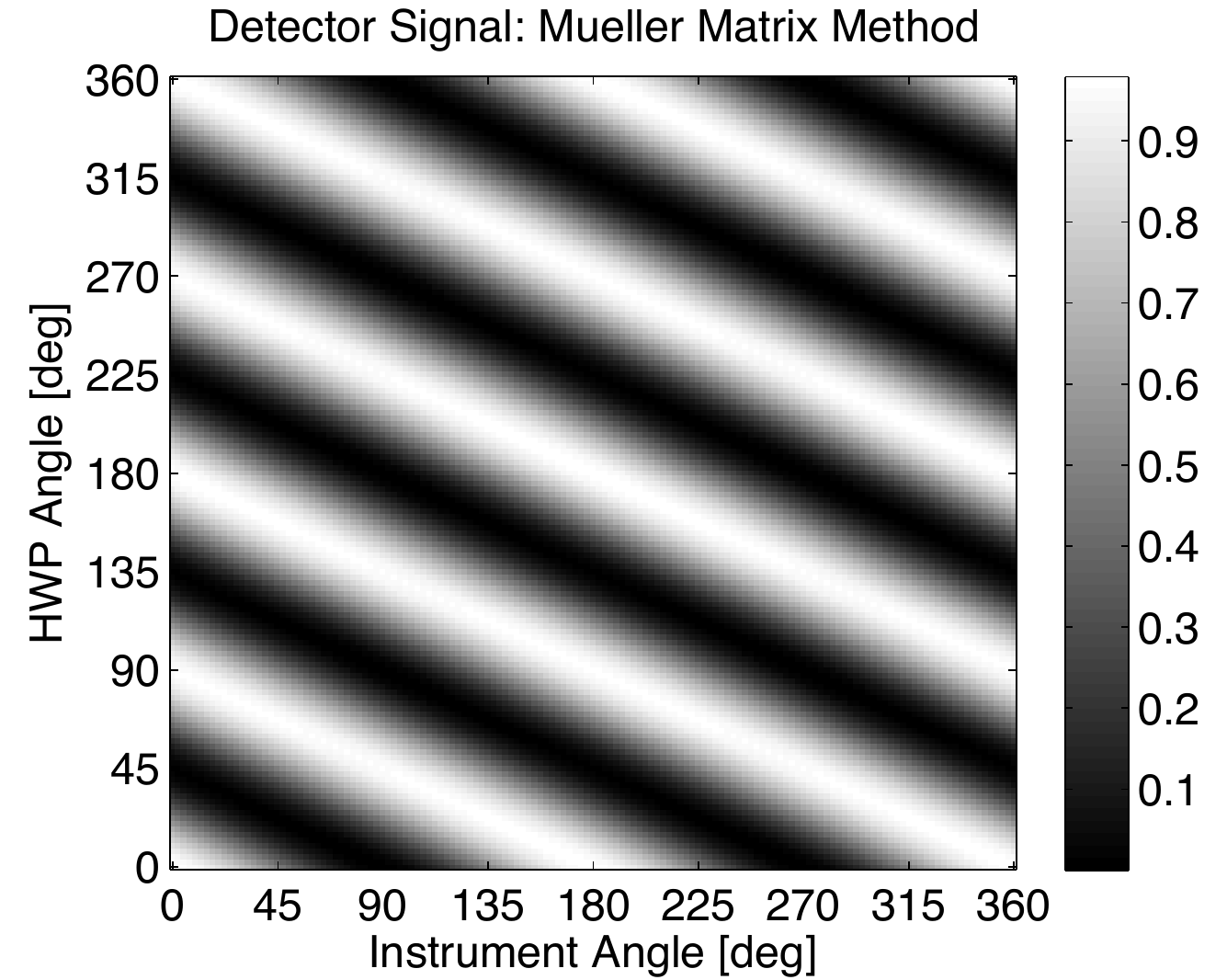}} 
\subfigure{\includegraphics[width=0.51\textwidth]{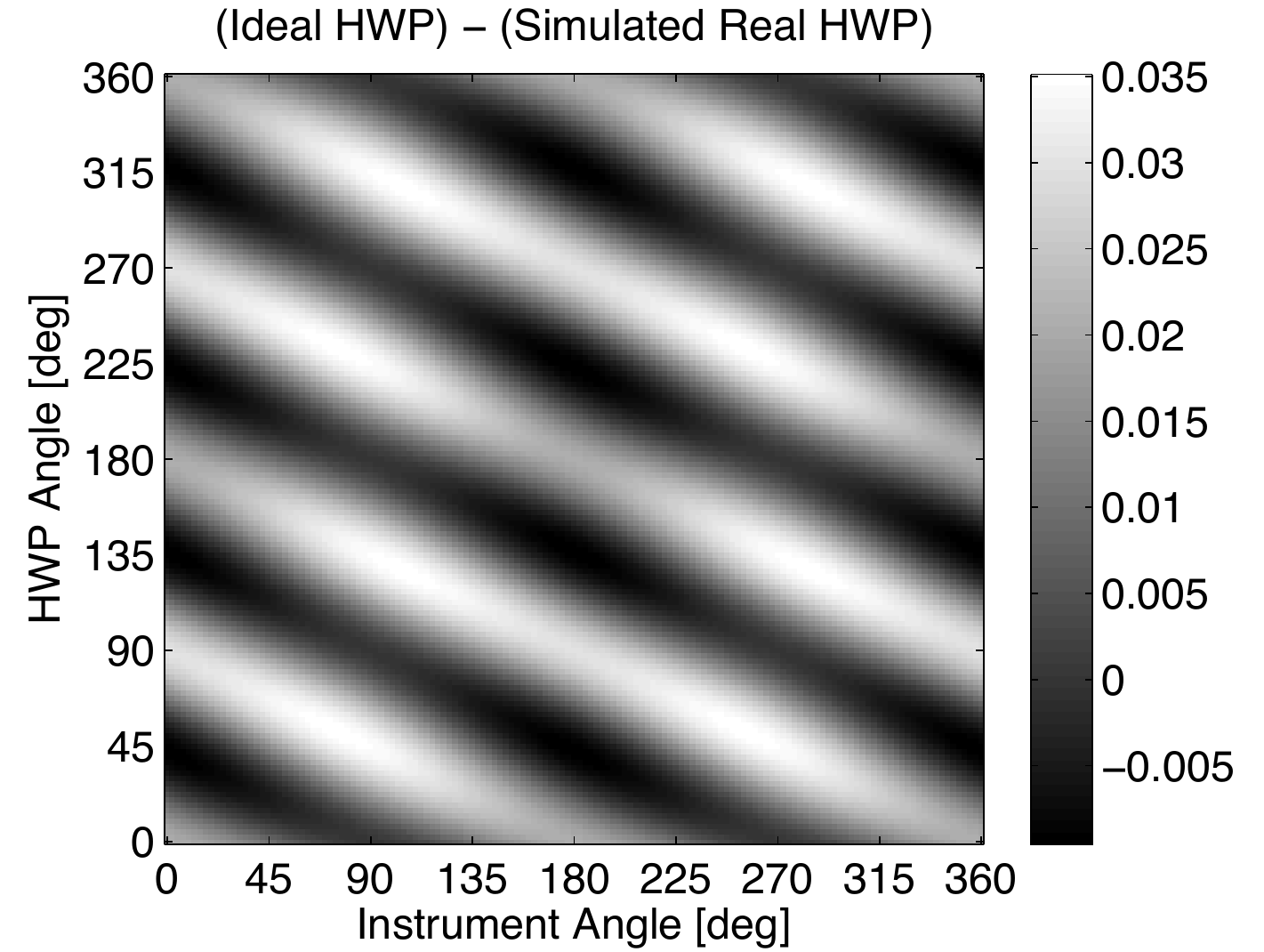}} 
\caption[Simulated detector output for the Spider instrument as a function of HWP and instrument angle for a $Q$-polarized CMB $dB/dT$ source.]{Simulated detector output for the Spider instrument as a function of HWP and instrument angle for a $Q$-polarized CMB $dB/dT$ source. The left panel shows the detector output calculated for a Spider 150 GHz HWP using our band-averaged Mueller matrix formalism.
The right panel shows the percent-level differences between an ideal HWP and the simulated non-ideal HWP. 
\label{cmb_sim}}
\end{center}
\end{figure}

An example of the output from our model, along with the percent-level differences between that and a naiive treatment, is shown in Figure~\ref{cmb_sim} for the specific case of the Spider 150 GHz HWP.  

\section{Detector Pair Summing and Differencing}

Many CMB polarization experiments, including Spider, use pairs of Polarization-Sensitive Bolometers (PSBs) located at the same point on the focal plane \cite{jones02}. Both detectors in a pair view the same patch of the sky, but detector B is oriented at $90^\circ$ with respect to detector A. Detector differencing within a pair reduces common-mode noise while retaining sensitivity 
to linear polarization. The sum of a pair measures the $I$ Stokes parameter. To form the sum and differences, the gain of each detector needs to be measured and corrected for. In Spider, the absolute calibration of the instrument will be obtained by measuring temperature fluctuations, and cross-correlating with previous measurements. Temporal variations of each detector's gain will be monitored by periodically injecting small electrical bias steps, allowing these gain drifts to be corrected for post-flight. Here, for simplicity we assume that the absolute calibration of the experiment only needs to be obtained once, and the bias steps will be used to track gain variations over the observations at all HWP angles. This means that in all the detector differencing formulas that follow, no gain terms need to be explicitly included. On the other hand, the situation would be more complicated if it turned out that it was necessary to repeat the absolute calibration of the experiment more frequently, say once per HWP angle. Because the $M_{II}$ term in Equation~\ref{hwp_matrix_elements} depends on HWP angle, this calibration process itself would yield slightly different results depending on the HWP angle. Percent level correction factors would need to be introduced to treat this effect in the model.

We take Equation~\ref{eqn_one} with $\mathsf{M}$ given by Equation~\ref{mat_multiply} with $\xi_{det} = 0$ as a model for the A detector timestream $d_i^A$. The B detector timestream $d_i^B$ can be similarly calculated by setting
$\xi_{det} = 90^\circ$;  we note however that the Jones matrix of a horizontal polarizer,
\begin{equation}\label{j_pol_b}
\mathsf{J}_{pol}^B = \left[ \begin{array}{cc} \delta & 0  \\ 0 & \eta \end{array} \right],
\end{equation}
is related to that of a vertical polarizer (Equation~\ref{j_pol_a}) via the substitutions
$\eta \rightarrow \delta$ and $\delta \rightarrow \eta$.  We can thus just make those substitutions in Equations~\ref{start_of_big_equation} through \ref{end_of_big_equation} to find a model for the B detector timestream. We then construct the sum and difference timestreams
\begin{eqnarray}
d_i^{sum} &\equiv&  d^A_i + d^B_i , \nonumber\\
d_i^{diff} &\equiv&  d^A_i - d^B_i.
\end{eqnarray}
Only  $(\delta^2 + \eta^2)$ terms will remain in the sum timestream and only $(\delta^2 - \eta^2)$ terms will remain in the difference timestream. The matrix elements for the sum timestream with a single-plate HWP are therefore
\begin{eqnarray}
M_{II}^{sum} &=& T (\eta^2 + \delta^2) \nonumber \\
M_{IQ}^{sum}  &=&  \rho \cos(2 (\psi_{inst} + \theta_{hwp}) ) (\eta^2 + \delta^2) \nonumber \\
M_{IU}^{sum}  &=&  \rho \sin(2 (\psi_{inst} + \theta_{hwp}) ) (\eta^2 + \delta^2) \nonumber \\
M_{IV}^{sum}  &=& 0.
\end{eqnarray}
Even with the non-idealities of the HWP, the sum timestream coupling to intensity is independent of HWP angle. There is also a small coupling to linear polarization. The matrix elements for the difference timestream are
\begin{eqnarray}
M_{II}^{diff} &=& \rho \cos(2 \theta_{hwp}) (\eta^2 - \delta^2) \nonumber \\
M_{IQ}^{diff}  &=&  \textcolor{white}{-}\left[ \mathcal{F}'   \sin(2 \psi_{inst}) + \mathcal{G}'  \cos(2 \psi_{inst}) \right] (\eta^2 - \delta^2) \nonumber \\
M_{IU}^{diff}  &=&  \left[-\mathcal{F}' \cos(2 \psi_{inst}) + \mathcal{G}'\sin(2 \psi_{inst}) \right] (\eta^2 - \delta^2) \nonumber \\
M_{IV}^{diff}  &=& s \sin(2 \theta_{hwp}) (\eta^2 - \delta^2) ,
\end{eqnarray}
where $\mathcal{F}'$ and $\mathcal{G}'$ are
\begin{eqnarray}
\mathcal{F}' &\equiv&    -\frac{1}{2} (T-c) \sin(4 \theta_{hwp}) \\
\mathcal{G}' &\equiv& \frac{1}{2} \left[T + c + (T-c) \cos(4 \theta_{hwp}) \right],
\end{eqnarray}
and nominal calculated values for $T$, $\rho$, $c$, and $s$ are given in Tables~\ref{hwp_params} and \ref{hwp_params_90}. The difference timestream unfortunately has small couplings to intensity and circular polarization. Note that for both the sum and difference timestreams, detector cross-polarization only shows up as an overall factor of $(\eta^2~\pm~\delta^2)$, and will not result in leakage between the estimates of $Q$ and $U$.




\chapter{Modeling Polarized Ghosting}\label{ghosting-chapter}

\label{ghost_chapter}

In Spider, reflections from the HWP and/or filters reflect again from the focal plane to create a $\sim$percent level coupling of the detector to another part of the sky, called a ghost beam. Simulations suggest that corrupting the polarized signal with a ghost beam with a temperature coupling at the $10\%$ level would be a systematic error that would just start to contaminate the B-mode signal. Measurements (presented in Chapter~\ref{texas_results}) show that the ghost coupling is at the few percent level, which is already good enough for Spider. Still, to better understand this potential problem, we developed a polarimetric model of ghosting generated both by the HWP and by unpolarized reflections from a filter. Our model shows that by combining maps made at several selected HWP angles, the temperature coupling of the ghost caused by the HWP itself exactly cancels out. Ghosting caused by filters is also calculated to have minimal impact on the final maps in Spider.

\section{Modeling the Ghost Beam}

\begin{figure}
\begin{center}
\includegraphics[width=0.80\textwidth]{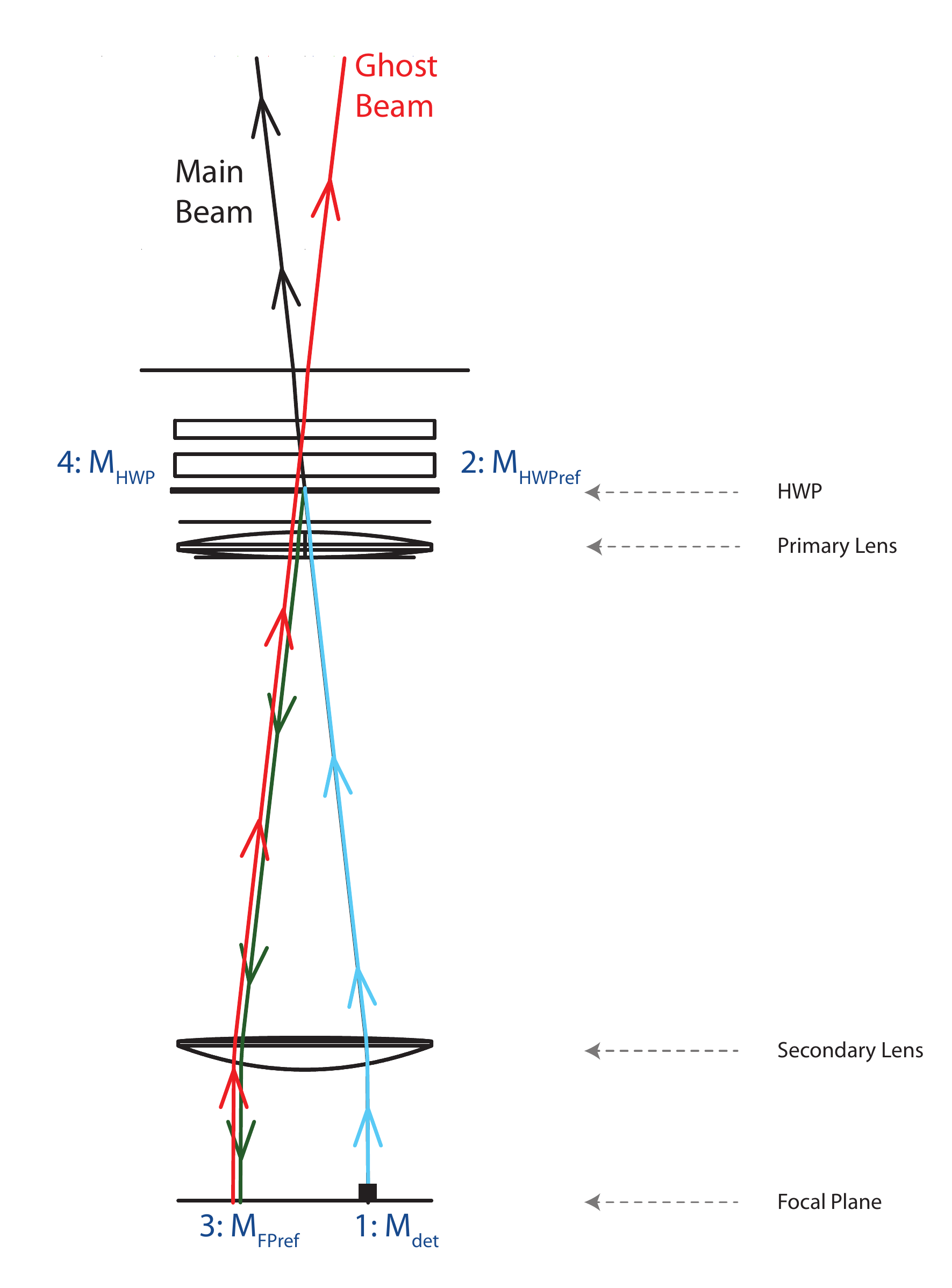}
\caption[Sketch illustrating the prime ray of the ghost beam.]{Sketch illustrating the prime ray of the ghost beam of an off-axis detector. Thinking in broadcast mode, the detector's sensitivity is launched from the focal plane (cyan ray). Most of this sensitivity continues through the HWP and goes out to the sky forming the main beam (black ray). However, some of this sensitivity is reflected from the HWP (green ray), bounces off the focal plane (red ray, shifted to the left for clarity), and leaves the telescope as a second beam of sensitivity called the ghost beam. Each of these interactions has a corresponding Mueller matrix to model the polarization of ghost beam. \label{ghost_ray}}
\end{center}
\end{figure}

The ray picture of how a ghost beam forms is shown in Figure~\ref{ghost_ray}. Thinking in broadcast mode, a percentage of the power in the parallel ray bundle at the HWP is sent back through the telescope and comes to a focus again at the point on the focal plane diametrically opposite from the detector that launched the rays. After a large fraction of that power reflects from the focal plane, this cone gets reformed by the telescope back into a parallel bundle and goes out to the sky. This means that in the ray picture, the ghost beam is the same size as the main beam.

As the main beam modeling in Chapter~\ref{zemax_chapter} indicates, the ray picture does not model all of the effects of the main beam. Ideally a Zemax model with multiple reflections would be used to calculate the electric field as it undergoes multiple reflections, but Zemax does not have this capability. Instead, to get a qualitative sense of the electric field distribution of the ghost beam, we use the Gaussian beam formalism. First we multiply ray transfer matrices to project the detector beam up to the HWP. Then, in reverse order, we multiply out the ray transfer matrices to go back to the focal plane. Here it is important to note that the lens surfaces are in reverse order and have reversed signs for their curvature radii. Then, we propagate the Gaussian beam using the usual ray transfer matrices of the entire optical system. The results of this calculation are shown in Table~\ref{ghosting_gaussbeam}. The far-field width of the ghosted Gaussian beam and the unghosted Gaussian beam are calculated to be different by roughly a factor of two. Far-field beam measurements in Spider show that the ghost and main beam widths are more similar than this. Also, the main beam width calculated with Gaussian beams is different from measurements and Zemax. This shows that the Gaussian beam approach is only qualitatively accurate for Spider.

\begin{table}
\begin{center}
\noindent\makebox[\textwidth]{%
\footnotesize
\begin{tabular}{c|c|c|c|c|c|c|}
 & \textbf{ } & \textbf{ } & \textbf{Back at} & \textbf{Vacuum} & & \textit{Unghosted Beam at}\\
 & \textbf{Focal Plane} & \textbf{HWP} & \textbf{Focal Plane} & \textbf{Window} & & \textit{Vacuum Window} \\ \hline \hline
$w$  & 3.51 mm & \phantom{31,}106.99 mm & \phantom{-3}9.15 mm & \phantom{10,}113.77 mm & & \phantom{31,}109.11 mm  \\ \hline
$R$  & $\infty$  & 14,645.72 mm & -54.56 mm & \phantom{1}5,231.00 mm & & 14,748.86 mm  \\ \hline \hline
$w_0$  & 3.51 mm & \phantom{31,1}67.56 mm & \phantom{3-}3.51 mm & \phantom{10,1}28.35 mm & & \phantom{31,1}67.57 mm  \\ \hline
$z$  & 0\phantom{.00} mm  & \phantom{1}8,804.42 mm & -46.55 mm & \phantom{1}4,906.24 mm & & \phantom{1}9,092.78 mm  \\ \hline
$\theta_{FWHM}$  & $24.47^\circ$\phantom{000..}  & \phantom{1.}$1.27^\circ$ & $24.47^\circ$\phantom{0..} &  \phantom{1.}$3.03^\circ$  & & \phantom{1.}$1.27^\circ$  \\ \hline
\end{tabular}}
\caption[Calculated Gaussian beam widths and curvature $R$ for a ghost beam formed by a reflection from the HWP at 150 GHz.]{Calculated Gaussian beam widths and curvature $R$ for a ghost beam formed by a reflection from the HWP at 150 GHz. The beam waist $w_0$ and the distance $z$ of the waist from that surface are also shown. In this calculation, the ghosted and unghosted beams have far-field expansion angles that differ by roughly a factor of two. This contrasts somewhat with the measured far-field beam patterns, which are similar in size to the main beam. Also, the unghosted beam width both from measurements and Zemax is $0.5^\circ$, showing the limitations of the Gaussian beam model for Spider. \label{ghosting_gaussbeam}}
\end{center}
\end{table}

The polarization action of each of the interactions shown in Figure~\ref{ghost_ray} that form the ghost can be represented as a Mueller matrix. Multiplying them together yields the end-to-end Mueller matrix $\mathsf{M_A}$ that couples the ghost beam to the A detector timestream.
\begin{equation}\label{M_A_eqn}
\mathsf{M_A} = \mathsf{M}_{pol}~\mathsf{M}_{\theta}~\mathsf{M}_{ref\_HWP}~\mathsf{M}_{-\theta}~\mathsf{M}_{ref\_FP}~\mathsf{M}_{-\theta}~\mathsf{M}_{HWP}~\mathsf{M}_{\theta}~\mathsf{M}_{\psi}
\end{equation}
Here $\mathsf{M}_{HWP}$ is the HWP transmission from Equation~\ref{m_hwp}
\begin{equation}
\mathsf{M}_{HWP} \equiv \left[ \begin{array}{cccc} T & \rho & 0 & 0 \\ \rho & T & 0 & 0  \\ 0 & 0 & c & -s \\ 0 & 0 & s & c \end{array} \right],
\end{equation}
with nominal values of $T = .97683$, $\rho = .00962$, $c = -.95992$, and $s = -.01622$ for a CMB source and a Spider 150 GHz HWP.
$\mathsf{M}_{ref\_FP}$ is the focal plane reflection
\begin{equation}\label{m_fp_ref}
\mathsf{M}_{ref\_FP} = \left[ \begin{array}{cccc} R_{FP} & 0 & 0 & 0 \\ 0 & R_{FP} & 0 & 0  \\ 0 & 0 & -R_{FP} & 0 \\ 0 & 0 & 0 & -R_{FP} \end{array} \right].
\end{equation}
Here we somewhat conservatively assume $R_{FP} = 0.5$. Here we also assume that the reflection from the focal plane is due to conductive structures of the detector feeds and/or the conductive niobium ground plane. As shown in \cite{fymat71}, a reflection from a metallic surface reflects the coordinate system without modulating the electric field vector, which has the effect of flipping the sign of $U$ and $V$ in the reflected coordinate system. $\mathsf{M}_{\psi}$ and $\mathsf{M}_{\theta}$ are the instrument and HWP rotation matrices, and $\mathsf{M}_{pol}$ is the A-detector Mueller matrix. The Jones matrix of an A-polarized detector is
\begin{equation}\label{j_pol_a}
\mathsf{J}_{pol} = \left[ \begin{array}{cc} \eta & 0  \\ 0 & \delta \end{array} \right],
\end{equation}
with $\eta \sim 1$ and $\delta \sim 0$. The corresponding Mueller matrix is
\begin{eqnarray}\label{m_pol}
\mathsf{M}_{pol} = \left[ \begin{array}{cccc} \frac{1}{2}(\eta^2 + \delta^2) & \frac{1}{2}(\eta^2 - \delta^2) & 0 & 0 \\ \frac{1}{2}(\eta^2 - \delta^2) & \frac{1}{2}(\eta^2 + \delta^2) & 0 & 0  \\ 0 & 0 & \eta \delta & 0 \\ 0 & 0 & 0 & \eta \delta \end{array} \right].
\end{eqnarray}

\begin{figure}
\begin{center}
\includegraphics[width=1.0\textwidth]{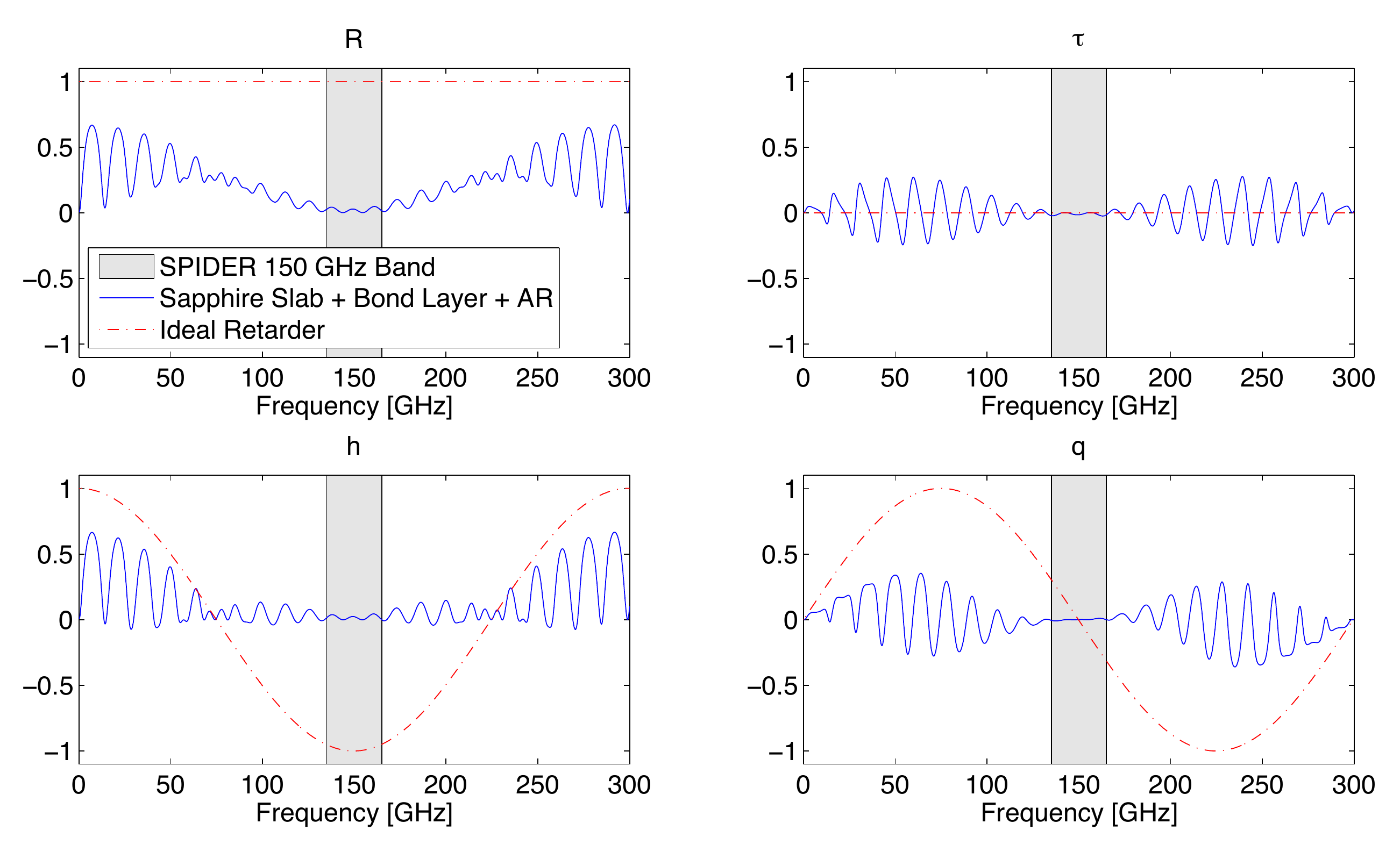}
\caption[HWP reflection Mueller matrix elements across the 150 GHz passband.]{HWP reflection Mueller matrix elements across the 150 GHz passband (shaded grey). The matrix elements are defined in Equation~\ref{m_ref_hwp}. The calculated reflection frequency response (blue) is markedly different from an ideal waveplate in transmission (dotted-red). \label{hwp_reflection_mueller_matrix}}
\end{center}
\end{figure}

$\mathsf{M}_{ref\_HWP}$ is the HWP reflection. Similar to HWP transmission, in the HWP crystal axis coordinate system, incident x-polarization can only reflect to x-polarization, and incident y-polarization can only reflect to y-polarization. This means that the HWP reflection Mueller matrix has the same functional form as the HWP transmission Mueller matrix, just with different numerical coefficients.
\begin{equation}\label{m_ref_hwp}
\mathsf{M}_{ref\_HWP} \equiv \left[ \begin{array}{cccc} R & \tau & 0 & 0 \\ \tau & R & 0 & 0  \\ 0 & 0 & h & -q \\ 0 & 0 & q & h \end{array} \right]
\end{equation}
For the Spider 150 GHz HWPs and a CMB source spectrum, these parameters are nominally $R = .02313$, $\tau = -.00952$, $h = .01908$, and $q = .00024$. Note that since $R$ and $h$ have the same sign, a reflection off the HWP does not rotate linear polarization. Unlike the reflection from the surface of a metal modeled with the Mueller matrix in Equation~\ref{m_fp_ref}, reflection from dielectric material does induce a phase shift. This cancels out the minus sign on the $U$-to-$U$ and $V$-to-$V$ terms of this Mueller matrix that comes from the flip of the coordinate system. This means that $R$ and $h$ having the same sign is not surprising. The birefringence of the HWP does not dramatically affect this either. This is because a ray bouncing off the cold side of the HWP is simply reflected, and a ray bouncing off the warm side has gone through the HWP twice. This means 2 half-wave delays, or one full wave, which is the same as no relative phase delay at all. The $R$ and $h$ values above include reflections from both the cold and warm sides of the HWP, since they are calculated using the full E\&M scattering-matrix code. The coordinate definitions used in the scattering matrix code implicitly account for the flip in the reflected coordinate system. The fact that $\tau$ and $q$ terms are nonzero affects polarization, as does the fact that $R \neq h$.

Multiplying out Equation~\ref{M_A_eqn} yields the Mueller matrix $\mathsf{M_A}$ for the A detector. The signal from the ghost on the A detector will be
\begin{equation}
d_{A} = I  M_{A}^{II} + Q  M_{A}^{IQ}+ U  M_{A}^{IU} + V M_{A}^{IV},
\end{equation}
where $I$, $Q$, $U$, and $V$ are the beam-averaged Stokes parameters of the light at the ghost beam location on the sky. The relevant elements of the matrix  $\mathsf{M}_{A}$ for the ghost beam are
\begin{eqnarray}\label{M_diff}
M_{A}^{II} &=&  \frac{R_{FP}}{2}   \left[ (R T + \rho \tau) (\eta^2 + \delta^2) + (R \rho + T \tau) (\eta^2 - \delta^2) \cos{(2 \theta_{hwp})} \right] \nonumber \\
M_{A}^{IQ} &=& \frac{R_{FP}}{4} \left[ (c h + R T - q s + \rho \tau) (\eta^{2} - \delta^{2}) \cos{(2 \psi_{inst})} \right. \nonumber \\
                                      &+& (2 T \rho + 2 T \tau )  (\eta^{2} + \delta^{2})  \cos{(2 \psi_{inst} + 2 \theta_{hwp})} \nonumber \\
                                      &+& \left. (R T - c h + q s + \rho \tau )  (\eta^{2} - \delta^{2})  \cos{(2 \psi_{inst} + 4 \theta_{hwp})} \right] \nonumber \\
M_{A}^{IU} &=& \frac{R_{FP}}{4} \left[ (c h + R T - q s + \rho \tau) (\eta^{2} - \delta^{2}) \sin{(2 \psi_{inst})} \right. \nonumber \\
                                      &+& (2 T \rho + 2 T \tau )  (\eta^{2} + \delta^{2})  \sin{(2 \psi_{inst} + 2 \theta_{hwp})} \nonumber \\
                                      &+& \left. (R T - c h + q s + \rho \tau )  (\eta^{2} - \delta^{2})  \sin{(2 \psi_{inst} + 4 \theta_{hwp})} \right] \nonumber \\
M_{A}^{IV} &=& \frac{R_{FP}}{2} \sin{(2 \theta_{hwp})} (\eta^2 - \delta^2) (c q + h s).
\end{eqnarray}
The nominal numerical value of the $I$ term above is at most .0081, the $Q$ and $U$ couplings are at most .0103, and the $V$ coupling is at most .0001. Measurements in Chapter~\ref{texas_results} show that the actual ghost beam coupling is 2 to 6 times this calculated value, but this is still below the $10\%$ ghost that simulations suggest would still be tolerable.

As a check on the functional form of this model, we can set the values of the variables to use this formula to model reflection from an unpolarized filter in an instrument with no HWP. To do this, we set the reflection matrix parameters to $h = R$ and $\tau = q = 0$, and remove the HWP from the model by setting $T = c = 1$ and $\rho = s = 0$. As expected for an unpolarized reflection in a system with no HWP, in this limit the ghost Mueller matrix is the same as that of a partially-polarized detector.
\begin{eqnarray}\label{M_diff_nohwp}
M_{A,noHWP}^{II} &=&  \frac{R_{FP}}{2} R (\eta^{2} + \delta^{2}) \nonumber \\
M_{A,noHWP}^{IQ} &=&\frac{R_{FP}}{2} R (\eta^{2} - \delta^{2}) \cos{(2 \psi_{inst})} \nonumber \\
M_{A,noHWP}^{IU} &=&\frac{R_{FP}}{2} R (\eta^{2} - \delta^{2}) \sin{(2 \psi_{inst})} \nonumber \\
M_{A,noHWP}^{IV} &=& 0
\end{eqnarray}

Assuming its sensitivity is rotated by exactly 90 degrees, and that the A and B detectors have identical cross-pol, we make the substitutions $\eta \rightarrow \delta$ and $\delta \rightarrow \eta$ to get the Mueller matrix $\mathsf{M_B}$ of the B detector. Also, measurements of the ghost beam in Spider indicate that the actual focal plane reflection experienced by each detector in the focal plane varies considerably, by up to a factor of two. To handle this effect, the focal plane reflection experienced by the A detector is labeled $R_{FP}^A$, and the reflection for the B detector is $R_{FP}^B$. For detector differencing, we want the difference of the two Mueller matrices, $\mathsf{M}_{diff} \equiv \mathsf{M_A} - \mathsf{M_B}$. The detector difference signal will then be
\begin{equation}
d_{diff} \equiv d_{A} - d_{B} = I  M_{diff}^{II} + Q  M_{diff}^{IQ}+ U  M_{diff}^{IU} + V M_{diff}^{IV}.
\end{equation}

\section{Averaging Down the Ghost}
In Spider, the HWP will be stepped to a different angle once per day. With maps taken at multiple HWP angles, the ghost beam will be averaged down as daily maps from a single detector pair are coadded. For Spider the HWP angle schedule for the $Q$ maps taken with a single detector pair is
\begin{equation}
\{\theta_{hwp}^{i}\}^{Q} = \{0^{\circ},45^{\circ},90^{\circ},135^{\circ} \}
\end{equation}
and the schedule for the $U$ maps from the same detector pair is
\begin{equation}
\{\theta_{hwp}^{i}\}^{U} = \{22.5^{\circ},67.5^{\circ},112.5^{\circ},157.5^{\circ} \} 
\end{equation}
Because the Spider scan strategy repeats every sidereal day, in a given RA/dec map pixel the set of instrument angles between Spider and the sky $Q/U$ coordinate system will be the same for each day's map. To model this effect here, the same $\psi_{inst}$ angle variable is used in each day's ghost Mueller matrix.

To generate a map of $Q$, the sub-maps are added together such that the sign of $Q_{cmb}$ is always positive, and the signs of HWP systematics, beam shape systematics, and AB beam offset systematics, alternate so they each add to zero. The $U$ map is generated in the same way. The set of HWP angles was chosen to always have the instrument sensitive to $Q$ alone or $U$ alone, and also chosen so the temperature leakage term caused by HWP non-idealities would cancel out when the individual maps are combined. This approach significantly reduces the requirements on lab measurements of HWP non-idealities and detector crosspol, which is important since those measurements have proven to be difficult in the integrated instrument.

For modeling the ghost beam, adding together the ghost beam Mueller matrices $\mathsf{M}^{\theta}_{diff}$ corresponding to angles in each sub-map yields a schedule-averaged ghost beam Mueller matrix $\mathsf{M}_{ave}^{Q}$ for the $Q$ observations. Doing this for the $U$ schedule yields a schedule-averaged Mueller matrix $\mathsf{M}_{ave}^{U} $for the $U$ observations.
\begin{eqnarray}\label{m_ave}
\mathsf{M}_{ave}^{Q} &\equiv& \mathsf{M}^{0}_{diff} - \mathsf{M}^{45}_{diff} + \mathsf{M}^{90}_{diff} - \mathsf{M}^{135}_{diff} \nonumber \\
\mathsf{M}_{ave}^{U} &\equiv& \mathsf{M}^{22.5}_{diff} - \mathsf{M}^{67.5}_{diff} + \mathsf{M}^{112.5}_{diff} - \mathsf{M}^{157.5}_{diff}
\end{eqnarray}
These Mueller matrices describe the coupling of the sky signal in the ghost beam all the way to the final coadded map for a given detector A-B pair. 

Applying this HWP angle schedule results in some very significant cancellations in the ghost beam coupling. Coadding maps in this way for these particular angles cancels out the constant and $2 \theta_{hwp}$ terms, and the $4 \theta_{hwp}$ trig functions are always $\pm 1$ for these angles. The $Q$ angles cancel out the $\sin{4 \theta_{hwp}}$ term, so the schedule-averaged Mueller matrix elements for $Q$ are
\begin{eqnarray}
(\mathsf{M}_{ave}^{Q})^{II} &=& 0 \nonumber \\
(\mathsf{M}_{ave}^{Q})^{IQ} &=& \phantom{-}(R_{FP}^A + R_{FP}^B) \cos(\psi_{inst}) (R T - c h + q s + \rho \tau ) (\eta^2 - \delta^2)\nonumber \\
(\mathsf{M}_{ave}^{Q})^{IU} &=& \phantom{-}(R_{FP}^A + R_{FP}^B) \sin(\psi_{inst}) (R T - c h + q s + \rho \tau ) (\eta^2 - \delta^2) \nonumber \\
(\mathsf{M}_{ave}^{Q})^{IV} &=& 0.
\end{eqnarray}
 The $U$ angles cancel out the $\cos{4 \theta_{hwp}}$ term, so for the $U$ angles they are
\begin{eqnarray}
(\mathsf{M}_{ave}^{U})^{II} &=& 0 \nonumber \\
(\mathsf{M}_{ave}^{U})^{IQ} &=& -(R_{FP}^A + R_{FP}^B) \sin(\psi_{inst}) (R T - c h + q s + \rho \tau ) (\eta^2 - \delta^2) \nonumber \\
(\mathsf{M}_{ave}^{U})^{IU} &=& \phantom{-}(R_{FP}^A + R_{FP}^B) \cos(\psi_{inst}) (R T - c h + q s + \rho \tau ) (\eta^2 - \delta^2) \nonumber \\
(\mathsf{M}_{ave}^{U})^{IV} &=& 0. 
\end{eqnarray}
The numerical value of these terms (with $\psi_{inst} = 0$ and $R_{FP}^A = R_{FP}^B = 0.5$) is nominally .0408. Since four maps were combined using Equation~\ref{m_ave} to form this Mueller matrix, this represents $1\%$ of the main beam coupling to $Q$ or $U$. Remarkably, once the detectors are differenced and the daily maps are coadded, the temperature ghosting caused by the HWP cancels out leaving only a small $Q$-to-$Q$ and a $U$-to-$U$ polarization ghost. These polarization ghost beam couplings have the functional form of instrument $Q$ to instrument $Q$ coupling, and instrument $U$ to instrument $U$ coupling, so the ghost is not cross-polarized. The fact that the temperature ghost cancels out represents a further improvement over a measured ghost beam amplitude that is already acceptable for Spider according to simulations.

\begin{table}
\begin{center}
\noindent\makebox[\textwidth]{
\scriptsize
\begin{tabular}{|c|c|c|c|c|c|c|c|c|} \hline
\textbf{HWP} & \multicolumn{4}{c|}{\textbf{Ghost Beam}} & \multicolumn{4}{c|}{\textbf{Main Beam}} \\ 
\textbf{Angle} & $I$ & $Q$ & $U$ & $V$ & $I$ & $Q$ & $U$ & $V$ \\ \hline
$\mathit{0^{\circ}}$ & -0.004538   & \phantom{-}0.011251& 0 & 0 & \phantom{-}0.009620 &  \phantom{-}0.976830 & 0  & 0 \\ \hline
$\mathit{90^{\circ}}$ & \phantom{-}0.004538 &  \phantom{-}0.011251 & 0 & 0 & -0.009620 & \phantom{-}0.976830 & 0 & 0 \\ \hline
$\mathit{45^{\circ}}$ & 0 & -0.009160 & 0 & -0.000040 & 0 & -0.959920&  0 &  -0.016220 \\ \hline
$\mathit{135^{\circ}}$ & 0 &  -0.009160 & 0 &  \phantom{-}0.000040 & 0 & -0.959920 &  0 &   \phantom{-}0.016220 \\ \hline
$\mathit{22.5^{\circ}}$ & -0.003223 & \phantom{-}0.000915 & \phantom{-}0.010188 & -0.000051 & \phantom{-}0.006802 & \phantom{-}0.008455 &  \phantom{-}0.968375 &  -0.011469 \\ \hline
$\mathit{157.5^{\circ}}$ & -0.003223 & \phantom{-}0.000915 & -0.010188 & \phantom{-}0.000051 & \phantom{-}0.006802 &  \phantom{-}0.008455 & -0.968375 & \phantom{-}0.011469 \\ \hline
$\mathit{67.5^{\circ}}$ & \phantom{-}0.003223 & \phantom{-}0.000915 &  -0.010188 & -0.000051 & -0.006802 & \phantom{-}0.008455 & -0.968375 & -0.011469 \\ \hline
$\mathit{112.5^{\circ}}$ & \phantom{-}0.003223 &  \phantom{-}0.000915 & \phantom{-}0.010188 & \phantom{-}0.000051 &  -0.006802 & \phantom{-}0.008455 & \phantom{-}0.968375 & \phantom{-}0.011469 \\ \hline
\end{tabular}}
\caption[Main and ghost beam couplings to different source polarizations at different HWP angles.]{Main and ghost beam couplings of a differenced detector pair to different source polarizations (columns) at different HWP angles (rows). The couplings are band-averaged for a CMB source spectrum, nominal 150 GHz HWP and AR-coat thicknesses, no cross-polarization in the detector, and the focal plane reflection for both the A and B detector is assumed to be 0.5. Enough digits are displayed in this table such that distinct table entries appear distinct, and table entries that appear equal actually are equal. The instrument angle is $0^{\circ}$. \label{hwp_no_rot_schedule}}
\end{center}
\end{table}

\section{Ghosting from a Filter Skyward of the HWP}

If a filter skyward of the HWP is reflective, this will also form a ghost beam. Unlike a beam systematic originating from the optics on the cold side of the HWP, this will be modulated by the HWP multiple times in a way that could in principle be problematic, so here we model its polarization properties. Assuming the filter reflection has the polarization properties of a reflection from a dielectric, the Mueller matrix of this reflection is
\begin{equation}\label{m_filt_ref}
\mathsf{M}_{ref\_filter} = \left[ \begin{array}{cccc} R_{filter} & 0 & 0 & 0 \\ 0 & R_{filter} & 0 & 0  \\ 0 & 0 & R_{filter} & 0 \\ 0 & 0 & 0 & R_{filter} \end{array} \right].
\end{equation}
Thinking in broadcast mode, the sensitivity will leave the polarized detector and pass through the HWP. Then, the sensitivity reflects from the filter, and goes back through the HWP. The rotation matrices for this second pass through the HWP will appear in reverse order because the beam is going through the other direction. Then, just as before the beam will bounce off the focal plane, pass through the HWP in the forward direction, and go to the sky.

The Mueller matrix product for this ghost beam is
\begin{equation}\label{M_A_eqn}
\mathsf{M_A} = \mathsf{M}_{pol}~\mathsf{M}_{-\theta}~\mathsf{M}_{HWP}~\mathsf{M}_{\theta}~\mathsf{M}_{ref\_filter}~\mathsf{M}_{\theta}~\mathsf{M}_{HWP}~\mathsf{M}_{-\theta}~\mathsf{M}_{ref\_FP}~\mathsf{M}_{-\theta}~\mathsf{M}_{HWP}~\mathsf{M}_{\theta}~\mathsf{M}_{\psi}.
\end{equation}
With this modified form of $\mathsf{M_A}$, we rerun the calculation all the way through to the schedule-averaged Mueller matrices for $Q$ and $U$ maps for a detector pair, a new version of Equation~\ref{m_ave}. The result for $Q$ is
\begin{eqnarray}
(\mathsf{M}_{ave}^{Q})^{II} &=& 4 R_{filter} T \rho^2 (R_{FP}^A - R_{FP}^B) (\eta^2 + \delta^2)\nonumber \\
(\mathsf{M}_{ave}^{Q})^{IQ} &=& \cos{(2 \psi_{inst})} R_{filter} (R_{FP}^A + R_{FP}^B) (T^3 + 3 T \rho^2 - c^3 - c s^2) (\eta^2 - \delta^2)\nonumber \\
(\mathsf{M}_{ave}^{Q})^{IU} &=& \sin{(2 \psi_{inst})} R_{filter} (R_{FP}^A + R_{FP}^B) (T^3 + 3 T \rho^2 - c^3 - c s^2) (\eta^2 - \delta^2) \nonumber \\
(\mathsf{M}_{ave}^{Q})^{IV} &=& 0,
\end{eqnarray}
and the result for $U$ is
\begin{eqnarray}
(\mathsf{M}_{ave}^{U})^{II} &=& 0 \nonumber \\
(\mathsf{M}_{ave}^{U})^{IQ} &=& -\sin{(2 \psi_{inst})} R_{filter} (R_{FP}^A + R_{FP}^B) \mathcal{H} (\eta^2 - \delta^2)\nonumber \\
(\mathsf{M}_{ave}^{U})^{IU} &=& \phantom{-}\cos{(2 \psi_{inst})} R_{filter} (R_{FP}^A + R_{FP}^B) \mathcal{H} (\eta^2 - \delta^2)\nonumber \\
(\mathsf{M}_{ave}^{U})^{IV} &=& 4 R_{filter} (R_{FP}^A - R_{FP}^B) (\eta^2 + \delta^2),
\end{eqnarray}
where $\mathcal{H}$ is
\begin{equation}
\mathcal{H} = T^2 c - T c^2 + 2 T \rho^2 + T s^2 + c \rho^2 - 2 c s^2.
\end{equation}
The temperature ghost in the $Q$ maps is non-zero, which could be a problem. However, the coupling is small because it goes as the third power of parameters, $\sim \rho^2 R_{filter}$, that are each of order $1\%$. Assuming that the filter reflection is $1\%$, and pessimistically assuming that the focal plane reflections experienced by the A and B detectors in the pair are 0.50 and 0.25 respectively, the nominal numerical value of this coupling is $2 \times 10^{-7}$, very small. Setting the instrument angle to zero, the $Q$ ghost in the $Q$ map is nominally $0.4\%$, and the $U$ ghost in the $U$ map is very small, nominally $0.006\%$.

\section{Anomalous Focal Plane Reflection}
So far, we have assumed that the reflection from the focal plane is a reflection from a perfect mirror. For the focal plane reflection caused by the niobium ground plane, this should be approximately correct. However, reflection from and interactions with the detector feeds could in principle cause horizontal and vertical polarization to reflect differently. To model this, we can modify the Mueller matrix of the focal plane reflection in Equation~\ref{m_fp_ref} to be a weak partial polarizer. To avoid changing variables away from $R_{FP}$, the Mueller matrix of a metal reflective partial polarizer co-oriented with the A detector can be re-written as
\begin{eqnarray}\label{m_weird_fp}
\mathsf{M}_{ref\_FP} = \left[ \begin{array}{cccc} R_{FP} & a & 0 & 0 \\ a & R_{FP} & 0 & 0  \\ 0 & 0 & -\sqrt{(R_{FP} + a)(R_{FP} - a)} & 0 \\ 0 & 0 & 0 & -\sqrt{(R_{FP} + a)(R_{FP} - a)} \end{array} \right].
\end{eqnarray}
Here $a$ is a parameter related to the difference between the reflection of horizontal and vertical polarization. Setting $a$ to zero recovers the result for a mirror-like focal plane reflection. Allowing it to be nonzero models a focal plane that reflects horizontal and vertical polarizations with different amplitudes, but still with no phase delay as is appropriate for a metallic surface reflection. Setting $a=R_{FP}$ recovers the result for a completely polarizing reflector.

Using this new form of $\mathsf{M}_{ref\_FP}$, we rerun the same calculation all the way through to the schedule-averaged Mueller matrices of Equation~\ref{m_ave} for a detector pair.  For simplicity, here we assume that the focal plane reflection experienced by both the $A$ and $B$ detectors is the same, and that the non-ideal reflection parameter $a$ is also the same for both detectors. For the Spider schedule, the schedule-averaged Mueller matrix elements for $Q$ are
\begin{eqnarray}
(\mathsf{M}_{ave}^{Q})^{II} &=& 2 a (\eta^{2} - \delta^{2}) (R T - T h + \rho \tau)  \nonumber \\
(\mathsf{M}_{ave}^{Q})^{IQ} &=& 2 (\eta^{2} - \delta^{2}) \cos{(2 \psi_{inst})} \left(R R_{FP} T - R_{FP} c h + R_{FP} \rho \tau + q s \sqrt{R_{FP}^2 - a^2} \right) \nonumber \\
(\mathsf{M}_{ave}^{Q})^{IU} &=& 2 (\eta^{2} - \delta^{2}) \sin{(2 \psi_{inst})} \left(R R_{FP} T - R_{FP} c h + R_{FP} \rho \tau + q s \sqrt{R_{FP}^2 - a^2} \right) \nonumber \\
(\mathsf{M}_{ave}^{Q})^{IV} &=& 0,
\end{eqnarray}
and for $U$ they are
\begin{eqnarray}
(\mathsf{M}_{ave}^{U})^{II} &=& 0 \nonumber \\
(\mathsf{M}_{ave}^{U})^{IQ} &=& - (\eta^{2} - \delta^{2}) \sin{(2\psi_{inst})} \left(R_{FP} (R  T - R c + T h - c h + 2 \rho \tau) \phantom{\sqrt{R_{FP}^2}} \right. \nonumber \\
&+& \left. (R T + R c - T h - c h + 2 q s) \sqrt{R_{FP}^2 - a^2} \right) \nonumber \\
(\mathsf{M}_{ave}^{U})^{IU} &=& \phantom{-}(\eta^{2} - \delta^{2}) \cos{(2\psi_{inst})} \left(R_{FP} (R  T - R c + T h - c h + 2 \rho \tau) \phantom{\sqrt{R_{FP}^2}} \right. \nonumber \\
&+& \left. (R T + R c - T h - c h + 2 q s) \sqrt{R_{FP}^2 - a^2} \right) \nonumber \\
(\mathsf{M}_{ave}^{U})^{IV} &=& 2 a s \tau (\eta^{2} - \delta^{2}). 
\end{eqnarray}
Anomalous focal plane reflection makes the schedule-averaged instrument $Q$ and instrument $U$ ghosts slightly unequal, and causes a small intensity ghost in the instrument $Q$ map. The intensity ghost is roughly $a \times 0.2\%$ of the amplitude of the main beam, so if the non-ideality $a$ of the focal plane reflections is at the several percent level or lower this ghost coupling will be very small.


\chapter{HWP Performance in Spider}
\label{texas_results}

Before deploying the instrument to fly from Antarctica, in summer 2013 we integrated the instrument at the Columbia Scientific Balloon Facility in Palestine, Texas, to perform initial testing of the instrument as a whole. Here some optical testing results relevant for HWP performance are described.

\section{Optical Response}

The HWPs are fabricated from materials that have low mm-wave loss, and AR coats have been bonded to both sides of the sapphire to reduce reflection. This means that the HWPs should not significantly reduce the optical response of the detectors. The overall band-averaged transmission of the HWP is the $T$ parameter of Equation~\ref{m_hwp}, calculated to be $97.6\%$ for the 150 GHz HWPs, and $96.8\%$ for the 95 GHz HWPs.

\begin{figure}
\begin{center}
\includegraphics[width=0.85\textwidth]{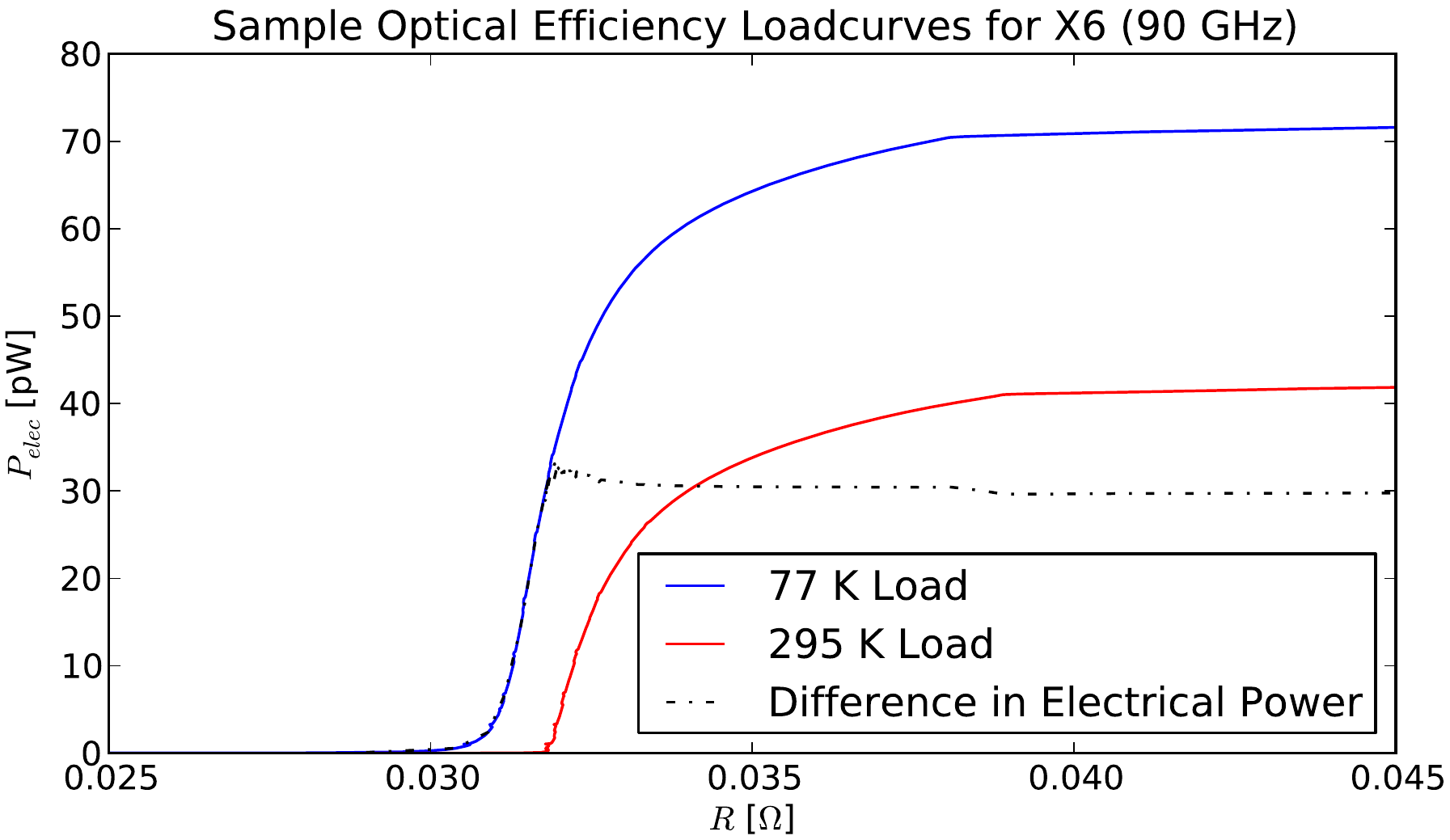}
\caption[Sample loadcurves from X6.]{Sample loadcurves from X6. The blue curve shows electrical power as a function of device resistance when the telescope beam is filled with a room-temperature black HR-10 load. The red curve is the same but with a liquid-nitrogen cold load filling the beam. Only the aluminum TES transition is observed for this range of optical loading. The difference in electrical power between the two loads is shown as a dashed black line, which is roughly constant in the power ranges when both devices are active. \label{sample_loadcurve}}
\end{center}
\end{figure}

In Spider, we measure the optical response of the detectors by taking load curves of the TES bolometers with different temperature beam-filling loads placed outside the cryostat. The goal is to measure how much optical power is deposited onto the detectors, per Kelvin of temperature change of the load outside the cryostat. Sample load curves for a single detector in the X6 receiver are shown in Figure~\ref{sample_loadcurve}. This receiver had no HWP installed during the run in Texas. The measurement begins by placing a room temperature (295 K) microwave absorbing sheet in front of the aperture of the telescope. The detectors are then biased high enough such that the electrical power deposited in the TES is enough to heat it well above its superconducting transition temperature. Then, the bias is decreased until the TES is cool enough to just start to transition, the bias continues to decrease until it cools through its superconducting transition, and the bias continues to drop until the device goes fully superconducting. This sweep through biases is called a load curve. The raw dataset consists of a measurement of the current $I$ through the TES as a function of the bias voltage $V$. A useful change of variables is to instead plot $R = V/I$ vs $P = IV$. For the device shown in Figure~\ref{sample_loadcurve}, the electrical power to just begin to drop into the aluminum TES transition for the 295~K load was 40.6~pW

For an optical response measurement, the aperture is then filled with a colder load of known temperature, and a second loadcurve is taken. For the data shown in Figure~\ref{sample_loadcurve}, the second loadcurve was taken with a microwave absorbing sheet cooled with liquid nitrogen to 77 K. For this load, the electrical power to transition was 70.0~pW. This is interpreted as a measurement that an extra $(295~\mathrm{K}~-~77~\mathrm{K})$ of loading outside the cryostat deposits an extra $(70.0~\mathrm{pW}~-~40.6~\mathrm{pW})$ of optical power on the TES. Therefore, the optical response of this device is
\begin{equation}
\frac{70.0~\mathrm{pW}~-~40.6~\mathrm{pW}}{295~\mathrm{K}~-~77~\mathrm{K}} = 0.135~\mathrm{pW/K}.
\end{equation}

For most of the detector tiles in the receivers that had waveplates installed for the cryostat run in Texas, the detector optical efficiency had already been measured in a separate cold run without a HWP installed. The no-waveplate results for receivers X1, X3, and X4 are from earlier runs in the Spider test cryostat, and the no-waveplate results for X2 are from an earlier run in the flight cryostat. Comparing the optical response measurements of these two runs gives a sense of the percentage of light entering the cryostat that HWP absorbs or reflects. The HWP transmission estimates are simply dividing the median optical response from the with-waveplate run by the optical response from the no-waveplate run. These are estimates only, because different filters were used at VCS1 and VCS2 between running at Caltech and Texas, and even X2 which was run previously in the flight cryostat had its VCS1 filter replaced.  The results of this are shown in Table~\ref{hwp_opt_eff}. Both of the 150 GHz HWPs have high transmission, and the HWP mounted in the X2 95 GHz receiver also has high transmission. The measured optical response in the X4 95 GHz receiver dropped by a factor of 79\% to 85\% in the run in Texas. 

\begin{table}
\begin{center}
\noindent\makebox[\textwidth]{%
\small
    \begin{tabular}{r|l|l|l|l|}
    ~                   & \textbf{Tile 1}     & \textbf{Tile 2}     & \textbf{Tile 3}     & \textbf{Tile 4}     \\ \hline
    {X1 (150 GHz) with HWP}         & 0.203 pW/K & 0.189 pW/K & 0.219 pW/K & 0.208 pW/K \\ \hline
    {X1 (150 GHz) no HWP}           & 0.215 pW/K & 0.197 pW/K & ~          & 0.214 pW/K \\ \hline
    \textbf{\textit{X1 HWP Transmission}} & 94.4\%      & 95.9\%      & ~          & 97.2\%      \\ \hline \hline
    {X3 (150 GHz) with HWP}         & 0.169 pW/K & 0.170 pW/K & 0.182 pW/K & 0.164 pW/K \\ \hline
    {X3 (150 GHz) no HWP}           & ~          & 0.170 pW/K & 0.184 pW/K & 0.168 pW/K \\ \hline
    \textbf{\textit{X3 HWP Transmission}} & ~          & 100.0\%     & 98.9\%      & 97.6\%      \\ \hline \hline
    {X2 (95 GHz) with HWP}         & 0.108 pW/K & 0.104 pW/K & 0.111 pW/K & 0.103 pW/K \\ \hline
    {X2 (95 GHz) no HWP}           & 0.103 pW/K & 0.099 pW/K & 0.103 pW/K & 0.098 pW/K \\ \hline
    \textbf{\textit{X2 HWP Transmission}} & 104.9\%     & 105.1\%     & 107.8\%     & 105.1\%     \\ \hline \hline
    {X4 (95 GHz) with HWP}         & 0.098 pW/K & 0.088 pW/K & 0.108 pW/K & 0.097 pW/K \\ \hline
    {X4 (95 GHz) no HWP}           & 0.124 pW/K & 0.103 pW/K & ~          & 0.118 pW/K \\ \hline
    \textbf{\textit{X4 HWP Transmission}} & 79.0\%      & 85.4\%      & ~          & 82.2\%      \\ \hline
    \end{tabular}}
    \caption[HWP transmission estimates from detector optical response measurements.]{HWP transmission estimates from detector optical response measurements. Each entry in the table has the median optical response across all bolometers that were working in that detector tile in that run, and the measured HWP transmission is estimated by dividing the two results by each other. The expected HWP transmission is calculated to be $97.6\%$ for the 150 GHz HWPs, and $96.8\%$ for the 95 GHz HWP. In addition to adding the HWP, there were significant changes in the filter stack between the with-HWP and no-HWP runs. \label{hwp_opt_eff}}
    \end{center}
\end{table}

\section{Beam Performance}

\subsection{Ghost Amplitude}

As shown in Chapter~\ref{ghost_chapter}, a ghost beam forms when detector sensitivity reflects from the HWP or the filters. The amount of power in this ghost beam is approximately the focal plane reflectance $R_{FP}$ multiplied by the reflectance of the HWP and/or filter. This means that the ghost beam power relative to the main beam power is a good way to confirm that the AR coatings are performing well across the band.

We took a beam map in Texas using an unpolarized hot thermal source chopped at 13 Hz located approximately 40 m away from the cryostat. We demodulated the data with a time-domain software lock-in. The source distance we chose is just entering the far field of the telescope. One definition of the far field is to require that the phase difference across the telescope aperture $d$ (which is 290 mm for Spider) be less than $\lambda / 8$ for a wavefront launched from a point source a distance $D$ away. This yields the relation $D > d^2 / \lambda$ = 42 m for 150 GHz, and 27 m for 95 GHz.

For the beam map, we scanned the gondola in azimuth, stepping in elevation at the end of each scan. We chose a scan strategy that would cover most of the field of view of all of the telescopes, but difficulties in realtime pointing reconstruction prevented us from having enough control over the range of the scan to do this. Still, we covered a wide enough range to see several detectors and their ghost beams, which allows for a measurement of the ghost amplitude of those detectors.

\begin{figure}
\begin{center}
\includegraphics[width=0.94\textwidth]{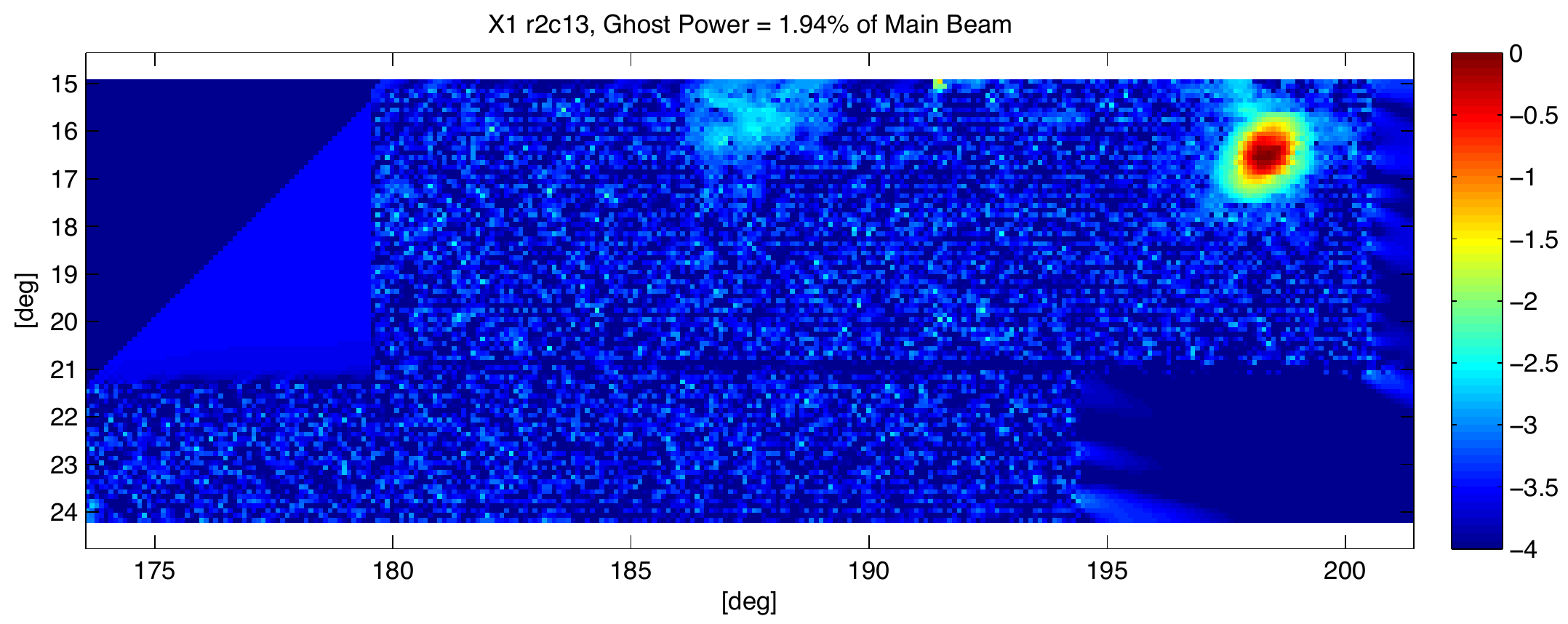} \\
\includegraphics[width=0.94\textwidth]{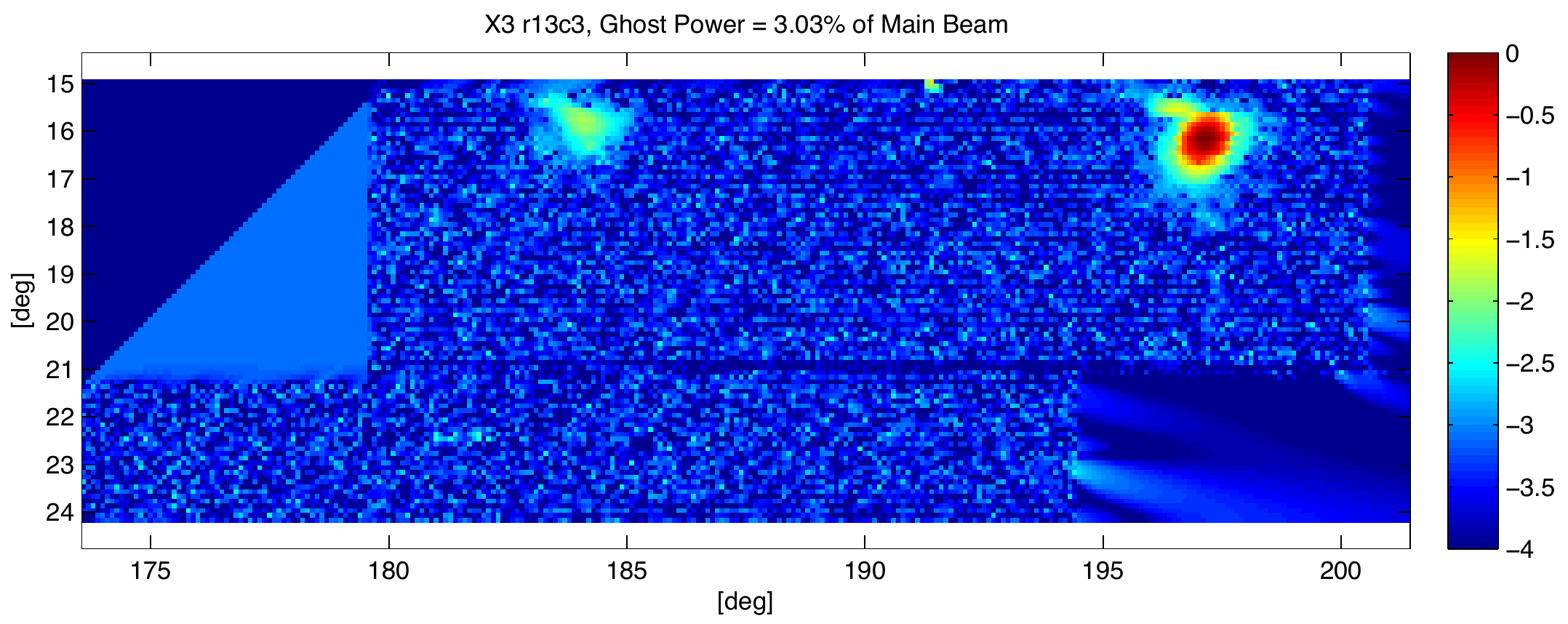} \\
\includegraphics[width=0.94\textwidth]{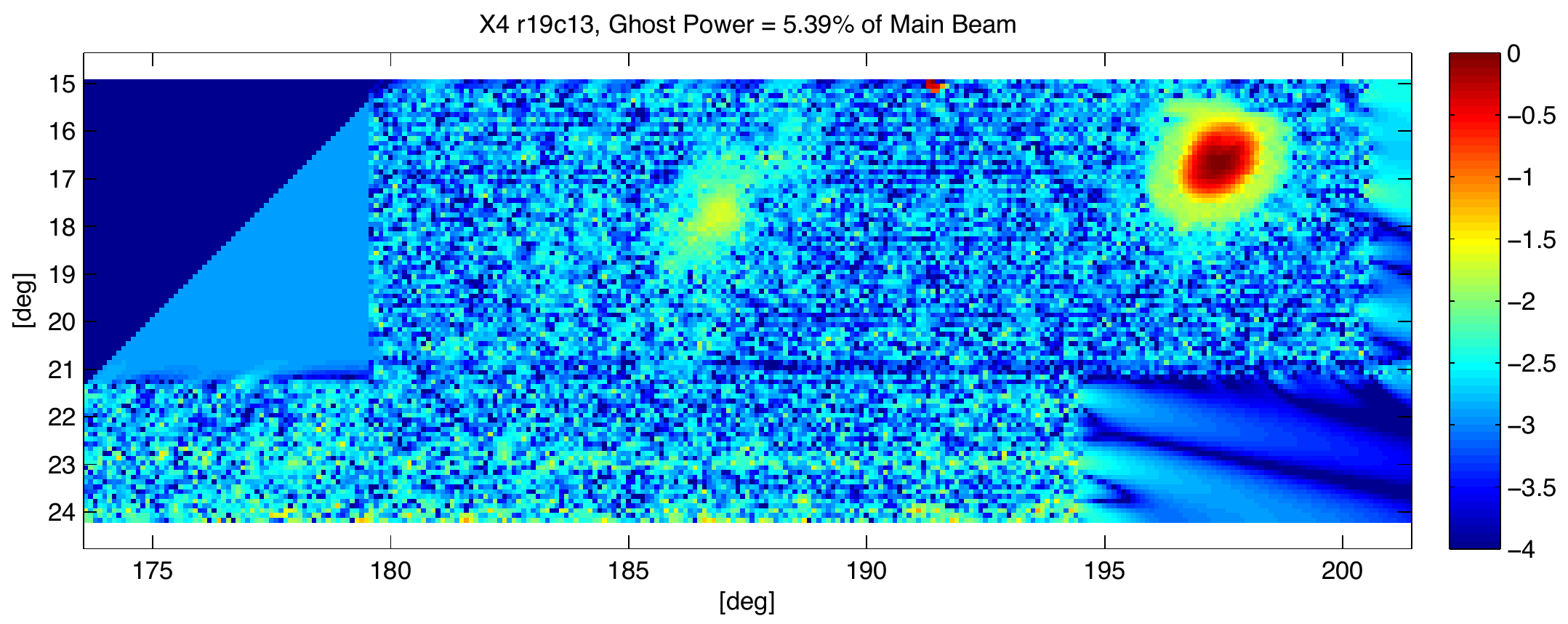}
\caption[Beam maps from Texas showing ghost beam power.]{Beam maps from Texas showing ghost beam power. These maps are not noise-biased, because they were made from the lock-in demodulated timestreams. For presentation on the logarithmic color scale, the image pixels are rectified, yielding a noise floor just below $10^{-3}$ on the top (X1) and middle (X3) maps, and just below $10^{-2}$ on the bottom map (X4). The colorscale is in $\log_{10}($Power$)$ relative to the peak of the main beam. The main beam is the red structure to the right in all three images, and the ghost beam is the light yellow/light blue structure to the left. The integrated ghost beam power is shown above each plot as a percentage of the integrated main beam power. The features above and left of $(180^\circ,21^\circ)$ and below and right of $(195^\circ,21^\circ)$ are software artifacts from outside the scan area. Data courtesy of Rebecca Tucker. \label{ghost_amplitude_images}}
\end{center}
\end{figure}

Figure~\ref{ghost_amplitude_images} shows beam maps from the run in Texas. One detector from each of three different receivers is shown, with both main and ghost beams visible. The detector from X1 had a ghost beam coupling of $1.94\%$, the detector in X3 had $3.03\%$, and the detector in X4 had $5.39\%$. These numbers come from integrating the measured beam power within a $2^\circ$ radius of the center of the main beam, integrating the measured beam power within a $2^\circ$ radius of the center of the ghost beam, and taking the ratio of the two.

X4 is the receiver that had the largest decrease in optical response when the HWP was added. Because of the changes in the filter stack between the with-HWP and no-HWP measurements, it is possible that this decrease is not related to the HWP at all. Still, the larger ghost coupling in this receiver compared to the others could indicate some of the decrease in optical response is due to HWP reflection. The ghost beam coupling is approximately the focal plane reflectance $R_{FP}$ in Equation~\ref{m_fp_ref}, multiplied by the reflectance of the HWP or filter causing the ghosting. Ignoring the filter stack changes and pessimistically assuming the $79\% - 85\%$ optical efficiency decrease in X4 is all due to HWP reflection, this measurement of the ghost beam coupling suggests that the focal plane reflectance is not $1/2$ as was conservatively assumed for the ghost beam modeling in Chapter~\ref{ghost_chapter}, but is approximately $1/3$ to $1/4$.

\subsection{Beam Unchanged vs. HWP Angle}

In Texas, we were not able to control the beam maps well enough to obtain maps of the same detectors at two or more HWP angles. However, there was a run in the Spider test cryostat at Caltech with the X3 receiver with a 150 GHz HWP mounted outside the cryostat where this dataset was taken. To get a quantitative sense of the degree to which the HWPs leave beam systematics unchanged, we consider the AB beam offsets, the offset in the location of the A beam centroid relative to the B beam centroid within a detector pair. The detectors in X3 have AB beam offsets of roughly 0.3 arcminutes, compared to a beam FWHM of 29 arcminutes. Each pair has a different seemingly random direction and magnitude of its offset, and they presumably originate from slight fabrication non-uniformities. The magnitude of the AB beam offset has been larger than this in previously fabricated prototype detector tiles, and a significant development effort by the detector group at Caltech/JPL reduced the offsets of the deployed detector tiles to these lower levels.

As a back-of-the-envelope estimate of how much this matters, consider if the AB beam offset were as large as the angular scale of the first acoustic peak ($\sim1^\circ = 60'$) in the temperature anisotropies. When the AB pair is differenced, that could couple roughly all of the power of the first acoustic peak into the difference timestream, say 100 $\mu$K. This means that, very roughly, a 0.3 arcminute AB beam offset could create a signal of $(0.3' / 60') \times 100~\mu$K $ = 0.5~\mu$K. Having that signal unpredictably changing as the sky rotates would corrupt the AB difference maps roughly at the level of the E mode signal. 

\begin{figure}
\begin{center}
\includegraphics[width=0.85\textwidth]{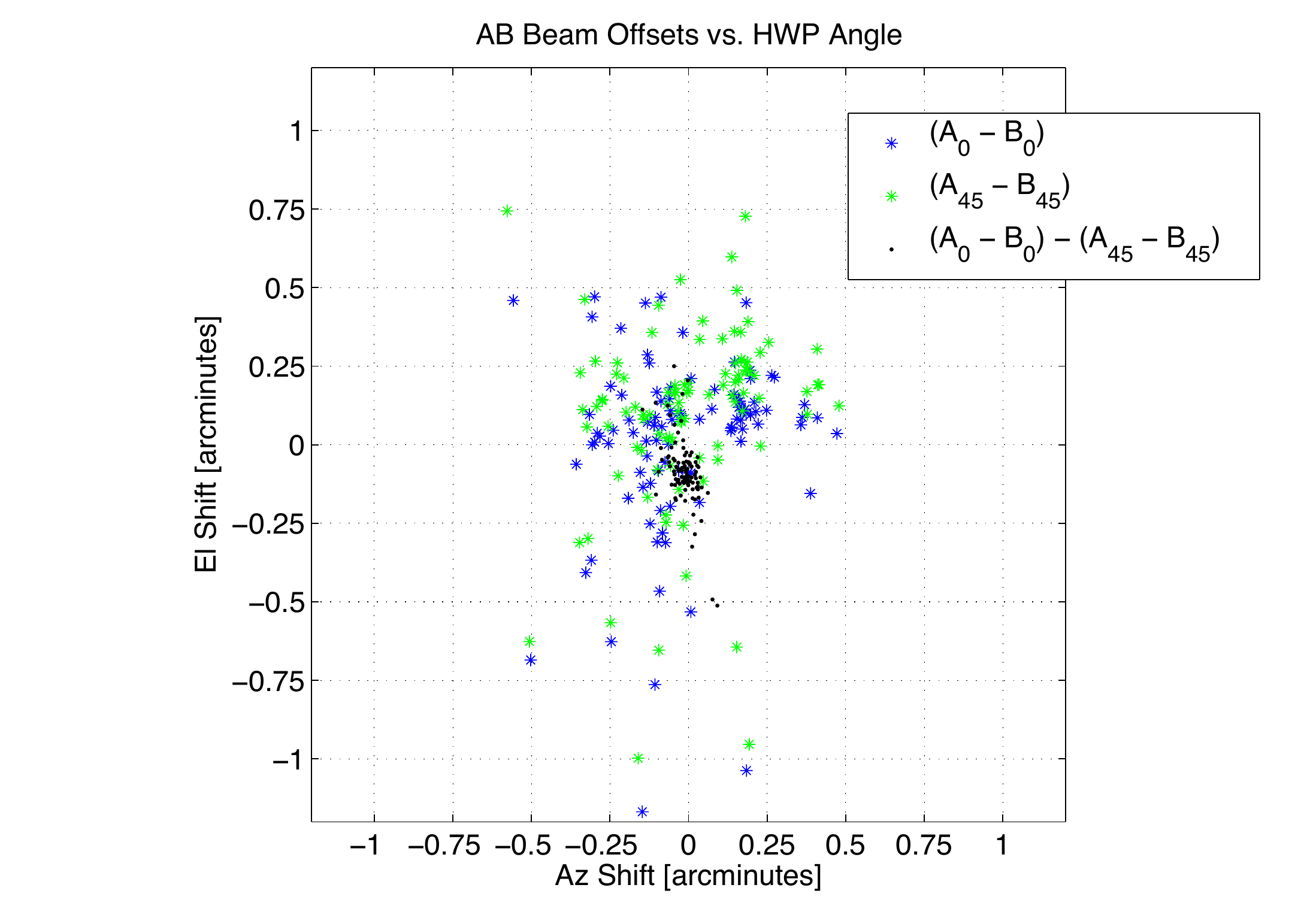}
\caption[AB beam offsets from X3 at two HWP angles.]{AB beam offsets  from X3 at two HWP angles. The scatter in the AB beam offsets (blue and green stars) is well above the detector noise level, and is the result of slight fabrication non-uniformities. The scatter expected in the AB beam offset shifts (black dots) from noise alone is $\pm 0.05$ arcminutes, roughly consistent with what is observed. Data courtesy of Rebecca Tucker. \label{beam_shift_scatterplot}}
\end{center}
\end{figure}

As shown in Figure~\ref{beam_shift_scatterplot}, rotating the HWP does not measurably change the AB beam offsets. The mean and scatter of the az shifts are $-0.014 \pm 0.039$ arcminutes. The mean and RMS scatter of the el shifts are $-0.085 \pm 0.108$ arcminutes. The scatter expected from detector noise in this measurement is roughly 0.05 arcminutes which is consistent with the observed scatter in the az shifts. The elevation steps are 6 arcminutes so there may be systematics that cause scatter at some fraction of that level, which may explain the slightly higher scatter in the el shifts.

\begin{figure}
\includegraphics[width=1.0\textwidth]{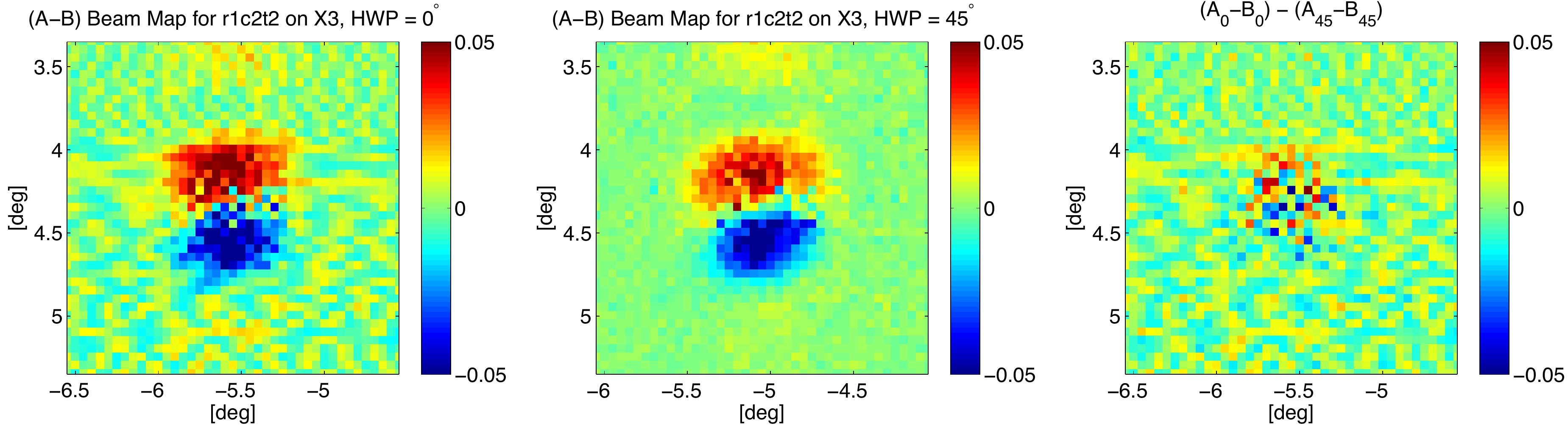} \\
\includegraphics[width=1.0\textwidth]{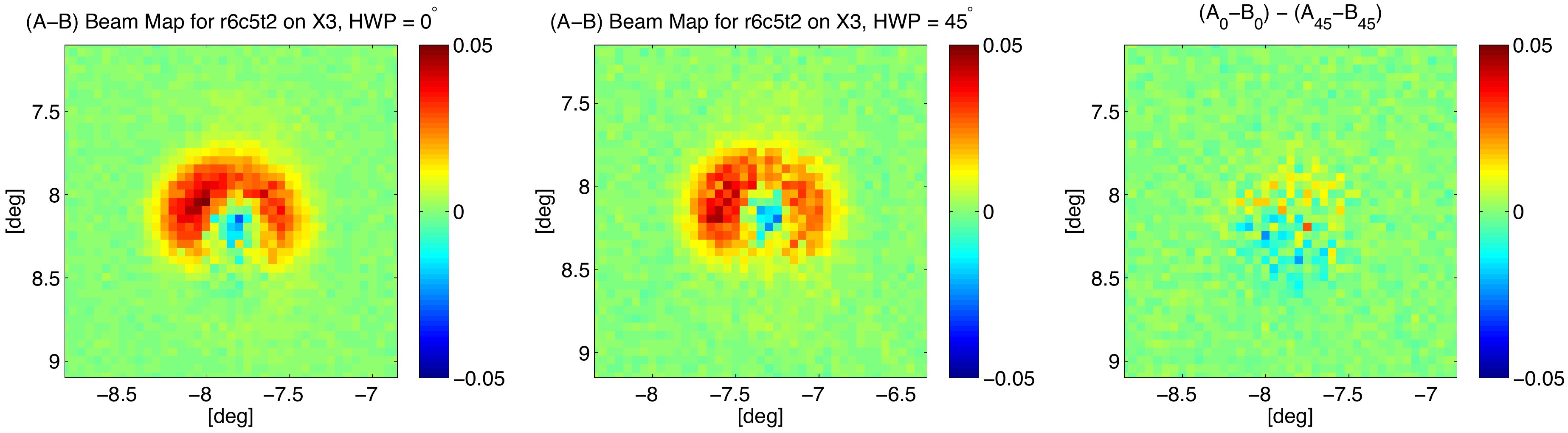} \\
\includegraphics[width=1.0\textwidth]{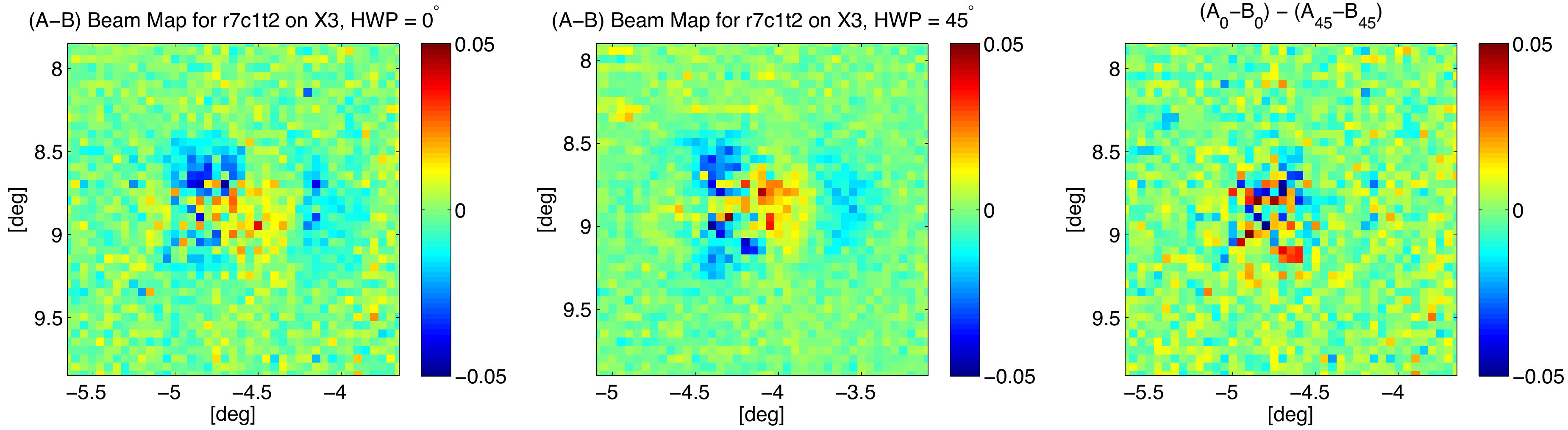}
\begin{center}
\caption[Sample beam maps from three detector pairs in the X3 receiver at two HWP angles.]{Sample beam maps from three detector pairs in the X3 receiver at two HWP angles. The top and bottom detector pairs show a dipole feature in their AB beam difference map, indicating an offset in the beam centers of the A and B pixels. The middle detector pair shows a ring structure, likely caused by a slight difference in the beam sizes of A and B. The left images are all taken with the HWP at a certain angle, the center images are all taken with the HWP rotated by $45^\circ$, and the right images are the difference between the HWP angles. The difference images were formed after shifting the $45^\circ$ data to the left in az by $0.5^\circ$ to correct for an offset in the az encoder between the measurements.  As the difference maps show, the non-idealities of the beam remain basically unchanged under this HWP rotation. Data courtesy of Rebecca Tucker. \label{beam_shift_sample_images}}
\end{center}
\end{figure}

The beams are not symmetric gaussians. They have a slightly elliptical shape, and there are also near sidelobes. These are likely caused by a combination of non-ideal feeds at the focal plane, and the stops and lenses in the telescope itself. As expected, the beam map images themselves do not appear to change significantly with HWP angle, as shown in Figure~\ref{beam_shift_sample_images}.

\section{Polarization Modulation}

The detector feed structures are designed to have high polarization efficiency, and the HWP thicknesses were chosen to have high modulation efficiency in each band. To measure the polarization efficiency, we placed a chopped hot thermal source with a high efficiency polarizing grid in front of it approximately 10~m away from the aperture of the telescope. The grid was mounted in a precision rotation stage with an encoder, and the mount also had a tiltmeter. The grid absolute orientation relative to gravity was determined by diffracting a laser pointer off the wire grid, and rotating the grid until the diffraction pattern aligned with the line of a plumb bob. The encoder and tiltmeter readings for this state were then logged away as an absolute reference. Because the source is chopped, if the grid is perpendicular to the detector orientation and the polarization efficiency is high, the demodulated detector signal really will go to zero.

With the source chopping at roughly 13 Hz and the grid at a particular orientation angle, we slewed the telescope until the signal was maximized on a single detector pair in the X1 150 GHz receiver. The source lit up the edge of the main beam of some nearby detectors as well, but with much lower signal-to-noise. We then stepped the HWP in 5 degree increments, spending 10-20 seconds at each HWP angle. The demodulated data was fit to a model with a $4\theta_{hwp}$ term and a DC term. This allows a best-fit estimate of the minimum and maximum signal level, and taking the ratio yields a measurement of the polarization leakage. This also allows a best-fit estimate of the HWP angle (measured with the HWP absolute angle encoder) at which the polarized detector signal is maximized.

\begin{figure}
\begin{center}
\includegraphics[width=0.78\textwidth]{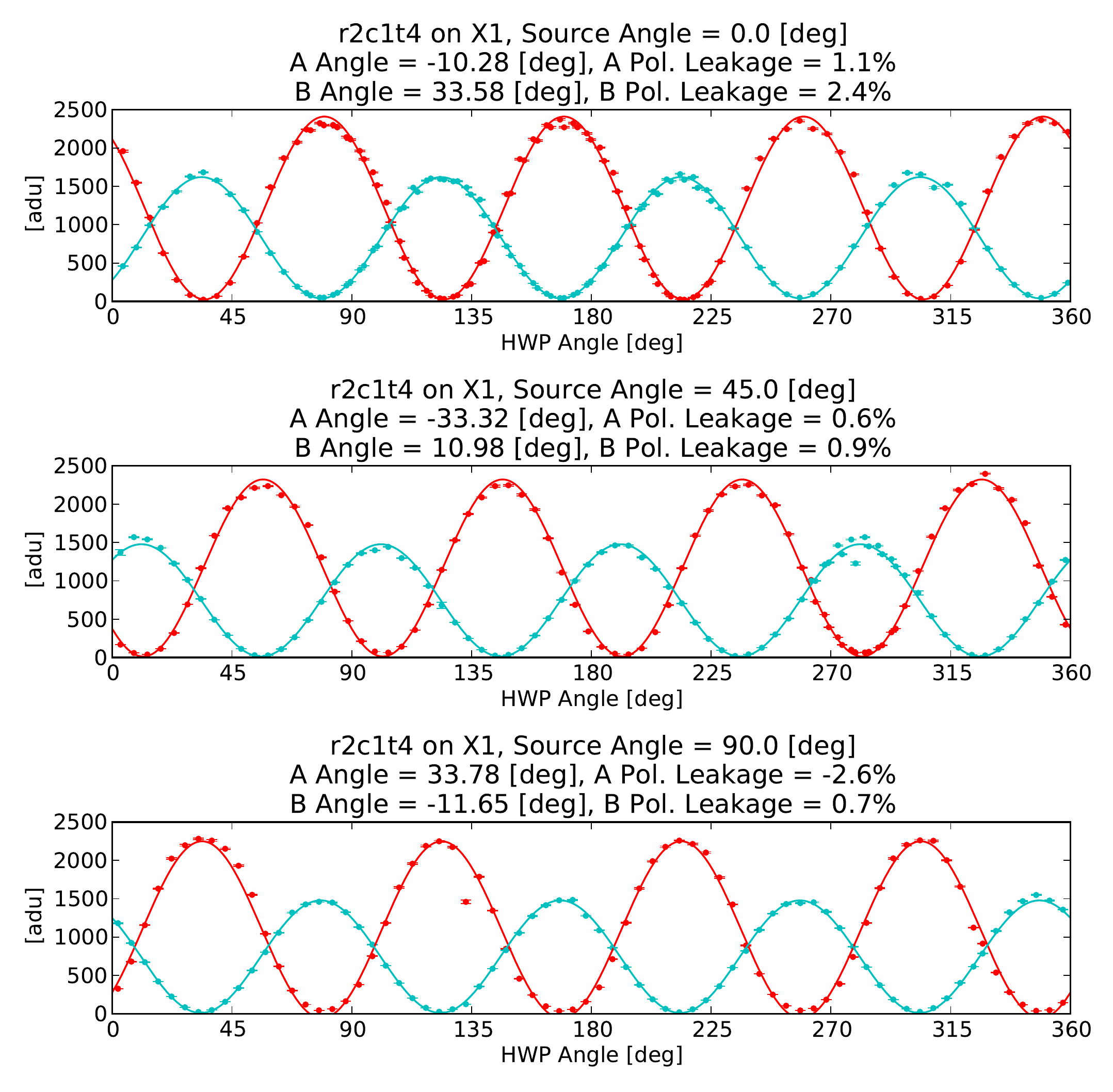} \\
\includegraphics[width=0.78\textwidth]{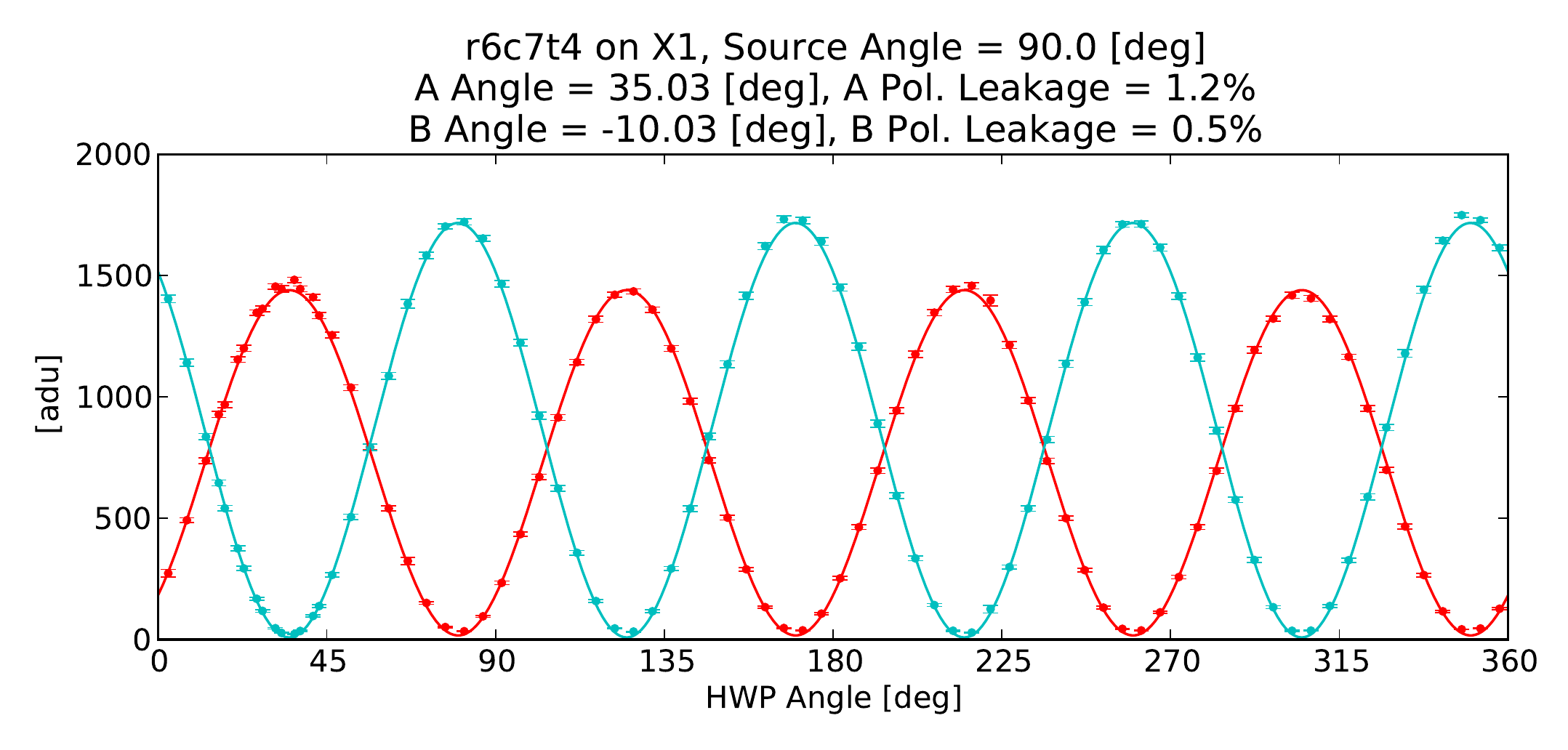}
\caption[HWP polarization modulation measurement in X1. ]{HWP polarization modulation measurement in X1 (150 GHz). The A detector is in red and B is in blue. The top panel shows data and fit with the telescope pointed up on a detector near the outside corner of the focal plane. The second panel is the same detector with the source polarization rotated by $45^\circ$, and the third panel is rotated by $90^\circ$. The bottom panel is the same source angle as the third panel, but the telescope was pointed up on a detector near the center of the focal plane. Throughout this set of measurements, the modulation efficiency is high, as predicted from the HWP model. The HWP angles of the peak detector response are compared in Table~\ref{hwp_angle_best_fits_x1} \label{hwp_pol_mod_x1}}
\end{center}
\end{figure}

\begin{figure}
\begin{center}
\includegraphics[width=0.78\textwidth]{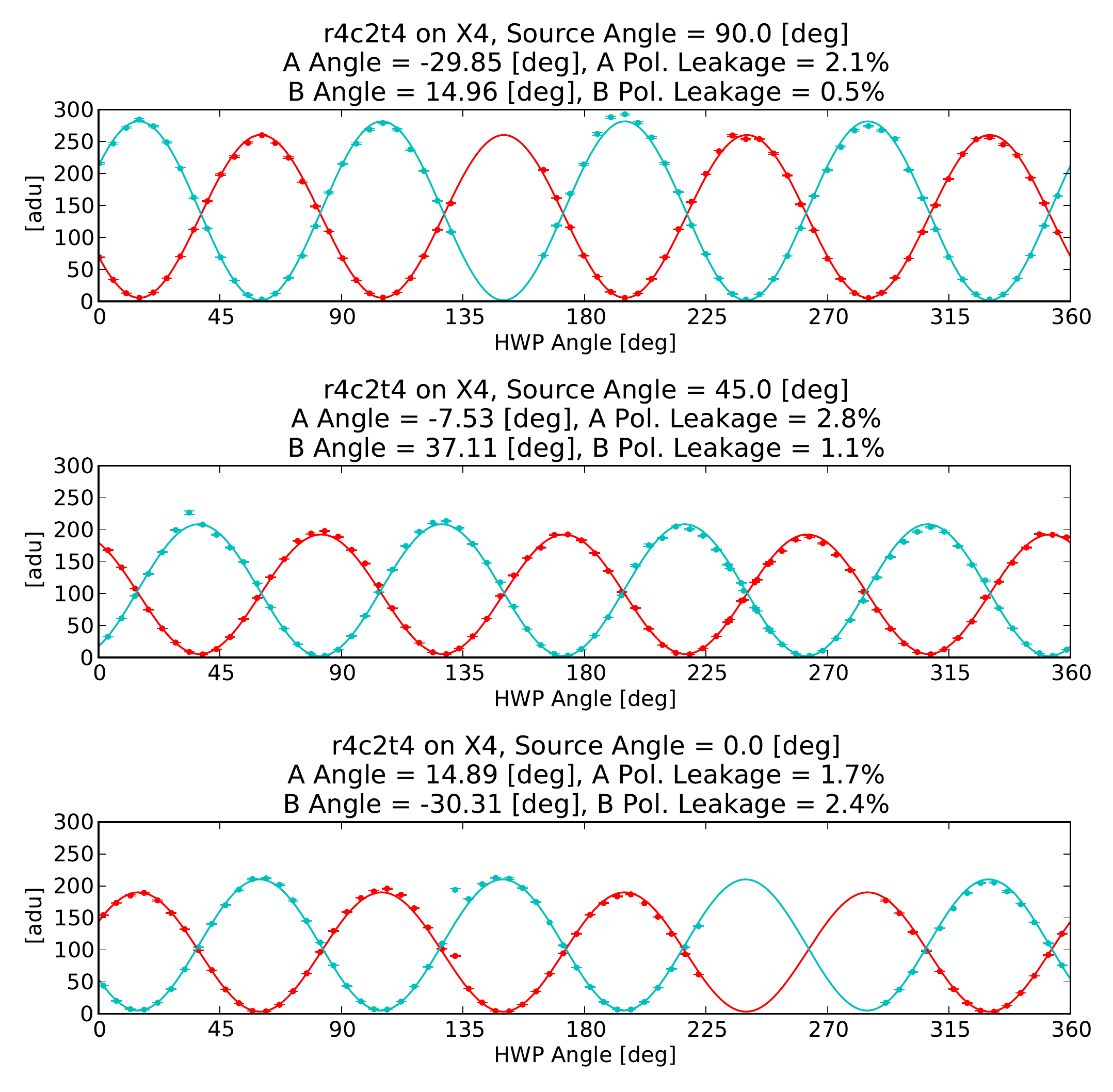}
\caption[HWP polarization modulation measurement in X4. ]{HWP polarization modulation measurement in X4 (95 GHz), similar to Figure~\ref{hwp_pol_mod_x1}. The detector was located near the center of a tile on that focal plane, and the three panels are data taken at three different source polarization angles. This data shows high polarization efficiency. \label{hwp_pol_mod_x4}}
\end{center}
\end{figure}

The results of the polarization efficiency measurements in X1 are shown in Figure~\ref{hwp_pol_mod_x1} and Table~\ref{hwp_angle_best_fits_x1}, and results from the X4 receiver are shown in Figure~\ref{hwp_pol_mod_x4} and Table~\ref{hwp_angle_best_fits_x4}. To confirm that the peak HWP angle changed as expected with the source polarization angle, in X1 we repeated the measurement with the grid at $45^\circ$ and $90^\circ$ relative to gravity. In X1, with the source grid remaining at a $90^\circ$ orientation, we slewed the telescope to maximize the signal on a second detector pair located elsewhere on the focal plane. The signal on this detector pair peaked at a HWP angle rotated a little over $1^\circ$ relative to the previous pair. Because we moved across several degrees of the telescope's field of view to get between the two detectors, the projected polarization sensitivity of the two different detector pairs onto the source is calculated to change by roughly the amount observed. Also, fabrication non-uniformities across the different detectors in the focal plane may also result in polarization sensitivity angles that vary by up to a degree. Finally, we repeated this measurement with a single detector pair on X4, a 95 GHz receiver, and the results are shown in Figure~\ref{hwp_pol_mod_x4} and Table~\ref{hwp_angle_best_fits_x4}. Both of these inserts were mounted in the cryostat rotated $22.5^\circ$ relative to the zero of the source encoder angle coordinate system. Simulations \cite{fraisse13} show that we need to understand the polarization angles of the receivers to $\pm1^\circ$, and these measurements demonstrate that we can do that in Spider.

Using the single-detector model of the detector timestream in Equation~\ref{hwp_matrix_elements}, fixing the instrument angle to zero, and assuming a perfectly-polarized source and a detector with $1\%$ crosspol aligned with the source yields an estimate for how HWP non-idealities should show up in this measurement.
\begin{equation}
d = \frac{1}{2} T + \frac{1}{4} (T + c) (\eta^2 - \delta^2) + \left[ \frac{1}{2} \rho (\eta^2 - \delta^2) + \frac{1}{2} \rho \right]\cos(2 \theta_{hwp}) +  \frac{1}{4} (T-c) (\eta^2 - \delta^2) \cos(4 \theta_{hwp})
\end{equation}
Since the $\rho$ term only provides a small correction to polarization leakage estimation, we only fit to DC and $4 \theta_{hwp}$. Under the definition of polarization leakage used in this analysis, this means the HWP model predicts that with aligned source and detector polarizations
\begin{eqnarray}
\mathrm{Polarization~Leakage} &=& \frac{\frac{1}{2} T + \left[\frac{1}{4} (T + c) -  \frac{1}{4} (T - c)\right](\eta^2 - \delta^2)}{\frac{1}{2} T + \left[\frac{1}{4} (T + c) +  \frac{1}{4} (T - c)\right] (\eta^2 - \delta^2)} \\
&=& 2.29\%~\mathrm{for~150~GHz} \nonumber\\
&=& 2.39\%~\mathrm{for~95~GHz}. \nonumber
\end{eqnarray}
Repeating the calculation for crossed source and detector polarization predicts polarization leakage of $0.59\%$ at 150 GHz and $0.55\%$ at 95 GHz. The measured polarization leakage at the different source-detector orientations varies from $0.6\%$ to $2.8\%$, in good agreement with expectations.

\begin{table}
    {\scriptsize
    \begin{tabular}{c|c|c|c|c|c|c|c|c|}
    \textbf{Detector}             & \textbf{Source}            & \textit{Change from}          &  & \textbf{A Detector}   & \textit{Change from}          &  & \textbf{B Detector}   & \textit{Change from} ~                    \\
              \textbf{Pair} & \textbf{Angle} & \textit{Previous} &  & \textbf{Peak Angle} & \textit{Previous} &  & \textbf{Peak Angle} & \textit{Previous} \\ \hline \hline
    r2c1t4        & $\phantom{0}0^\circ$  & ~                    &  & $-10.28^\circ$ & ~                    &  & $\phantom{-}33.58^\circ$ & ~                    \\ \hline
    r2c1t4        & $45^\circ$ & $45^\circ$         & & $-33.32^\circ$ & $23.04^\circ$      &  & $\phantom{-}10.98^\circ$ & $22.60^\circ$      \\ \hline
    r2c1t4        & $90^\circ$ & $45^\circ$         & & $\phantom{-}33.78^\circ$ & $22.90^\circ$      &  & $-11.65^\circ$ & $22.93^\circ$      \\ \hline
    r6c7t4        & $90^\circ$ & $\phantom{0}0^\circ$          & & $\phantom{-}35.03^\circ$ & $\phantom{2}1.25^\circ$       &  & $-10.03^\circ$ & $\phantom{2}1.62^\circ$       \\ \hline
    \end{tabular}}
    \caption[HWP angles of peak detector response from the polarization modulation measurement in X1]{HWP angles of peak detector response from the polarization modulation measurement in X1 (150 GHz) shown in Figure~\ref{hwp_pol_mod_x1}. The best-fit HWP angles with maximum detector signal are shown for both the A and B detectors. As the source changed by steps of $45^\circ$, this HWP angle should also shift by $22.5^\circ$ which it does to within $\pm0.5^\circ$. Between the last two measurements, the source angle did not change at all but we did slew the telescope to point up on a different detector pair. The peak HWP angles should not have shifted at all, but they did by a little over $1^\circ$. This is due to projection effects between the telescope and the source, fabrication non-uniformities in the detector array, or possibly a combination of both. \label{hwp_angle_best_fits_x1}}
\end{table}

\begin{table}
    {\scriptsize
    \begin{tabular}{c|c|c|c|c|c|c|c|c|}
    \textbf{Detector}             & \textbf{Source}            & \textit{Change from}          &  & \textbf{A Detector}   & \textit{Change from}          &  & \textbf{B Detector}   & \textit{Change from} ~                    \\
              \textbf{Pair} & \textbf{Angle} & \textit{Previous} &  & \textbf{Peak Angle} & \textit{Previous} &  & \textbf{Peak Angle} & \textit{Previous} \\ \hline \hline
    r4c2t4        & $\phantom{0}0^\circ$  & ~                    &  & $-29.85^\circ$ & ~                    &  & $\phantom{-}14.96^\circ$ & ~                    \\ \hline
    r4c2t4        & $45^\circ$ & $45^\circ$         & & $\phantom{.}-7.53^\circ$ & $22.32^\circ$      &  & $\phantom{-}37.11^\circ$ & $22.15^\circ$      \\ \hline
    r4c2t4        & $90^\circ$ & $45^\circ$         & & $\phantom{-}14.89^\circ$ & $22.42^\circ$      &  & $-30.31^\circ$ & $22.58^\circ$      \\ \hline
    \end{tabular}}
    \caption[HWP angles of peak detector response from the polarization modulation measurement in X4]{HWP angles of peak detector response from the polarization modulation measurement in X4 (95 GHz), similar to Table~\ref{hwp_angle_best_fits_x1}. Just as with the measurement in X1, the peak HWP angle shifted between source angles by $22.5^\circ$ as expected, within $\pm0.5^\circ$. \label{hwp_angle_best_fits_x4}}
\end{table}

\chapter{Conclusion}

In lab testing, the HWP system for the Spider instrument performs well. The hot-press Cirlex bonded AR coats used for the 150 GHz HWPs have high optical transmission and are robust enough to survive several thermal cycles to cryogenic temperatures. The fused-quartz glued AR coats on the 95 GHz HWPs perform similarly. The quartz AR coats on the 95 GHz are more fragile than the bonded Cirlex ARs used at 150 GHz, so it could be desirable in the future to switch the 95 GHz HWPs to Cirlex AR coats as well. Future work could include determining why Cirlex bonding failed to adhere when used in the 95 GHz HWPs, and also determining why the Cirlex ARs only survive a finite number of cryogenic thermal cycles before the bond fails. This is not critical for the success of the Spider program, since we have already built working HWPs for the instrument.

The cryogenic rotation mechanisms deliver reliable, precise rotation at cryogenic temperatures. The power dissipation is low enough to have a minimal impact on liquid helium consumption, and the precision of the encoder readout is $\pm0.1^\circ$, well below our polarization angle error budget of $\pm1^\circ$. The absolute encoder repeatably measures the absolute angle of the HWP bearing. Fitting the template to interpret the angle encoder waveforms after a turn is slightly computationally costly, running somewhat slower than real-time even on a fast multicore computer. While we already have the bearing performance necessary for Spider, a possible future operational improvement could be to use encoder ticks that are finer than $0.5^\circ$ spacing and design a LED/photodiode optical system with correspondingly finer resolution. This could allow a digital tick-counting algorithm to yield the precision necessary with simpler postprocessing.

We developed and presented a polarimetric model for the HWP transmission. Lab testing has shown that this model correctly describes the polarization modulation of bare sapphire, and that it correctly predicts the observed high polarization modulation efficiency of the AR-coated HWPs. Lab measurements of the AR-coated HWPs have proven difficult to connect very precisely with the model so far. Our spectral data from the integrated instrument has always been of the HWP+detector system, and because the system was under active development we have always made significant changes between HWP+detector runs and detector-only runs. This has made determining the HWP-only response difficult. Dedicating two runs of the instrument, one with detectors only and a second cold run with HWP+detectors could yield spectral data about the AR-coated HWPs, but it is likely that our schedule will not have room for this.

We also derived a polarimetric model for the ghost beam of the instrument with a HWP, caused by the HWP itself or caused by a filter. This model and the model for the main beam HWP response show that the HWP non-idealities and ghost coupling are calculated to have minimal impact on the Spider science results, which means the reduction of the impact of beam systematics provided by the HWP is calculated to come without introducing other problems. A beam map campaign to study the properties of the ghost beam at many different HWP angles could yield data to confirm our ghost beam model and confirm that the ghosting at different HWP angles really will cancel out of the science results as the calculations suggest.

In the integrated instrument, the measured beam patterns are unchanged with HWP angle. This is a crucial verification of the ability of the HWPs to reduce the impact of beam systematics.

Spider will make its first flight in the austral summer of 2014-2015. Lab testing and modeling of the HWP system show that it will reduce the impact of beam systematics on the Spider science results and that we will be able to use its polarization modulation as a consistency check on the data.

\bibliographystyle{unsrt}
\bibliography{bibliography}

\end{document}